\documentclass[12pt]{article}
\usepackage{graphicx}
\usepackage{epsfig}
\usepackage{epstopdf}
\DeclareGraphicsExtensions{.pdf,.eps,.png,.jpg,.mps,.tif}
\usepackage{amssymb,graphicx}

\usepackage{graphicx}
\usepackage{wrapfig}
\usepackage{lscape}
\usepackage{rotating}

\usepackage[
singlelinecheck=false 
]{caption}

\usepackage{cite}

\def\lsim{\mathrel{\rlap{\lower3pt\hbox{\hskip0pt$\sim$}}
     \raise1pt\hbox{$<$}}}         
\def\gsim{\mathrel{\rlap{\lower4pt\hbox{\hskip1pt$\sim$}}
     \raise1pt\hbox{$>$}}}         

\usepackage{amsmath}
\usepackage{amsfonts}

\begin{document}
\begin{titlepage}

\centerline{\Large \bf Mutation Clusters from Cancer Exome}
\medskip

\centerline{Zura Kakushadze$^\S$$^\dag$\footnote{\, Zura Kakushadze, Ph.D., is the President of Quantigic$^\circledR$ Solutions LLC,
and a Full Professor at Free University of Tbilisi. Email: zura@quantigic.com} and Willie Yu$^\sharp$\footnote{\, Willie Yu, Ph.D., is a Research Fellow at Duke-NUS Medical School. Email: willie.yu@duke-nus.edu.sg}}
\bigskip

\centerline{\em $^\S$ Quantigic$^\circledR$ Solutions LLC}
\centerline{\em 1127 High Ridge Road \#135, Stamford, CT 06905\,\,\footnote{\, DISCLAIMER: This address is used by the corresponding author for no
purpose other than to indicate his professional affiliation as is customary in
publications. In particular, the contents of this paper
are not intended as an investment, legal, tax or any other such advice,
and in no way represent views of Quantigic$^\circledR$ Solutions LLC,
the website \underline{www.quantigic.com} or any of their other affiliates.
}}
\centerline{\em $^\dag$ Free University of Tbilisi, Business School \& School of Physics}
\centerline{\em 240, David Agmashenebeli Alley, Tbilisi, 0159, Georgia}
\centerline{\em $^\sharp$ Centre for Computational Biology, Duke-NUS Medical School}
\centerline{\em 8 College Road, Singapore 169857}
\medskip
\centerline{(March 31, 2017)}

\bigskip
\medskip

\begin{abstract}
{}We apply our statistically deterministic machine learning/clustering algorithm *K-means (recently developed in https://ssrn.com/abstract=2908286) to 10,656 published exome samples for 32 cancer types. A majority of cancer types exhibit mutation clustering structure. Our results are in-sample stable. They are also out-of-sample stable when applied to 1,389 published genome samples across 14 cancer types. In contrast, we find in- and out-of-sample instabilities in cancer signatures extracted from exome samples via nonnegative matrix factorization (NMF), a computationally costly and non-deterministic method. Extracting stable mutation structures from exome data could have important implications for speed and cost, which are critical for early-stage cancer diagnostics such as novel blood-test methods currently in development.
\end{abstract}
\medskip
{\bf Keywords:} Clustering, K-Means, Nonnegative Matrix Factorization, Somatic Mutation, Cancer Signatures, Genome, Exome, DNA, eRank, Correlation, Covariance, Machine Learning, Sample, Matrix, Source Code, Quantitative Finance, Statistical Risk Model, Industry Classification

\end{titlepage}

\newpage
\section{Introduction and Summary}

{}Unless humanity finds a cure, about a billion people alive today will die of cancer. Unlike other diseases, cancer occurs at the DNA level via somatic alterations in the genome. A common type of such mutations found in cancer is due to alterations to single bases in the genome (single nucleotide variations, or SNVs). These alterations are accumulated throughout the lifespan of an individual via various mutational processes such as imperfect DNA replication during cell division or spontaneous cytosine deamination \cite{Goodman}, \cite{Lindahl}, or due to exposures to chemical insults or ultraviolet radiation \cite{Loeb}, \cite{Ananthaswamy}, etc. The footprint left by these mutations in the cancer genome is characterized by distinctive alteration patterns known as cancer signatures.

{}Identifying all cancer signatures would greatly facilitate progress in understanding the origins of cancer and its development. Therapeutically, if there are common underlying structures across different cancer types, then treatment for one cancer type might be applicable to other cancer types, which would be a great news. From a diagnostic viewpoint, identification of all underlying cancer signatures would aid cancer detection and identification methodologies, including vital early detection\footnote{\, See, e.g., \cite{Cho}. A goal of early detection (via blood tests) is behind Grail, Inc.'s recent $\sim$\$1B series B funding round -- see, e.g., \cite{Nasdaq}.} -- according to American Cancer Society, late stage metastatic cancers of unknown origin represent about 2\% of all cancers \cite{ACS} and can make treatment almost impossible.  Another practical application is prevention by pairing the signatures extracted from cancer samples with those caused by known carcinogens (e.g., tobacco, aflatoxin, UV radiation, etc.). At the end of the day, it all boils down to the question of usefulness: is there a small enough number of cancer signatures underlying all (100+) known cancer types, or is this number too large to be meaningful/useful? Thus, if we focus on 96 mutation categories of SNVs,\footnote{\, In brief, DNA is a double helix of two strands, and each strand is a string of letters A, C, G, T corresponding to adenine, cytosine, guanine and thymine, respectively. In the double helix, A in one strand always binds with T in the other, and G always binds with C. This is known as base complementarity. Thus, there are six possible base mutations C $>$ A, C $>$ G, C $>$ T, T $>$ A, T $>$ C, T $>$ G, whereas the other six base mutations are equivalent to these by base complementarity. Each of these 6 possible base mutations is flanked by 4 possible bases on each side thereby producing $4 \times 6 \times 4 = 96$ distinct mutation categories.} we cannot have more than 96 signatures.\footnote{\, A priori, nonlinearities could alter this conclusion. However, such nonlinearities may also render cancer signatures essentially useless...} Even if the number of true underlying signatures is, say, of order 50, it is unclear whether they would be useful, especially within practical applications. On the other hand, if there are only about a dozen underlying cancer signatures, then a hope for an order of magnitude simplification may well be warranted.

{}The commonly used method for extracting cancer signatures \cite{Alexandrov.NMF} is based on nonnegative matrix factorization (NMF) \cite{Paatero}, \cite{LeeSeung}. Thus, one analyzes SNV patterns in a cohort of DNA sequenced whole cancer genomes, and organizes the data into a matrix $G_{i\mu}$, where the rows correspond to the $N=96$ mutation categories, the columns correspond to $d$ samples, and each element is a nonnegative occurrence count of a given mutation category in a given sample.  Under NMF the matrix $G$ is then approximated via $G \approx W~H$, where $W_{iA}$ is an $N\times K$ matrix, $H_{A\mu}$ is a $K\times d$ matrix, and both $W$ and $H$ are nonnegative. The appeal of NMF is its biologic interpretation whereby the $K$ columns of the matrix $W$ are interpreted as the weights with which the $K$ cancer signatures contribute into the $N=96$ mutation categories, and the columns of the matrix $H$ are interpreted as the exposures to these $K$ signatures in each sample. The price to pay for this is that NMF, which is an iterative procedure, is computationally costly and depending on the number of samples $d$ it can take days or even weeks to run it. Furthermore, NMF does not fix the number of signatures $K$, which must be either guessed or obtained via trial and error, thereby further adding to the computational cost. Perhaps most importantly, NMF is a nondeterministic algorithm and produces a different matrix $W$ in each run.\footnote{\, Each $W$ corresponds to one in myriad local minima of the NMF objective function.} This is dealt with by averaging over many such $W$ matrices obtained via multiple NMF runs (or samplings). However, each run generally produces a weights matrix $W_{iA}$ with columns (i.e., signatures) not aligned with those in other runs. Aligning or matching the signatures across different runs (before averaging over them) is typically achieved via nondeterministic clustering such as k-means. Therefore, the result, even after averaging, generally is both noisy\footnote{\, By ``noise" we mean the statistical errors in the weighs obtained by averaging. Usually, such error bars are not reported in the literature on cancer signatures. Typically, they are large.} and nondeterministic! I.e., if this computationally costly procedure (which includes averaging) is run again and again on the same data, generally it will yield different looking cancer signatures every time! Simply put, the NMF-based method for extracting cancer signatures is not designed to be even in-sample stable. Under these circumstances, out-of-sample stability cannot even be dreamt about...\footnote{\, I.e., cancer signatures obtained from non-overlapping sets of samples can be dramatically different. And out-of-sample stability is crucial for practical usefulness, e.g., diagnostically.}

{}Without in- and out-of-sample stability practical therapeutic and diagnostic applications of cancer signatures would be challenging. For instance, suppose one sequences genome (or exome -- see below) data from a patient sample.\footnote{\, Be it via a liquid biopsy, a blood test, or some other (potentially novel) method.} Let us focus on SNVs. We have a vector of occurrence counts for 96 mutation categories. We need a quick computational test to determine with high enough confidence level whether i) there is a cancer signature present in this data, and ii) which cancer type this cancer signature corresponds to (i.e., which organ the cancer originated in). If cancer signatures are not even in-sample stable, then we cannot trust them. They could simply be noise. Indeed, there is always {\em somatic mutational noise} present in such data and must be factored out of the data before extracting cancer signatures. A simple way to understand somatic mutational noise is to note that mutations (i) are already present in humans unaffected by cancer, and (ii) such mutations, which are unrelated to cancer, are further exacerbated when cancer occurs as it disrupts the normal operation of various processes (including repair) in the DNA. At the level of the data matrix $G$, in \cite{BioFM} we discussed a key component of the somatic mutational noise and gave a prescription for removing it.\footnote{\, This is achieved by cross-sectionally (i.e., across the 96 mutation categories) demeaning ``log-counts". This ``de-noising" dramatically improved NMF-based signatures we extracted from genome data in \cite{BioFM} and cut computational cost (these savings would scale nonlinearly for larger datasets) by a factor of about 10 on a genome dataset for 1,389 samples in 14 cancer types. In \cite{BioFM}, by adapting the methods used in statistical risk models in quantitative finance \cite{StatRM}, we also proposed a simple method for fixing the number of cancer signatures based on eRank (effective rank) \cite{RV}.\label{fn.eRank}} However, there likely exist other, deeper sources of somatic mutational noise, which must be further identified and carefully factored out. Simply put, somatic mutational noise unequivocally is a substantial source of systematic error in cancer signatures.

{}However, then there is also the statistical error, which is large and due to the nondeterministic nature of NMF discussed above. This statistical error is exacerbated by the somatic mutational noise but would be present even if this noise was somehow completely factored out. Therefore, the in-sample instability must somehow be addressed. We emphasize that, a priori, this does not automatically address out-of-sample stability, without which any therapeutic or diagnostic applications would still be farfetched. However, without in-sample stability nothing is clear...

{}The problem at hand is nontrivial and requires a step-by-step approach, including identification of various sources of in-sample instability. One simple observation of \cite{BioFM} is that, if we work directly with occurrence counts $G_{i\mu}$ for individual samples, (i) the data is very noisy, and (ii) the number of signatures is bound to be too large to be meaningful/useful if the number of samples is large. A simple way to deal with this is to aggregate samples by cancer types. In doing so, we have a matrix $G_{is}$, where $s$ now labels cancer types, which is (i) less noisy, and (ii) much smaller ($96 \times n$, where $n$ is the number of cancer types), so the number of resultant signatures is much more reasonable.\footnote{\, In aggregating samples by cancer types, for some cancer types pertinent information may be muddled up as there may be biologic factors one may wish to understand, e.g., mutational spectra of liver cancers can have substantial regional dependence as they are mutagenized by exposures to different chemicals (alcohol, aflatoxin, tobacco, etc.). In such cases,
aggregation by regions (or other applicable characteristics, as the case may be) within a cancer type may still be warranted to reduce noise (or else, without any aggregation, there are simply too many cancer signatures -- see, e.g., Table 7 in \cite{BioFM}). However, not to get ahead of ourselves -- one step at a time -- in this paper we will work with (exome) data aggregated by cancer types (see below).\label{fn.liver}} Thus, such aggregation is helpful.

{}Still, even with aggregation, we must address nondeterminism (of NMF). To circumvent this, in \cite{*K-means} we proposed an alternative approach which bypasses NMF altogether. As we argue in \cite{*K-means}, NMF is -- at least to a certain degree -- clustering in disguise. E.g., many COSMIC cancer signatures \cite{COSMIC} obtained via NMF\footnote{\, Augmented with additional heuristics based on biologic intuition and empirical observations.} exhibit clustering substructure, i.e., in many of these signatures there are mutation categories with high weights (``peaks" or ``tall mountain landscapes") with other mutation categories having small weights likely well within statistical and systematic errors. For all practical purposes such low weights could be set to zero. Then, many cancer signatures would start looking like clusters, albeit some clusters could be overlapping between different signatures. Considering that various signatures may be somatic mutational noise artifacts in the first instance and statistical error bars are large, it is natural to wonder whether there are some robust underlying clustering structures present in the data -- with the understanding that such structures may not be present for all cancer types. However, even if they are present for a substantial number of cancer types, unveiling them would amount to a major step forward in understanding cancer signature structure.

{}To address this question, in \cite{*K-means} we proposed a new clustering algorithm termed {\em *K-means}. Its basic building block is the vanilla k-means algorithm, which computationally is very cheap. However, it is also nondeterministic. *K-means uses 2 machine learning levels on top of k-means to achieve {\em statistical determinism} -- see Section \ref{sec.2} for details\footnote{\, There is virtually no way to make this paper self-contained without essentially copying all the technical details over from \cite{*K-means}. We will not do so here. Instead, readers interested in technical details should read this paper together with \cite{*K-means}.} -- without any initialization of the centers, etc.\footnote{\, It also fixes the number of clusters $K$: it fixes the {\em target} number of clusters $K_1$ via an eRank based method (see fn. \ref{fn.eRank}); then the final number of clusters $K \leq K_1$ follows via machine learning.} Once *K-means fixes the clustering, it turns out that the weights and exposures can be computed using (normalized) regressions \cite{*K-means}, thereby altogether bypassing computationally costly NMF. In \cite{*K-means} we applied this method to cancer genome data corresponding to 1,389 published samples for 14 cancer types. We found that clustering works well for 10 out the 14 cancer types -- the metrics include within-cluster correlations and overall fit quality. This suggests that there is indeed clustering substructure present in underlying cancer genome data, at least for most cancer types!\footnote{\, One of the cancer types for which clustering does not appear to work well -- completely consistently with and expectedly from the results of \cite{BioFM} -- is Liver Cancer. In particular, the dominant (with a 96\% contribution) NMF-based cancer signature we found in \cite{BioFM} for Liver Cancer does not have ``peaks" (``rolling hills landscape"), with no resemblance to a clustering substructure. In this regard, note our comments in fn. \ref{fn.liver}.} This is exciting!

{}In this paper we apply the method of \cite{*K-means} to {\em exome} data consisting of 10,656 published samples (sample IDs with sources are in Appendix \ref{app.IDs}) aggregated by 32 cancer types. *K-means produces a robustly stable clustering (11 clusters) from this data. One motivation for using exome data is that exome is a small subset ($\sim$1\%) of full genome containing only protein coding regions of genome \cite{Ng}. Exome is much cheaper and less time consuming to sequence -- which can be especially important for early stage diagnostics -- than whole genome, yet it encodes important information about cancer signatures. As we discuss in the subsequent sections, our method appears to work well on exome data for most cancer types. In fact, overall it appears to work better than COSMIC signatures, including out-of-sample, when applying clusters derived from our exome data to genome data.

\newpage
\section{*K-means}\label{sec.2}

{}In \cite{*K-means}, by extending a prior work \cite{StatIndClass} in quantitative finance on building statistical industry classifications using clustering algorithms, we developed a clustering method termed {\em *K-means} (``Star K-means") and applied it to extraction of cancer signatures from genome data. *K-means anchors on the standard k-means algorithm\footnote{\, See \cite{Steinhaus}, \cite{Lloyd1957}, \cite{Forgy}, \cite{MacQueen}, \cite{Hartigan}, \cite{HartWong}, \cite{Lloyd1982}.} as its basic building block. However, k-means is not deterministic. *K-means is {\em statistically deterministic}, without specifying initial centers, etc. This is achieved via 2 machine learning levels sitting on top of k-means. At the first level we aggregate a large number $M$ of k-means clusterings with randomly initialized centers (and the number of {\em target} clusters fixed using eRank) via a nontrivial aggregation procedure -- see \cite{*K-means} for details. This aggregation is based on clustering (again, using k-means) the centers produced in the $M$ clusterings, so the resultant aggregated clustering is nondeterministic. However, it is a lot less nondeterministic than vanilla k-means clusterings as aggregation dramatically reduces the degree of nondeterminism. At the second level, we take a large number $P$ of such aggregated clusterings and determine the ``ultimate" clustering with the maximum occurrence count (among the $P$ aggregations). For sufficiently large $M$ and $P$ the ``ultimate" clustering is stable, i.e., if we run the algorithm over and over again, we will get the same ``ultimate" clustering every time, even though the occurrence counts within different $P$ aggregations are going to be different for various aggregations. What is important here is that the most frequently occurring (``ultimate") aggregation remains the same run after run.

\section{Empirical Results}\label{sec.3}
\subsection{Data Summary}\label{data.summary}

{}In this paper we apply *K-means to exome data.\footnote{\, In \cite{*K-means} we applied it to published genome data.} We use data consisting of 10,656 published exome samples aggregated by 32 cancer types listed in Table \ref{table.exome.summary}, which summarizes total occurrence counts, numbers of samples, and data sources. Appendix \ref{app.IDs} provides sample IDs together with references for the data sources. Occurrence counts for the 96 mutation categories for each cancer type are given in Tables \ref{table.aggr.data.11}-\ref{table.aggr.data.22}.

\subsubsection{Structure of Data}

{}The underlying data consists of matrices $[G(s)]_{i\mu(s)}$ whose elements are occurrence counts of mutation categories labeled by $i = 1,\dots, N = 96$ in samples labeled by $\mu(s) = 1,\dots, d(s)$. Here $s = 1,\dots,n$ labels $n$ different cancer types (in our case $n = 32$). We can choose to work with individual matrices $[G(s)]_{i\mu(s)}$, or with the $N \times d_{tot}$ ``big matrix" $\Gamma$ obtained by appending (i.e., bootstrapping) the matrices $[G(s)]_{i\mu(s)}$ together column-wise (so $d_{tot} = \sum_{s=1}^n d(s)$). Alternatively, we can aggregate samples by cancer types and work with the so-aggregated matrix
\begin{equation}
 G_{is} = \sum_{\mu(s) = 1}^{d(s)} [G(s)]_{i\mu(s)}
\end{equation}
Generally, individual matrices $[G(s)]_{i\mu(s)}$ and, thereby, the ``big matrix" $\Gamma$ contain a lot of noise. For some cancer types we can have a relatively small number of samples. We can also have ``sparsely populated" data, i.e., with many zeros for some mutation categories. In fact, different samples are not even necessarily uniformly normalized. Etc. The bottom line is that the data is noisy. To mitigate the aforementioned issues, following \cite{BioFM}, here we work with the $N\times n$ matrix $G_{is}$ with samples aggregated by cancer types. Below we apply *K-means to $G_{is}$.

\subsection{Exome Data Results}\label{sub.res}

{}The $96 \times 32$ matrix $G_{is}$ given in Tables \ref{table.aggr.data.11}-\ref{table.aggr.data.22} is what we pass into the function {\tt\small bio.cl.sigs()} in Appendix A of \cite{*K-means} as the input matrix {\tt\small x}. We use: {\tt\small iter.max = 100} (this is the maximum number of iterations used in the built-in R function {\tt\small kmeans()} -- we note that there was not a single instance in our 30 million runs of {\tt\small kmeans()} where more iterations were required);\footnote{\, The R function {\tt\small kmeans()} produces a warning if it does not converge within {\tt\small iter.max}.} {\tt\small num.try = 1000} (this is the number of individual k-means samplings we aggregate every time); and {\tt\small num.runs = 30000} (which is the number of aggregated clusterings we use to determine the ``ultimate" -- that is, the most frequently occurring -- clustering). More precisely, we ran 3 batches with {\tt\small num.runs = 10000} as a sanity check, to make sure that the final result based on 30000 aggregated clusterings was consistent with the results based on smaller batches, i.e., that it was stable from batch to batch.\footnote{\, We ran these 3 batches consecutively, and each batch produced slightly different top-10 (by occurrence counts) clusterings with varying occurrence counts across the batches, etc. However, Clustering-E1 invariably had the highest occurrence count by a large margin. See Table \ref{table.occurrence.cts}.} Based on Table \ref{table.occurrence.cts}, we identify Clustering-E1 as the ``ultimate" clustering\footnote{\, It is evident that the top-10 clusterings in Table \ref{table.occurrence.cts} essentially are variations of each other.} (see Section \ref{sec.2}).

\begin{figure}[ht]
\centering
\includegraphics[scale=0.7]{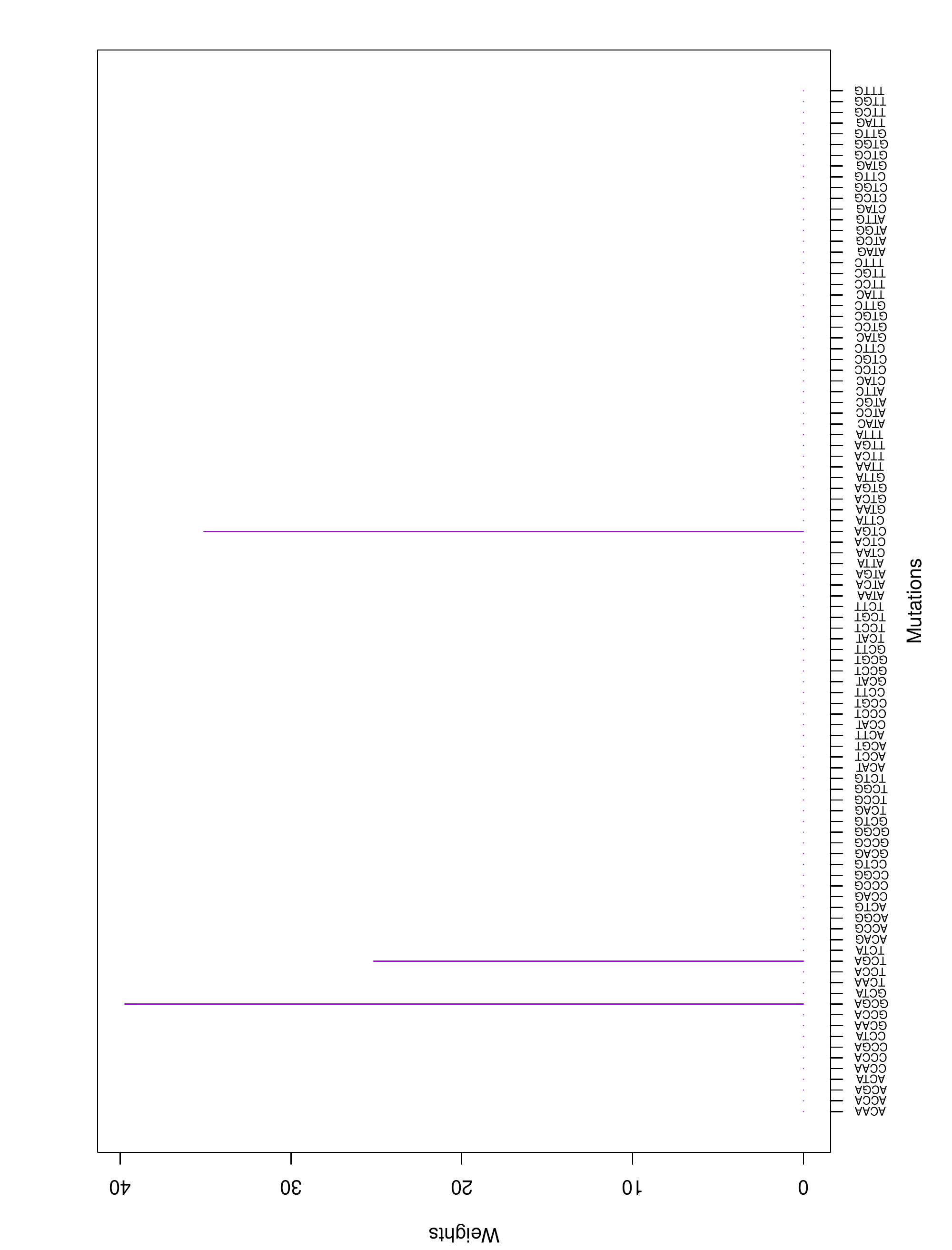}
\caption{Cluster Cl-1 in Clustering-E1 with weights based on unnormalized regressions with arithmetic means.
}
\label{Figure1}
\end{figure}

\begin{figure}[ht]
\centering
\includegraphics[scale=0.7]{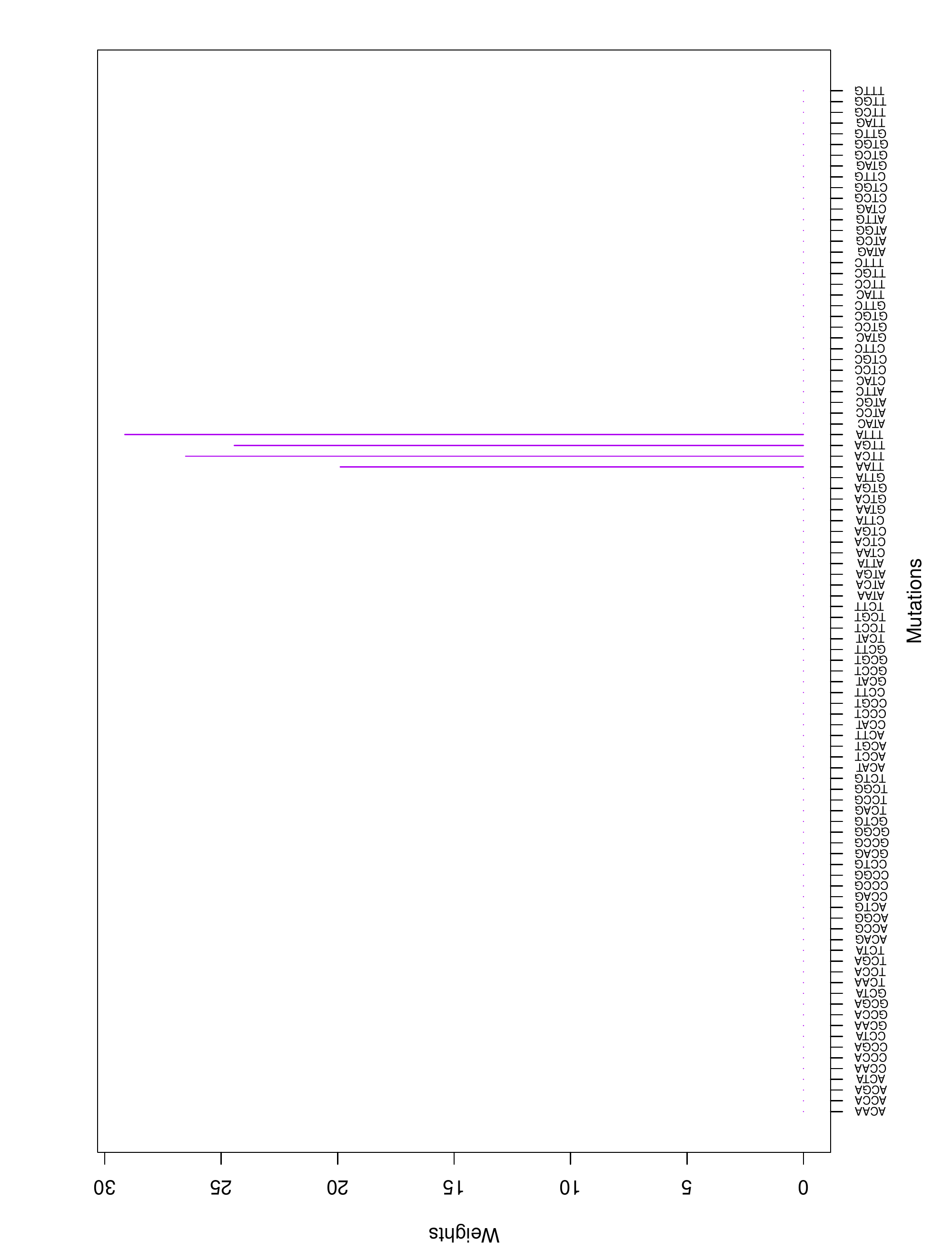}
\caption{Cluster Cl-2 in Clustering-E1 with weights based on unnormalized regressions with arithmetic means.
}
\label{Figure2}
\end{figure}

\begin{figure}[ht]
\centering
\includegraphics[scale=0.7]{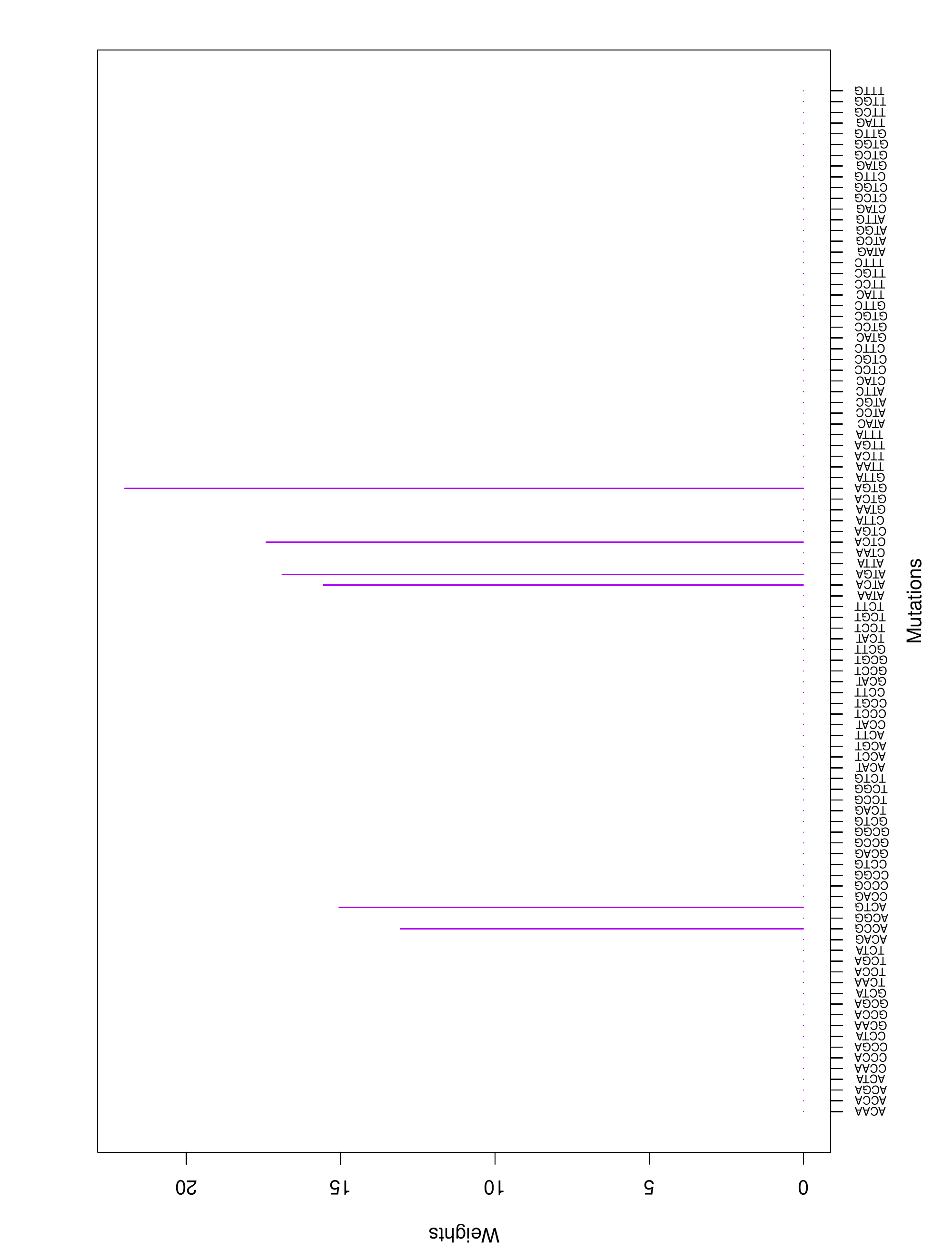}
\caption{Cluster Cl-3 in Clustering-E1 with weights based on unnormalized regressions with arithmetic means.
}
\label{Figure3}
\end{figure}

\begin{figure}[ht]
\centering
\includegraphics[scale=0.7]{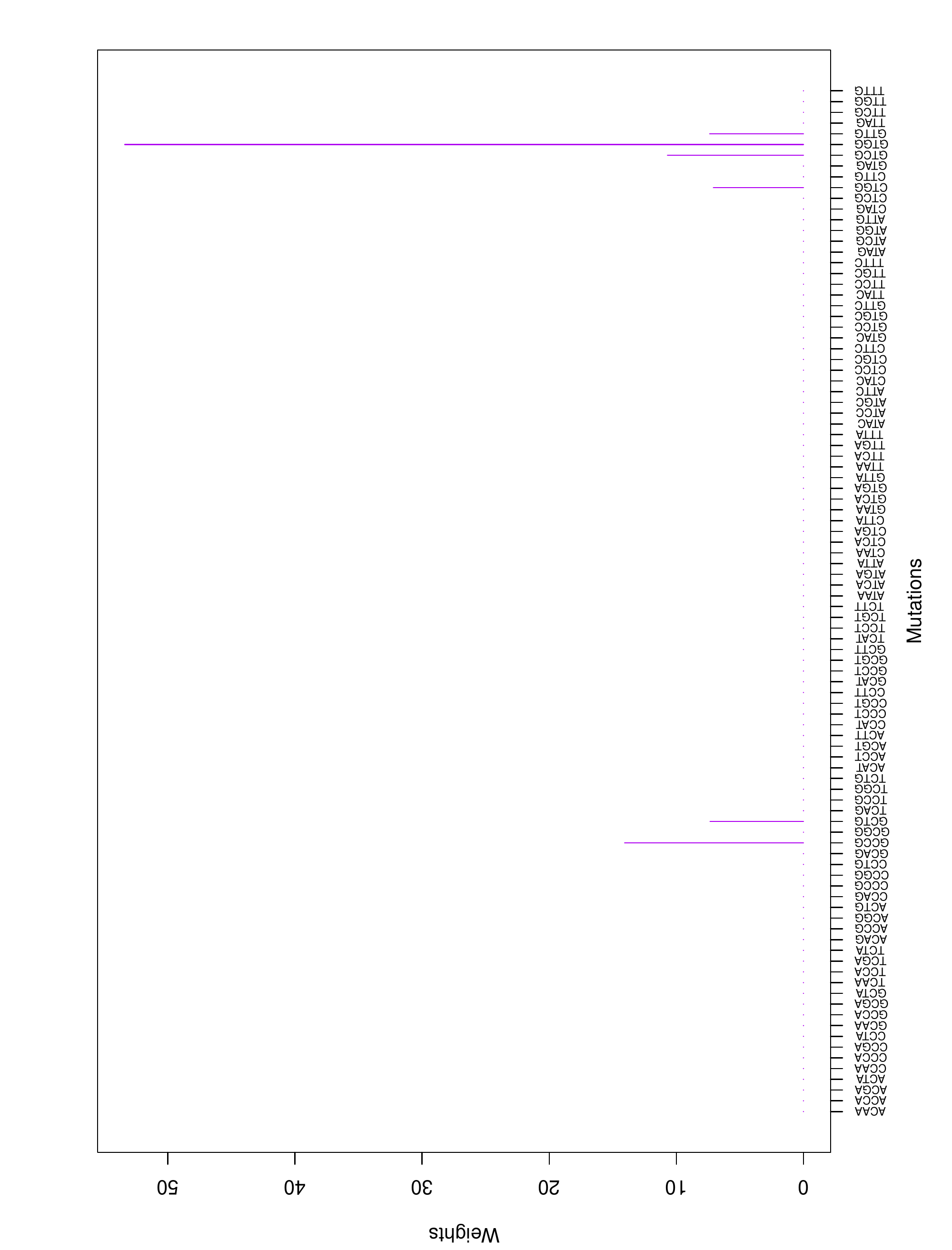}
\caption{Cluster Cl-4 in Clustering-E1 with weights based on unnormalized regressions with arithmetic means.
}
\label{Figure4}
\end{figure}

\begin{figure}[ht]
\centering
\includegraphics[scale=0.7]{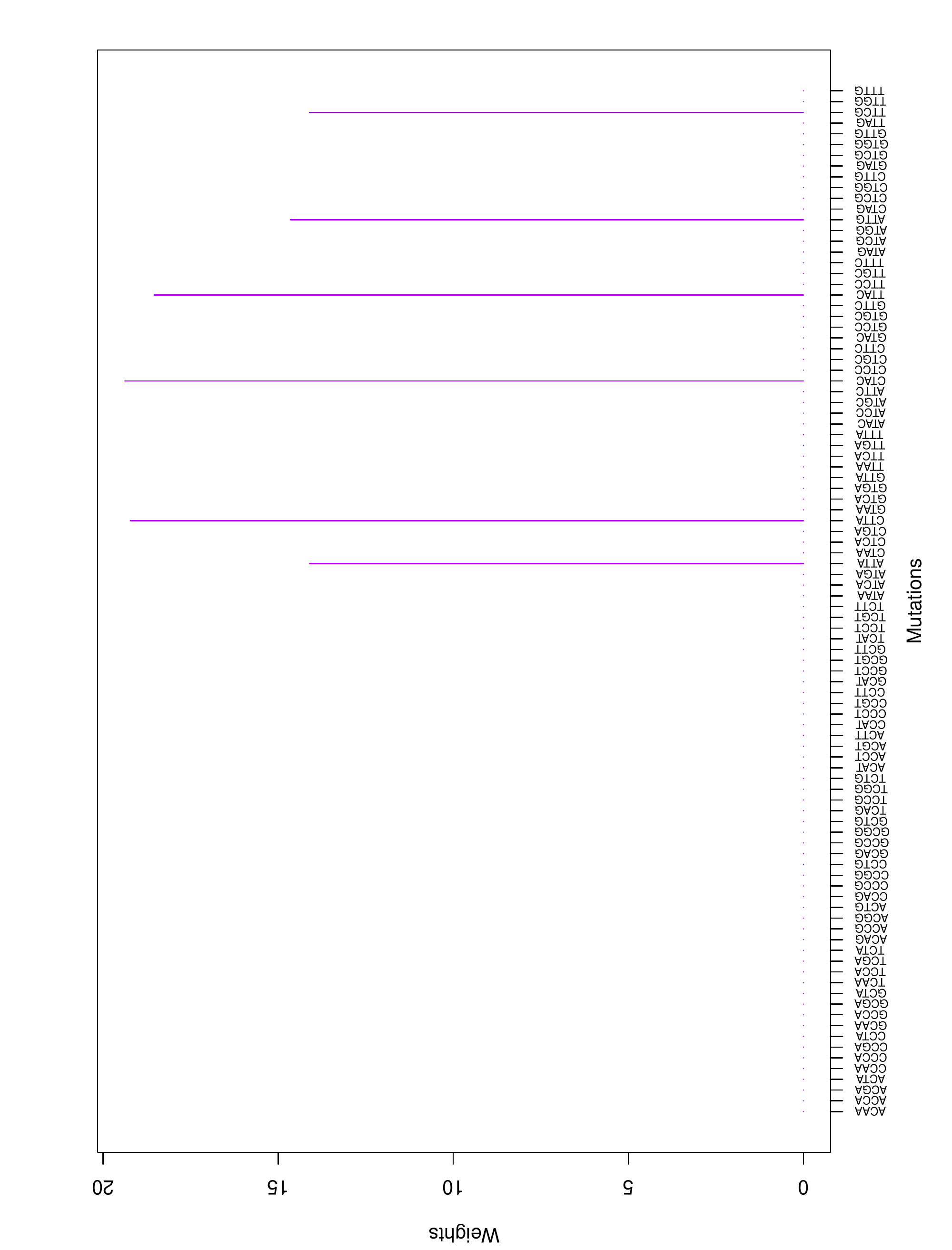}
\caption{Cluster Cl-5 in Clustering-E1 with weights based on unnormalized regressions with arithmetic means.
}
\label{Figure5}
\end{figure}

\begin{figure}[ht]
\centering
\includegraphics[scale=0.7]{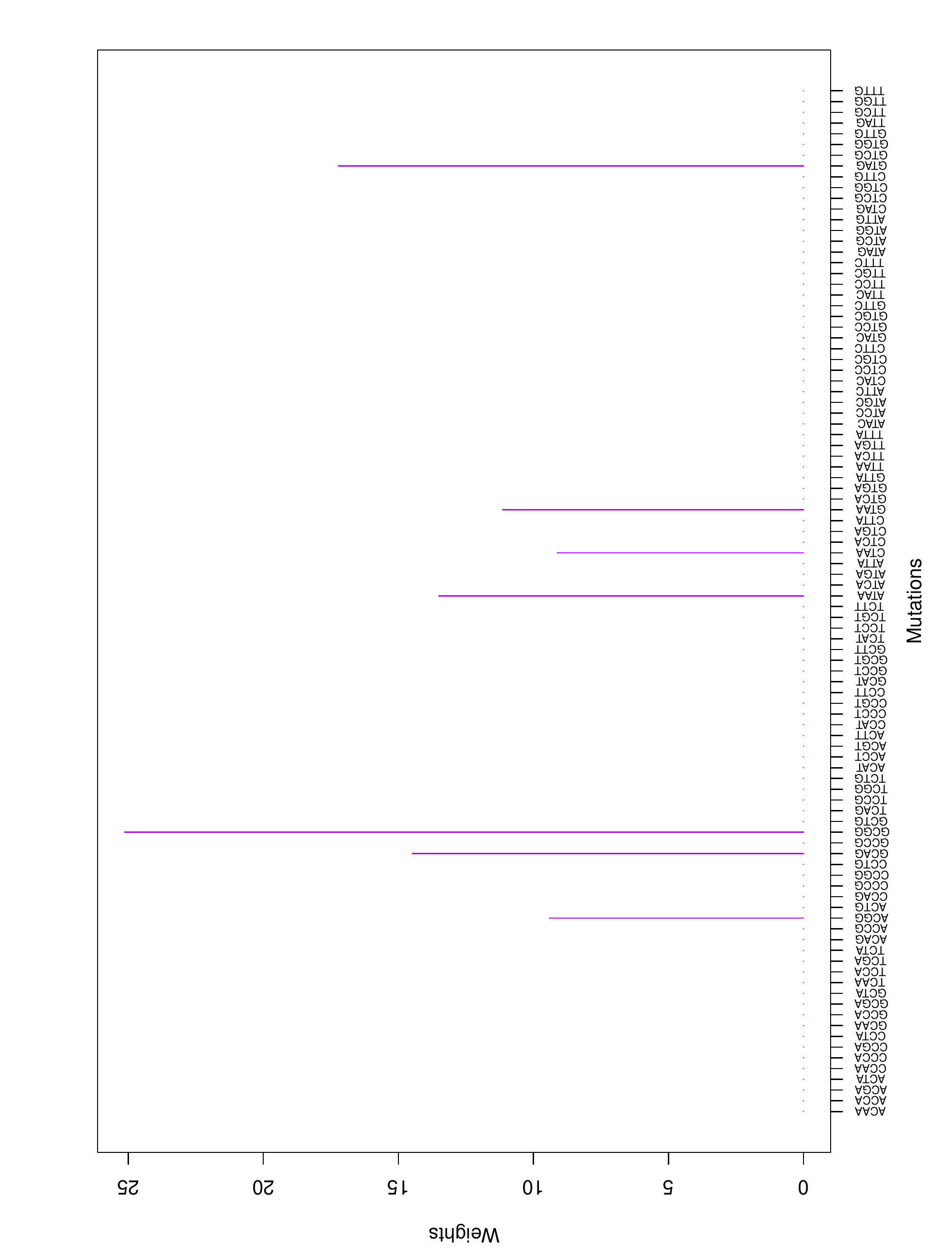}
\caption{Cluster Cl-6 in Clustering-E1 with weights based on unnormalized regressions with arithmetic means.
}
\label{Figure6}
\end{figure}

\begin{figure}[ht]
\centering
\includegraphics[scale=0.7]{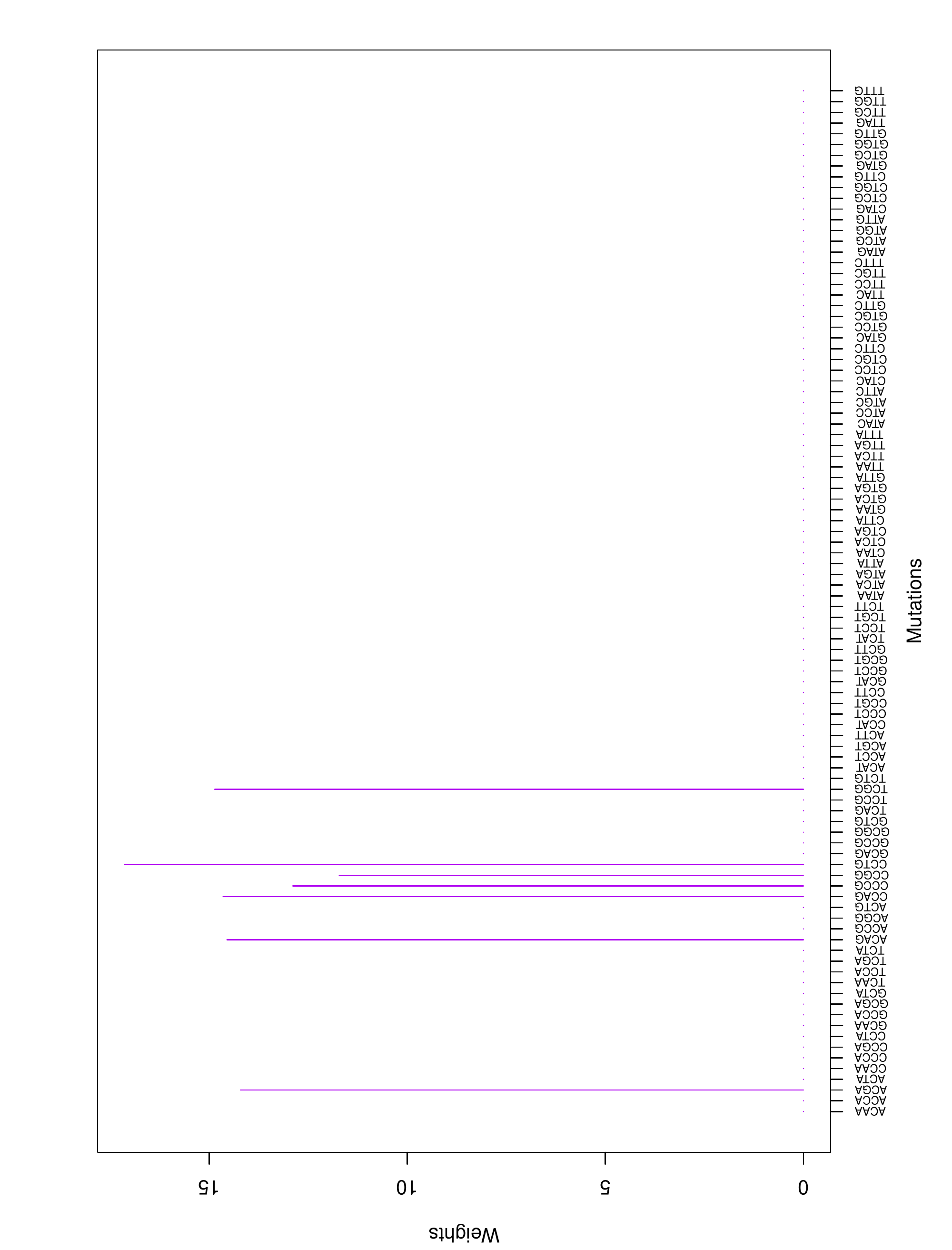}
\caption{Cluster Cl-7 in Clustering-E1 with weights based on unnormalized regressions with arithmetic means.
}
\label{Figure7}
\end{figure}

\begin{figure}[ht]
\centering
\includegraphics[scale=0.7]{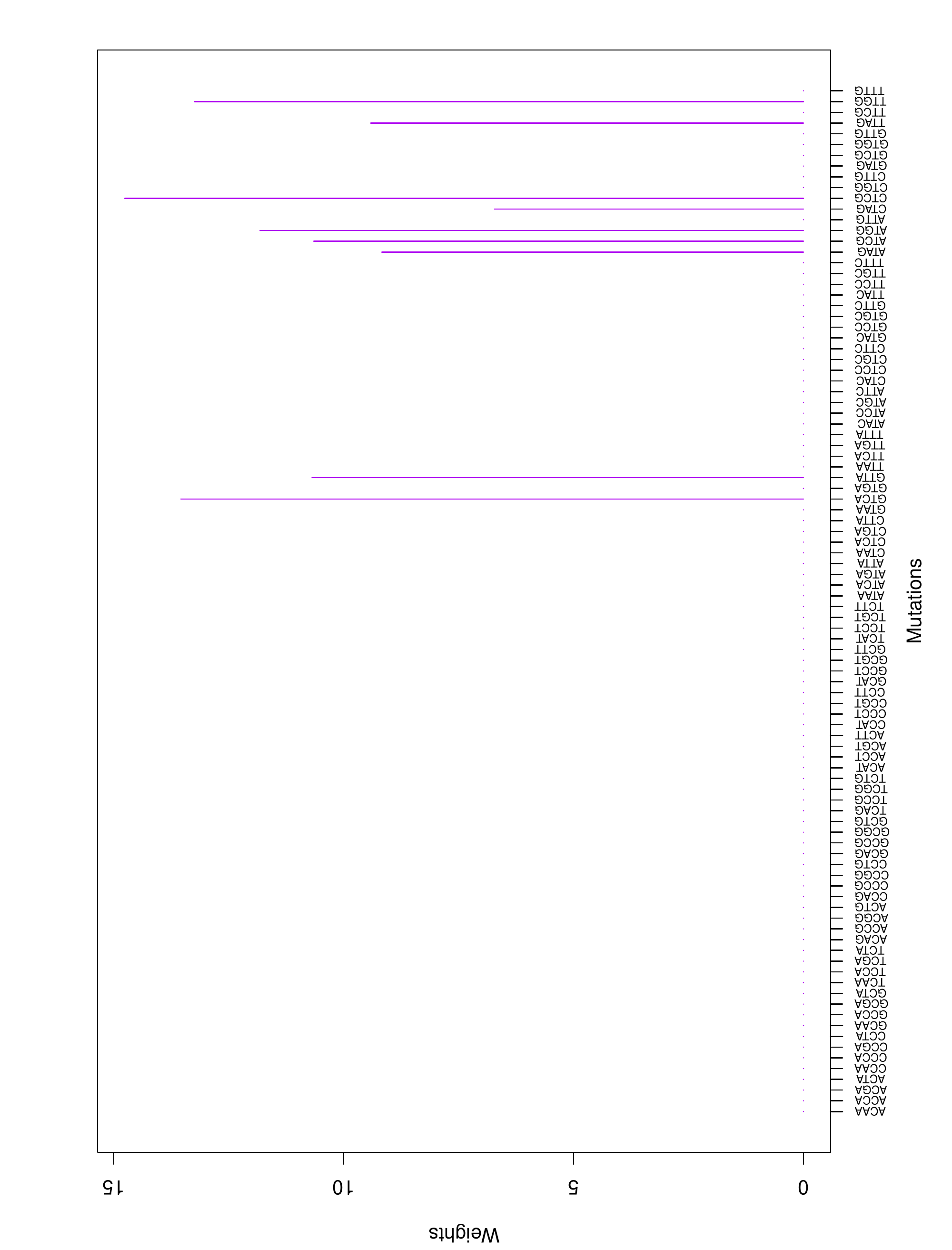}
\caption{Cluster Cl-8 in Clustering-E1 with weights based on unnormalized regressions with arithmetic means.
}
\label{Figure8}
\end{figure}

\begin{figure}[ht]
\centering
\includegraphics[scale=0.7]{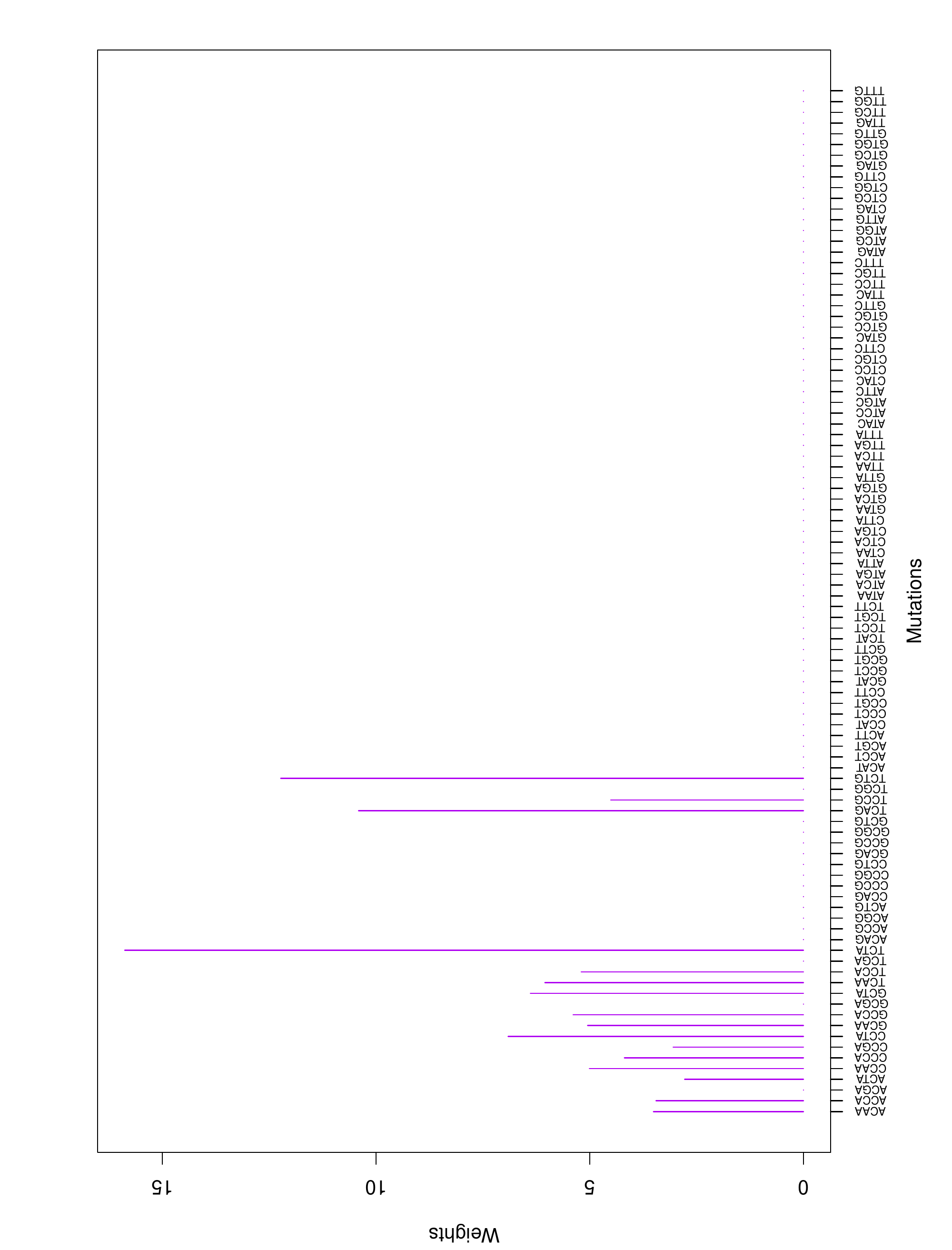}
\caption{Cluster Cl-9 in Clustering-E1 with weights based on unnormalized regressions with arithmetic means.
}
\label{Figure9}
\end{figure}

\begin{figure}[ht]
\centering
\includegraphics[scale=0.7]{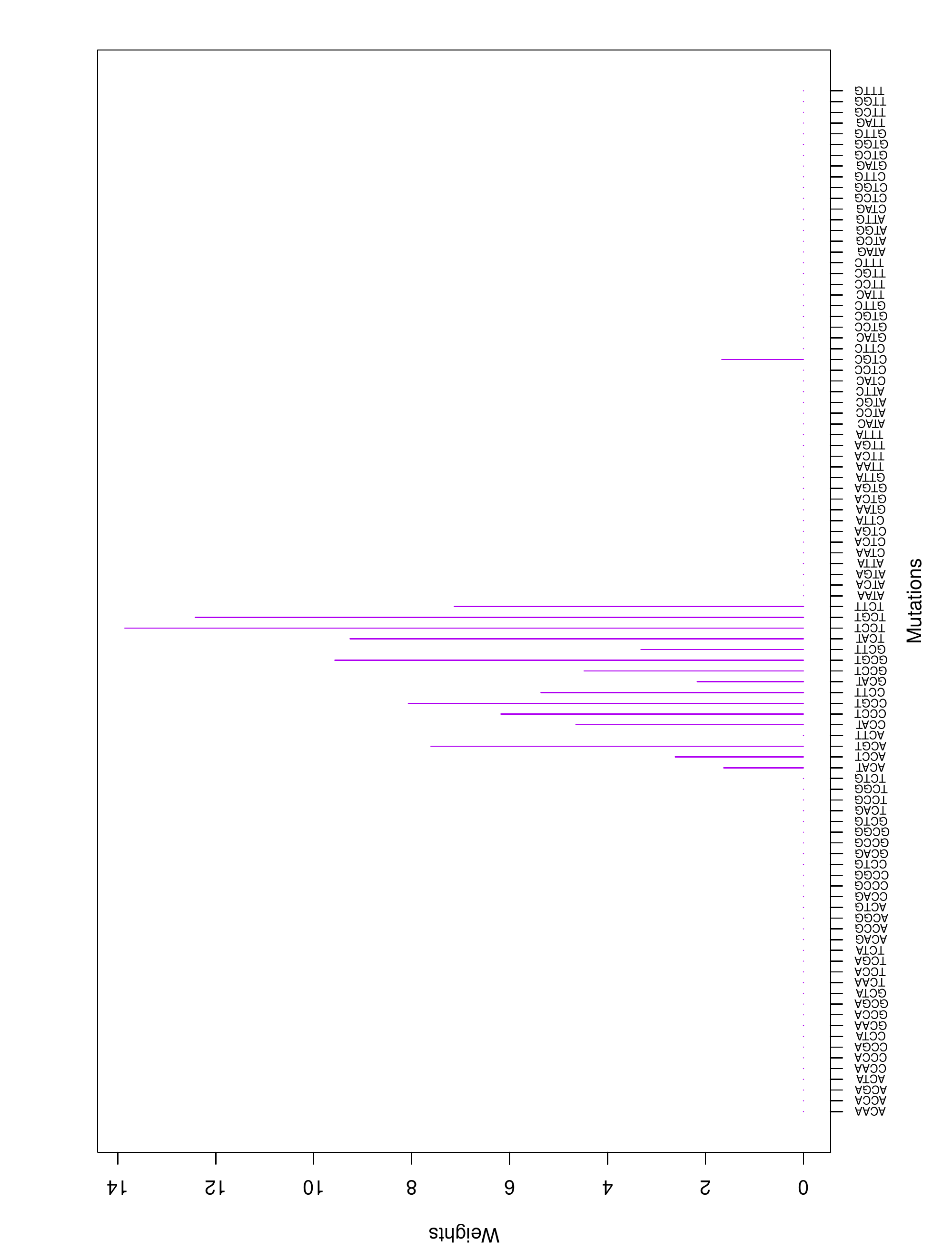}
\caption{Cluster Cl-10 in Clustering-E1 with weights based on unnormalized regressions with arithmetic means.
}
\label{Figure10}
\end{figure}

\begin{figure}[ht]
\centering
\includegraphics[scale=0.7]{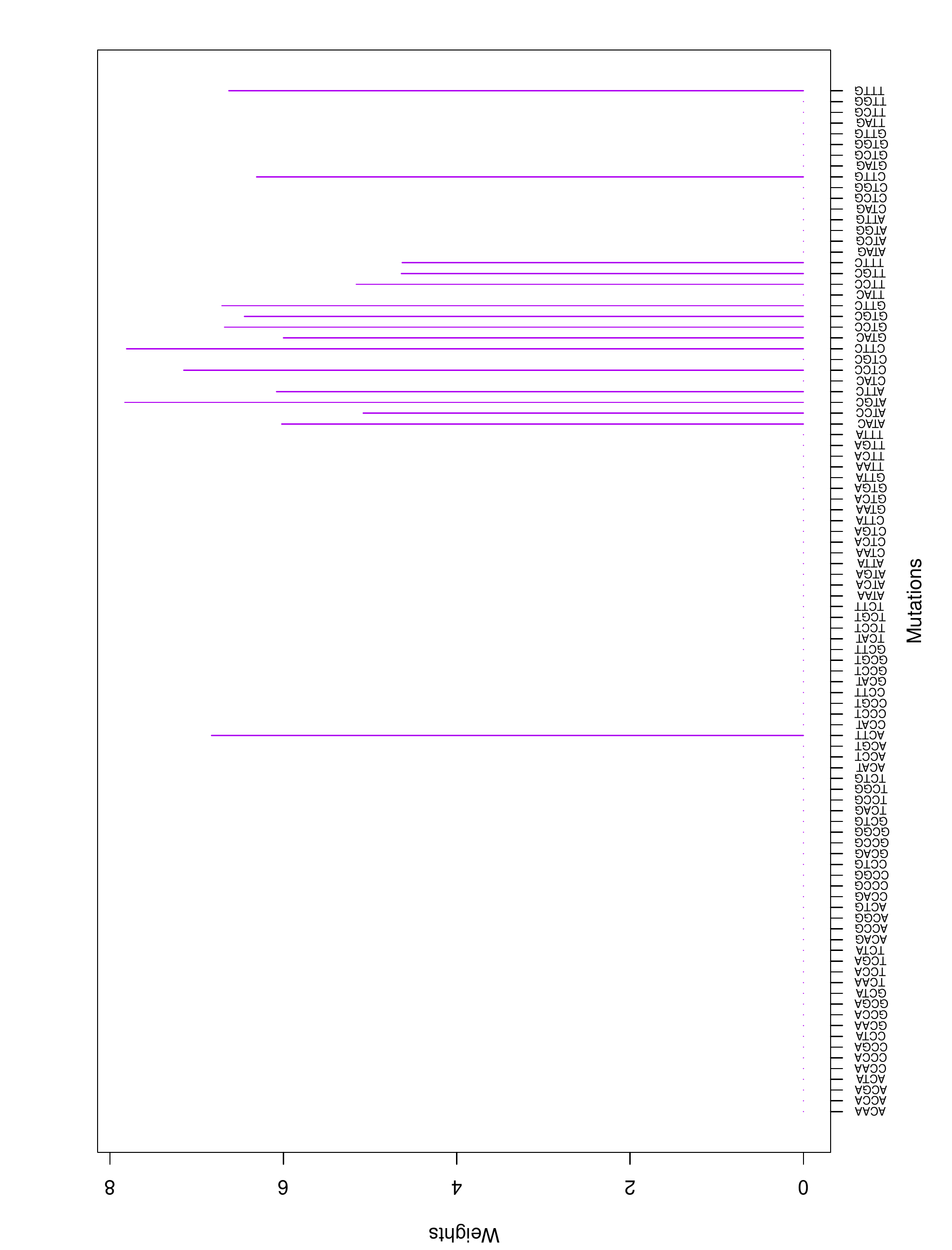}
\caption{Cluster Cl-11 in Clustering-E1 with weights based on unnormalized regressions with arithmetic means.
}
\label{Figure11}
\end{figure}

\begin{figure}[ht]
\centering
\includegraphics[scale=0.7]{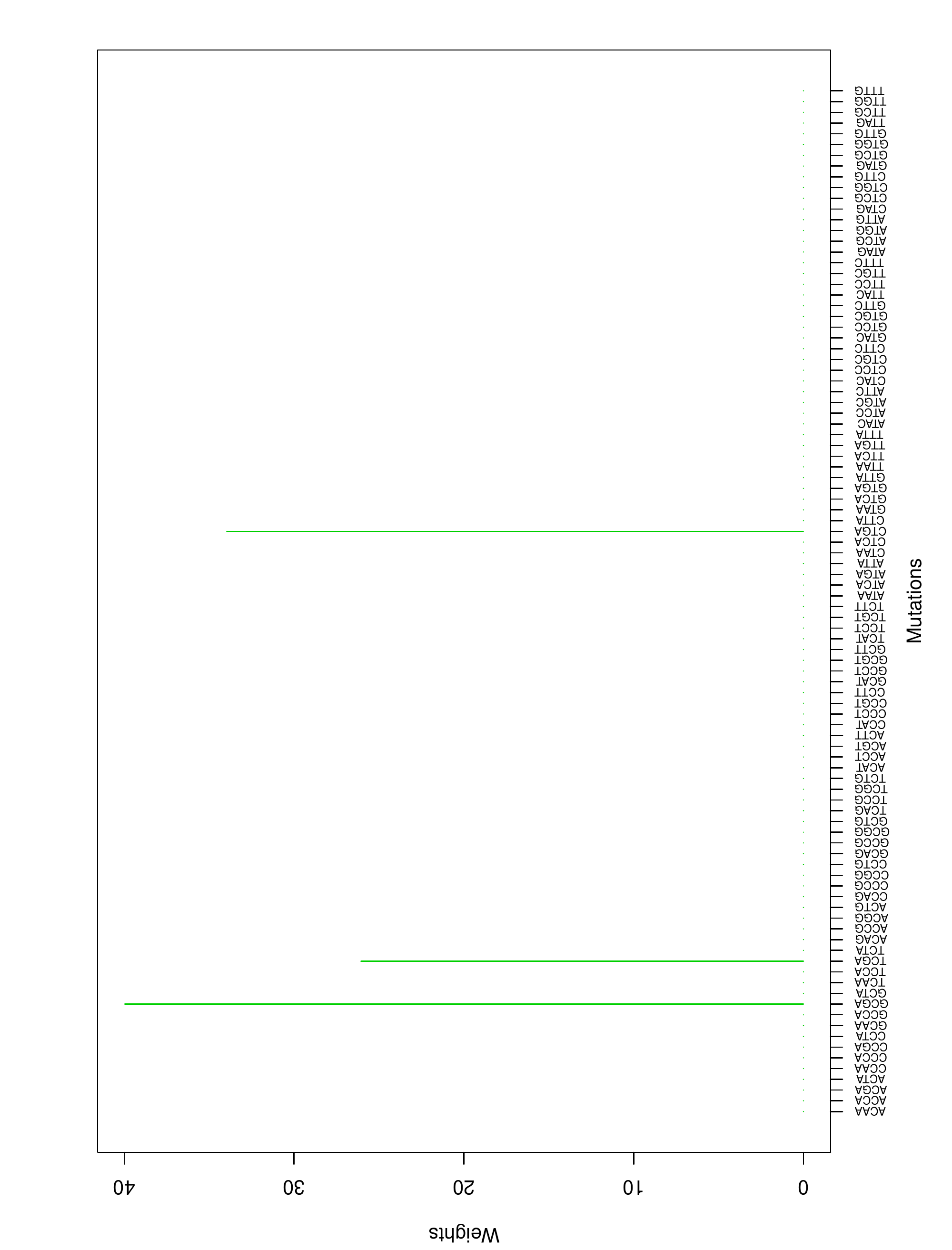}
\caption{Cluster Cl-1 in Clustering-E1 with weights based on normalized regressions with arithmetic means.
}
\label{FigureNorm1}
\end{figure}

\begin{figure}[ht]
\centering
\includegraphics[scale=0.7]{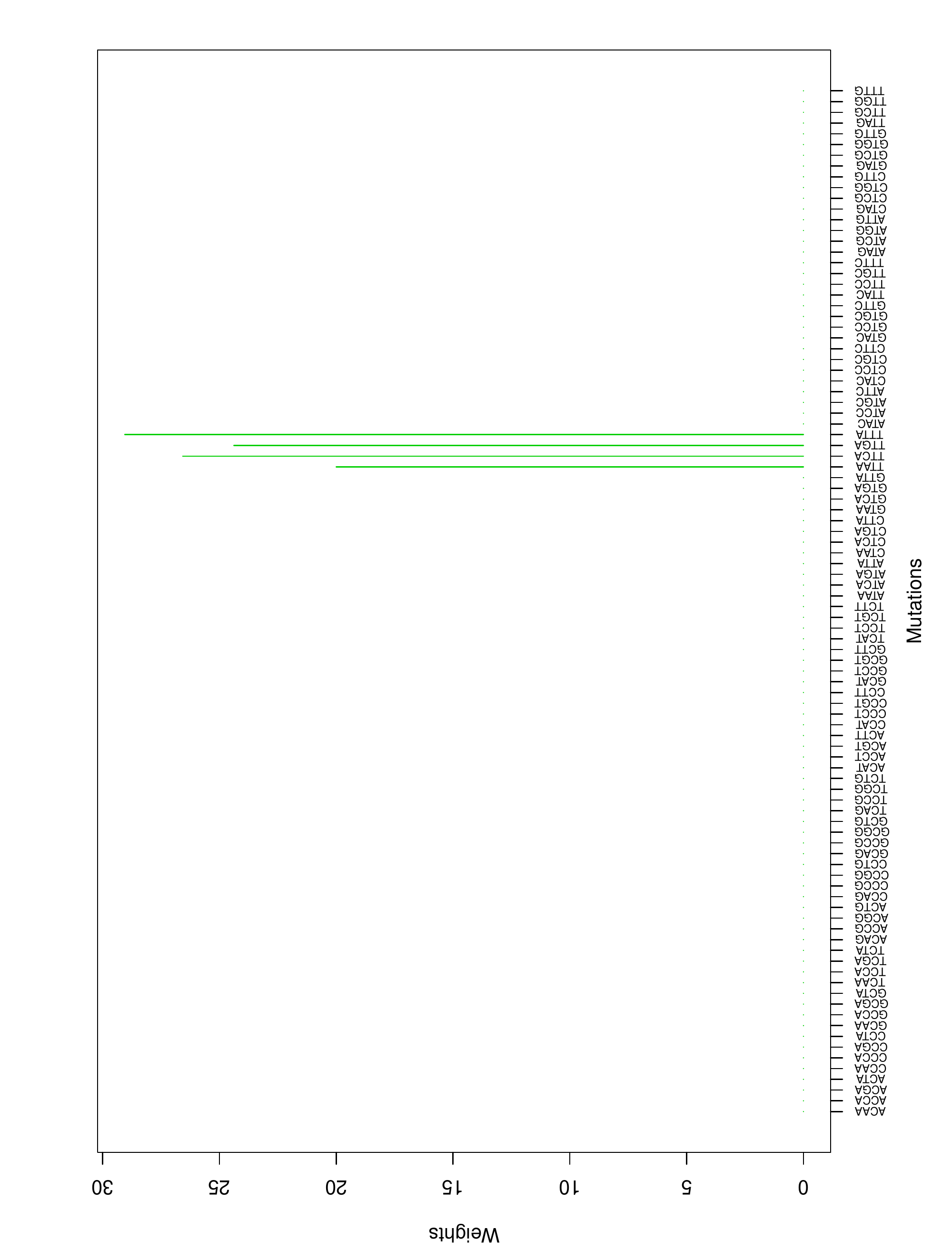}
\caption{Cluster Cl-2 in Clustering-E1 with weights based on normalized regressions with arithmetic means.
}
\label{FigureNorm2}
\end{figure}

\begin{figure}[ht]
\centering
\includegraphics[scale=0.7]{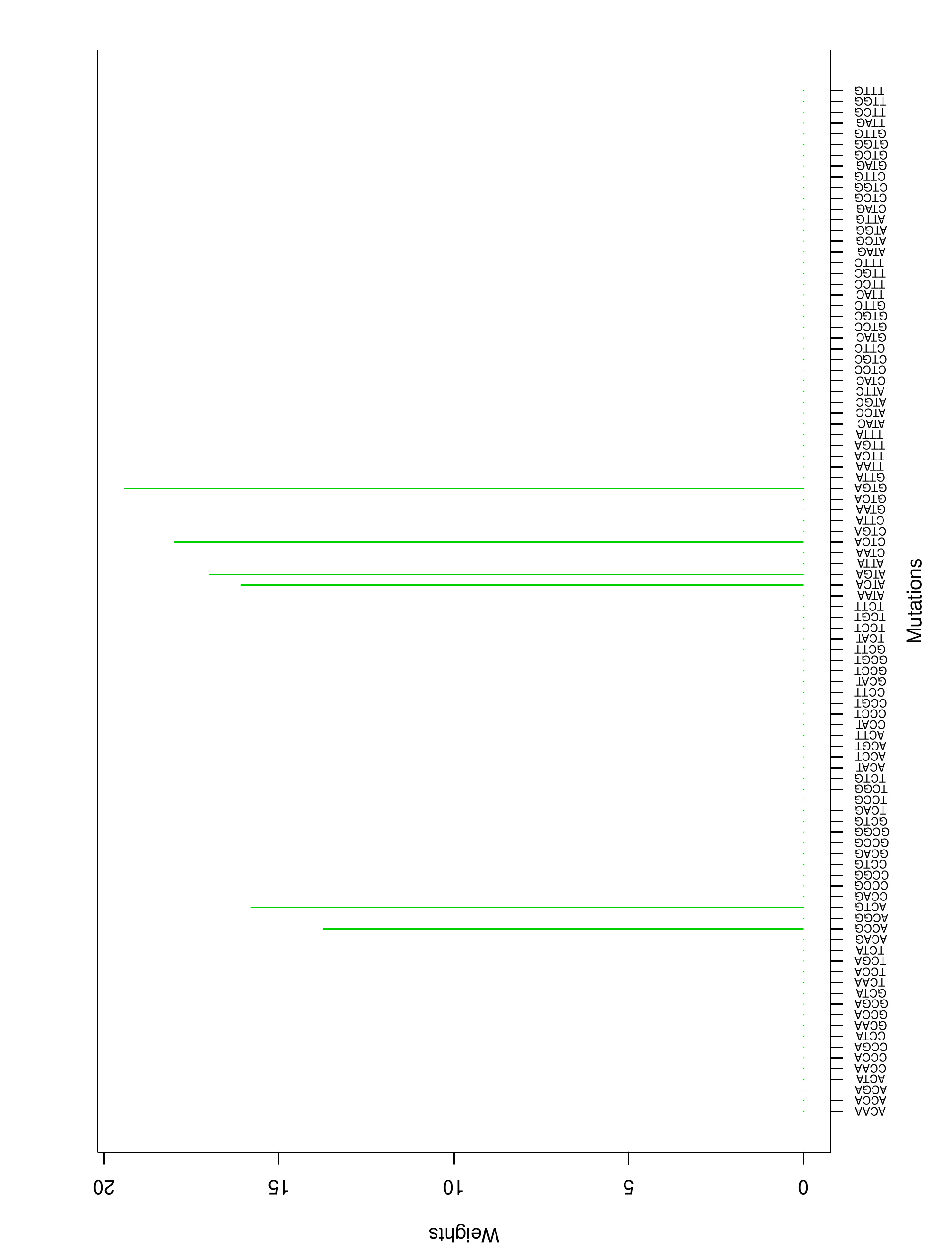}
\caption{Cluster Cl-3 in Clustering-E1 with weights based on normalized regressions with arithmetic means.
}
\label{FigureNorm3}
\end{figure}

\begin{figure}[ht]
\centering
\includegraphics[scale=0.7]{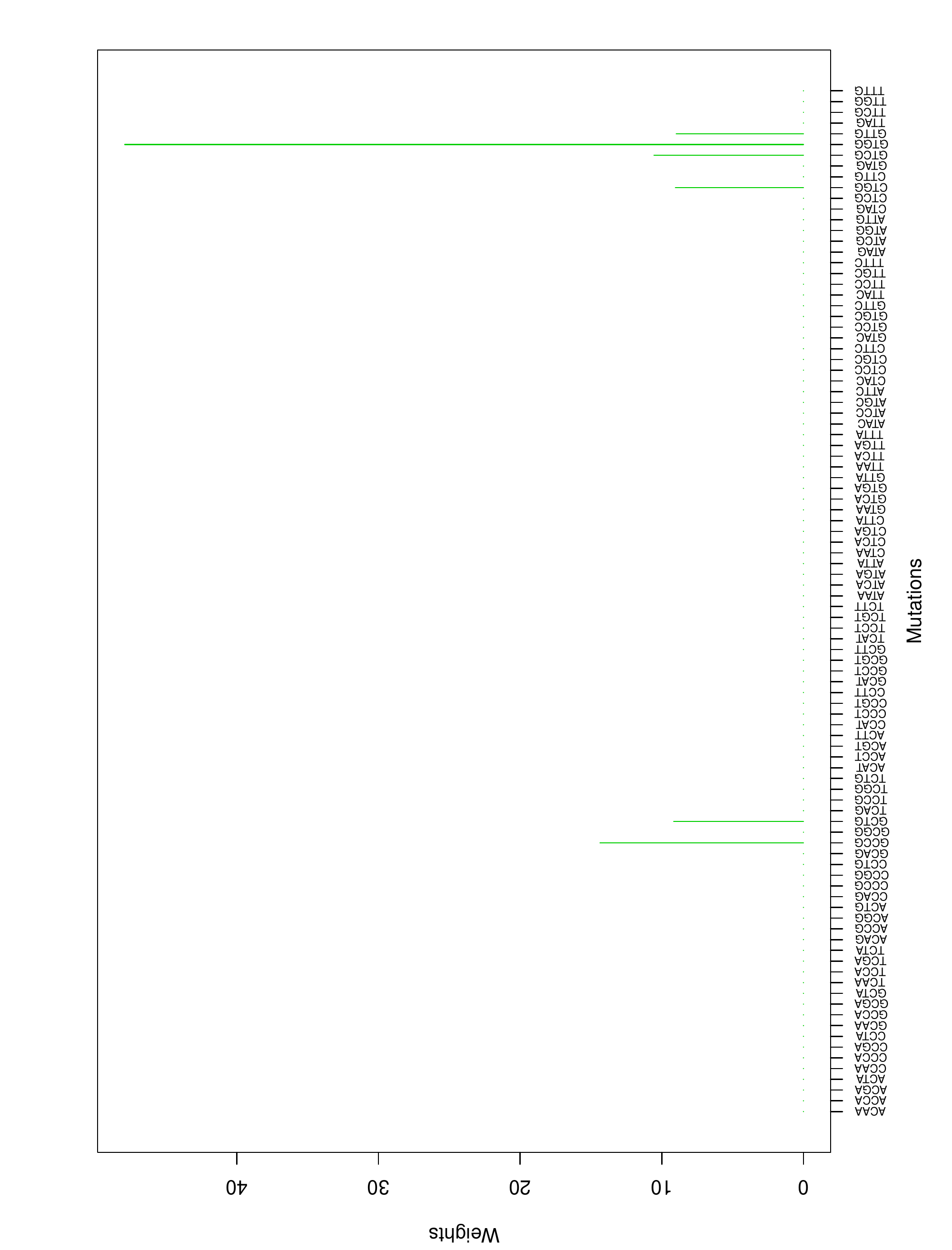}
\caption{Cluster Cl-4 in Clustering-E1 with weights based on normalized regressions with arithmetic means.
}
\label{FigureNorm4}
\end{figure}

\begin{figure}[ht]
\centering
\includegraphics[scale=0.7]{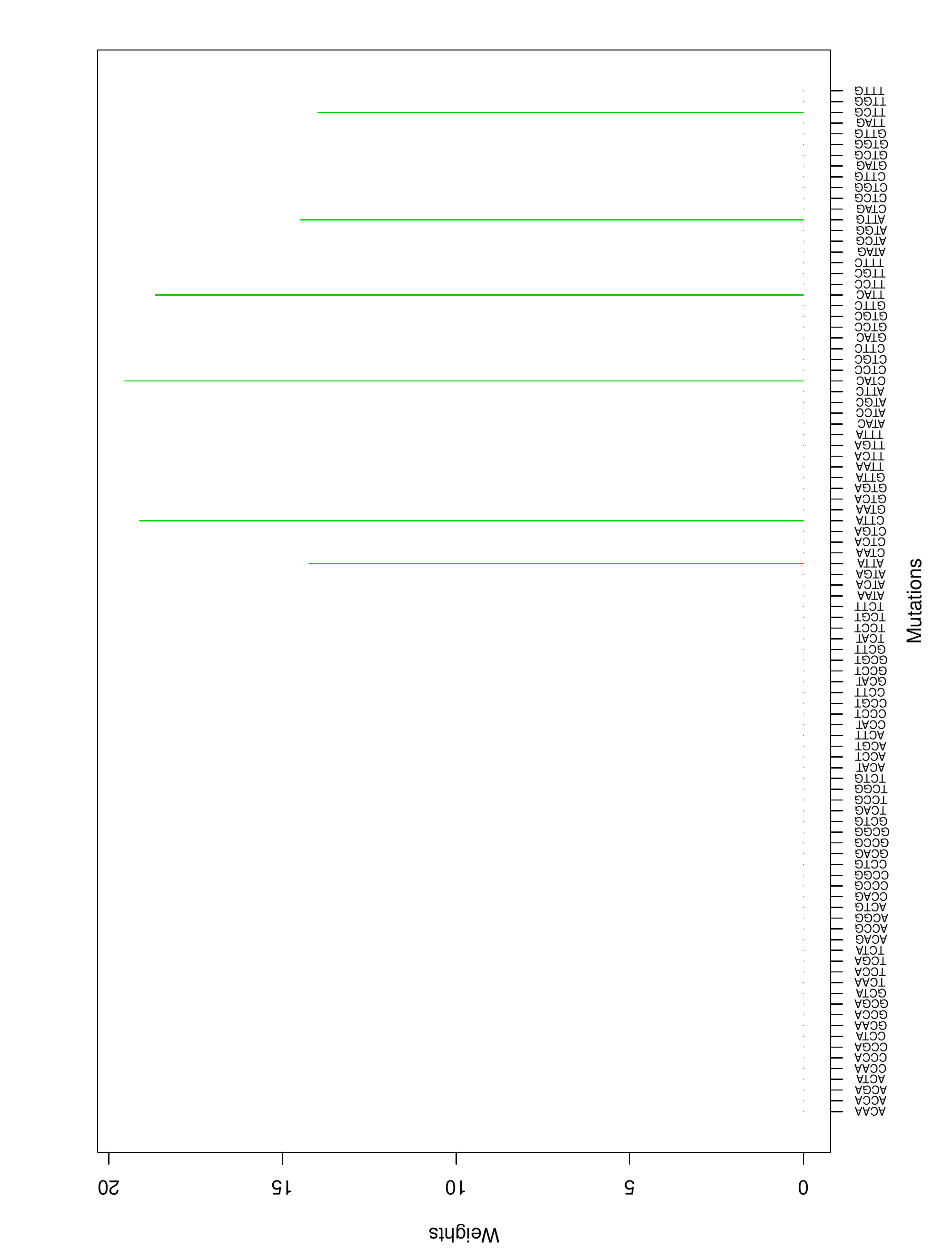}
\caption{Cluster Cl-5 in Clustering-E1 with weights based on normalized regressions with arithmetic means.
}
\label{FigureNorm5}
\end{figure}

\begin{figure}[ht]
\centering
\includegraphics[scale=0.7]{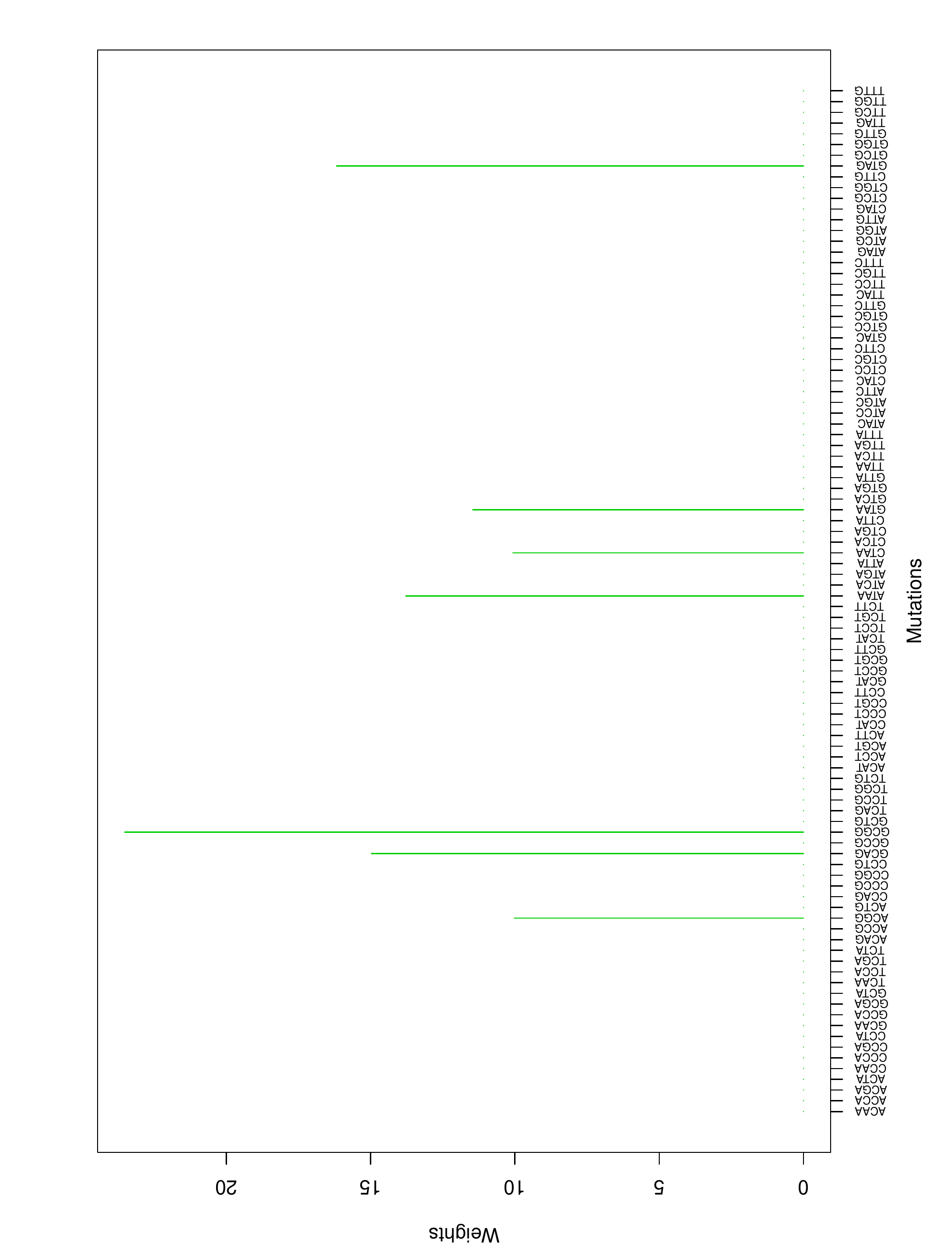}
\caption{Cluster Cl-6 in Clustering-E1 with weights based on normalized regressions with arithmetic means.
}
\label{FigureNorm6}
\end{figure}

\begin{figure}[ht]
\centering
\includegraphics[scale=0.7]{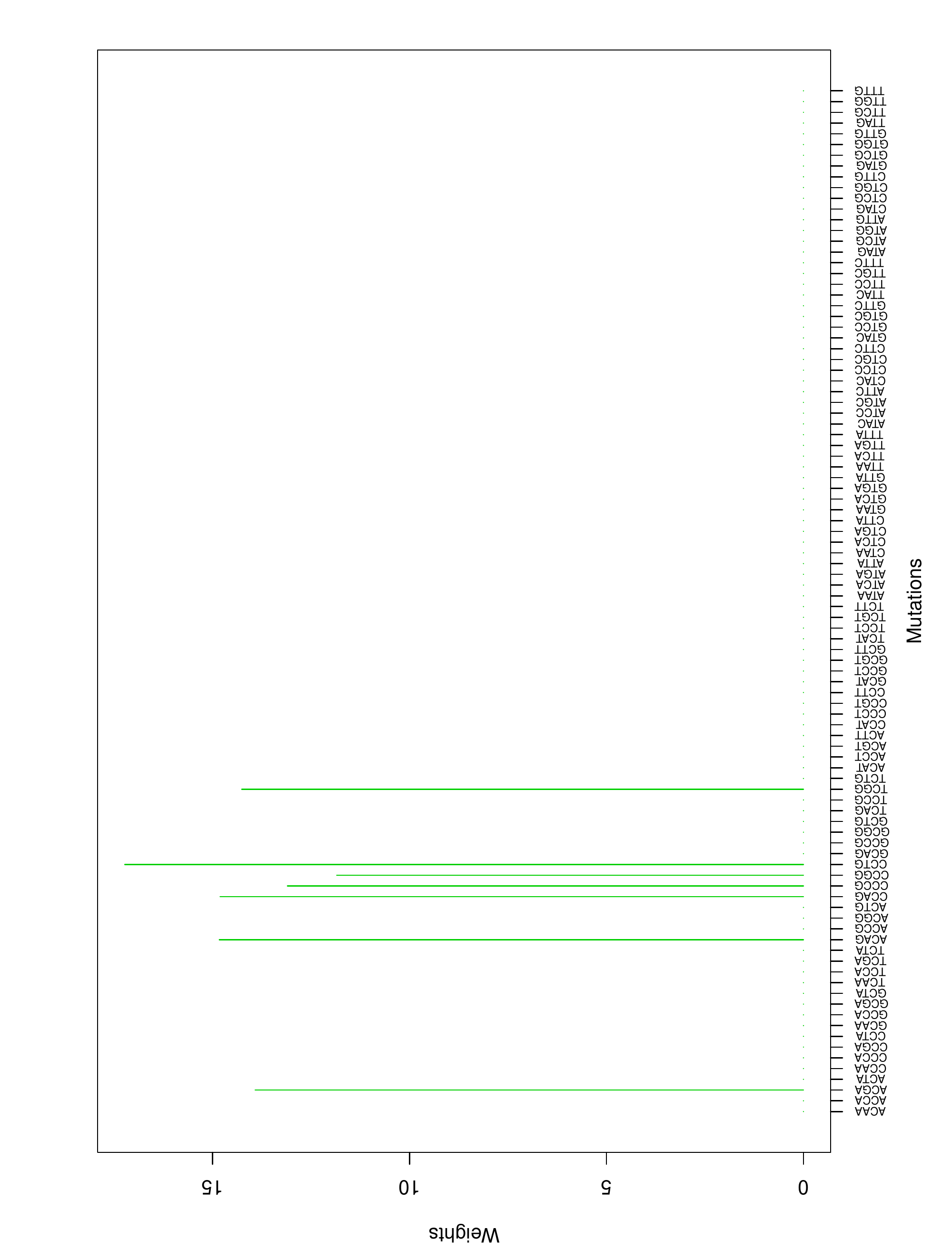}
\caption{Cluster Cl-7 in Clustering-E1 with weights based on normalized regressions with arithmetic means.
}
\label{FigureNorm7}
\end{figure}

\begin{figure}[ht]
\centering
\includegraphics[scale=0.7]{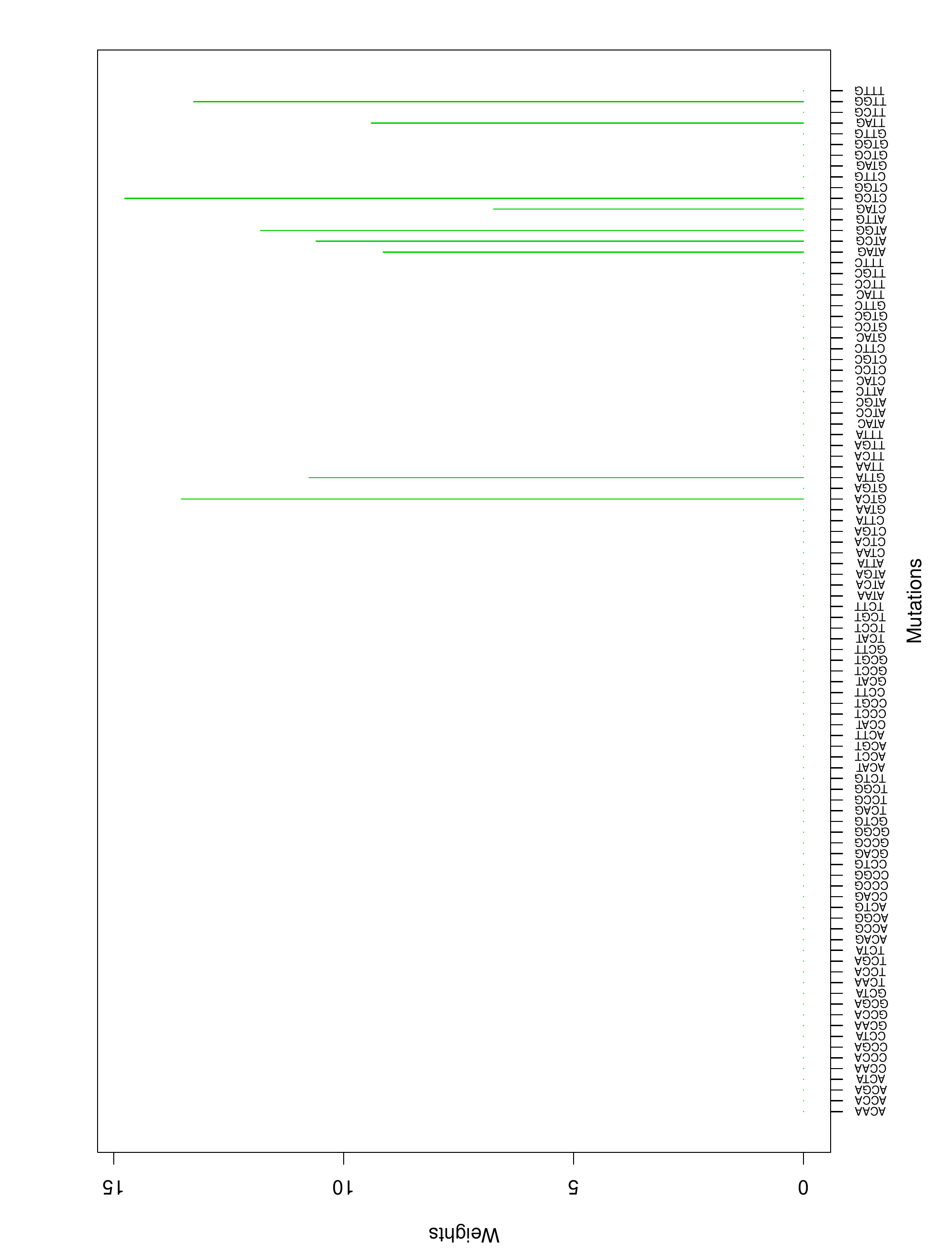}
\caption{Cluster Cl-8 in Clustering-E1 with weights based on normalized regressions with arithmetic means.
}
\label{FigureNorm8}
\end{figure}

\begin{figure}[ht]
\centering
\includegraphics[scale=0.7]{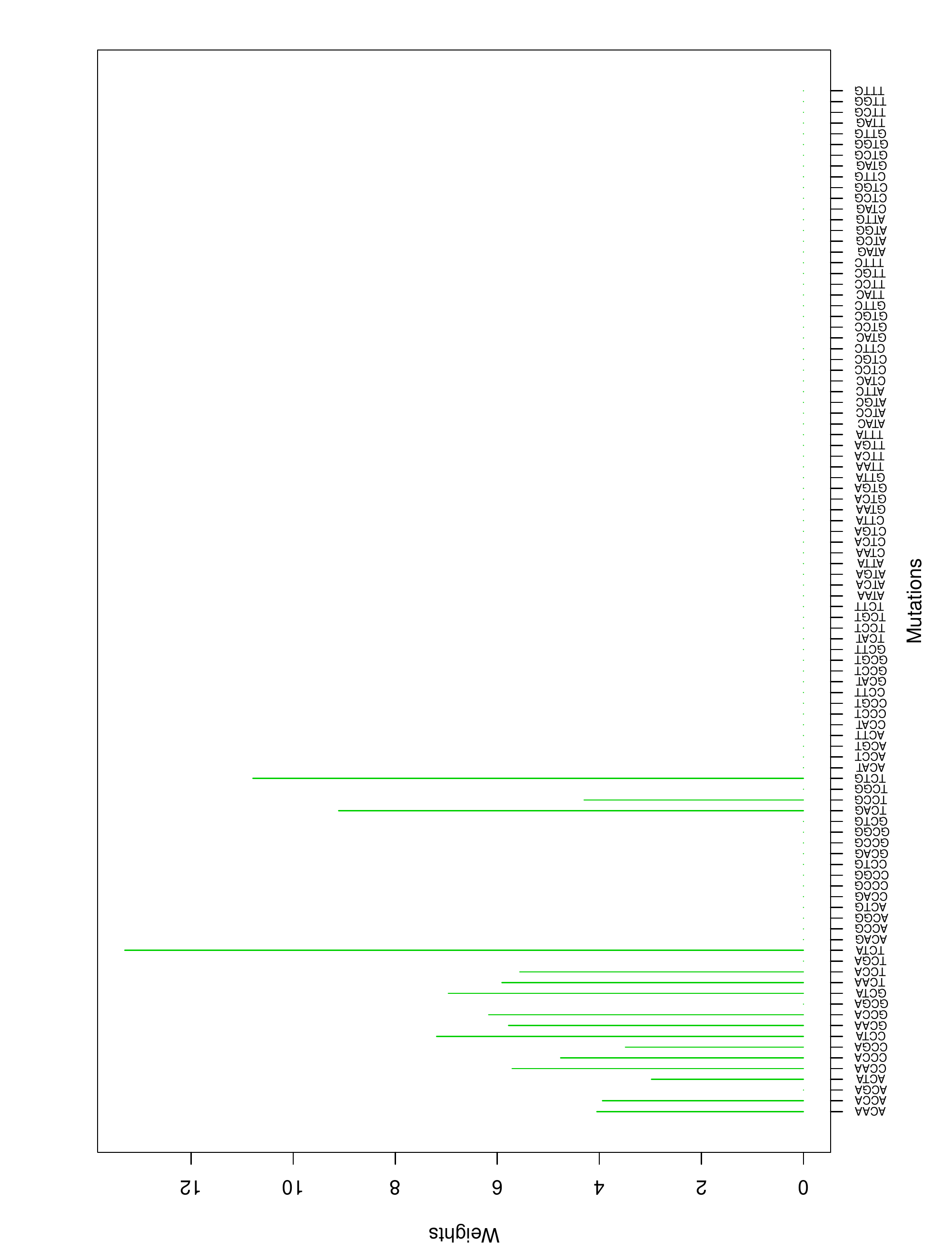}
\caption{Cluster Cl-9 in Clustering-E1 with weights based on normalized regressions with arithmetic means.
}
\label{FigureNorm9}
\end{figure}

\begin{figure}[ht]
\centering
\includegraphics[scale=0.7]{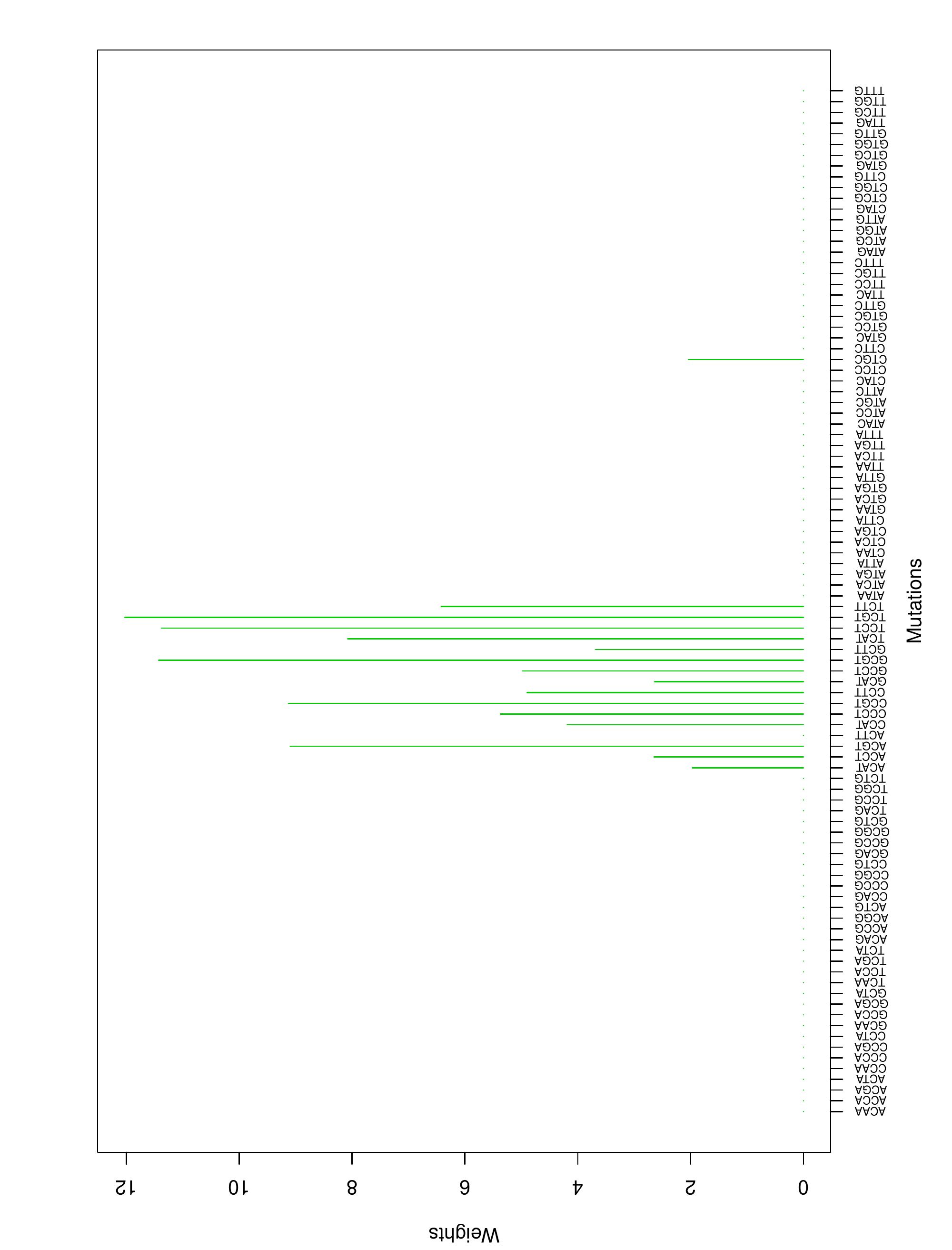}
\caption{Cluster Cl-10 in Clustering-E1 with weights based on normalized regressions with arithmetic means.
}
\label{FigureNorm10}
\end{figure}

\begin{figure}[ht]
\centering
\includegraphics[scale=0.7]{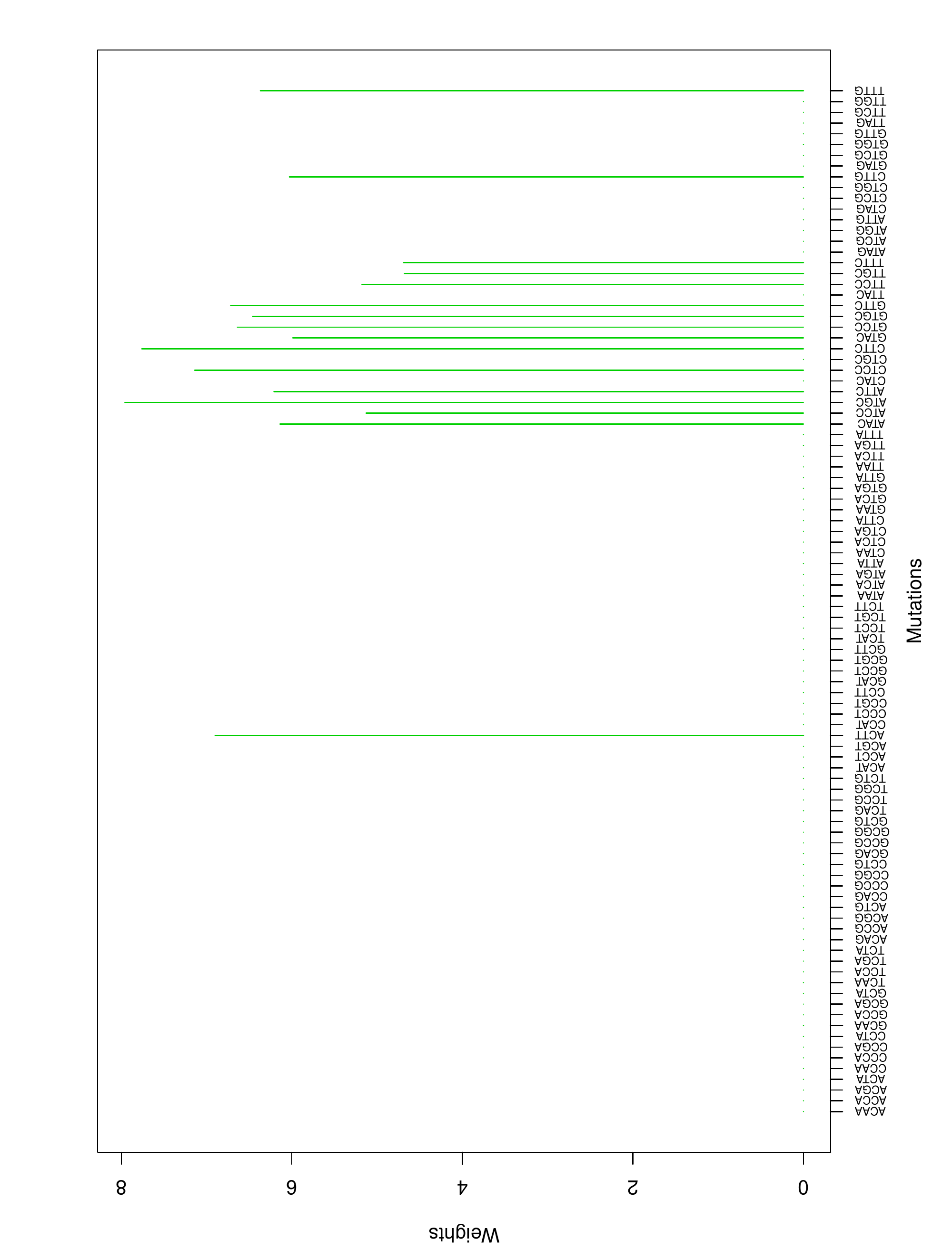}
\caption{Cluster Cl-11 in Clustering-E1 with weights based on normalized regressions with arithmetic means.
}
\label{FigureNorm11}
\end{figure}

{}For Clustering-E1, as in \cite{*K-means}, we compute the within-cluster weights based on unnormalized regressions (via Eqs. (13), (14) and (15) thereof) and normalized regressions (via Eqs. (17), (14) and (16) thereof) with exposures calculated based on arithmetic averages (see Subsection 2.6 of \cite{*K-means} for details). We give the within-cluster weights for Clustering-E1 in Tables \ref{table.weights.E.1} and \ref{table.weights.E.2} and plot them in Figures \ref{Figure1} through \ref{Figure11} for unnormalized regressions, and in Tables \ref{table.weights.E.N1} and \ref{table.weights.E.N2} and Figures \ref{FigureNorm1} through \ref{FigureNorm11} for normalized regressions. The actual mutation categories in each cluster can be read off the aforesaid Tables \ref{table.weights.E.1} and \ref{table.weights.E.2} with the weights (thus, the mutation categories with nonzero weights belong to a given cluster), or from the horizontal axis labels in the aforesaid Figures \ref{Figure1}-\ref{Figure11}.

\subsection{Reconstruction and Correlations}\label{sub.cor}

\subsubsection{Within-cluster Correlations}\label{sub.within}

{}We have our data matrix $G_{is}$. We are approximating this matrix via the following factorized matrix:
\begin{equation}\label{fac.st}
 G^*_{is} = \sum_{A=1}^K W_{iA}~H_{As} = w_i~H_{Q(i),s}
\end{equation}
where $W_{iA}$ are the within-cluster weights ($i=1,\dots,N$; $A=1\dots,K$), $H_{As}$ are the exposures ($s=1,\dots,n=32$ labels the cancer types), $Q:\{1,\dots,N\}\mapsto\{1,\dots,K\}$ is the map between the $N=96$ mutations and $K=11$ clusters in Clustering-E1, and we have\footnote{\, Due to a binary clustering structure, the within-cluster weights $W_{iA}$ are encoded in an $N$-vector $w_i$. This is because all but $N$ elements of the matrix $W_{iA}$ are zero.} $W_{iA} = w_i~\delta_{Q(i), A}$. It is the matrix $W_{iA}$ that is given in Tables \ref{table.weights.E.1} and \ref{table.weights.E.2} for the unnormalized regressions, and Tables \ref{table.weights.E.N1} and \ref{table.weights.E.N2} for the normalized regressions.

{}We can now compute an $n\times K$ matrix $\Theta_{sA}$ of {\em within-cluster} cross-sectional correlations between $G_{is}$ and $G^*_{is}$ defined via ($\mbox{xCor}(\cdot,\cdot)$ stands for ``cross-sectional correlation", i.e., ``correlation across the index $i$")\footnote{\, Due to the factorized structure (\ref{fac.st}), these correlations do not directly depend on $H_{As}$.}
\begin{equation}
 \Theta_{sA} = \left.\mbox{xCor}(G_{is}, G^*_{is})\right|_{i\in J(A)} = \left.\mbox{xCor}(G_{is}, w_i)\right|_{i\in J(A)}
\end{equation}
Here $J(A) = \{i|Q(i) = A\}$ is the set of mutations labeled by $i$ that belong to a given cluster labeled by $A$. We give the matrix $\Theta_{sA}$ for Clustering-E1 for weights based on unnormalized regressions in Table \ref{table.fit.theta}, and weights based on normalized regressions in Table \ref{table.fit.theta.N}. As for genome data \cite{*K-means}, the fit for normalized regressions is somewhat better than that for unnormalized regressions.

\subsubsection{Overall Correlations}\label{ov.cor}

{}Another useful metric, which we use as a sanity check, is this. For each value of $s$ (i.e., for each cancer type), we can run a linear cross-sectional regression (without the intercept) of $G_{is}$ over the matrix $W_{iA}$. So, we have $n=32$ of these regressions. Each regression produces multiple $R^2$ and adjusted $R^2$, which we give in Tables \ref{table.fit.theta}-\ref{table.fit.theta.N}. Furthermore, we can compute the {\em fitted} values ${\widehat G}^*_{is}$ based on these regressions, which are given by
\begin{equation}\label{overall.reg}
 {\widehat G}^*_{is} = \sum_{A=1}^K W_{iA}~F_{As} = w_i~F_{G(i),s}
\end{equation}
where (for each value of $s$) $F_{As}$ are the regression coefficients. We can now compute the {\em overall} cross-sectional correlations (i.e., the index $i$ runs over all $N=96$ mutation categories)
\begin{equation}
 \Xi_s = \mbox{xCor}(G_{is}, {\widehat G}^*_{is})
\end{equation}
These correlations are also given in Tables \ref{table.fit.theta}-\ref{table.fit.theta.N} and measure the overall fit quality.

\subsubsection{Interpretation}

{}Looking at Table \ref{table.fit.theta.N}, a few things jump out. First, most -- 24 out of 32 -- cancer types have high (80\%+) within-cluster correlations with at least one cluster. Out of the other 8 cancer types, 6 have reasonably high (70\%+) within-cluster correlations with at least one cluster. The remaining 2 cancer types are X9 (Cervical Cancer) and X17 (Liver Cancer). In \cite{*K-means}, based on genome data, we already observed that Liver Cancer does not have clustering structure, so this is not surprising. On the other hand, with Cervical Cancer the story appears to be trickier. According to \cite{COSMIC}, we should expect COSMIC signatures CSig2+13 and CSig26 (see Section \ref{sec.4} for more details) to appear in Cervical Cancer. According to Table \ref{table.cosmic.exome.cor.1} (see Section \ref{sec.4}), CSig2+13 indeed have high correlations with X9 (but not CSig26). On the other hand, the dominant part of CSig2 (C $>$ T mutations in TCA, TCC, TCG, TCT) is subsumed in Cluster Cl-10 (see Figure \ref{FigureNorm10}), and the dominant part of CSig13 (C $>$ G mutations in TCA, TCC, TCT) is subsumed in Cluster Cl-9 (see Figure \ref{FigureNorm9}). Basically, it appears that the large (each with 16 mutation categories) clusters Cl-9, Cl-10 and Cl-11 probably could be split into smaller clusters. In fact, Cl-9 and Cl-11 do not have 80\%+ correlations with any cancer types (they do have 70\%+ correlations with one cancer type each). This is another indication that these clusters might be ``oversized". The same was observed with the largest cluster (with 21 mutation categories) in \cite{*K-means} in the context of genome data. Simply put, these ``oversized" clusters may have to be dealt with via appropriately tweaking the underlying clustering algorithm.\footnote{\, This is outside of the scope hereof and will be dealt with elsewhere.}

{}The last 3 columns in Table \ref{table.fit.theta.N} provide metrics for the overall fit for each cancer type. The overall correlations (between the original data $G_{is}$ and the model-fitted values ${\widehat G}^*_{is}$ -- see Subsection \ref{ov.cor}) in the last column of Table \ref{table.fit.theta.N} are above 80\% for 16 (out of the 32) cancer types, and above 70\% for 26 cancer types. These high correlations indicate a good in-sample agreement between the original and reconstructed (model-fitted) data for each of these 26 cancer types. The remaining 6 cancer types, which all have overall correlations above 60\%, are: X4 (B-Cell Lymphoma), X6 (Bladder Cancer), X8 (Breast Cancer), X9 (Cervical Cancer), X26 (Rectum Adenocarcinoma), and X29 (Testicular Germ Cell Tumor). We already discussed Cervical Cancer above. We address Breast Cancer in Section \ref{sec.4} hereof. Now, the X4 data is sparsely populated: there are 24 samples and the total number of counts is 706, so there are many zeros in the underlying sample data, albeit only 2 zeros in the aggregated data. According to \cite{COSMIC}, we should expect CSig9 and CSig17 in B-Cell Lymphoma. However, according to Table \ref{table.cosmic.exome.cor.1} (see Section \ref{sec.4}) these signatures do not have high correlations with X4. Note that clustering worked well for B-Cell Lymphoma for the genome data in \cite{*K-means}, but there the genome data was well-populated. Therefore, it is reasonable to assume that here the ``underperformance" is likely due to the sparsity of the underlying data. For X6 (Bladder Cancer) the situation is similar to X9 (Cervical Cancer) above: according to \cite{COSMIC}, we should expect CSig2+13 in Bladder Cancer, and Table \ref{table.cosmic.exome.cor.1} is consistent with this. However, as mentioned above CSig2 and CSig13 are subsumed in Clusters Cl-10 and Cl-9, respectively (``oversizing"). According to Table \ref{table.cosmic.exome.cor.2}, we should expect CSig10 in X26. CSig10 is dominated by the C $>$ A mutation in TCT (which is subsumed in Cluster Cl-9) and C $>$ T mutation in TCG (which is subsumed in Cluster Cl-10). Again, here we are dealing with ``oversizing" of these clusters. X29 has high within-cluster correlations with Clusters Cl-4 and Cl-5. The overall fit correlation apparently is lowered by the high negative correlation with Cluster Cl-3. To summarize, ``oversizing" is one potential ``shortcoming" here.

\section{Concluding Remarks}\label{sec.4}

{}In order to understand the significance of our results, let us compare them to the fit that COSMIC signatures\footnote{\, For details, see \cite{COSMIC}. For references, see \cite{Nik-Zainal}, \cite{Alexandrov.NMF}, \cite{Alexandrov}, \cite{Helleday}, \cite{AlexStrat}.} provide for our exome data. We can do this by computing the following $p \times n$ cross-sectional correlation matrix
\begin{equation}
 \Delta_{\alpha s} = \mbox{xCor}(U_{i\alpha}, G_{is})
\end{equation}
where $U_{i\alpha}$ ($\alpha = 1,\dots, p$) is the $N \times p$ matrix of weights for $p = 30$ COSMIC signatures, which for brevity we will refer to as CSig1, \dots, CSig30.\footnote{\, See http://cancer.sanger.ac.uk/cancergenome/assets/signatures\_probabilities.txt; note that the ordering of mutation categories in this file is not the same as ours.} The matrix $\Delta_{\alpha s}$ is given in Tables \ref{table.cosmic.exome.cor.1} and \ref{table.cosmic.exome.cor.2}. Let us look at the 80\%+ correlations (which are in bold font in Tables \ref{table.cosmic.exome.cor.1} and \ref{table.cosmic.exome.cor.2}).\footnote{\, Relaxing this cut-off to 70\% (see Tables \ref{table.cosmic.exome.cor.1} and \ref{table.cosmic.exome.cor.2}) does not alter our conclusions below.} Only 6 out 30 COSMIC signatures, to wit, CSig1,2,6,7,10,15 have 80\%+ correlations with the exome data for the 32 cancer types. Aetiology of these signatures is known \cite{COSMIC}. CSig1 is the result of an endogenous mutational process initiated by spontaneous 5-methylcytosine deamination, hence the ubiquity of its high correlations with many cancer types. CSig2 (which usually appears in tandem with CSig13) is due to APOBEC mediated cytosine deamination, hence its high correlations with some cancer types. CSig6 is associated with defective DNA mismatch repair, hence its high correlations with several cancer types. CSig7 is due to ultraviolet light exposure, so its high correlation with X19 (Melanoma) is spot on.\footnote{\, However, there is no magic here. Apparently there is a large overlap between the exome data we use here and that used by \cite{COSMIC}. Furthermore, caution is in order when it comes to any NMF-based signature that dominates a given cancer type. What this means is that the signature is close to the properly normalized underlying occurrence counts data (either aggregated or appropriately averaged over all samples), and NMF samplings fail to find a local minimum substantially different along this particular direction from the local minima that include this cancer signature. Such a signature indicates that the corresponding cancer type is of a ``stand-alone" type and has little in common with other cancer types. An example of such a signature is the Liver Cancer dominant NMF-based cancer signature found in \cite{BioFM}.\label{fn.dom.sigs}} CSig10 is associated with recurrent error-prone polymerase POLE somatic mutations.\footnote{\, Its high correlations with X26 (Rectum Adenocarcinoma) and X32 (Uterine Cancer) are consistent with \cite{COSMIC} and, once again, apparently are due to a large overlap between the exome data we use here and that used by \cite{COSMIC}.} CSig15 is associated with defective DNA mismatch repair; the significance of its high correlation with X23 (Pancreatic Cancer) is unclear. So, only a handful of COSMIC signatures -- all associated with known mutational processes -- do well on our exome data.\footnote{\, Note that considering the overall fit quality for COSMIC signatures by running overall regressions (of $G_{is}$ over $U_{i\alpha}$ without the intercept) as we did above for clusters would not be meaningful. The regression coefficients $F_{As}$ in (\ref{overall.reg}) in the case of clusters are guaranteed to be nonnegative. This is because the $N$-vectors corresponding to the columns in the cluster weights matrix $W_{iA}$ are orthogonal to each other. The $N$-vectors corresponding to the columns in the COSMIC weights matrix $U_{i\alpha}$ are not orthogonal, unacceptably resulting in many negative regression coefficients $F_{\alpha s}$.} Others do not fit well.

{}This is the out-of-sample stability issue emphasized in \cite{BioFM}. It traces to the fact that NMF is an intrinsically unstable method, both in- and out-of-sample. In-sample instability relates to the fact that NMF is nondeterministic and produces different looking signatures from one run to another. In fact, we attempted running NMF on our exome data. We ran 3 batches with 800 sampling in each batch -- a computationally time-consuming procedure.\footnote{\, Thus, to run one batch of NMF with 800 samplings on a 4-CPU (8 cores each, 2.60GHz) machine with 529Gb of RAM and hyperthreading (Operating System: Debian 3.2.84-2 x86\_64 GNU/Linux), it took 6-7 days (and 3-4 days when the input data was ``de-noised" following \cite{BioFM}). In contrast, to run each of our 3 batches of *K-means with 10 million instances of k-means in each batch (see Subsection \ref{sub.res}), it only took under 24 hours on a single CPU (quad-core, 3.1GHz) machine with 16GB of RAM (Operating System: 64-bit Windows Server 2008 R2 Standard). From this data it is evident that *K-means computationally is much cheaper than NMF, even if NMF is improved via ``de-noising" \cite{BioFM}.} The 3 batches produced different looking results, which with a lot of manual curation could only be partially matched to some COSMIC signatures, but this matching was different and highly unstable across the 3 batches. Simply put, NMF failed to produce any meaningful results on our exome data. Furthermore, the above discussion illustrates that most COSMIC signatures (extracted using NMF from exome and genome data) apparently are unstable out-of-sample, e.g., when applied to our exome data aggregated by cancer types. Here one may argue that exome data contains only partial information and NMF should not be used on it. However, the COSMIC signatures are in fact based on 10,952 exomes and 1,048 whole-genomes across 40 cancer types \cite{COSMIC}.\footnote{\, Also, see, e.g., \cite{Schulze}.} The difference here is that we are aggregating samples by cancer types and most COSMIC signatures apparently do not apply, which means that COSMIC signatures are highly sample-set-specific (that is, unstable out-of-sample). Furthermore, as mentioned above, CSig7 (UV exposure) is spot on in that it has 99.66\% correlation with X19 (Melanoma).\footnote{\, Albeit one should keep in mind the comments in fn. \ref{fn.dom.sigs}.} So, one can argue that the culprit is not the exome data but the method (NMF) itself. To quantify this, let us look at correlations of COSMIC signatures with {\em genome} data for 14 cancer types used in \cite{BioFM} and \cite{*K-means}. The results are given in Table \ref{table.cosmic.fit.genome}. As in the case of exome data, here too we have high correlations only for a handful of COSMIC signatures corresponding to known mutational processes, to wit, CSig1,4,6,13. So, most COSMIC signatures do not appear to have explanatory power on genome data aggregated by cancer types, a further indication that most COSMIC signatures lack out-of-sample stability.

{}What about out-of-sample stability for our clusters we obtained from exome data? One way to test it is to look at within-cluster correlations and the overall fit metrics as in Table \ref{table.fit.theta.N} but for the aforesaid genome data for 14 cancer types used in \cite{BioFM} and \cite{*K-means}. The results are given in Table \ref{table.fit.theta.N.genome}. Unsurprisingly, the quality of the fit for genome data (out-of-sample) is not as good as for exome data (in-sample). However, it is i) reasonably good, and ii) unequivocally much better than the fit provided by the COSMIC signatures (Table \ref{table.cosmic.fit.genome}). Furthermore, the exome based 11 clusters have a poor overall fit for G.X4 (Breast Cancer), G.X8 (Liver Cancer), G.X9 (Lung Cancer), and G.X14 (Renal Cell Carcinoma), the same 4 cancer types for which genome based 7 clusters in \cite{*K-means} produced a poor overall fit, and for a good reason too (see \cite{*K-means} for details). It is less clear why the exome based 11 clusters do not have a better fit for G.X7 (Gastric Cancer) considering the in-sample fits for this cancer type based on exome data (X15, Table \ref{table.fit.theta.N} hereof) and genome data (row 7, Table 15 of \cite{*K-means}) are petty good.

{}So, unlike NMF, *K-means clustering, being a {\em statistically deterministic} method, is in-sample stable. Here we can ask, what if we apply to NMF the same 2 machine learning levels as those that sit on top of k-means in *K-means to make it statistically deterministic? The answer is that when applying NMF, one already uses one machine learning method, which is a form of aggregation of a large number of samplings (i.e., individual NMF runs).\footnote{\, Thus, as mentioned above, we ran 3 batches of 800 NMF samplings. In each batch, 800 samplings are aggregated via nondeterministic clustering (e.g., via k-means -- see, e.g., \cite{*K-means} for a detailed discussion). The net result -- by design -- is nondeterministic.} This is conceptually similar to the first machine learning level in *K-means. So, then we can ask, what if we add to NMF the second machine learning level as in *K-means, to wit, by comparing a large number of such ``averagings"? A simple, prosaic answer is that it would make NMF -- which is already computationally costly as is and much more so with the first machine learning level -- {\em computationally prohibitive}. The reason why *K-means is computationally much cheaper is that the basic building block of *K-means -- on top of which we add the two machine learning methods -- is vanilla k-means, which is much, much cheaper than NMF. And that is what makes all the difference.\footnote{\, Furthermore, as was argued in \cite{*K-means}, NMF, at least to some degree, is clustering in disguise. In fact, visual inspection of COSMIC signatures makes it evident that many of them -- albeit possibly not all -- have clustering substructure. This will be discussed in more detail in a forthcoming paper. Also, it would be interesting to understand the relation between ``R-mutations" \cite{Vog} (also see references therein) and somatic mutational noise.}

{}Finally, let us mention that exome data for Chronic Myeloid Disorders (121 samples, 175 total counts) was published in \cite{Papaemmanuil}, \cite{Malcovati}, and for Neuroblastoma (13 samples, 298 total counts) in \cite{Sausen}. However, this data is so sparsely populated (too many zeros even after aggregation) that we specifically excluded it from our analysis. Much more unpublished data is available for the cancer types we analyze here as well as other cancer types, and it would be very interesting to apply our methods to this data, including to (still embargoed) extensive genome data of the International Cancer Genome Consortium.

\section*{Acknowledgments} The results published here are in whole or part based upon data generated by the TCGA Research Network: http://cancergenome.nih.gov/.

\appendix

\section{Exome Sample IDs}\label{app.IDs}

{}In this Appendix we give the sample IDs with the corresponding publication references for the exome data we use. We label these references as H1, Z1, etc., and use these labels in Table \ref{table.exome.summary} in the Sources column.\\
{\tiny
$\blacklozenge$ {\bf Acute Lymphoblastic Leukemia} (86 samples):\\
$\bullet$ Source H1 = \cite{Holmfeldt}. Sample IDs are of the form SJHYPO*, where * is:\\
001-D,
002-D,
004-D,
005-D,
006-D,
009-D,
009-R,
012-D,
013-D,
014-D,
016-D,
019-D,
020-D,
022-D,
024-D,
026-D,
029-D,
032-D,
036-D,
037-D,
037-R,
039-D,
040-D,
041-D,
042-D,
044-D,
045-D,
046-D,
047-D,
051-D,
052-D,
052-R,
055-D,
056-D,
116-D,
117-D,
119-D,
120-D,
123-D,
124-D,
125-D,
126-D.\\
$\bullet$ Source Z1 = \cite{Zhang2012}. Sample IDs are of the form SJTALL*, where * is:\\
001,
002,
003,
004,
005,
006,
007,
008,
009,
011,
012,
013,
169,
192,
208.\\
$\bullet$ Source D1 = \cite{De Keersmaecker}:\\
TBR01,
TBR03,
TBR05,
TBR06,
TBR08,
TLE02,
TLE10,
TLE109,
TLE31,
TLE33,
TLE34,
TLE38,
TLE39,
TLE41,
TLE42,
TLE43,
TLE50,
TLE51,
TLE54,
TLE55,
TLE57,
TLE60,
TLE61,
TLE63,
TLE64,
TLE65,
TLE66,
TLE67,
TLE68.\\
$\blacklozenge$ {\bf Acute Myeloid Leukemia} (190 samples):\\
$\bullet$ Source T1 = TCGA (see Acknowledgments). Sample IDs are of the form TCGA-AB-*, where * is:\\
2802,
2803,
2804,
2805,
2806,
2807,
2808,
2809,
2810,
2811,
2812,
2813,
2814,
2816,
2817,
2818,
2819,
2820,
2821,
2822,
2824,
2825,
2826,
2827,
2828,
2829,
2830,
2831,
2832,
2833,
2835,
2836,
2837,
2838,
2839,
2841,
2842,
2843,
2844,
2845,
2846,
2847,
2849,
2850,
2851,
2853,
2854,
2855,
2857,
2858,
2859,
2860,
2861,
2862,
2863,
2864,
2865,
2866,
2867,
2868,
2869,
2870,
2871,
2872,
2873,
2874,
2875,
2876,
2877,
2878,
2879,
2880,
2881,
2882,
2883,
2884,
2885,
2886,
2887,
2888,
2889,
2890,
2891,
2892,
2893,
2894,
2895,
2896,
2897,
2898,
2899,
2900,
2901,
2904,
2905,
2906,
2907,
2908,
2910,
2911,
2912,
2913,
2914,
2915,
2916,
2917,
2918,
2919,
2920,
2921,
2922,
2923,
2924,
2925,
2926,
2927,
2928,
2929,
2930,
2931,
2932,
2933,
2934,
2935,
2936,
2937,
2938,
2939,
2940,
2941,
2943,
2945,
2946,
2947,
2948,
2949,
2950,
2952,
2954,
2955,
2956,
2957,
2959,
2963,
2964,
2965,
2966,
2967,
2968,
2969,
2970,
2971,
2972,
2973,
2974,
2975,
2976,
2977,
2978,
2979,
2980,
2981,
2982,
2983,
2984,
2985,
2986,
2987,
2988,
2989,
2990,
2991,
2992,
2993,
2994,
2995,
2996,
2997,
2998,
2999,
3000,
3001,
3002,
3005,
3006,
3007,
3008,
3009,
3011,
3012.\\
$\blacklozenge$ {\bf Adrenocortical Carcinoma} (91 samples):\\
$\bullet$ Source T2 = TCGA (see Acknowledgments). Sample IDs are of the form TCGA-*, where * is:\\
OR-A5J1,
OR-A5J2,
OR-A5J3,
OR-A5J4,
OR-A5J5,
OR-A5J6,
OR-A5J7,
OR-A5J8,
OR-A5J9,
OR-A5JA,
OR-A5JB,
OR-A5JC,
OR-A5JD,
OR-A5JE,
OR-A5JF,
OR-A5JG,
OR-A5JH,
OR-A5JI,
OR-A5JJ,
OR-A5JK,
OR-A5JL,
OR-A5JM,
OR-A5JO,
OR-A5JP,
OR-A5JQ,
OR-A5JR,
OR-A5JS,
OR-A5JT,
OR-A5JU,
OR-A5JV,
OR-A5JW,
OR-A5JX,
OR-A5JY,
OR-A5JZ,
OR-A5K0,
OR-A5K1,
OR-A5K2,
OR-A5K3,
OR-A5K4,
OR-A5K5,
OR-A5K6,
OR-A5K8,
OR-A5K9,
OR-A5KB,
OR-A5KO,
OR-A5KP,
OR-A5KQ,
OR-A5KS,
OR-A5KT,
OR-A5KU,
OR-A5KV,
OR-A5KW,
OR-A5KX,
OR-A5KY,
OR-A5KZ,
OR-A5L1,
OR-A5L2,
OR-A5L3,
OR-A5L4,
OR-A5L5,
OR-A5L6,
OR-A5L8,
OR-A5L9,
OR-A5LA,
OR-A5LB,
OR-A5LC,
OR-A5LD,
OR-A5LE,
OR-A5LF,
OR-A5LG,
OR-A5LH,
OR-A5LI,
OR-A5LJ,
OR-A5LK,
OR-A5LL,
OR-A5LN,
OR-A5LO,
OR-A5LP,
OR-A5LR,
OR-A5LS,
OR-A5LT,
OU-A5PI,
P6-A5OF,
P6-A5OG,
P6-A5OH,
PA-A5YG,
PK-A5H8,
PK-A5H9,
PK-A5HA,
PK-A5HB,
PK-A5HC.\\
$\blacklozenge$ {\bf B-Cell Lymphoma} (24 samples):\\
$\bullet$ Source M1 = \cite{Morin}. In DLBCL sample IDs * runs from A though M (e.g., DLBCL-PatientC):\\
07-35482,
DLBCL-Patient*,
FL-PatientA,
FL009.\\
$\bullet$ Source L1 = \cite{Love1}:\\
1060,
1061,
1065,
1093,
1096,
1102,
515,
EB2.\\
$\blacklozenge$ {\bf Benign Liver Tumor} (40 samples):\\
$\bullet$ Source P1 = \cite{Pilati}. Sample IDs are of the form CHC*, where * is:\\
1023T,
1124T,
1315T,
1328T,
1329T,
1337T,
1382T,
1383T,
1424T,
1425T,
1428T,
1432T,
1434T,
1439T,
1488T,
1489T,
1665T,
1666T,
1854T,
1916T,
340T,
361TB,
462T,
463T,
464T,
470T,
471T,
517T,
575T,
578T,
603T,
605T,
623T,
624T,
674T,
687T,
689T,
846T,
918T,
976T.\\
$\blacklozenge$ {\bf Bladder Cancer} (341 samples):\\
$\bullet$ Source G1 = \cite{Guo}. Sample IDs are of the form TCC+AF8-B**+AC0-Tumor, where ** is (below * stands for +AC0-, e.g., 104*0 = 104+AC0-0, and the full sample ID is TCC+AF8-B104+AC0-0+AC0-Tumor):\\
10,
100,
101,
102,
103,
104*0,
104,
105*0,
105*1,
105,
106,
107,
109,
11,
110,
111,
112,
114,
13,
14,
15,
16,
17,
18,
19,
2,
20,
21,
22,
23,
24,
25,
34,
35,
37,
41,
43,
45,
47,
5,
50,
52,
54,
55,
56,
57,
58,
59*0,
59*1,
59*3,
59,
60,
61,
62*0,
63,
64,
65,
66*0,
66,
68,
70,
71,
73,
74,
77,
78,
79,
8,
80*0,
80*1,
80*11,
80*13,
80*3,
80*4,
80*5,
80*7,
80*8,
80,
81*1,
81*2,
81,
82,
83,
84,
85*0,
85*2,
86,
87,
88,
89*1,
89*10,
89*11,
89*12,
89*16,
89*3,
89*4,
89*5,
9,
90,
92,
96,
98,
99.\\
$\bullet$ Source T3 = TCGA (see Acknowledgments). Sample IDs are of the form TCGA+AC0-**, where ** is (below * stands for +AC0-A, e.g., BL*0C8 = BL+AC0-A0C8, and the full sample ID is TCGA+AC0-BL+AC0-A0C8; also, below $\star$ = 3OO = 3-double-O):\\
BL*0C8,
BL*13I,
BL*13J,
BL*3JM,
BL*5ZZ,
BT*0S7,
BT*0YX,
BT*20J,
BT*20N,
BT*20O,
BT*20P,
BT*20Q,
BT*20R,
BT*20T,
BT*20U,
BT*20V,
BT*20W,
BT*20X,
BT*2LA,
BT*2LB,
BT*2LD,
BT*3PH,
BT*3PJ,
BT*3PK,
BT*42B,
BT*42C,
BT*42E,
BT*42F,
C4*0EZ,
C4*0F0,
C4*0F1,
C4*0F6,
C4*0F7,
CF*1HR,
CF*1HS,
CF*27C,
CF*3MF,
CF*3MG,
CF*3MH,
CF*3MI,
CF*47S,
CF*47T,
CF*47V,
CF*47W,
CF*47X,
CF*47Y,
CF*5U8,
CF*5UA,
CU*0YN,
CU*0YO,
CU*0YR,
CU*3KJ,
CU*3QU,
CU*3YL,
CU*5W6,
CU*72E,
DK*1A3,
DK*1A5,
DK*1A6,
DK*1A7,
DK*1AA,
DK*1AB,
DK*1AC,
DK*1AD,
DK*1AE,
DK*1AF,
DK*1AG,
DK*2HX,
DK*2I1,
DK*2I2,
DK*2I4,
DK*2I6,
DK*3IK,
DK*3IL,
DK*3IM,
DK*3IN,
DK*3IQ,
DK*3IS,
DK*3IT,
DK*3IU,
DK*3IV,
DK*3WW,
DK*3WX,
DK*3WY,
DK*3X1,
DK*3X2,
DK*6AV,
DK*6AW,
DK*6B0,
DK*6B1,
DK*6B2,
DK*6B5,
DK*6B6,
E5*2PC,
E5*4TZ,
E5*4U1,
E7*3X6,
E7*3Y1,
E7*4IJ,
E7*4XJ,
E7*541,
E7*5KE,
E7*5KF,
E7*677,
E7*678,
E7*6ME,
E7*6MF,
E7*7DU,
E7*7DV,
FD*3B3,
FD*3B4,
FD*3B5,
FD*3B6,
FD*3B7,
FD*3B8,
FD*3N5,
FD*3N6,
FD*3NA,
FD*3SJ,
FD*3SL,
FD*3SM,
FD*3SN,
FD*3SO,
FD*3SP,
FD*3SQ,
FD*3SR,
FD*3SS,
FD*43N,
FD*43P,
FD*43S,
FD*43U,
FD*43X,
FD*5BR,
FD*5BS,
FD*5BU,
FD*5BV,
FD*5BX,
FD*5BY,
FD*5BZ,
FD*5C0,
FD*5C1,
FD*62N,
FD*62O,
FD*62P,
FD*62S,
FD*6TA,
FD*6TB,
FD*6TC,
FD*6TD,
FD*6TE,
FD*6TF,
FD*6TG,
FD*6TH,
FD*6TI,
FD*6TK,
FJ*3Z7,
FJ*3Z9,
FJ*3ZE,
FJ*3ZF,
FT*3EE,
FT*61P,
G2*2EC,
G2*2EF,
G2*2EJ,
G2*2EK,
G2*2EL,
G2*2EO,
G2*2ES,
G2*3IB,
G2*3IE,
G2*3VY,
GC*3BM,
GC*3I6,
GC*$\star$,
GC*3RB,
GC*3RC,
GC*3RD,
GC*3WC,
GC*3YS,
GC*6I1,
GC*6I3,
GD*2C5,
GD*3OP,
GD*3OQ,
GD*3OS,
GD*6C6,
GD*76B,
GU*42P,
GU*42Q,
GU*42R,
GU*762,
GU*763,
GU*766,
GU*767,
GV*3JV,
GV*3JW,
GV*3JX,
GV*3JZ,
GV*3QF,
GV*3QG,
GV*3QH,
GV*3QI,
GV*3QK,
GV*40E,
GV*40G,
GV*6ZA,
H4*2HO,
H4*2HQ,
HQ*2OE,
HQ*2OF,
HQ*5ND,
HQ*5NE,
K4*3WS,
K4*3WU,
K4*3WV,
K4*4AB,
K4*4AC,
K4*54R,
K4*5RH,
K4*5RI,
K4*5RJ,
K4*6FZ,
K4*6MB,
KQ*41N,
KQ*41P,
KQ*41Q,
KQ*41S,
LC*66R,
LT*5Z6,
MV*51V,
PQ*6FI,
PQ*6FN,
R3*69X,
S5*6DX,
UY*78K,
UY*78L,
UY*78N,
UY*78O.\\
$\blacklozenge$ {\bf Brain Lower Grade Glioma} (465 samples):\\
$\bullet$ Source T4 = TCGA (see Acknowledgments). Sample IDs are of the form TCGA-*, where * is:\\
CS-4938,
CS-4941,
CS-4942,
CS-4943,
CS-4944,
CS-5390,
CS-5393,
CS-5394,
CS-5395,
CS-5396,
CS-5397,
CS-6186,
CS-6188,
CS-6290,
CS-6665,
CS-6666,
CS-6667,
CS-6668,
CS-6669,
CS-6670,
DB-5270,
DB-5273,
DB-5274,
DB-5275,
DB-5276,
DB-5277,
DB-5278,
DB-5279,
DB-5280,
DB-5281,
DB-A4X9,
DB-A4XA,
DB-A4XB,
DB-A4XC,
DB-A4XD,
DB-A4XE,
DB-A4XF,
DB-A4XG,
DB-A4XH,
DB-A64L,
DB-A64O,
DB-A64P,
DB-A64Q,
DB-A64R,
DB-A64S,
DB-A64U,
DB-A64V,
DB-A64W,
DB-A64X,
DB-A75K,
DB-A75L,
DB-A75M,
DB-A75O,
DB-A75P,
DH-5140,
DH-5141,
DH-5142,
DH-5143,
DH-5144,
DH-A669,
DH-A66B,
DH-A66D,
DH-A66F,
DH-A66G,
DH-A7UR,
DH-A7US,
DH-A7UT,
DH-A7UU,
DH-A7UV,
DU-5847,
DU-5849,
DU-5851,
DU-5852,
DU-5853,
DU-5854,
DU-5855,
DU-5870,
DU-5871,
DU-5872,
DU-5874,
DU-6392,
DU-6393,
DU-6394,
DU-6395,
DU-6396,
DU-6397,
DU-6399,
DU-6400,
DU-6401,
DU-6402,
DU-6403,
DU-6404,
DU-6405,
DU-6406,
DU-6407,
DU-6408,
DU-6410,
DU-6542,
DU-7006,
DU-7007,
DU-7008,
DU-7009,
DU-7010,
DU-7011,
DU-7012,
DU-7013,
DU-7014,
DU-7015,
DU-7018,
DU-7019,
DU-7290,
DU-7292,
DU-7294,
DU-7298,
DU-7299,
DU-7300,
DU-7301,
DU-7302,
DU-7304,
DU-7306,
DU-7309,
DU-8158,
DU-8161,
DU-8162,
DU-8163,
DU-8164,
DU-8165,
DU-8166,
DU-8167,
DU-8168,
DU-A5TP,
DU-A5TR,
DU-A5TS,
DU-A5TT,
DU-A5TU,
DU-A5TW,
DU-A5TY,
DU-A6S2,
DU-A6S3,
DU-A6S6,
DU-A6S7,
DU-A6S8,
DU-A76K,
DU-A76L,
DU-A76O,
DU-A76R,
DU-A7T6,
DU-A7T8,
DU-A7TA,
DU-A7TB,
DU-A7TC,
DU-A7TD,
DU-A7TG,
DU-A7TJ,
E1-5302,
E1-5303,
E1-5304,
E1-5305,
E1-5307,
E1-5311,
E1-5318,
E1-5319,
E1-5322,
E1-A7YD,
E1-A7YE,
E1-A7YH,
E1-A7YI,
E1-A7YJ,
E1-A7YK,
E1-A7YL,
E1-A7YM,
E1-A7YN,
E1-A7YO,
E1-A7YQ,
E1-A7YS,
E1-A7YU,
E1-A7YV,
E1-A7YW,
E1-A7YY,
E1-A7Z2,
E1-A7Z3,
E1-A7Z4,
E1-A7Z6,
EZ-7264,
FG-5962,
FG-5963,
FG-5964,
FG-5965,
FG-6688,
FG-6689,
FG-6690,
FG-6691,
FG-6692,
FG-7634,
FG-7636,
FG-7637,
FG-7638,
FG-7641,
FG-7643,
FG-8181,
FG-8182,
FG-8185,
FG-8186,
FG-8187,
FG-8188,
FG-8189,
FG-8191,
FG-A4MT,
FG-A4MU,
FG-A4MW,
FG-A4MX,
FG-A4MY,
FG-A60J,
FG-A60K,
FG-A60L,
FG-A6IZ,
FG-A6J1,
FG-A6J3,
FG-A70Y,
FG-A70Z,
FG-A710,
FG-A711,
FG-A713,
FN-7833,
HT-7467,
HT-7468,
HT-7469,
HT-7470,
HT-7471,
HT-7472,
HT-7473,
HT-7474,
HT-7475,
HT-7476,
HT-7477,
HT-7478,
HT-7479,
HT-7480,
HT-7481,
HT-7482,
HT-7483,
HT-7485,
HT-7601,
HT-7602,
HT-7603,
HT-7604,
HT-7605,
HT-7606,
HT-7607,
HT-7608,
HT-7609,
HT-7610,
HT-7611,
HT-7616,
HT-7620,
HT-7676,
HT-7677,
HT-7680,
HT-7681,
HT-7684,
HT-7686,
HT-7687,
HT-7688,
HT-7689,
HT-7690,
HT-7691,
HT-7692,
HT-7693,
HT-7694,
HT-7695,
HT-7854,
HT-7855,
HT-7856,
HT-7857,
HT-7858,
HT-7860,
HT-7873,
HT-7874,
HT-7875,
HT-7877,
HT-7879,
HT-7880,
HT-7881,
HT-7882,
HT-7884,
HT-7902,
HT-8010,
HT-8011,
HT-8012,
HT-8013,
HT-8015,
HT-8018,
HT-8019,
HT-8104,
HT-8105,
HT-8106,
HT-8107,
HT-8108,
HT-8109,
HT-8110,
HT-8111,
HT-8113,
HT-8114,
HT-8558,
HT-8563,
HT-8564,
HT-A4DS,
HT-A4DV,
HT-A5R5,
HT-A5R7,
HT-A5R9,
HT-A5RA,
HT-A5RB,
HT-A5RC,
HT-A614,
HT-A615,
HT-A616,
HT-A617,
HT-A618,
HT-A619,
HT-A61A,
HT-A61B,
HT-A61C,
HT-A74H,
HT-A74J,
HT-A74K,
HT-A74L,
HT-A74O,
HW-7486,
HW-7487,
HW-7489,
HW-7490,
HW-7491,
HW-7493,
HW-7495,
HW-8319,
HW-8320,
HW-8321,
HW-8322,
HW-A5KJ,
HW-A5KK,
HW-A5KL,
HW-A5KM,
IK-7675,
IK-8125,
KT-A74X,
KT-A7W1,
P5-A5ET,
P5-A5EU,
P5-A5EV,
P5-A5EW,
P5-A5EX,
P5-A5EY,
P5-A5EZ,
P5-A5F0,
P5-A5F1,
P5-A5F2,
P5-A5F4,
P5-A5F6,
P5-A72U,
P5-A72W,
P5-A72X,
P5-A72Z,
P5-A730,
P5-A731,
P5-A733,
P5-A735,
P5-A736,
P5-A737,
P5-A77W,
P5-A77X,
P5-A780,
P5-A781,
QH-A65R,
QH-A65S,
QH-A65V,
QH-A65X,
QH-A65Z,
QH-A6CS,
QH-A6CU,
QH-A6CV,
QH-A6CW,
QH-A6CX,
QH-A6CY,
QH-A6CZ,
QH-A6X3,
QH-A6X4,
QH-A6X5,
QH-A6X8,
QH-A6X9,
QH-A6XA,
QH-A6XC,
R8-A6MK,
R8-A6ML,
R8-A6MO,
R8-A6YH,
R8-A73M,
S9-A6TS,
S9-A6TU,
S9-A6TV,
S9-A6TW,
S9-A6TX,
S9-A6TY,
S9-A6TZ,
S9-A6U0,
S9-A6U1,
S9-A6U2,
S9-A6U5,
S9-A6U6,
S9-A6U8,
S9-A6U9,
S9-A6UA,
S9-A6UB,
S9-A6WD,
S9-A6WE,
S9-A6WG,
S9-A6WH,
S9-A6WI,
S9-A6WL,
S9-A6WM,
S9-A6WN,
S9-A6WO,
S9-A6WP,
S9-A6WQ,
S9-A7IQ,
S9-A7IS,
S9-A7IX,
S9-A7IY,
S9-A7IZ,
S9-A7J0,
S9-A7J1,
S9-A7J2,
S9-A7J3,
S9-A7QW,
S9-A7QX,
S9-A7QY,
S9-A7QZ,
S9-A7R1,
S9-A7R2,
S9-A7R3,
S9-A7R4,
S9-A7R7,
S9-A7R8,
TM-A7C3,
TM-A7C4,
TM-A7C5,
TM-A7CA,
TM-A7CF,
TQ-A7RF,
TQ-A7RG,
TQ-A7RH,
TQ-A7RI,
TQ-A7RJ,
TQ-A7RK,
TQ-A7RM,
TQ-A7RN,
TQ-A7RO,
TQ-A7RP,
TQ-A7RQ,
TQ-A7RR,
TQ-A7RS,
TQ-A7RU,
TQ-A7RV,
TQ-A7RW,
VW-A7QS.\\
$\blacklozenge$ {\bf Breast Cancer} (1182 samples):\\
$\bullet$ Source N1 = \cite{Nik-Zainal2012}. Sample IDs are of the form CGP\_specimen\_*, where * is:\\
1096043,
1142475,
1142532,
1142534,
1192095,
1192097,
1192099,
1192101,
1192103,
1192105,
1192107,
1192111,
1192113,
1192115,
1192117,
1192119,
1192121,
1192123,
1192125,
1192127,
1192129,
1192131,
1192133,
1192135,
1192137,
1195364,
1195366,
1195368,
1212804,
1212810,
1212816,
1212822,
1212825,
1212828,
1215490,
1215532,
1215535,
1215553,
1215559,
1215561,
1215563,
1215565,
1215567,
1215573,
1223855,
1223858,
1223861,
1227889,
1227916,
1227918,
1227920,
1227922,
1227924,
1227926,
1227928,
1227951,
1227953,
1227955,
1227957,
1227959,
1227961,
1227963,
1227965,
1227969,
1227971,
1241537,
1241539,
1241541,
1241543,
1241545,
1241547,
1241549,
1241551,
1241553,
1241555,
1241557,
1241559,
1241562,
1241565,
1241568,
1241571,
1241574,
1241579,
1241581,
1261287,
1261291,
1261293,
1261295,
1261297,
1261299,
1261301,
1261303,
1261305,
1261307,
1261309,
1261311,
1261313,
1261337,
1261382,
1261391,
1266549,
1266551,
1266553,
1266561,
1266563,
1266565,
1266567,
1343241,
1343244,
1343247,
1380057,
1380059,
1380061,
1380063,
1380065,
1380067.\\
$\bullet$ Source S1 = \cite{Stephens}. Sample IDs are of the form PD*a, where * is:\\
4842,
4843,
4844,
4934,
4935,
4936,
4937,
4938,
4939,
5961,
7206,
7211,
7316,
9193.\\
$\bullet$ Source S2 = \cite{Shah}. Sample IDs are of the form SA*, where * is:\\
018,
029,
030,
031,
051,
052,
053,
054,
055,
063,
065,
067,
068,
069,
071,
072,
073,
074,
075,
076,
077,
080,
083,
084,
085,
089,
090,
092,
093,
094,
096,
097,
098,
101,
102,
103,
106,
208,
210,
212,
213,
214,
215,
216,
217,
218,
219,
220,
222,
223,
224,
225,
226,
227,
228,
229,
230,
231,
233,
234,
235,
236,
237.\\
$\bullet$ Source T5 = TCGA (see Acknowledgments). Sample IDs are of the form TCGA-*, where * is:\\
A1-A0SB,
A1-A0SD,
A1-A0SE,
A1-A0SF,
A1-A0SG,
A1-A0SH,
A1-A0SI,
A1-A0SJ,
A1-A0SK,
A1-A0SM,
A1-A0SN,
A1-A0SO,
A1-A0SP,
A1-A0SQ,
A2-A04N,
A2-A04P,
A2-A04Q,
A2-A04R,
A2-A04T,
A2-A04U,
A2-A04V,
A2-A04W,
A2-A04X,
A2-A04Y,
A2-A0CK,
A2-A0CL,
A2-A0CM,
A2-A0CO,
A2-A0CP,
A2-A0CQ,
A2-A0CR,
A2-A0CS,
A2-A0CT,
A2-A0CU,
A2-A0CV,
A2-A0CW,
A2-A0CX,
A2-A0CZ,
A2-A0D0,
A2-A0D1,
A2-A0D2,
A2-A0D3,
A2-A0D4,
A2-A0EM,
A2-A0EN,
A2-A0EO,
A2-A0EP,
A2-A0EQ,
A2-A0ER,
A2-A0ES,
A2-A0ET,
A2-A0EU,
A2-A0EV,
A2-A0EW,
A2-A0EX,
A2-A0EY,
A2-A0ST,
A2-A0SU,
A2-A0SV,
A2-A0SW,
A2-A0SX,
A2-A0SY,
A2-A0T0,
A2-A0T1,
A2-A0T2,
A2-A0T3,
A2-A0T4,
A2-A0T5,
A2-A0T6,
A2-A0T7,
A2-A0YC,
A2-A0YD,
A2-A0YE,
A2-A0YF,
A2-A0YG,
A2-A0YH,
A2-A0YI,
A2-A0YJ,
A2-A0YK,
A2-A0YL,
A2-A0YM,
A2-A0YT,
A2-A1FV,
A2-A1FW,
A2-A1FX,
A2-A1FZ,
A2-A1G0,
A2-A1G1,
A2-A1G4,
A2-A1G6,
A2-A259,
A2-A25A,
A2-A25B,
A2-A25C,
A2-A25D,
A2-A25E,
A2-A25F,
A2-A3KC,
A2-A3KD,
A2-A3XS,
A2-A3XT,
A2-A3XU,
A2-A3XV,
A2-A3XW,
A2-A3XX,
A2-A3XY,
A2-A3XZ,
A2-A3Y0,
A2-A4RW,
A2-A4RX,
A2-A4RY,
A2-A4S0,
A2-A4S1,
A2-A4S2,
A2-A4S3,
A7-A0CD,
A7-A0CE,
A7-A0CG,
A7-A0CH,
A7-A0CJ,
A7-A0D9,
A7-A0DA,
A7-A0DB,
A7-A0DC,
A7-A13D,
A7-A13E,
A7-A13F,
A7-A13G,
A7-A13H,
A7-A26E,
A7-A26F,
A7-A26G,
A7-A26H,
A7-A26I,
A7-A26J,
A7-A2KD,
A7-A3IY,
A7-A3IZ,
A7-A3J0,
A7-A3J1,
A7-A3RF,
A7-A425,
A7-A426,
A7-A4SA,
A7-A4SB,
A7-A4SC,
A7-A4SD,
A7-A4SE,
A7-A4SF,
A7-A56D,
A7-A5ZV,
A7-A5ZW,
A7-A5ZX,
A8-A06N,
A8-A06O,
A8-A06P,
A8-A06Q,
A8-A06R,
A8-A06T,
A8-A06U,
A8-A06X,
A8-A06Y,
A8-A06Z,
A8-A075,
A8-A076,
A8-A079,
A8-A07B,
A8-A07C,
A8-A07E,
A8-A07F,
A8-A07G,
A8-A07I,
A8-A07J,
A8-A07L,
A8-A07O,
A8-A07P,
A8-A07R,
A8-A07U,
A8-A07W,
A8-A07Z,
A8-A081,
A8-A082,
A8-A083,
A8-A084,
A8-A085,
A8-A086,
A8-A08A,
A8-A08B,
A8-A08F,
A8-A08G,
A8-A08H,
A8-A08I,
A8-A08J,
A8-A08L,
A8-A08O,
A8-A08P,
A8-A08R,
A8-A08S,
A8-A08T,
A8-A08X,
A8-A08Z,
A8-A090,
A8-A091,
A8-A092,
A8-A093,
A8-A094,
A8-A095,
A8-A096,
A8-A097,
A8-A099,
A8-A09A,
A8-A09B,
A8-A09C,
A8-A09D,
A8-A09E,
A8-A09G,
A8-A09I,
A8-A09K,
A8-A09M,
A8-A09N,
A8-A09Q,
A8-A09R,
A8-A09T,
A8-A09V,
A8-A09W,
A8-A09X,
A8-A09Z,
A8-A0A1,
A8-A0A2,
A8-A0A4,
A8-A0A6,
A8-A0A7,
A8-A0A9,
A8-A0AB,
A8-A0AD,
AC-A23C,
AC-A23E,
AC-A23G,
AC-A23H,
AC-A2B8,
AC-A2BK,
AC-A2BM,
AC-A2FB,
AC-A2FE,
AC-A2FF,
AC-A2FG,
AC-A2FK,
AC-A2FM,
AC-A2FO,
AC-A2QH,
AC-A2QI,
AC-A2QJ,
AC-A3BB,
AC-A3EH,
AC-A3HN,
AC-A3OD,
AC-A3QP,
AC-A3TM,
AC-A3TN,
AC-A3W5,
AC-A3W6,
AC-A3W7,
AC-A3YI,
AC-A3YJ,
AC-A5EH,
AC-A5EI,
AC-A5XS,
AC-A5XU,
AC-A62X,
AC-A62Y,
AN-A03X,
AN-A03Y,
AN-A041,
AN-A046,
AN-A049,
AN-A04A,
AN-A04C,
AN-A04D,
AN-A0AJ,
AN-A0AK,
AN-A0AL,
AN-A0AM,
AN-A0AR,
AN-A0AS,
AN-A0AT,
AN-A0FD,
AN-A0FF,
AN-A0FJ,
AN-A0FK,
AN-A0FL,
AN-A0FN,
AN-A0FS,
AN-A0FT,
AN-A0FV,
AN-A0FW,
AN-A0FX,
AN-A0FY,
AN-A0FZ,
AN-A0G0,
AN-A0XL,
AN-A0XN,
AN-A0XO,
AN-A0XP,
AN-A0XR,
AN-A0XS,
AN-A0XT,
AN-A0XU,
AN-A0XV,
AN-A0XW,
AO-A03L,
AO-A03M,
AO-A03N,
AO-A03O,
AO-A03P,
AO-A03R,
AO-A03T,
AO-A03U,
AO-A03V,
AO-A0J2,
AO-A0J3,
AO-A0J4,
AO-A0J5,
AO-A0J6,
AO-A0J7,
AO-A0J8,
AO-A0J9,
AO-A0JA,
AO-A0JB,
AO-A0JC,
AO-A0JD,
AO-A0JE,
AO-A0JF,
AO-A0JG,
AO-A0JI,
AO-A0JJ,
AO-A0JL,
AO-A0JM,
AO-A124,
AO-A125,
AO-A126,
AO-A128,
AO-A129,
AO-A12A,
AO-A12B,
AO-A12D,
AO-A12E,
AO-A12F,
AO-A12G,
AO-A12H,
AO-A1KO,
AO-A1KP,
AO-A1KR,
AO-A1KS,
AO-A1KT,
AQ-A04H,
AQ-A04J,
AQ-A04L,
AQ-A0Y5,
AQ-A1H2,
AQ-A1H3,
AQ-A54N,
AQ-A54O,
AR-A0TP,
AR-A0TQ,
AR-A0TR,
AR-A0TS,
AR-A0TT,
AR-A0TU,
AR-A0TV,
AR-A0TW,
AR-A0TX,
AR-A0TY,
AR-A0TZ,
AR-A0U0,
AR-A0U1,
AR-A0U2,
AR-A0U3,
AR-A0U4,
AR-A1AH,
AR-A1AI,
AR-A1AJ,
AR-A1AK,
AR-A1AL,
AR-A1AM,
AR-A1AN,
AR-A1AO,
AR-A1AP,
AR-A1AQ,
AR-A1AR,
AR-A1AS,
AR-A1AT,
AR-A1AU,
AR-A1AV,
AR-A1AW,
AR-A1AX,
AR-A1AY,
AR-A24H,
AR-A24K,
AR-A24L,
AR-A24M,
AR-A24N,
AR-A24O,
AR-A24P,
AR-A24Q,
AR-A24R,
AR-A24S,
AR-A24T,
AR-A24U,
AR-A24V,
AR-A24W,
AR-A24X,
AR-A24Z,
AR-A250,
AR-A251,
AR-A252,
AR-A254,
AR-A255,
AR-A256,
AR-A2LE,
AR-A2LH,
AR-A2LJ,
AR-A2LK,
AR-A2LL,
AR-A2LM,
AR-A2LN,
AR-A2LO,
AR-A2LQ,
AR-A2LR,
AR-A5QM,
AR-A5QN,
AR-A5QP,
AR-A5QQ,
B6-A0I1,
B6-A0I2,
B6-A0I5,
B6-A0I6,
B6-A0I8,
B6-A0I9,
B6-A0IA,
B6-A0IB,
B6-A0IC,
B6-A0IE,
B6-A0IG,
B6-A0IH,
B6-A0IJ,
B6-A0IK,
B6-A0IM,
B6-A0IN,
B6-A0IO,
B6-A0IP,
B6-A0IQ,
B6-A0RE,
B6-A0RG,
B6-A0RH,
B6-A0RI,
B6-A0RL,
B6-A0RM,
B6-A0RN,
B6-A0RO,
B6-A0RP,
B6-A0RQ,
B6-A0RS,
B6-A0RT,
B6-A0RU,
B6-A0RV,
B6-A0WS,
B6-A0WT,
B6-A0WV,
B6-A0WW,
B6-A0WX,
B6-A0WY,
B6-A0WZ,
B6-A0X0,
B6-A0X1,
B6-A0X4,
B6-A0X5,
B6-A0X7,
B6-A1KC,
B6-A1KF,
B6-A1KI,
B6-A1KN,
B6-A2IU,
B6-A3ZX,
B6-A400,
B6-A401,
B6-A402,
B6-A408,
B6-A409,
B6-A40B,
B6-A40C,
BH-A0AU,
BH-A0AV,
BH-A0AW,
BH-A0AY,
BH-A0AZ,
BH-A0B0,
BH-A0B1,
BH-A0B3,
BH-A0B4,
BH-A0B5,
BH-A0B6,
BH-A0B7,
BH-A0B8,
BH-A0B9,
BH-A0BA,
BH-A0BC,
BH-A0BD,
BH-A0BF,
BH-A0BG,
BH-A0BJ,
BH-A0BL,
BH-A0BM,
BH-A0BO,
BH-A0BP,
BH-A0BQ,
BH-A0BR,
BH-A0BS,
BH-A0BT,
BH-A0BV,
BH-A0BW,
BH-A0BZ,
BH-A0C0,
BH-A0C1,
BH-A0C3,
BH-A0C7,
BH-A0DD,
BH-A0DE,
BH-A0DG,
BH-A0DH,
BH-A0DI,
BH-A0DK,
BH-A0DL,
BH-A0DO,
BH-A0DP,
BH-A0DQ,
BH-A0DS,
BH-A0DT,
BH-A0DV,
BH-A0DX,
BH-A0DZ,
BH-A0E0,
BH-A0E1,
BH-A0E2,
BH-A0E6,
BH-A0E7,
BH-A0E9,
BH-A0EA,
BH-A0EB,
BH-A0EE,
BH-A0EI,
BH-A0GY,
BH-A0GZ,
BH-A0H0,
BH-A0H3,
BH-A0H5,
BH-A0H6,
BH-A0H7,
BH-A0H9,
BH-A0HA,
BH-A0HB,
BH-A0HF,
BH-A0HI,
BH-A0HK,
BH-A0HL,
BH-A0HN,
BH-A0HO,
BH-A0HP,
BH-A0HQ,
BH-A0HU,
BH-A0HW,
BH-A0HX,
BH-A0HY,
BH-A0RX,
BH-A0W3,
BH-A0W4,
BH-A0W5,
BH-A0W7,
BH-A0WA,
BH-A18F,
BH-A18G,
BH-A18H,
BH-A18I,
BH-A18J,
BH-A18K,
BH-A18L,
BH-A18M,
BH-A18N,
BH-A18P,
BH-A18Q,
BH-A18R,
BH-A18S,
BH-A18T,
BH-A18U,
BH-A18V,
BH-A1EN,
BH-A1EO,
BH-A1ES,
BH-A1ET,
BH-A1EU,
BH-A1EV,
BH-A1EW,
BH-A1EX,
BH-A1EY,
BH-A1F0,
BH-A1F2,
BH-A1F5,
BH-A1F6,
BH-A1F8,
BH-A1FC,
BH-A1FD,
BH-A1FE,
BH-A1FG,
BH-A1FH,
BH-A1FJ,
BH-A1FL,
BH-A1FM,
BH-A1FN,
BH-A1FR,
BH-A1FU,
BH-A201,
BH-A202,
BH-A203,
BH-A204,
BH-A208,
BH-A209,
BH-A28O,
BH-A28Q,
BH-A2L8,
BH-A42T,
BH-A42U,
BH-A42V,
BH-A5IZ,
BH-A5J0,
C8-A12K,
C8-A12L,
C8-A12M,
C8-A12N,
C8-A12O,
C8-A12P,
C8-A12Q,
C8-A12T,
C8-A12U,
C8-A12V,
C8-A12W,
C8-A12X,
C8-A12Y,
C8-A12Z,
C8-A130,
C8-A131,
C8-A132,
C8-A133,
C8-A134,
C8-A135,
C8-A137,
C8-A138,
C8-A1HE,
C8-A1HF,
C8-A1HG,
C8-A1HI,
C8-A1HJ,
C8-A1HK,
C8-A1HL,
C8-A1HM,
C8-A1HN,
C8-A1HO,
C8-A26V,
C8-A26W,
C8-A26X,
C8-A26Y,
C8-A26Z,
C8-A273,
C8-A274,
C8-A275,
C8-A278,
C8-A27A,
C8-A27B,
C8-A3M7,
C8-A3M8,
D8-A13Y,
D8-A13Z,
D8-A140,
D8-A141,
D8-A142,
D8-A143,
D8-A145,
D8-A146,
D8-A147,
D8-A1J8,
D8-A1J9,
D8-A1JA,
D8-A1JB,
D8-A1JC,
D8-A1JD,
D8-A1JE,
D8-A1JF,
D8-A1JG,
D8-A1JH,
D8-A1JI,
D8-A1JJ,
D8-A1JK,
D8-A1JL,
D8-A1JM,
D8-A1JN,
D8-A1JP,
D8-A1JS,
D8-A1JT,
D8-A1JU,
D8-A1X5,
D8-A1X6,
D8-A1X7,
D8-A1X8,
D8-A1X9,
D8-A1XA,
D8-A1XB,
D8-A1XC,
D8-A1XF,
D8-A1XG,
D8-A1XJ,
D8-A1XK,
D8-A1XL,
D8-A1XM,
D8-A1XO,
D8-A1XQ,
D8-A1XR,
D8-A1XS,
D8-A1XT,
D8-A1XU,
D8-A1XV,
D8-A1XW,
D8-A1XY,
D8-A1XZ,
D8-A1Y0,
D8-A1Y1,
D8-A1Y2,
D8-A1Y3,
D8-A27E,
D8-A27F,
D8-A27G,
D8-A27H,
D8-A27I,
D8-A27K,
D8-A27L,
D8-A27M,
D8-A27N,
D8-A27P,
D8-A27R,
D8-A27T,
D8-A27V,
D8-A27W,
D8-A3Z5,
D8-A3Z6,
D8-A4Z1,
E2-A105,
E2-A107,
E2-A108,
E2-A109,
E2-A10A,
E2-A10B,
E2-A10C,
E2-A10E,
E2-A10F,
E2-A14N,
E2-A14O,
E2-A14P,
E2-A14Q,
E2-A14R,
E2-A14S,
E2-A14T,
E2-A14U,
E2-A14V,
E2-A14W,
E2-A14X,
E2-A14Y,
E2-A14Z,
E2-A150,
E2-A152,
E2-A153,
E2-A154,
E2-A155,
E2-A156,
E2-A158,
E2-A159,
E2-A15A,
E2-A15C,
E2-A15D,
E2-A15E,
E2-A15F,
E2-A15G,
E2-A15H,
E2-A15I,
E2-A15J,
E2-A15K,
E2-A15L,
E2-A15M,
E2-A15O,
E2-A15P,
E2-A15R,
E2-A15S,
E2-A15T,
E2-A1AZ,
E2-A1B0,
E2-A1B1,
E2-A1B4,
E2-A1B5,
E2-A1B6,
E2-A1BC,
E2-A1BD,
E2-A1IE,
E2-A1IF,
E2-A1IG,
E2-A1IH,
E2-A1II,
E2-A1IJ,
E2-A1IK,
E2-A1IL,
E2-A1IN,
E2-A1IO,
E2-A1IU,
E2-A1L6,
E2-A1L7,
E2-A1L8,
E2-A1L9,
E2-A1LA,
E2-A1LB,
E2-A1LE,
E2-A1LG,
E2-A1LH,
E2-A1LI,
E2-A1LK,
E2-A1LL,
E2-A1LS,
E2-A2P5,
E2-A2P6,
E2-A3DX,
E2-A56Z,
E2-A570,
E2-A573,
E2-A574,
E9-A1N3,
E9-A1N4,
E9-A1N5,
E9-A1N8,
E9-A1N9,
E9-A1NA,
E9-A1NC,
E9-A1ND,
E9-A1NE,
E9-A1NF,
E9-A1NG,
E9-A1NH,
E9-A1NI,
E9-A1QZ,
E9-A1R0,
E9-A1R2,
E9-A1R3,
E9-A1R4,
E9-A1R5,
E9-A1R6,
E9-A1R7,
E9-A1RA,
E9-A1RB,
E9-A1RC,
E9-A1RD,
E9-A1RE,
E9-A1RF,
E9-A1RG,
E9-A1RH,
E9-A1RI,
E9-A226,
E9-A227,
E9-A228,
E9-A229,
E9-A22A,
E9-A22B,
E9-A22D,
E9-A22E,
E9-A22G,
E9-A22H,
E9-A243,
E9-A244,
E9-A245,
E9-A247,
E9-A248,
E9-A249,
E9-A24A,
E9-A295,
E9-A2JS,
E9-A2JT,
E9-A3HO,
E9-A3Q9,
E9-A3QA,
E9-A3X8,
E9-A54X,
E9-A54Y,
E9-A5FK,
E9-A5FL,
E9-A5UO,
E9-A5UP,
EW-A1IW,
EW-A1IX,
EW-A1IY,
EW-A1IZ,
EW-A1J1,
EW-A1J2,
EW-A1J3,
EW-A1J5,
EW-A1J6,
EW-A1OV,
EW-A1OX,
EW-A1OY,
EW-A1OZ,
EW-A1P0,
EW-A1P1,
EW-A1P3,
EW-A1P4,
EW-A1P5,
EW-A1P6,
EW-A1P7,
EW-A1P8,
EW-A1PA,
EW-A1PB,
EW-A1PC,
EW-A1PD,
EW-A1PE,
EW-A1PG,
EW-A1PH,
EW-A2FR,
EW-A2FS,
EW-A2FV,
EW-A2FW,
EW-A3E8,
EW-A3U0,
EW-A423,
GI-A2C8,
GI-A2C9,
GM-A2D9,
GM-A2DA,
GM-A2DB,
GM-A2DC,
GM-A2DD,
GM-A2DF,
GM-A2DH,
GM-A2DI,
GM-A2DK,
GM-A2DL,
GM-A2DM,
GM-A2DN,
GM-A2DO,
GM-A3NW,
GM-A3NY,
GM-A3XG,
GM-A3XL,
GM-A3XN,
GM-A4E0,
GM-A5PV,
GM-A5PX,
HN-A2NL,
HN-A2OB,
JL-A3YW,
JL-A3YX,
LL-A440,
LL-A441,
LL-A50Y,
LL-A5YL,
LL-A5YM,
LL-A5YN,
LL-A5YO,
LL-A5YP,
LQ-A4E4,
MS-A51U,
OK-A5Q2,
OL-A5D6,
OL-A5D7,
OL-A5D8,
OL-A5DA,
OL-A5RU,
OL-A5RV,
OL-A5RW,
OL-A5RX,
OL-A5RY,
OL-A5RZ,
OL-A5S0,
OL-A66H,
OL-A66I,
OL-A66J,
OL-A66K,
PE-A5DC,
PE-A5DD,
PE-A5DE.\\
$\blacklozenge$ {\bf Cervical Cancer} (197 samples):\\
$\bullet$ Source T6 = TCGA (see Acknowledgments). Sample IDs are of the form TCGA-*, where * is:\\
BI-A0VR,
BI-A0VS,
BI-A20A,
C5-A0TN,
C5-A1BE,
C5-A1BF,
C5-A1BI,
C5-A1BJ,
C5-A1BK,
C5-A1BL,
C5-A1BM,
C5-A1BN,
C5-A1BQ,
C5-A1M5,
C5-A1M6,
C5-A1M7,
C5-A1M8,
C5-A1M9,
C5-A1ME,
C5-A1MF,
C5-A1MH,
C5-A1MI,
C5-A1MJ,
C5-A1MK,
C5-A1ML,
C5-A1MN,
C5-A1MP,
C5-A1MQ,
C5-A2LS,
C5-A2LT,
C5-A2LV,
C5-A2LX,
C5-A2LY,
C5-A2LZ,
C5-A2M1,
C5-A2M2,
C5-A3HD,
C5-A3HE,
C5-A3HF,
C5-A3HL,
C5-A7CG,
C5-A7CH,
C5-A7CJ,
C5-A7CK,
C5-A7CL,
C5-A7CM,
C5-A7CO,
C5-A7UC,
C5-A7UE,
C5-A7UH,
C5-A7X3,
DG-A2KH,
DG-A2KJ,
DG-A2KK,
DG-A2KL,
DG-A2KM,
DR-A0ZL,
DR-A0ZM,
DS-A0VK,
DS-A0VL,
DS-A0VM,
DS-A0VN,
DS-A1OA,
DS-A3LQ,
DS-A5RQ,
DS-A7WF,
DS-A7WH,
DS-A7WI,
EA-A1QS,
EA-A1QT,
EA-A3HQ,
EA-A3HR,
EA-A3HT,
EA-A3HU,
EA-A3QD,
EA-A3QE,
EA-A3Y4,
EA-A410,
EA-A411,
EA-A439,
EA-A43B,
EA-A44S,
EA-A4BA,
EA-A50E,
EA-A556,
EA-A5FO,
EA-A5O9,
EA-A5ZD,
EA-A5ZE,
EA-A5ZF,
EA-A6QX,
EA-A78R,
EK-A2GZ,
EK-A2H0,
EK-A2H1,
EK-A2IP,
EK-A2PG,
EK-A2PI,
EK-A2PK,
EK-A2PL,
EK-A2PM,
EK-A2R7,
EK-A2R8,
EK-A2R9,
EK-A2RA,
EK-A2RB,
EK-A2RC,
EK-A2RD,
EK-A2RE,
EK-A2RJ,
EK-A2RK,
EK-A2RL,
EK-A2RM,
EK-A2RN,
EK-A2RO,
EK-A3GJ,
EK-A3GK,
EK-A3GM,
EK-A3GN,
EX-A1H5,
EX-A1H6,
EX-A3L1,
EX-A449,
EX-A69L,
EX-A69M,
FU-A23K,
FU-A23L,
FU-A2QG,
FU-A3EO,
FU-A3HY,
FU-A3HZ,
FU-A3NI,
FU-A3TQ,
FU-A3TX,
FU-A3WB,
FU-A3YQ,
FU-A40J,
FU-A57G,
FU-A5XV,
FU-A770,
HG-A2PA,
HM-A3JJ,
HM-A3JK,
HM-A4S6,
HM-A6W2,
IR-A3L7,
IR-A3LA,
IR-A3LB,
IR-A3LC,
IR-A3LF,
IR-A3LH,
IR-A3LI,
IR-A3LK,
IR-A3LL,
JW-A5VG,
JW-A5VH,
JW-A5VI,
JW-A5VJ,
JW-A5VK,
JW-A5VL,
JW-A69B,
JW-A852,
JX-A3PZ,
JX-A3Q0,
JX-A3Q8,
JX-A5QV,
LP-A4AU,
LP-A4AV,
LP-A4AW,
LP-A4AX,
LP-A5U2,
LP-A5U3,
LP-A7HU,
MU-A51Y,
MU-A5YI,
MY-A5BD,
MY-A5BE,
MY-A5BF,
Q1-A5R1,
Q1-A5R2,
Q1-A5R3,
Q1-A6DT,
Q1-A6DV,
Q1-A6DW,
Q1-A73O,
Q1-A73P,
Q1-A73Q,
Q1-A73R,
Q1-A73S,
R2-A69V,
RA-A741,
UC-A7PD,
UC-A7PF,
WL-A834,
DS-A1OB,
DS-A1OC,
DS-A1OD.\\
$\blacklozenge$ {\bf Cholangiocarcinoma} (139 samples):\\
$\bullet$ Source Z2 = \cite{Zou}:\\
1,
10,
100,
101,
107,
108,
109,
110,
111,
112,
113,
115,
116,
118,
119,
120,
121,
122,
123,
125,
127,
128,
129,
13,
130,
131,
132,
133,
134,
135,
137,
139,
140,
141,
142,
143,
144,
145,
146,
147,
16,
17,
18,
19,
2,
20,
24,
25,
26,
28,
29,
3,
33,
34,
35,
39,
41,
42,
44,
46,
48,
5,
50,
51,
52,
53,
56,
58,
59,
6,
60,
61,
63,
64,
66,
67,
69,
7,
70,
71,
74,
79,
8,
8\_1,
8\_2,
8\_4,
8\_6,
80,
81,
82,
85,
86,
87,
88,
89,
9,
90,
91,
94,
95,
97,
98,
99.\\
$\bullet$ Source T7 = TCGA (see Acknowledgments). Sample IDs are of the form TCGA-*, where * is:\\
3X-AAV9,
3X-AAVA,
3X-AAVB,
3X-AAVC,
3X-AAVE,
4G-AAZO,
4G-AAZT,
W5-AA2G,
W5-AA2H,
W5-AA2I,
W5-AA2O,
W5-AA2Q,
W5-AA2R,
W5-AA2T,
W5-AA2U,
W5-AA2W,
W5-AA2X,
W5-AA2Z,
W5-AA30,
W5-AA31,
W5-AA33,
W5-AA34,
W5-AA36,
W5-AA38,
W5-AA39,
W6-AA0S,
WD-A7RX,
YR-A95A,
ZD-A8I3,
ZH-A8Y1,
ZH-A8Y2,
ZH-A8Y4,
ZH-A8Y5,
ZH-A8Y6,
ZH-A8Y8,
ZU-A8S4.\\
$\blacklozenge$ {\bf Chronic Lymphocytic Leukemia} (80 samples):\\
$\bullet$ Source Q1 = \cite{Quesada}:\\
170,
171,
172,
173,
174,
175,
18,
181,
182,
184,
185,
186,
188,
189,
19,
191,
193,
194,
195,
197,
20,
22,
23,
264,
266,
267,
27,
270,
272,
273,
274,
275,
276,
278,
279,
280,
29,
290,
30,
319,
32,
321,
322,
323,
324,
325,
326,
328,
33,
375,
39,
40,
41,
42,
43,
44,
45,
48,
49,
5,
51,
52,
53,
54,
6,
618,
63,
64,
642,
680,
7,
758,
761,
785,
8,
82,
83,
9,
90,
91.\\
$\blacklozenge$ {\bf Colorectal Cancer} (581 samples):\\
$\bullet$ Source S3 = \cite{Seshagiri}:\\
587220,
587222,
587224,
587226,
587228,
587230,
587232,
587234,
587238,
587242,
587246,
587254,
587256,
587260,
587262,
587264,
587268,
587270,
587276,
587278,
587282,
587284,
587286,
587288,
587290,
587292,
587294,
587298,
587300,
587302,
587304,
587306,
587316,
587318,
587322,
587328,
587330,
587332,
587334,
587336,
587338,
587340,
587342,
587344,
587346,
587348,
587350,
587352,
587354,
587356,
587358,
587360,
587362,
587364,
587368,
587370,
587372,
587374,
587376,
587378,
587380,
587382,
587384,
587386,
587388,
587390,
587392,
587394,
587398,
587400.\\
$\bullet$ Source T8 = TCGA (see Acknowledgments). Sample IDs are of the form TCGA-*, where * is:\\
A6-2670,
A6-2671,
A6-2672,
A6-2674,
A6-2675,
A6-2676,
A6-2677,
A6-2678,
A6-2683,
A6-3807,
A6-3808,
A6-3809,
A6-3810,
A6-4105,
A6-5656,
A6-5657,
A6-5659,
A6-5660,
A6-5661,
A6-5662,
A6-5664,
A6-5665,
A6-5666,
A6-5667,
A6-6137,
A6-6138,
A6-6140,
A6-6141,
A6-6142,
A6-6648,
A6-6649,
A6-6650,
A6-6651,
A6-6652,
A6-6653,
A6-6654,
A6-6780,
A6-6781,
A6-6782,
AA-3489,
AA-3492,
AA-3496,
AA-3502,
AA-3510,
AA-3511,
AA-3514,
AA-3516,
AA-3517,
AA-3518,
AA-3519,
AA-3520,
AA-3521,
AA-3522,
AA-3524,
AA-3525,
AA-3526,
AA-3527,
AA-3529,
AA-3531,
AA-3532,
AA-3534,
AA-3538,
AA-3542,
AA-3543,
AA-3544,
AA-3548,
AA-3549,
AA-3552,
AA-3553,
AA-3554,
AA-3555,
AA-3556,
AA-3558,
AA-3560,
AA-3561,
AA-3562,
AA-3655,
AA-3660,
AA-3662,
AA-3663,
AA-3664,
AA-3666,
AA-3667,
AA-3672,
AA-3673,
AA-3678,
AA-3679,
AA-3680,
AA-3681,
AA-3684,
AA-3685,
AA-3688,
AA-3692,
AA-3693,
AA-3695,
AA-3696,
AA-3697,
AA-3710,
AA-3712,
AA-3713,
AA-3715,
AA-3811,
AA-3812,
AA-3814,
AA-3815,
AA-3818,
AA-3819,
AA-3821,
AA-3831,
AA-3833,
AA-3837,
AA-3842,
AA-3844,
AA-3845,
AA-3846,
AA-3848,
AA-3850,
AA-3851,
AA-3852,
AA-3854,
AA-3855,
AA-3856,
AA-3858,
AA-3860,
AA-3861,
AA-3864,
AA-3866,
AA-3867,
AA-3869,
AA-3870,
AA-3872,
AA-3875,
AA-3877,
AA-3930,
AA-3939,
AA-3941,
AA-3947,
AA-3949,
AA-3950,
AA-3952,
AA-3955,
AA-3956,
AA-3966,
AA-3968,
AA-3971,
AA-3972,
AA-3973,
AA-3975,
AA-3977,
AA-3979,
AA-3980,
AA-3982,
AA-3984,
AA-3986,
AA-3989,
AA-3994,
AA-A004,
AA-A00A,
AA-A00D,
AA-A00E,
AA-A00F,
AA-A00J,
AA-A00K,
AA-A00L,
AA-A00N,
AA-A00O,
AA-A00Q,
AA-A00R,
AA-A00U,
AA-A00W,
AA-A00Z,
AA-A010,
AA-A017,
AA-A01D,
AA-A01F,
AA-A01G,
AA-A01I,
AA-A01K,
AA-A01P,
AA-A01Q,
AA-A01R,
AA-A01S,
AA-A01T,
AA-A01V,
AA-A01X,
AA-A01Z,
AA-A022,
AA-A024,
AA-A029,
AA-A02F,
AA-A02H,
AA-A02J,
AA-A02K,
AA-A02O,
AA-A02W,
AA-A02Y,
AA-A03F,
AA-A03J,
AD-5900,
AD-6548,
AD-6888,
AD-6889,
AD-6890,
AD-6895,
AD-6899,
AD-6901,
AD-6963,
AD-6964,
AD-6965,
AF-2687,
AF-2689,
AF-2691,
AF-2692,
AF-2693,
AF-3400,
AF-3913,
AF-4110,
AF-5654,
AF-6136,
AF-6655,
AF-6672,
AG-3574,
AG-3575,
AG-3578,
AG-3580,
AG-3581,
AG-3582,
AG-3583,
AG-3584,
AG-3586,
AG-3587,
AG-3593,
AG-3594,
AG-3598,
AG-3599,
AG-3600,
AG-3601,
AG-3602,
AG-3605,
AG-3608,
AG-3609,
AG-3611,
AG-3612,
AG-3726,
AG-3727,
AG-3731,
AG-3732,
AG-3742,
AG-3878,
AG-3881,
AG-3882,
AG-3883,
AG-3885,
AG-3887,
AG-3890,
AG-3892,
AG-3893,
AG-3894,
AG-3896,
AG-3898,
AG-3901,
AG-3902,
AG-3909,
AG-3999,
AG-4001,
AG-4005,
AG-4007,
AG-4008,
AG-4015,
AG-A002,
AG-A008,
AG-A00C,
AG-A00H,
AG-A00Y,
AG-A011,
AG-A014,
AG-A015,
AG-A016,
AG-A01L,
AG-A01W,
AG-A01Y,
AG-A020,
AG-A025,
AG-A026,
AG-A02G,
AG-A02N,
AG-A02X,
AG-A032,
AG-A036,
AH-6544,
AH-6547,
AH-6549,
AH-6643,
AH-6644,
AM-5820,
AM-5821,
AU-3779,
AU-6004,
AY-4070,
AY-4071,
AY-5543,
AY-6196,
AY-6197,
AY-6386,
AZ-4315,
AZ-4323,
AZ-4615,
AZ-4616,
AZ-4681,
AZ-4682,
AZ-5403,
AZ-5407,
AZ-6598,
AZ-6599,
AZ-6600,
AZ-6601,
AZ-6603,
AZ-6605,
AZ-6606,
AZ-6607,
AZ-6608,
CA-5254,
CA-5255,
CA-5796,
CA-5797,
CA-6715,
CA-6716,
CA-6717,
CA-6718,
CA-6719,
CI-6619,
CI-6620,
CI-6621,
CI-6622,
CI-6624,
CK-4947,
CK-4948,
CK-4950,
CK-4952,
CK-5912,
CK-5913,
CK-5914,
CK-5915,
CK-5916,
CK-6746,
CK-6747,
CK-6748,
CK-6751,
CL-5917,
CL-5918,
CM-4743,
CM-4744,
CM-4746,
CM-4747,
CM-4748,
CM-4750,
CM-4752,
CM-5341,
CM-5344,
CM-5348,
CM-5349,
CM-5860,
CM-5861,
CM-5862,
CM-5863,
CM-5864,
CM-5868,
CM-6161,
CM-6162,
CM-6163,
CM-6164,
CM-6165,
CM-6166,
CM-6167,
CM-6168,
CM-6169,
CM-6170,
CM-6171,
CM-6172,
CM-6674,
CM-6675,
CM-6676,
CM-6677,
CM-6678,
CM-6679,
CM-6680,
D5-5537,
D5-5538,
D5-5539,
D5-5540,
D5-5541,
D5-6529,
D5-6531,
D5-6532,
D5-6533,
D5-6534,
D5-6535,
D5-6536,
D5-6537,
D5-6538,
D5-6539,
D5-6540,
D5-6541,
D5-6898,
D5-6920,
D5-6922,
D5-6923,
D5-6924,
D5-6926,
D5-6927,
D5-6928,
D5-6929,
D5-6930,
D5-6931,
D5-6932,
D5-7000,
DC-5337,
DC-5869,
DC-6155,
DC-6157,
DC-6158,
DC-6160,
DC-6681,
DC-6682,
DC-6683,
DM-A0X9,
DM-A0XD,
DM-A0XF,
DM-A1D0,
DM-A1D4,
DM-A1D6,
DM-A1D7,
DM-A1D8,
DM-A1D9,
DM-A1DA,
DM-A1DB,
DM-A1HA,
DM-A1HB,
DM-A282,
DM-A285,
DM-A28C,
DM-A28E,
DM-A28F,
DM-A28G,
DM-A28H,
DM-A28K,
DM-A28M,
DT-5265,
DY-A0XA,
DY-A1DC,
DY-A1DD,
DY-A1DF,
DY-A1DG,
DY-A1H8,
EF-5830,
EI-6506,
EI-6507,
EI-6508,
EI-6510,
F4-6459,
F4-6460,
F4-6461,
F4-6463,
F4-6569,
F4-6570,
F4-6703,
F4-6704,
F4-6805,
F4-6806,
F4-6807,
F4-6808,
F4-6809,
F4-6854,
F4-6855,
F4-6856,
F4-6857,
F5-6464,
F5-6465,
F5-6571,
F5-6702,
F5-6811,
F5-6812,
F5-6813,
G4-6293,
G4-6294,
G4-6295,
G4-6297,
G4-6298,
G4-6299,
G4-6302,
G4-6303,
G4-6304,
G4-6306,
G4-6307,
G4-6309,
G4-6310,
G4-6311,
G4-6314,
G4-6315,
G4-6317,
G4-6320,
G4-6321,
G4-6322,
G4-6323,
G4-6586,
G4-6588,
G4-6625,
G4-6626,
G4-6628,
G5-6235,
G5-6641.\\
$\blacklozenge$ {\bf Esophageal Cancer} (329 samples):\\
$\bullet$ Source D2 = \cite{Dulak}. Sample IDs are of the form ESO-*-Tumor, where * is:\\
0001,
0009,
0013,
0015,
0019,
0023,
0025,
0029,
003,
005,
0053,
0059,
0061,
0067,
007,
0071,
0079,
0103,
0115,
0123,
0125,
0129,
0133,
0149,
0167,
017,
0176,
021,
0255,
027,
0280,
037,
043,
045,
0459,
049,
051,
0590,
075,
077,
083,
085,
0950,
105,
1059,
1060,
107,
1096,
111,
1130,
1133,
114,
1145,
1154,
116,
1163,
117,
118,
119,
120,
122,
130,
131,
135,
137,
139,
141,
1427,
143,
147,
1481,
1488,
151,
152,
153,
155,
157,
159,
1594,
160,
1608,
161,
164,
165,
167,
1670,
169,
171,
173,
1733,
1748,
175,
177,
179,
184,
185,
187,
1872,
189,
191,
2143,
224,
2472,
250,
251,
2536,
327,
408,
409,
512,
536,
539,
555,
580,
582,
601,
610,
632,
640,
669,
682,
683,
708,
718,
720,
721,
732,
752,
805,
837,
838,
859,
864,
866,
874,
887,
913,
916,
931,
963,
D76,
H01,
H63,
K08,
R61,
S41.\\
$\bullet$ Source T9 = TCGA (see Acknowledgments). Sample IDs are of the form TCGA-*, where * is:\\
2H-A9GF,
2H-A9GH,
2H-A9GI,
2H-A9GJ,
2H-A9GK,
2H-A9GL,
2H-A9GM,
2H-A9GN,
2H-A9GO,
2H-A9GQ,
2H-A9GR,
IC-A6RE,
IC-A6RF,
IG-A3I8,
IG-A3QL,
IG-A3Y9,
IG-A3YA,
IG-A3YB,
IG-A3YC,
IG-A4P3,
IG-A4QS,
IG-A4QT,
IG-A50L,
IG-A51D,
IG-A5B8,
IG-A5S3,
IG-A625,
IG-A6QS,
IG-A7DP,
IG-A8O2,
IG-A97H,
IG-A97I,
JY-A6F8,
JY-A6FA,
JY-A6FB,
JY-A6FD,
JY-A6FE,
JY-A6FG,
JY-A6FH,
JY-A938,
JY-A939,
JY-A93C,
JY-A93D,
JY-A93E,
JY-A93F,
KH-A6WC,
L5-A43C,
L5-A43E,
L5-A43H,
L5-A43I,
L5-A43J,
L5-A43M,
L5-A4OE,
L5-A4OF,
L5-A4OG,
L5-A4OH,
L5-A4OI,
L5-A4OJ,
L5-A4OM,
L5-A4ON,
L5-A4OO,
L5-A4OP,
L5-A4OQ,
L5-A4OR,
L5-A4OS,
L5-A4OT,
L5-A4OU,
L5-A4OW,
L5-A4OX,
L5-A88S,
L5-A88T,
L5-A88V,
L5-A88W,
L5-A88Y,
L5-A88Z,
L5-A891,
L5-A893,
L5-A8NE,
L5-A8NF,
L5-A8NG,
L5-A8NH,
L5-A8NI,
L5-A8NJ,
L5-A8NK,
L5-A8NL,
L5-A8NM,
L5-A8NN,
L5-A8NQ,
L5-A8NR,
L5-A8NS,
L5-A8NT,
L5-A8NU,
L5-A8NV,
L5-A8NW,
L7-A56G,
L7-A6VZ,
LN-A49K,
LN-A49L,
LN-A49M,
LN-A49N,
LN-A49O,
LN-A49P,
LN-A49R,
LN-A49S,
LN-A49U,
LN-A49V,
LN-A49W,
LN-A49X,
LN-A49Y,
LN-A4A1,
LN-A4A2,
LN-A4A3,
LN-A4A4,
LN-A4A5,
LN-A4A6,
LN-A4A8,
LN-A4A9,
LN-A4MQ,
LN-A4MR,
LN-A5U5,
LN-A5U6,
LN-A5U7,
LN-A7HV,
LN-A7HW,
LN-A7HX,
LN-A7HY,
LN-A7HZ,
LN-A8HZ,
LN-A8I0,
LN-A8I1,
LN-A9FO,
LN-A9FP,
LN-A9FQ,
LN-A9FR,
M9-A5M8,
Q9-A6FU,
Q9-A6FW,
R6-A6DN,
R6-A6DQ,
R6-A6KZ,
R6-A6L4,
R6-A6L6,
R6-A6XG,
R6-A6XQ,
R6-A6Y0,
R6-A6Y2,
R6-A8W5,
R6-A8W8,
R6-A8WC,
R6-A8WG,
RE-A7BO,
S8-A6BV,
S8-A6BW,
V5-A7RB,
V5-A7RC,
V5-A7RE,
V5-AASV,
V5-AASW,
V5-AASX,
VR-A8EO,
VR-A8EP,
VR-A8EQ,
VR-A8ER,
VR-A8ET,
VR-A8EU,
VR-A8EW,
VR-A8EX,
VR-A8EY,
VR-A8EZ,
VR-A8Q7,
VR-AA4D,
VR-AA4G,
VR-AA7B,
VR-AA7D,
X8-AAAR,
XP-A8T6,
XP-A8T7,
XP-A8T8,
Z6-A8JD,
Z6-A8JE,
Z6-A9VB,
Z6-AAPN,
ZR-A9CJ.\\
$\blacklozenge$ {\bf Gastric Cancer} (401 samples):\\
$\bullet$ Source Z3 = \cite{Zang}:\\
2000362,
31231321,
76629543,
970010,
98748381,
990089,
990097,
990172,
990300,
990396,
990475,
990515,
TGH,
TWH.\\
$\bullet$ Source W1 = \cite{Wang}. Sample IDs are of the form pfg*T, where * is:\\
001,
002,
003,
005,
006,
007,
008,
009,
010,
011,
014,
015,
016,
017,
018,
019,
020,
021,
022,
024,
025,
029.\\
$\bullet$ Source T10 = TCGA (see Acknowledgments). Sample IDs are of the form TCGA-*, where * is:\\
B7-5816,
B7-5818,
B7-A5TI,
B7-A5TJ,
B7-A5TK,
B7-A5TN,
BR-4183,
BR-4184,
BR-4186,
BR-4187,
BR-4188,
BR-4190,
BR-4191,
BR-4194,
BR-4195,
BR-4197,
BR-4199,
BR-4200,
BR-4201,
BR-4205,
BR-4253,
BR-4255,
BR-4256,
BR-4257,
BR-4259,
BR-4261,
BR-4263,
BR-4264,
BR-4265,
BR-4267,
BR-4271,
BR-4273,
BR-4276,
BR-4277,
BR-4278,
BR-4279,
BR-4280,
BR-4281,
BR-4283,
BR-4284,
BR-4286,
BR-4288,
BR-4291,
BR-4292,
BR-4294,
BR-4298,
BR-4357,
BR-4361,
BR-4362,
BR-4363,
BR-4366,
BR-4368,
BR-4369,
BR-4370,
BR-4371,
BR-4375,
BR-4376,
BR-6452,
BR-6453,
BR-6454,
BR-6455,
BR-6456,
BR-6457,
BR-6458,
BR-6563,
BR-6564,
BR-6565,
BR-6566,
BR-6705,
BR-6706,
BR-6707,
BR-6709,
BR-6801,
BR-6802,
BR-6803,
BR-6852,
BR-7196,
BR-7197,
BR-7703,
BR-7704,
BR-7707,
BR-7715,
BR-7716,
BR-7717,
BR-7722,
BR-7723,
BR-7851,
BR-7901,
BR-7957,
BR-7958,
BR-7959,
BR-8058,
BR-8059,
BR-8060,
BR-8077,
BR-8078,
BR-8080,
BR-8081,
BR-8284,
BR-8285,
BR-8286,
BR-8289,
BR-8291,
BR-8295,
BR-8296,
BR-8297,
BR-8360,
BR-8361,
BR-8363,
BR-8364,
BR-8365,
BR-8366,
BR-8367,
BR-8368,
BR-8369,
BR-8370,
BR-8371,
BR-8372,
BR-8373,
BR-8380,
BR-8381,
BR-8382,
BR-8384,
BR-8483,
BR-8484,
BR-8485,
BR-8486,
BR-8487,
BR-8588,
BR-8589,
BR-8590,
BR-8591,
BR-8592,
BR-8676,
BR-8677,
BR-8678,
BR-8679,
BR-8680,
BR-8682,
BR-8683,
BR-8686,
BR-8687,
BR-8690,
BR-A44T,
BR-A44U,
BR-A452,
BR-A453,
BR-A4CQ,
BR-A4CR,
BR-A4CS,
BR-A4IU,
BR-A4IV,
BR-A4IY,
BR-A4IZ,
BR-A4J1,
BR-A4J2,
BR-A4J4,
BR-A4J5,
BR-A4J6,
BR-A4J7,
BR-A4J8,
BR-A4PD,
BR-A4PE,
BR-A4PF,
BR-A4QI,
BR-A4QL,
BR-A4QM,
CD-5798,
CD-5799,
CD-5800,
CD-5801,
CD-5802,
CD-5803,
CD-5804,
CD-5813,
CD-8524,
CD-8525,
CD-8526,
CD-8527,
CD-8528,
CD-8529,
CD-8530,
CD-8531,
CD-8532,
CD-8533,
CD-8534,
CD-8535,
CD-8536,
CD-A486,
CD-A487,
CD-A489,
CD-A48A,
CD-A48C,
CD-A4MG,
CD-A4MH,
CD-A4MI,
CD-A4MJ,
CG-4300,
CG-4301,
CG-4304,
CG-4305,
CG-4306,
CG-4436,
CG-4437,
CG-4438,
CG-4440,
CG-4441,
CG-4442,
CG-4443,
CG-4444,
CG-4449,
CG-4455,
CG-4460,
CG-4462,
CG-4465,
CG-4466,
CG-4469,
CG-4474,
CG-4475,
CG-4476,
CG-4477,
CG-5716,
CG-5717,
CG-5718,
CG-5719,
CG-5720,
CG-5721,
CG-5722,
CG-5723,
CG-5724,
CG-5725,
CG-5726,
CG-5727,
CG-5728,
CG-5730,
CG-5732,
CG-5733,
CG-5734,
D7-5577,
D7-5578,
D7-5579,
D7-6518,
D7-6519,
D7-6520,
D7-6521,
D7-6522,
D7-6524,
D7-6525,
D7-6526,
D7-6527,
D7-6528,
D7-6815,
D7-6817,
D7-6818,
D7-6820,
D7-6822,
D7-8570,
D7-8572,
D7-8573,
D7-8574,
D7-8575,
D7-8576,
D7-8578,
D7-8579,
D7-A4YT,
D7-A4YU,
D7-A4YV,
D7-A4YX,
D7-A4YY,
D7-A4Z0,
D7-A6ET,
D7-A6EV,
D7-A6EX,
D7-A6EY,
D7-A6EZ,
D7-A6F0,
D7-A6F2,
D7-A747,
D7-A748,
D7-A74A,
D7-A74B,
EQ-5647,
EQ-8122,
EQ-A4SO,
F1-6177,
F1-6874,
F1-6875,
F1-A448,
F1-A72C,
FP-7735,
FP-7829,
FP-7916,
FP-7998,
FP-8099,
FP-8209,
FP-8210,
FP-8211,
FP-8631,
FP-A4BE,
FP-A4BF,
HF-7131,
HF-7132,
HF-7133,
HF-7134,
HF-7136,
HF-A5NB,
HJ-7597,
HU-8238,
HU-8243,
HU-8244,
HU-8245,
HU-8249,
HU-8602,
HU-8604,
HU-8608,
HU-8610,
HU-A4G2,
HU-A4G3,
HU-A4G6,
HU-A4G8,
HU-A4G9,
HU-A4GC,
HU-A4GD,
HU-A4GF,
HU-A4GH,
HU-A4GJ,
HU-A4GN,
HU-A4GP,
HU-A4GQ,
HU-A4GT,
HU-A4GU,
HU-A4GX,
HU-A4GY,
HU-A4H0,
HU-A4H2,
HU-A4H3,
HU-A4H4,
HU-A4H5,
HU-A4H6,
HU-A4H8,
HU-A4HB,
HU-A4HD,
IN-7806,
IN-7808,
IN-8462,
IN-8663,
IN-A6RI,
IN-A6RJ,
IN-A6RL,
IN-A6RN,
IN-A6RO,
IN-A6RP,
IN-A6RR,
IP-7968,
KB-A6F5,
KB-A6F7,
MX-A5UG,
MX-A5UJ,
MX-A663,
MX-A666,
R5-A7O7,
RD-A7BS,
RD-A7BT,
RD-A7BW,
RD-A7C1.\\
$\blacklozenge$ {\bf Glioblastoma Multiforme} (359 samples):\\
$\bullet$ Source P2 = \cite{Parsons}. Sample IDs are of the form Br*, where * is:\\
001X,
018X,
019X,
02X,
03X,
04X,
05X,
06X,
07X,
08X,
102X,
103X,
104X,
10P,
112X,
116X,
117X,
118X,
11P,
128X,
12P,
132X,
133X,
136X,
13X,
143X,
148X,
14X,
15X,
16X,
17X,
20P,
21PT,
229T,
230T,
237T,
238T,
23X,
247T,
248T,
25X,
26X,
27P,
29P,
301T,
302T,
303T,
306T,
401X,
9PT.\\
$\bullet$ Source T11 = TCGA (see Acknowledgments). Sample IDs are of the form TCGA-*, where * is:\\
02-0003,
02-0007,
02-0010,
02-0014,
02-0015,
02-0021,
02-0028,
02-0033,
02-0043,
02-0047,
02-0055,
02-0083,
02-0089,
02-0099,
02-0107,
02-0114,
02-0115,
02-2470,
02-2483,
02-2485,
02-2486,
06-0119,
06-0122,
06-0124,
06-0125,
06-0126,
06-0128,
06-0129,
06-0130,
06-0132,
06-0137,
06-0138,
06-0139,
06-0140,
06-0141,
06-0142,
06-0143,
06-0145,
06-0147,
06-0148,
06-0151,
06-0152,
06-0154,
06-0155,
06-0157,
06-0158,
06-0165,
06-0166,
06-0167,
06-0168,
06-0169,
06-0171,
06-0173,
06-0174,
06-0176,
06-0178,
06-0184,
06-0185,
06-0188,
06-0189,
06-0190,
06-0192,
06-0195,
06-0201,
06-0209,
06-0210,
06-0211,
06-0213,
06-0214,
06-0216,
06-0219,
06-0221,
06-0237,
06-0238,
06-0240,
06-0241,
06-0644,
06-0645,
06-0646,
06-0648,
06-0649,
06-0650,
06-0686,
06-0743,
06-0744,
06-0745,
06-0747,
06-0749,
06-0750,
06-0875,
06-0876,
06-0877,
06-0878,
06-0879,
06-0881,
06-0882,
06-0939,
06-1804,
06-1806,
06-2557,
06-2558,
06-2559,
06-2561,
06-2562,
06-2563,
06-2564,
06-2565,
06-2567,
06-2569,
06-2570,
06-5408,
06-5410,
06-5411,
06-5412,
06-5413,
06-5414,
06-5415,
06-5417,
06-5418,
06-5856,
06-5858,
06-5859,
06-6388,
06-6389,
06-6390,
06-6391,
06-6693,
06-6694,
06-6695,
06-6697,
06-6698,
06-6699,
06-6700,
06-6701,
08-0244,
08-0345,
08-0352,
08-0353,
08-0360,
08-0373,
08-0375,
08-0385,
08-0386,
12-0615,
12-0616,
12-0618,
12-0619,
12-0688,
12-0692,
12-0821,
12-1597,
12-3649,
12-3650,
12-3652,
12-3653,
12-5295,
12-5299,
12-5301,
14-0740,
14-0781,
14-0786,
14-0787,
14-0789,
14-0790,
14-0813,
14-0817,
14-0862,
14-0871,
14-1034,
14-1043,
14-1395,
14-1450,
14-1456,
14-1823,
14-1825,
14-1829,
14-2554,
14-3476,
14-4157,
15-0742,
15-1444,
16-0846,
16-0861,
16-1045,
16-1048,
19-1390,
19-1790,
19-2619,
19-2620,
19-2623,
19-2624,
19-2625,
19-2629,
19-2631,
19-4068,
19-5953,
26-1439,
26-1442,
26-5132,
26-5133,
26-5134,
26-5135,
26-5136,
26-5139,
26-6173,
26-6174,
27-1830,
27-1831,
27-1832,
27-1833,
27-1834,
27-1835,
27-1836,
27-1837,
27-1838,
27-2518,
27-2519,
27-2521,
27-2523,
27-2524,
27-2526,
27-2527,
27-2528,
28-1747,
28-1753,
28-2499,
28-2501,
28-2502,
28-2509,
28-2510,
28-2513,
28-2514,
28-5204,
28-5207,
28-5208,
28-5209,
28-5211,
28-5213,
28-5214,
28-5215,
28-5216,
28-5218,
28-5219,
28-5220,
28-6450,
32-1970,
32-1977,
32-1979,
32-1980,
32-1982,
32-1986,
32-1991,
32-2491,
32-2494,
32-2495,
32-2498,
32-2615,
32-2632,
32-2634,
32-2638,
32-4208,
32-4209,
32-4210,
32-4211,
32-4213,
32-4719,
32-5222,
41-2571,
41-2572,
41-2573,
41-2575,
41-3392,
41-3393,
41-3915,
41-4097,
41-5651,
41-6646,
74-6573,
74-6575,
74-6577,
74-6578,
74-6584,
76-4925,
76-4926,
76-4927,
76-4928,
76-4929,
76-4931,
76-4932,
76-4934,
76-4935,
76-6191,
76-6192,
76-6193,
76-6280,
76-6282,
76-6283,
76-6285,
76-6286,
76-6656,
76-6657,
76-6660,
76-6661,
76-6662,
76-6663,
76-6664,
81-5910,
81-5911,
87-5896.\\
$\blacklozenge$ {\bf Head and Neck Cancer} (591 samples):\\
$\bullet$ Source A1 = \cite{Agrawal}:\\
139,
266,
325,
347,
388,
478,
91.\\
$\bullet$ Source S4 = \cite{Stransky}:\\
HN12PT,
HN22PT,
HN27PT,
HN32PT,
HN33PT.\\
Remaining sample IDs are of the form HN\_*-Tumor, where * is:\\
0-046,
0-064,
00076,
00122,
00190,
00313,
00338,
00361,
00378,
00443,
00466,
00761,
01000,
62237,
62298,
62318,
62338,
62374,
62415,
62417,
62421,
62426,
62469,
62481,
62493,
62505,
62506,
62515,
62532,
62539,
62601,
62602,
62624,
62646,
62652,
62671,
62672,
62686,
62699,
62739,
62740,
62741,
62755,
62756,
62807,
62814,
62825,
62832,
62854,
62857\_2,
62860,
62861,
62863,
62897,
62921,
62926,
62984,
62996,
63007,
63021,
63027,
63039,
63048,
63058,
63080,
63081,
63095,
63114.\\
$\bullet$ Source T12 = TCGA (see Acknowledgments). Sample IDs are of the form TCGA-*, where * is:\\
BA-4074,
BA-4075,
BA-4076,
BA-4077,
BA-4078,
BA-5149,
BA-5151,
BA-5152,
BA-5153,
BA-5555,
BA-5556,
BA-5557,
BA-5558,
BA-5559,
BA-6868,
BA-6869,
BA-6870,
BA-6871,
BA-6872,
BA-6873,
BA-7269,
BA-A4IF,
BA-A4IG,
BA-A4IH,
BA-A4II,
BA-A6D8,
BA-A6DA,
BA-A6DB,
BA-A6DD,
BA-A6DE,
BA-A6DF,
BA-A6DG,
BA-A6DI,
BA-A6DJ,
BA-A6DL,
BB-4217,
BB-4223,
BB-4224,
BB-4225,
BB-4227,
BB-4228,
BB-7861,
BB-7862,
BB-7863,
BB-7864,
BB-7866,
BB-7870,
BB-7871,
BB-7872,
BB-8596,
BB-8601,
BB-A5HU,
BB-A5HY,
BB-A5HZ,
BB-A6UM,
BB-A6UO,
C9-A47Z,
C9-A480,
CN-4723,
CN-4725,
CN-4726,
CN-4727,
CN-4728,
CN-4729,
CN-4730,
CN-4731,
CN-4733,
CN-4734,
CN-4735,
CN-4736,
CN-4737,
CN-4738,
CN-4739,
CN-4740,
CN-4741,
CN-4742,
CN-5355,
CN-5356,
CN-5358,
CN-5359,
CN-5360,
CN-5361,
CN-5363,
CN-5364,
CN-5365,
CN-5366,
CN-5367,
CN-5369,
CN-5370,
CN-5373,
CN-5374,
CN-6010,
CN-6011,
CN-6012,
CN-6013,
CN-6016,
CN-6017,
CN-6018,
CN-6019,
CN-6020,
CN-6021,
CN-6022,
CN-6023,
CN-6024,
CN-6988,
CN-6989,
CN-6992,
CN-6994,
CN-6995,
CN-6996,
CN-6997,
CN-6998,
CN-A497,
CN-A498,
CN-A499,
CN-A49A,
CN-A49B,
CN-A49C,
CN-A63T,
CN-A63U,
CN-A63V,
CN-A63W,
CN-A63Y,
CN-A640,
CN-A641,
CN-A642,
CN-A6UY,
CN-A6V1,
CN-A6V3,
CN-A6V6,
CN-A6V7,
CQ-5323,
CQ-5324,
CQ-5325,
CQ-5326,
CQ-5327,
CQ-5329,
CQ-5330,
CQ-5331,
CQ-5332,
CQ-5333,
CQ-5334,
CQ-6218,
CQ-6219,
CQ-6220,
CQ-6221,
CQ-6222,
CQ-6223,
CQ-6224,
CQ-6225,
CQ-6227,
CQ-6228,
CQ-6229,
CQ-7063,
CQ-7064,
CQ-7065,
CQ-7067,
CQ-7068,
CQ-7069,
CQ-7071,
CQ-7072,
CQ-A4C6,
CQ-A4C7,
CQ-A4C9,
CQ-A4CB,
CQ-A4CD,
CQ-A4CE,
CQ-A4CG,
CQ-A4CH,
CQ-A4CI,
CR-5243,
CR-5247,
CR-5248,
CR-5249,
CR-5250,
CR-6467,
CR-6470,
CR-6471,
CR-6472,
CR-6473,
CR-6474,
CR-6477,
CR-6478,
CR-6480,
CR-6481,
CR-6482,
CR-6484,
CR-6487,
CR-6488,
CR-6491,
CR-6492,
CR-6493,
CR-7364,
CR-7365,
CR-7367,
CR-7368,
CR-7369,
CR-7370,
CR-7371,
CR-7372,
CR-7373,
CR-7374,
CR-7376,
CR-7377,
CR-7379,
CR-7380,
CR-7382,
CR-7383,
CR-7385,
CR-7386,
CR-7388,
CR-7389,
CR-7390,
CR-7391,
CR-7392,
CR-7393,
CR-7394,
CR-7395,
CR-7397,
CR-7398,
CR-7399,
CR-7401,
CR-7402,
CR-7404,
CV-5430,
CV-5431,
CV-5432,
CV-5434,
CV-5435,
CV-5436,
CV-5439,
CV-5440,
CV-5441,
CV-5442,
CV-5443,
CV-5444,
CV-5966,
CV-5970,
CV-5971,
CV-5973,
CV-5976,
CV-5977,
CV-5978,
CV-5979,
CV-6003,
CV-6433,
CV-6436,
CV-6441,
CV-6933,
CV-6934,
CV-6935,
CV-6936,
CV-6937,
CV-6938,
CV-6939,
CV-6940,
CV-6941,
CV-6942,
CV-6943,
CV-6945,
CV-6948,
CV-6950,
CV-6951,
CV-6952,
CV-6953,
CV-6954,
CV-6955,
CV-6956,
CV-6959,
CV-6960,
CV-6961,
CV-6962,
CV-7089,
CV-7090,
CV-7091,
CV-7095,
CV-7097,
CV-7099,
CV-7100,
CV-7101,
CV-7102,
CV-7103,
CV-7104,
CV-7177,
CV-7178,
CV-7180,
CV-7183,
CV-7235,
CV-7236,
CV-7238,
CV-7242,
CV-7243,
CV-7245,
CV-7247,
CV-7248,
CV-7250,
CV-7252,
CV-7253,
CV-7254,
CV-7255,
CV-7261,
CV-7263,
CV-7406,
CV-7407,
CV-7409,
CV-7410,
CV-7411,
CV-7413,
CV-7414,
CV-7415,
CV-7416,
CV-7418,
CV-7421,
CV-7422,
CV-7423,
CV-7424,
CV-7425,
CV-7427,
CV-7429,
CV-7430,
CV-7432,
CV-7433,
CV-7434,
CV-7435,
CV-7437,
CV-7438,
CV-7440,
CV-7446,
CV-7568,
CV-A45O,
CV-A45P,
CV-A45Q,
CV-A45R,
CV-A45T,
CV-A45U,
CV-A45V,
CV-A45W,
CV-A45X,
CV-A45Y,
CV-A45Z,
CV-A460,
CV-A461,
CV-A463,
CV-A464,
CV-A465,
CV-A468,
CV-A6JD,
CV-A6JE,
CV-A6JM,
CV-A6JN,
CV-A6JO,
CV-A6JT,
CV-A6JU,
CV-A6JY,
CV-A6JZ,
CV-A6K0,
CV-A6K1,
CV-A6K2,
CX-7082,
CX-7085,
CX-7086,
CX-7219,
CX-A4AQ,
D6-6515,
D6-6516,
D6-6517,
D6-6823,
D6-6824,
D6-6825,
D6-6826,
D6-6827,
D6-8568,
D6-8569,
D6-A4Z9,
D6-A4ZB,
D6-A6EK,
D6-A6EM,
D6-A6EN,
D6-A6EO,
D6-A6EP,
D6-A6EQ,
D6-A6ES,
D6-A74Q,
DQ-5624,
DQ-5625,
DQ-5629,
DQ-5630,
DQ-5631,
DQ-7588,
DQ-7589,
DQ-7590,
DQ-7591,
DQ-7592,
DQ-7593,
DQ-7594,
DQ-7595,
DQ-7596,
F7-7848,
F7-8489,
F7-A50G,
F7-A50I,
F7-A50J,
F7-A61S,
F7-A61V,
F7-A61W,
F7-A620,
F7-A622,
F7-A623,
F7-A624,
H7-7774,
H7-8501,
H7-A6C4,
H7-A6C5,
H7-A76A,
HD-7229,
HD-7753,
HD-7754,
HD-7831,
HD-7832,
HD-7917,
HD-8224,
HD-8314,
HD-8634,
HD-8635,
HD-A4C1,
HD-A633,
HD-A634,
HD-A6HZ,
HD-A6I0,
HL-7533,
IQ-7630,
IQ-7631,
IQ-7632,
IQ-A61E,
IQ-A61G,
IQ-A61H,
IQ-A61I,
IQ-A61J,
IQ-A61K,
IQ-A61L,
IQ-A61O,
IQ-A6SG,
IQ-A6SH,
KU-A66S,
KU-A66T,
KU-A6H7,
KU-A6H8,
MT-A51W,
MT-A51X,
MT-A67A,
MT-A67D,
MT-A67F,
MT-A67G,
MT-A7BN,
MZ-A5BI,
MZ-A6I9,
MZ-A7D7,
P3-A5Q6,
P3-A5QA,
P3-A5QE,
P3-A5QF,
P3-A6SW,
P3-A6SX,
P3-A6T0,
P3-A6T2,
P3-A6T3,
P3-A6T4,
P3-A6T5,
P3-A6T6,
P3-A6T7,
P3-A6T8,
QK-A64Z,
QK-A652,
QK-A6IF,
QK-A6IG,
QK-A6IH,
QK-A6II,
QK-A6IJ,
QK-A6V9,
QK-A6VB,
QK-A6VC,
RS-A6TO,
RS-A6TP,
T2-A6WX,
T2-A6WZ,
T2-A6X0,
T2-A6X2,
TN-A7HI,
TN-A7HJ,
TN-A7HL,
UF-A718,
UF-A719,
UF-A71A,
UF-A71B,
UF-A71D,
UF-A71E,
UF-A7J9,
UF-A7JA,
UF-A7JC,
UF-A7JD,
UF-A7JF,
UF-A7JH,
UF-A7JJ,
UF-A7JK,
UF-A7JO,
UF-A7JS,
UF-A7JT,
UF-A7JV,
UP-A6WW,
WA-A7GZ,
WA-A7H4.\\
$\blacklozenge$ {\bf Liver Cancer} (452 samples):\\
$\bullet$ Source S5 = \cite{Schulze}. Sample IDs are of the form BCB*, where * is:\\
109T,
111T,
151T,
157T,
167T,
231T,
301T,
307T,
325T.\\
Additional sample IDs are of the form BCM*, where * is:\\
229T,
257T,
265T,
269T,
275T,
321T,
325T,
329T,
337T,
339T,
371T,
375T,
397T,
399T,
423T,
439T,
455T,
483T,
489T,
501T,
529T,
531T,
543T,
545T,
565T,
567T,
617T,
643T,
671T,
683T,
689T,
695T,
703T,
711T,
723T,
735T,
739T,
759T,
769T,
783T,
791T.\\
Remaining sample IDs are of the form CHC*, where * is:\\
051T,
059T,
060T,
097T,
1010T,
1028T,
1035T,
1040T,
1041T,
1044T,
1052T,
1053T,
1055T,
1060T,
1061T,
1062T,
1065T,
1079T,
1081T,
1082T,
1083T,
1085T,
1089T,
1091T,
1097T,
1098T,
1137T,
1148T,
1152T,
1154T,
1162T,
1177T,
1180T,
1182T,
1183T,
1185T,
1186T,
1190T,
1191T,
1192T,
1201T,
1205T,
1207T,
1209T,
1210T,
1211T,
121T,
1530T,
1531T,
1534T,
1539T,
1545T,
1556T,
155T,
1566T,
1568T,
1569T,
1591T,
1592T,
1594T,
1595T,
1596T,
1597T,
1598T,
1600T,
1601T,
1602T,
1603T,
1604T,
1611T,
1616T,
1624T,
1626T,
1629T,
1700T,
1704T,
1708T,
1712T,
1714T,
1715T,
1717T,
1719T,
1720T,
1725T,
1731T,
1732T,
1734T,
1736T,
1737T,
1738T,
1739T,
1741T,
1742T,
1743T,
1744T,
1745T,
1746T,
1747T,
1749T,
1750T,
1751T,
1753T,
1754T,
1756T,
1757T,
1763T,
1774T,
1775T,
1915T,
197T,
2029T,
2034T,
2039Tbis,
2043T,
2048T,
2052T,
205T,
2098T,
2099T,
2103T,
2110Tbis,
2111T,
2112T,
2113T,
2115T,
2127T,
2128T,
2134T,
2141T,
218T,
2200T,
2202T,
2206T,
2208T,
2211T,
2213T,
2215T,
2216T,
2321T,
2351T,
2352T,
2358T,
2362T,
253T,
258T,
301T,
302T,
303T,
304T,
306T,
307T,
313T,
314T,
320T,
322T,
326T,
327T,
361TA,
429T,
432T,
433T,
434T,
437T,
451T,
465T,
469T,
510T,
609T,
614T,
703T,
734T,
736T,
789T,
793T,
794T,
796T,
798T,
799T,
801T,
805T,
879T,
884T,
889T,
891T,
892T,
896T,
898T,
902T,
909T,
912T,
917T,
923T,
961T.\\
$\bullet$ Source H2 = \cite{Huang}:\\
P47,
P48,
P51,
P52,
P55,
P56,
P929.\\
$\bullet$ Source T13 = TCGA (see Acknowledgments). Sample IDs are of the form TCGA-*, where * is:\\
BC-4073,
BC-A10Q,
BC-A10R,
BC-A10S,
BC-A10T,
BC-A10U,
BC-A10W,
BC-A10X,
BC-A10Y,
BC-A10Z,
BC-A110,
BC-A112,
BC-A216,
BC-A217,
BC-A3KF,
BC-A3KG,
BC-A5W4,
BC-A69H,
BC-A69I,
BD-A2L6,
BD-A3EP,
BD-A3ER,
BW-A5NO,
BW-A5NP,
BW-A5NQ,
CC-5258,
CC-5259,
CC-5260,
CC-5261,
CC-5262,
CC-5263,
CC-5264,
CC-A123,
CC-A1HT,
CC-A3M9,
CC-A3MA,
CC-A3MB,
CC-A3MC,
CC-A5UC,
CC-A5UD,
CC-A5UE,
CC-A7IF,
CC-A7IG,
CC-A7IH,
CC-A7II,
CC-A7IJ,
CC-A7IK,
CC-A7IL,
DD-A113,
DD-A114,
DD-A115,
DD-A116,
DD-A118,
DD-A119,
DD-A11A,
DD-A11B,
DD-A11C,
DD-A11D,
DD-A1E9,
DD-A1EA,
DD-A1EB,
DD-A1EC,
DD-A1ED,
DD-A1EF,
DD-A1EG,
DD-A1EH,
DD-A1EI,
DD-A1EJ,
DD-A1EK,
DD-A1EL,
DD-A39V,
DD-A39W,
DD-A39X,
DD-A39Y,
DD-A39Z,
DD-A3A0,
DD-A3A1,
DD-A3A2,
DD-A3A3,
DD-A3A4,
DD-A3A5,
DD-A3A6,
DD-A3A7,
DD-A3A8,
DD-A3A9,
DD-A4NA,
DD-A4NB,
DD-A4ND,
DD-A4NE,
DD-A4NF,
DD-A4NG,
DD-A4NH,
DD-A4NI,
DD-A4NJ,
DD-A4NK,
DD-A4NL,
DD-A4NN,
DD-A4NO,
DD-A4NP,
DD-A4NQ,
DD-A4NR,
DD-A4NS,
DD-A4NV,
DD-A73A,
DD-A73B,
DD-A73C,
DD-A73D,
DD-A73E,
DD-A73F,
DD-A73G,
ED-A459,
ED-A4XI,
ED-A5KG,
ED-A627,
ED-A66X,
ED-A66Y,
ED-A7PX,
ED-A7PY,
ED-A7PZ,
ED-A7XO,
ED-A7XP,
ED-A82E,
EP-A12J,
EP-A26S,
EP-A2KA,
EP-A2KB,
EP-A2KC,
EP-A3JL,
EP-A3RK,
ES-A2HS,
ES-A2HT,
FV-A23B,
FV-A2QQ,
FV-A2QR,
FV-A3I0,
FV-A3I1,
FV-A3R2,
FV-A3R3,
FV-A495,
FV-A496,
FV-A4ZP,
FV-A4ZQ,
G3-A25S,
G3-A25T,
G3-A25U,
G3-A25V,
G3-A25W,
G3-A25Y,
G3-A25Z,
G3-A3CG,
G3-A3CH,
G3-A3CI,
G3-A3CJ,
G3-A3CK,
G3-A5SI,
G3-A5SJ,
G3-A5SK,
G3-A5SL,
G3-A5SM,
G3-A6UC,
G3-A7M5,
G3-A7M6,
G3-A7M7,
G3-A7M8,
G3-A7M9,
GJ-A6C0,
HP-A5MZ,
HP-A5N0,
K7-A5RF,
K7-A5RG,
K7-A6G5,
KR-A7K0,
KR-A7K2,
KR-A7K7,
KR-A7K8,
LG-A6GG,
MI-A75C,
MI-A75E,
MI-A75G,
MI-A75H,
MI-A75I,
MR-A520,
NI-A4U2,
O8-A75V,
PD-A5DF,
QA-A7B7,
RC-A6M3,
RC-A6M4,
RC-A6M5,
RC-A6M6,
RC-A7S9,
RC-A7SB,
RC-A7SF,
RC-A7SK,
RG-A7D4,
T1-A6J8,
UB-A7MA,
UB-A7MB,
UB-A7MC,
UB-A7MD,
UB-A7ME,
UB-A7MF.\\
$\blacklozenge$ {\bf Lung Cancer} (1018 samples):\\
$\bullet$ Source D3 = \cite{Ding}:\\
16600,
16608,
16628,
16632,
16648,
16660,
16668,
16678,
16686,
16724,
16802,
16814,
16835,
16857,
16949,
17042,
17055,
17156,
17174,
17210,
17218,
17226,
17242,
17268,
17290,
17308,
17733,
17746,
17759,
17763.\\
$\bullet$ Source R1 = \cite{Rudin}:\\
113368,
134398,
134413,
134417,
134421,
134426,
134427,
134430,
2334187,
2334188,
2334189,
2334191,
2334193,
2334195,
2334196,
2334199,
2334201,
2334202,
585203,
585205,
585208,
585210,
585223,
585258,
585260,
585265,
585267,
585270,
585272,
585276,
631052,
631056,
631060,
631064,
631076,
631084,
631092,
98687,
98711,
98735.\\
$\bullet$ Source P3 = \cite{Peifer}:\\
H1672,
H2171,
S00022,
S00050,
S00356,
S00472,
S00501,
S00539,
S00827,
S00830,
S00833,
S00836,
S00837,
S00841,
S00932,
S00933,
S00935,
S00936,
S00943,
S00944,
S00945,
S00946,
S00947,
S01366,
S01453,
S01494,
S01512,
S01563,
S01728.\\
$\bullet$ Source S6 = \cite{Seo}. Sample IDs are of the form LC\_*, where * is:\\
C1,
C10,
C11,
C13,
C14,
C15,
C17,
C18,
C19,
C2,
C20,
C21,
C22,
C23,
C24,
C25,
C26,
C27,
C28,
C29,
C30,
C32,
C33,
C34,
C35,
C36,
C4,
C5,
C6,
C7,
C8,
C9,
S10,
S11,
S12,
S13,
S14,
S15,
S16,
S17,
S18,
S19,
S2,
S20,
S21,
S23,
S24,
S25,
S27,
S28,
S29,
S3,
S31,
S32,
S34,
S35,
S37,
S38,
S39,
S4,
S40,
S41,
S42,
S43,
S44,
S45,
S46,
S47,
S48,
S49,
S5,
S51,
S6,
S8,
S9.\\
$\bullet$ Source I1 = \cite{Imielinski}. Sample IDs are of the form LUAD.**.Tumor, where ** is (below * stands for NYU, e.g., *1021 = NYU1021 and the full sample ID is LUAD.NYU1021.Tumor):\\
5O6B5,
74TBW,
B00416,
B00523,
B00859,
B00915,
B01102,
B01145,
B01811,
B01970,
B02077,
B02216,
B02477,
B02515,
B02594,
D00147,
D01278,
D01603,
D01751,
D02085,
D02185,
E00163,
E00443,
E00897,
E00918,
F00018,
F00057,
F00089,
F00121,
F00134,
F00162,
F00170,
F00257,
F00282,
F00365,
F00368,
GU4I3,
LC15C,
LIP77,
*1021,
*1026,
*1027,
*1051S,
*1093,
*1096,
*1101,
*1142,
*1177,
*1195,
*1210,
*1219,
*160,
*184,
*195,
*201,
*213,
*252,
*259,
*263,
*282,
*284,
*287,
*315,
*330,
*408,
*508,
*574S,
*575,
*584S,
*608,
*627,
*669,
*689,
*696,
*704,
*739,
*796,
*802,
*803,
*846,
*847,
*848,
*947,
*994,
QCHM7,
QJN9L,
S00484,
S00486,
S00499,
S01304,
S01306,
S01315,
S01320,
S01354,
S01357,
S01362,
S01373,
S01409,
S01413,
S01482,
TLLGS,
UF7HM,
VUMN6,
YINHD,
YKER9.\\
Additional sample IDs are of the form LUAD.CHTN.*.Tumor, where * is:\\
3090346,
3090415,
3090416,
4090680,
MAD04.00674,
MAD06.00490,
MAD06.00668,
MAD06.00678,
MAD08.00104,
Z4716A.\\
Further sample IDs are of the form LUAD.RT.*.Tumor, where * is:\\
S01477,
S01487,
S01699,
S01700,
S01702,
S01703,
S01709,
S01711,
S01721,
S01769,
S01770,
S01771,
S01774,
S01777,
S01808,
S01810,
S01813,
S01818,
S01831,
S01832,
S01840,
S01852,
S01856,
S01866.\\
Remaining sample IDs are of the form LUAD\_*.Tumor, where * is:\\
E00522,
E00565,
E00623,
E00703,
E00945,
E01047,
E01086,
E01147,
E01166,
E01319,
E01419.\\
$\bullet$ Source T14 = TCGA (see Acknowledgments). Sample IDs are of the form TCGA.*, where * is:\\
05.4244,
05.4249,
05.4250,
05.4382,
05.4384,
05.4389,
05.4390,
05.4395,
05.4396,
05.4397,
05.4398,
05.4402,
05.4403,
05.4405,
05.4410,
05.4415,
05.4417,
05.4418,
05.4420,
05.4422,
05.4424,
05.4425,
05.4426,
05.4427,
05.4430,
05.4432,
05.4433,
05.4434,
05.5420,
05.5423,
05.5425,
05.5428,
05.5429,
05.5715,
17.Z000,
17.Z001,
17.Z002,
17.Z003,
17.Z004,
17.Z005,
17.Z007,
17.Z008,
17.Z009,
17.Z010,
17.Z011,
17.Z012,
17.Z013,
17.Z014,
17.Z015,
17.Z016,
17.Z017,
17.Z018,
17.Z019,
17.Z020,
17.Z021,
17.Z022,
17.Z023,
17.Z025,
17.Z026,
17.Z027,
17.Z028,
17.Z030,
17.Z031,
17.Z032,
17.Z033,
17.Z035,
17.Z036,
17.Z037,
17.Z040,
17.Z041,
17.Z042,
17.Z043,
17.Z044,
17.Z045,
17.Z046,
17.Z047,
17.Z048,
17.Z049,
17.Z050,
17.Z051,
17.Z052,
17.Z053,
17.Z054,
17.Z055,
17.Z056,
17.Z057,
17.Z058,
17.Z059,
17.Z060,
17.Z061,
17.Z062,
18.3406,
18.3407,
18.3408,
18.3409,
18.3410,
18.3411,
18.3412,
18.3414,
18.3415,
18.3416,
18.3417,
18.3419,
18.3421,
18.4083,
18.4086,
18.4721,
18.5592,
18.5595,
21.1070,
21.1071,
21.1076,
21.1077,
21.1078,
21.1081,
21.5782,
21.5784,
21.5786,
21.5787,
22.0944,
22.1002,
22.1011,
22.1012,
22.1016,
22.4591,
22.4593,
22.4595,
22.4599,
22.4601,
22.4604,
22.4607,
22.4613,
22.5471,
22.5472,
22.5473,
22.5474,
22.5477,
22.5478,
22.5480,
22.5482,
22.5485,
22.5489,
22.5491,
22.5492,
33.4532,
33.4533,
33.4538,
33.4547,
33.4566,
33.4582,
33.4583,
33.4586,
33.6737,
34.2596,
34.2600,
34.2608,
34.5231,
34.5232,
34.5234,
34.5236,
34.5239,
34.5240,
34.5927,
34.5928,
34.5929,
35.3615,
35.3621,
35.4122,
35.4123,
35.5375,
37.3783,
37.3789,
37.4133,
37.4135,
37.4141,
37.5819,
38.4625,
38.4626,
38.4627,
38.4628,
38.4629,
38.4630,
38.4631,
38.4632,
38.6178,
38.7271,
38.A44F,
39.5016,
39.5019,
39.5021,
39.5022,
39.5024,
39.5027,
39.5028,
39.5029,
39.5030,
39.5031,
39.5035,
39.5036,
39.5037,
39.5039,
43.2578,
43.3394,
43.3920,
43.5668,
43.6143,
43.6647,
43.6770,
43.6771,
44.2655,
44.2656,
44.2657,
44.2659,
44.2661,
44.2662,
44.2665,
44.2666,
44.2668,
44.3396,
44.3398,
44.3918,
44.3919,
44.4112,
44.5643,
44.5644,
44.5645,
44.6144,
44.6145,
44.6146,
44.6147,
44.6148,
44.6774,
44.6775,
44.6776,
44.6777,
44.6778,
44.6779,
44.7659,
44.7660,
44.7661,
44.7662,
44.7667,
44.7669,
44.7670,
44.7671,
44.7672,
44.8117,
44.8119,
44.8120,
44.A479,
44.A47A,
44.A47B,
44.A47F,
44.A47G,
44.A4SS,
44.A4SU,
46.3765,
46.3767,
46.3768,
46.3769,
46.6025,
46.6026,
49.4486,
49.4487,
49.4488,
49.4490,
49.4494,
49.4501,
49.4505,
49.4506,
49.4507,
49.4510,
49.4512,
49.4514,
49.6742,
49.6743,
49.6744,
49.6745,
49.6761,
49.6767,
50.5044,
50.5045,
50.5049,
50.5051,
50.5055,
50.5066,
50.5068,
50.5072,
50.5930,
50.5931,
50.5932,
50.5933,
50.5935,
50.5936,
50.5939,
50.5941,
50.5942,
50.5944,
50.5946,
50.6590,
50.6591,
50.6592,
50.6593,
50.6594,
50.6595,
50.6597,
50.6673,
50.7109,
50.8457,
50.8459,
50.8460,
51.4079,
51.4080,
51.4081,
53.7624,
53.7626,
53.7813,
53.A4EZ,
55.1592,
55.1594,
55.1595,
55.1596,
55.5899,
55.6543,
55.6642,
55.6712,
55.6968,
55.6969,
55.6970,
55.6971,
55.6972,
55.6975,
55.6978,
55.6979,
55.6980,
55.6981,
55.6982,
55.6983,
55.6984,
55.6985,
55.6986,
55.6987,
55.7227,
55.7281,
55.7283,
55.7284,
55.7570,
55.7573,
55.7574,
55.7576,
55.7724,
55.7725,
55.7726,
55.7727,
55.7728,
55.7815,
55.7816,
55.7903,
55.7907,
55.7910,
55.7911,
55.7913,
55.7914,
55.7994,
55.7995,
55.8085,
55.8087,
55.8089,
55.8090,
55.8091,
55.8092,
55.8094,
55.8096,
55.8097,
55.8203,
55.8204,
55.8205,
55.8206,
55.8207,
55.8208,
55.8299,
55.8301,
55.8302,
55.8505,
55.8506,
55.8507,
55.8508,
55.8510,
55.8511,
55.8512,
55.8513,
55.8514,
55.8614,
55.8615,
55.8616,
55.8619,
55.8620,
55.8621,
55.A48X,
55.A48Y,
55.A48Z,
55.A490,
55.A491,
55.A492,
55.A493,
55.A494,
55.A4DF,
55.A4DG,
56.1622,
56.5897,
56.5898,
56.6545,
56.6546,
60.2698,
60.2707,
60.2708,
60.2709,
60.2710,
60.2711,
60.2712,
60.2713,
60.2715,
60.2719,
60.2720,
60.2721,
60.2722,
60.2723,
60.2724,
60.2725,
60.2726,
62.8394,
62.8395,
62.8397,
62.8398,
62.8399,
62.8402,
62.A46O,
62.A46P,
62.A46R,
62.A46S,
62.A46U,
62.A46V,
62.A46Y,
62.A470,
62.A471,
62.A472,
63.5128,
63.5131,
63.6202,
64.1676,
64.1677,
64.1678,
64.1679,
64.1680,
64.1681,
64.5774,
64.5775,
64.5778,
64.5779,
64.5781,
64.5815,
66.2727,
66.2734,
66.2742,
66.2744,
66.2754,
66.2755,
67.3770,
67.3771,
67.3772,
67.3773,
67.3774,
67.4679,
67.6215,
67.6216,
67.6217,
69.7760,
69.7761,
69.7763,
69.7764,
69.7765,
69.7973,
69.7974,
69.7978,
69.7979,
69.7980,
69.8253,
69.8254,
69.8255,
69.A59K,
71.6725,
71.8520,
73.4658,
73.4659,
73.4662,
73.4666,
73.4668,
73.4670,
73.4675,
73.4676,
73.4677,
73.7498,
73.7499,
75.5122,
75.5125,
75.5126,
75.5146,
75.5147,
75.6203,
75.6205,
75.6206,
75.6207,
75.6211,
75.6212,
75.6214,
75.7025,
75.7027,
75.7030,
75.7031,
78.7143,
78.7145,
78.7146,
78.7147,
78.7148,
78.7149,
78.7150,
78.7152,
78.7153,
78.7154,
78.7155,
78.7156,
78.7158,
78.7159,
78.7160,
78.7161,
78.7162,
78.7163,
78.7166,
78.7167,
78.7220,
78.7535,
78.7536,
78.7537,
78.7539,
78.7540,
78.7542,
78.7633,
78.8640,
78.8648,
78.8655,
78.8660,
78.8662,
80.5607,
80.5608,
80.5611,
83.5908,
86.6562,
86.6851,
86.7701,
86.7711,
86.7713,
86.7714,
86.7953,
86.7954,
86.7955,
86.8054,
86.8055,
86.8056,
86.8073,
86.8074,
86.8075,
86.8076,
86.8278,
86.8279,
86.8280,
86.8281,
86.8358,
86.8359,
86.8585,
86.8668,
86.8669,
86.8671,
86.8672,
86.8673,
86.8674,
86.A456,
86.A4D0,
86.A4JF,
86.A4P7,
86.A4P8,
91.6828,
91.6829,
91.6830,
91.6831,
91.6835,
91.6836,
91.6840,
91.6847,
91.6848,
91.6849,
91.7771,
91.8496,
91.8497,
91.8499,
91.A4BC,
91.A4BD,
93.7347,
93.7348,
93.8067,
93.A4JN,
93.A4JO,
93.A4JP,
93.A4JQ,
95.7039,
95.7043,
95.7562,
95.7567,
95.7944,
95.7947,
95.7948,
95.8039,
95.8494,
95.A4VK,
95.A4VN,
95.A4VP,
97.7546,
97.7547,
97.7552,
97.7553,
97.7554,
97.7937,
97.7938,
97.7941,
97.8171,
97.8172,
97.8174,
97.8175,
97.8176,
97.8177,
97.8179,
97.8547,
97.8552,
97.A4LX,
97.A4M0,
97.A4M1,
97.A4M2,
97.A4M3,
97.A4M5,
97.A4M6,
97.A4M7,
99.7458,
99.8025,
99.8028,
99.8032,
99.8033,
J2.8192,
J2.8194,
J2.A4AD,
J2.A4AE,
J2.A4AG,
L4.A4E5,
L4.A4E6,
L9.A443,
L9.A444,
MN.A4N1,
MN.A4N4,
MN.A4N5,
MP.A4SV,
MP.A4SW,
MP.A4SY,
MP.A4T2,
MP.A4T4,
MP.A4T6,
MP.A4T7,
MP.A4T8,
MP.A4T9,
MP.A4TA,
MP.A4TC,
MP.A4TD,
MP.A4TE,
MP.A4TF,
MP.A4TH,
MP.A4TI,
MP.A4TK,
MP.A5C7,
NJ.A4YF,
NJ.A4YG,
NJ.A4YI,
NJ.A4YP,
NJ.A4YQ,
NJ.A55A,
NJ.A55O,
NJ.A55R,
O1.A52J.\\
$\blacklozenge$ {\bf Melanoma} (594 samples):\\
$\bullet$ Source S7 = \cite{Stark}:\\
A02,
A06,
D05,
D14,
D35,
D36,
D41,
D49.\\
$\bullet$ Source D4 = \cite{Davies}:\\
COLO-829.\\
$\bullet$ Source B1 = \cite{Berger}. Sample IDs are of the form ME*-Tumor, where * is:\\
001,
002,
009,
010,
011,
012,
014,
015,
016,
017,
018,
020,
021,
024,
029,
030,
032,
033,
034,
035,
037,
041,
043,
044,
045,
048,
049,
050.\\
Remaining sample IDs are:\\
Mel-BRAFi-03-Tumor,
Mel\_BRAFi\_02\_PRE-Tumor.\\
$\bullet$ Source A2 = \cite{Alexandrov}. Sample IDs are of the form PD*, where * is:\\
10020a,
10021a,
10022a,
9024a2,
9024b,
9025a,
9025b,
9026a,
9027a,
9027b,
9028a,
9028b,
9029a,
9030a,
9031a,
9032a,
9033a.\\
$\bullet$ Source H3 = \cite{Hodis}. Sample IDs are of the form SKCM-*-Tumor, where * is:\\
13447,
13456,
13463,
13468,
13473,
13531,
13537,
13543,
13549,
13560,
13561,
13567,
13575,
13591,
13600.\\
Additional sample IDs are of the form SKCM-JWCI-*-Tumor, where * is:\\
14,
27,
WGS-1,
WGS-11,
WGS-12,
WGS-13,
WGS-15,
WGS-18,
WGS-19,
WGS-2,
WGS-20,
WGS-21,
WGS-22,
WGS-23,
WGS-24,
WGS-25,
WGS-26,
WGS-29,
WGS-3,
WGS-32,
WGS-33,
WGS-34,
WGS-35,
WGS-36,
WGS-37,
WGS-38,
WGS-39,
WGS-4,
WGS-42,
WGS-43,
WGS-5,
WGS-6,
WGS-7,
WGS-8.\\
Further sample IDs are of the form SKCM-Ma-Mel-*-Tumor, where * is:\\
04,
05,
08a,
102,
103b,
105,
107,
108,
114,
119,
120,
122,
123,
15,
16,
19,
27,
28,
35,
36,
37,
48,
53,
54a,
55,
59,
62,
63,
65,
67,
71,
76,
79,
85,
86,
91,
92,
94.\\
Remaining sample IDs are:\\
SKCM-UKRV-Mel-20-Tumor,
SKCM-UKRV-Mel-24-Tumor,
SKCM-UKRV-Mel-6-Tumor.\\
$\bullet$ Source T15 = TCGA (see Acknowledgments). Sample IDs are of the form TCGA-*, where * is:\\
BF-A1PU,
BF-A1PV,
BF-A1PX,
BF-A1PZ,
BF-A1Q0,
BF-A3DJ,
BF-A3DL,
BF-A3DM,
BF-A3DN,
D3-A1Q1,
D3-A1Q3,
D3-A1Q4,
D3-A1Q5,
D3-A1Q6,
D3-A1Q7,
D3-A1Q8,
D3-A1Q9,
D3-A1QA,
D3-A1QB,
D3-A2J6,
D3-A2J7,
D3-A2J8,
D3-A2J9,
D3-A2JA,
D3-A2JB,
D3-A2JC,
D3-A2JD,
D3-A2JF,
D3-A2JG,
D3-A2JH,
D3-A2JK,
D3-A2JL,
D3-A2JN,
D3-A2JO,
D3-A2JP,
D3-A3BZ,
D3-A3C1,
D3-A3C3,
D3-A3C6,
D3-A3C7,
D3-A3C8,
D3-A3CB,
D3-A3CC,
D3-A3CE,
D3-A3CF,
D3-A3ML,
D3-A3MO,
D3-A3MR,
D3-A3MU,
D3-A3MV,
D9-A148,
D9-A149,
D9-A1JW,
D9-A1JX,
D9-A1X3,
DA-A1HV,
DA-A1HW,
DA-A1HY,
DA-A1I0,
DA-A1I1,
DA-A1I2,
DA-A1I4,
DA-A1I5,
DA-A1I7,
DA-A1I8,
DA-A1IA,
DA-A1IB,
DA-A1IC,
DA-A3F3,
DA-A3F5,
DA-A3F8,
EB-A1NK,
EB-A24C,
EB-A24D,
EB-A299,
EB-A3HV,
EE-A17X,
EE-A17Y,
EE-A17Z,
EE-A180,
EE-A181,
EE-A182,
EE-A183,
EE-A184,
EE-A185,
EE-A20B,
EE-A20C,
EE-A20F,
EE-A20H,
EE-A20I,
EE-A29A,
EE-A29B,
EE-A29C,
EE-A29D,
EE-A29E,
EE-A29G,
EE-A29H,
EE-A29L,
EE-A29M,
EE-A29N,
EE-A29P,
EE-A29Q,
EE-A29R,
EE-A29S,
EE-A29T,
EE-A29V,
EE-A29W,
EE-A29X,
EE-A2A0,
EE-A2A1,
EE-A2A2,
EE-A2A5,
EE-A2A6,
EE-A2GB,
EE-A2GC,
EE-A2GD,
EE-A2GE,
EE-A2GH,
EE-A2GI,
EE-A2GJ,
EE-A2GK,
EE-A2GL,
EE-A2GM,
EE-A2GN,
EE-A2GO,
EE-A2GP,
EE-A2GR,
EE-A2GS,
EE-A2GT,
EE-A2GU,
EE-A2M5,
EE-A2M6,
EE-A2M7,
EE-A2M8,
EE-A2MC,
EE-A2MD,
EE-A2ME,
EE-A2MF,
EE-A2MG,
EE-A2MH,
EE-A2MI,
EE-A2MJ,
EE-A2MK,
EE-A2ML,
EE-A2MM,
EE-A2MN,
EE-A2MP,
EE-A2MQ,
EE-A2MR,
EE-A2MS,
EE-A2MT,
EE-A2MU,
EE-A3AA,
EE-A3AB,
EE-A3AC,
EE-A3AD,
EE-A3AE,
EE-A3AF,
EE-A3AG,
EE-A3AH,
EE-A3J3,
EE-A3J4,
EE-A3J5,
EE-A3J7,
EE-A3J8,
EE-A3JA,
EE-A3JB,
EE-A3JD,
EE-A3JE,
EE-A3JH,
EE-A3JI,
ER-A193,
ER-A194,
ER-A195,
ER-A196,
ER-A197,
ER-A198,
ER-A199,
ER-A19A,
ER-A19B,
ER-A19C,
ER-A19D,
ER-A19E,
ER-A19F,
ER-A19G,
ER-A19H,
ER-A19J,
ER-A19K,
ER-A19L,
ER-A19N,
ER-A19O,
ER-A19P,
ER-A19Q,
ER-A19S,
ER-A19T,
ER-A1A1,
ER-A2NB,
ER-A2NC,
ER-A2ND,
ER-A2NE,
ER-A2NF,
ER-A2NG,
ER-A2NH,
ER-A3ES,
ER-A3ET,
ER-A3EV,
FR-A2OS,
FS-A1YX,
FS-A1YY,
FS-A1Z0,
FS-A1Z3,
FS-A1Z4,
FS-A1Z7,
FS-A1ZB,
FS-A1ZC,
FS-A1ZD,
FS-A1ZE,
FS-A1ZF,
FS-A1ZG,
FS-A1ZH,
FS-A1ZJ,
FS-A1ZK,
FS-A1ZM,
FS-A1ZN,
FS-A1ZP,
FS-A1ZQ,
FS-A1ZR,
FS-A1ZS,
FS-A1ZT,
FS-A1ZU,
FS-A1ZW,
FS-A1ZY,
FS-A1ZZ,
FW-A3I3,
GF-A2C7,
GN-A262,
GN-A263,
GN-A264,
GN-A265,
GN-A266,
GN-A267,
GN-A268,
GN-A269,
GN-A26A,
GN-A26C,
GN-A26D,
HR-A2OG,
HR-A2OH,
IH-A3EA,
D3-A5GT,
D9-A3Z4,
D9-A4Z2,
D9-A4Z3,
D9-A4Z5,
EB-A3XB,
EB-A3XC,
EB-A3XD,
EB-A3XE,
EB-A3Y6,
EB-A3Y7,
EB-A41A,
EB-A41B,
EB-A42Y,
EB-A42Z,
EB-A430,
EB-A431,
EB-A44N,
EB-A44O,
EB-A44P,
EB-A4IQ,
EB-A4IS,
EB-A4OY,
EB-A4OZ,
EB-A4P0,
EB-A551,
EB-A553,
EB-A57M,
EB-A5SE,
EB-A5SF,
EB-A5UM,
FR-A3R1,
FW-A5DX,
BF-A5EO,
BF-A5EP,
BF-A5EQ,
BF-A5ER,
BF-A5ES,
D3-A51E,
D3-A51F,
D3-A51G,
D3-A51H,
D3-A51J,
D3-A51K,
D3-A51N,
D3-A51R,
D3-A51T,
D3-A5GL,
D3-A5GN,
D3-A5GO,
D3-A5GR,
D3-A5GS,
D9-A3Z1,
D9-A3Z3,
D9-A6E9,
D9-A6EA,
D9-A6EC,
D9-A6EG,
DA-A3F2,
EB-A3XF,
EB-A44Q,
EB-A44R,
EB-A4XL,
EB-A5FP,
EB-A5KH,
EB-A5SG,
EB-A5SH,
EB-A5UL,
EB-A5UN,
EB-A5VU,
EB-A5VV,
EB-A6L9,
EB-A6QY,
EB-A6QZ,
EB-A6R0,
ER-A19M,
ER-A19W,
ER-A3PL,
ER-A42H,
ER-A42K,
ER-A42L,
FR-A3YN,
FR-A3YO,
FR-A44A,
FR-A69P,
FR-A726,
FR-A728,
FS-A1YW,
FS-A1ZA,
FS-A4F4,
FS-A4F5,
FS-A4F8,
FS-A4F9,
FS-A4FB,
FS-A4FC,
FS-A4FD,
FW-A3R5,
FW-A3TU,
FW-A3TV,
FW-A5DY,
GF-A3OT,
GF-A6C8,
GF-A6C9,
GF-A769,
GN-A4U3,
GN-A4U4,
GN-A4U5,
GN-A4U7,
GN-A4U8,
GN-A4U9,
OD-A75X,
QB-A6FS,
RP-A690,
RP-A693,
RP-A694,
RP-A695,
D3-A5GU,
FS-A4F0,
GF-A4EO,
RZ-AB0B,
V3-A9ZX,
V3-A9ZY,
V4-A9E5,
V4-A9E7,
V4-A9E8,
V4-A9E9,
V4-A9EA,
V4-A9EC,
V4-A9ED,
V4-A9EE,
V4-A9EF,
V4-A9EH,
V4-A9EI,
V4-A9EJ,
V4-A9EK,
V4-A9EL,
V4-A9EM,
V4-A9EO,
V4-A9EQ,
V4-A9ES,
V4-A9ET,
V4-A9EU,
V4-A9EV,
V4-A9EW,
V4-A9EX,
V4-A9EY,
V4-A9EZ,
V4-A9F0,
V4-A9F1,
V4-A9F2,
V4-A9F3,
V4-A9F4,
V4-A9F5,
V4-A9F7,
V4-A9F8,
VD-A8K7,
VD-A8K8,
VD-A8K9,
VD-A8KA,
VD-A8KB,
VD-A8KD,
VD-A8KE,
VD-A8KF,
VD-A8KG,
VD-A8KH,
VD-A8KI,
VD-A8KJ,
VD-A8KK,
VD-A8KL,
VD-A8KM,
VD-A8KN,
VD-A8KO,
VD-AA8M,
VD-AA8N,
VD-AA8O,
VD-AA8P,
VD-AA8Q,
VD-AA8R,
VD-AA8S,
VD-AA8T,
WC-A87T,
WC-A87U,
WC-A87W,
WC-A87Y,
WC-A880,
WC-A881,
WC-A882,
WC-A883,
WC-A884,
WC-A885,
WC-A888,
WC-A88A,
WC-AA9A,
WC-AA9E,
YZ-A980,
YZ-A982,
YZ-A983,
YZ-A984,
YZ-A985.\\
$\blacklozenge$ {\bf Nasopharyngeal Cancer} (11 samples):\\
$\bullet$ Source L2 = \cite{Lin}:\\
NPC088D,
NPC105D,
NPC29F,
NPC31F,
NPC34F,
NPC3F,
NPC42F,
NPC4D,
NPC4F,
NPC5D,
NPC5F.\\
$\blacklozenge$ {\bf Oral Cancer} (106 samples):\\
$\bullet$ Source I2 = \cite{India.ICGC}. Sample IDs are of the form OSCC-GB\_0*, where * is:\\
001011,
002011,
003011,
004011,
005011,
006011,
007011,
008011,
011011,
012011,
013011,
014011,
015011,
016011,
017011,
018011,
019011,
020011,
021011,
022011,
023011,
024011,
025011,
026011,
027011,
028011,
029011,
030011,
031011,
032011,
033011,
034011,
035011,
036011,
037011,
038011,
039011,
040011,
041011,
042011,
043011,
044011,
045011,
046011,
047011,
048011,
049011,
050011,
051011,
052011,
053011,
054011,
055011,
056011,
057011,
058011,
059011,
060011,
061011,
062011,
063011,
064011,
065011,
066011,
067011,
068011,
069011,
070011,
073011,
074011,
075011,
076011,
077011,
080011,
081011,
082011,
083011,
084011,
085011,
086011,
087011,
088011,
089011,
090011,
091011,
092011,
093011,
094011,
095011,
096011,
097011,
098011,
099011,
100011,
101011,
102011,
103011,
104011,
105011,
106011,
107011,
108011,
109011,
110011,
111011,
112011.\\
$\blacklozenge$ {\bf Ovarian Cancer} (471 samples):\\
$\bullet$ Source J1 = \cite{Jones1}:\\
OCC01PT,
OCC02PT,
OCC03PT,
OCC04PT,
OCC05PT,
OCC06PT,
OCC07PT,
OCC08PT.\\
$\bullet$ Source T16 = TCGA (see Acknowledgments). Sample IDs are of the form TCGA-*, where * is:\\
04-1331,
04-1332,
04-1336,
04-1337,
04-1338,
04-1342,
04-1343,
04-1346,
04-1347,
04-1348,
04-1349,
04-1350,
04-1351,
04-1353,
04-1356,
04-1357,
04-1361,
04-1362,
04-1364,
04-1365,
04-1367,
04-1369,
04-1514,
04-1516,
04-1517,
04-1519,
04-1525,
04-1530,
04-1542,
04-1638,
04-1644,
04-1646,
04-1648,
04-1649,
04-1651,
04-1652,
04-1655,
09-0364,
09-0365,
09-0366,
09-0367,
09-0369,
09-1659,
09-1661,
09-1662,
09-1664,
09-1665,
09-1666,
09-1669,
09-1670,
09-1672,
09-1673,
09-1674,
09-1675,
09-2044,
09-2045,
09-2049,
09-2050,
09-2051,
09-2053,
09-2056,
10-0926,
10-0927,
10-0928,
10-0930,
10-0931,
10-0933,
10-0934,
10-0935,
10-0937,
10-0938,
13-0714,
13-0717,
13-0720,
13-0723,
13-0724,
13-0726,
13-0727,
13-0730,
13-0751,
13-0755,
13-0758,
13-0760,
13-0761,
13-0762,
13-0765,
13-0791,
13-0792,
13-0793,
13-0795,
13-0800,
13-0801,
13-0804,
13-0807,
13-0883,
13-0884,
13-0885,
13-0886,
13-0887,
13-0889,
13-0890,
13-0891,
13-0893,
13-0894,
13-0897,
13-0899,
13-0900,
13-0901,
13-0903,
13-0904,
13-0905,
13-0906,
13-0910,
13-0911,
13-0912,
13-0913,
13-0916,
13-0919,
13-0920,
13-0923,
13-0924,
13-1403,
13-1404,
13-1405,
13-1407,
13-1408,
13-1409,
13-1410,
13-1411,
13-1412,
13-1477,
13-1481,
13-1482,
13-1483,
13-1484,
13-1487,
13-1488,
13-1489,
13-1491,
13-1492,
13-1494,
13-1495,
13-1496,
13-1497,
13-1498,
13-1499,
13-1501,
13-1504,
13-1505,
13-1506,
13-1507,
13-1509,
13-1510,
13-1512,
13-2057,
13-2059,
13-2060,
13-2061,
13-2065,
13-2066,
13-2071,
20-0987,
20-0990,
20-0991,
20-1682,
20-1683,
20-1684,
20-1685,
20-1686,
20-1687,
23-1021,
23-1022,
23-1023,
23-1024,
23-1026,
23-1027,
23-1028,
23-1029,
23-1030,
23-1031,
23-1032,
23-1109,
23-1110,
23-1111,
23-1114,
23-1116,
23-1117,
23-1118,
23-1119,
23-1120,
23-1122,
23-1123,
23-1124,
23-1809,
23-2072,
23-2077,
23-2078,
23-2079,
23-2081,
23-2641,
23-2643,
23-2645,
23-2647,
23-2649,
24-0966,
24-0968,
24-0970,
24-0975,
24-0979,
24-0980,
24-0982,
24-1103,
24-1104,
24-1105,
24-1413,
24-1416,
24-1417,
24-1418,
24-1419,
24-1422,
24-1423,
24-1424,
24-1425,
24-1426,
24-1427,
24-1428,
24-1431,
24-1434,
24-1435,
24-1436,
24-1463,
24-1464,
24-1466,
24-1469,
24-1470,
24-1471,
24-1474,
24-1544,
24-1545,
24-1546,
24-1548,
24-1549,
24-1551,
24-1552,
24-1553,
24-1555,
24-1556,
24-1557,
24-1558,
24-1560,
24-1562,
24-1563,
24-1564,
24-1565,
24-1567,
24-1603,
24-1604,
24-1614,
24-1616,
24-1842,
24-1843,
24-1844,
24-1845,
24-1846,
24-1847,
24-1849,
24-1850,
24-2019,
24-2024,
24-2030,
24-2035,
24-2038,
24-2254,
24-2260,
24-2261,
24-2262,
24-2267,
24-2271,
24-2280,
24-2281,
24-2288,
24-2289,
24-2290,
24-2293,
24-2298,
25-1313,
25-1315,
25-1316,
25-1317,
25-1318,
25-1319,
25-1320,
25-1321,
25-1322,
25-1324,
25-1325,
25-1326,
25-1328,
25-1329,
25-1623,
25-1625,
25-1626,
25-1627,
25-1628,
25-1630,
25-1631,
25-1632,
25-1633,
25-1634,
25-1635,
25-2042,
25-2391,
25-2392,
25-2393,
25-2396,
25-2398,
25-2399,
25-2400,
25-2401,
25-2404,
25-2408,
25-2409,
29-1688,
29-1690,
29-1691,
29-1693,
29-1694,
29-1695,
29-1696,
29-1697,
29-1698,
29-1699,
29-1701,
29-1702,
29-1703,
29-1705,
29-1707,
29-1710,
29-1711,
29-1761,
29-1762,
29-1763,
29-1764,
29-1766,
29-1768,
29-1769,
29-1770,
29-1771,
29-1774,
29-1775,
29-1776,
29-1777,
29-1778,
29-1781,
29-1783,
29-1784,
29-1785,
29-2427,
29-2429,
29-2431,
29-2432,
29-2434,
29-2436,
30-1714,
30-1718,
30-1853,
30-1855,
30-1856,
30-1857,
31-1950,
36-1568,
36-1569,
36-1570,
36-1571,
36-1574,
36-1575,
36-1576,
36-1577,
36-1578,
36-1580,
36-2530,
36-2532,
36-2533,
36-2534,
36-2537,
36-2538,
36-2539,
36-2540,
36-2542,
36-2543,
36-2544,
36-2545,
36-2547,
36-2548,
36-2551,
36-2552,
42-2582,
42-2587,
42-2588,
42-2589,
42-2590,
42-2591,
57-1582,
57-1584,
57-1586,
57-1993,
59-2348,
59-2350,
59-2351,
59-2352,
59-2354,
59-2355,
59-2363,
59-2372,
61-1722,
61-1725,
61-1727,
61-1728,
61-1730,
61-1733,
61-1736,
61-1737,
61-1738,
61-1740,
61-1741,
61-1895,
61-1899,
61-1900,
61-1901,
61-1903,
61-1904,
61-1906,
61-1907,
61-1910,
61-1911,
61-1913,
61-1914,
61-1915,
61-1995,
61-1998,
61-2000,
61-2002,
61-2003,
61-2008,
61-2009,
61-2012,
61-2016,
61-2092,
61-2094,
61-2095,
61-2097,
61-2101,
61-2102,
61-2104,
61-2109,
61-2110,
61-2111,
61-2113,
61-2610,
61-2611,
61-2612,
61-2613,
61-2614.\\
$\blacklozenge$ {\bf Pancreatic Cancer} (184 samples):\\
$\bullet$ Source W2 = \cite{Wu}:\\
IPMN 11,
IPMN 12,
IPMN 20,
IPMN 21,
IPMN 36,
IPMN 4,
IPMN 41,
MCN 162,
MCN 163,
MCN 164,
MCN 166,
MCN 168,
MCN 169,
MCN 170,
SCA 14,
SCA 23,
SCA 27,
SCA 35,
SCA 37,
SCA 38,
SCA 40,
SPN 8.\\
$\bullet$ Source J2 = \cite{Jiao}. Sample IDs are of the form PanNET*, where * is:\\
10PT,
21PT,
23PT,
24PT,
25PT,
31PT,
36PT,
3PT,
7PT,
93PT.\\
$\bullet$ Source T17 = TCGA (see Acknowledgments). Sample IDs are of the form TCGA-*, where * is:\\
2L-AAQA,
2L-AAQE,
2L-AAQI,
2L-AAQJ,
2L-AAQL,
2L-AAQM,
3A-A9I5,
3A-A9I7,
3A-A9I9,
3A-A9IB,
3A-A9IC,
3A-A9IH,
3A-A9IJ,
3A-A9IL,
3A-A9IN,
3A-A9IO,
3A-A9IR,
3A-A9IS,
3A-A9IU,
3E-AAAY,
3E-AAAZ,
F2-6879,
F2-6880,
F2-7273,
F2-7276,
F2-A44G,
F2-A44H,
F2-A7TX,
F2-A8YN,
FB-A4P5,
FB-A4P6,
FB-A545,
FB-A5VM,
FB-A78T,
FB-A7DR,
FB-AAPS,
FQ-6551,
FQ-6552,
FQ-6553,
FQ-6554,
FQ-6555,
FQ-6558,
FQ-6559,
FZ-5919,
FZ-5920,
FZ-5921,
FZ-5922,
FZ-5923,
FZ-5924,
FZ-5926,
H6-8124,
H6-A45N,
H8-A6C1,
HV-A5A3,
HV-A5A4,
HV-A5A5,
HV-A5A6,
HV-A7OL,
HV-A7OP,
HV-AA8X,
HZ-7289,
HZ-7918,
HZ-7919,
HZ-7920,
HZ-7922,
HZ-7923,
HZ-7924,
HZ-7925,
HZ-7926,
HZ-8001,
HZ-8002,
HZ-8003,
HZ-8005,
HZ-8315,
HZ-8317,
HZ-8636,
HZ-8637,
HZ-8638,
HZ-A49G,
HZ-A49H,
HZ-A49I,
HZ-A4BH,
HZ-A4BK,
HZ-A77O,
HZ-A77P,
HZ-A77Q,
HZ-A8P0,
HZ-A8P1,
IB-7644,
IB-7645,
IB-7646,
IB-7647,
IB-7649,
IB-7651,
IB-7652,
IB-7654,
IB-7885,
IB-7886,
IB-7887,
IB-7888,
IB-7889,
IB-7890,
IB-7891,
IB-7893,
IB-7897,
IB-8126,
IB-8127,
IB-A5SO,
IB-A5SP,
IB-A5SQ,
IB-A5SS,
IB-A5ST,
IB-A6UF,
IB-A6UG,
IB-A7LX,
IB-A7M4,
IB-AAUM,
IB-AAUN,
IB-AAUO,
IB-AAUP,
IB-AAUR,
IB-AAUS,
IB-AAUT,
IB-AAUU,
IB-AAUV,
IB-AAUW,
LB-A7SX,
LB-A8F3,
LB-A9Q5,
M8-A5N4,
OE-A75W,
PZ-A5RE,
Q3-A5QY,
Q3-AA2A,
RB-A7B8,
RB-AA9M,
RL-AAAS,
S4-A8RM,
S4-A8RO,
S4-A8RP,
US-A774,
US-A776,
US-A779,
US-A77E,
US-A77G,
US-A77J,
XD-AAUL,
XN-A8T3,
XN-A8T5,
YB-A89D,
YH-A8SY,
YY-A8LH.\\
$\blacklozenge$ {\bf Pheochromocytoma and Paraganglioma} (178 samples):\\
$\bullet$ Source T18 = TCGA (see Acknowledgments). Sample IDs are of the form TCGA-*, where * is:\\
P7-A5NX,
P7-A5NY,
P8-A5KC,
P8-A5KD,
P8-A6RX,
P8-A6RY,
PR-A5PF,
PR-A5PG,
PR-A5PH,
QR-A6GO,
QR-A6GR,
QR-A6GS,
QR-A6GT,
QR-A6GU,
QR-A6GW,
QR-A6GX,
QR-A6GY,
QR-A6GZ,
QR-A6H0,
QR-A6H1,
QR-A6H2,
QR-A6H3,
QR-A6H4,
QR-A6H5,
QR-A6H6,
QR-A6ZZ,
QR-A702,
QR-A703,
QR-A705,
QR-A706,
QR-A707,
QR-A708,
QR-A70A,
QR-A70C,
QR-A70D,
QR-A70E,
QR-A70G,
QR-A70H,
QR-A70I,
QR-A70J,
QR-A70K,
QR-A70M,
QR-A70N,
QR-A70O,
QR-A70P,
QR-A70Q,
QR-A70R,
QR-A70T,
QR-A70U,
QR-A70V,
QR-A70W,
QR-A70X,
QR-A7IN,
QR-A7IP,
QT-A5XJ,
QT-A5XK,
QT-A5XL,
QT-A5XM,
QT-A5XN,
QT-A5XO,
QT-A5XP,
QT-A69Q,
QT-A7U0,
RM-A68T,
RM-A68W,
RT-A6Y9,
RT-A6YA,
RT-A6YC,
RW-A67V,
RW-A67W,
RW-A67X,
RW-A67Y,
RW-A680,
RW-A681,
RW-A684,
RW-A685,
RW-A686,
RW-A688,
RW-A689,
RW-A68A,
RW-A68B,
RW-A68C,
RW-A68D,
RW-A68F,
RW-A68G,
RW-A7CZ,
RW-A7D0,
RW-A8AZ,
RX-A8JQ,
S7-A7WL,
S7-A7WM,
S7-A7WN,
S7-A7WO,
S7-A7WP,
S7-A7WQ,
S7-A7WR,
S7-A7WT,
S7-A7WU,
S7-A7WV,
S7-A7WW,
S7-A7WX,
S7-A7X0,
S7-A7X1,
S7-A7X2,
SA-A6C2,
SP-A6QC,
SP-A6QD,
SP-A6QF,
SP-A6QG,
SP-A6QH,
SP-A6QI,
SP-A6QJ,
SP-A6QK,
SQ-A6I4,
SQ-A6I6,
SR-A6MP,
SR-A6MQ,
SR-A6MR,
SR-A6MS,
SR-A6MT,
SR-A6MU,
SR-A6MV,
SR-A6MX,
SR-A6MY,
SR-A6MZ,
SR-A6N0,
TT-A6YJ,
TT-A6YK,
TT-A6YN,
TT-A6YO,
TT-A6YP,
W2-A7H5,
W2-A7H7,
W2-A7HA,
W2-A7HB,
W2-A7HC,
W2-A7HD,
W2-A7HE,
W2-A7HF,
W2-A7HH,
W2-A7UY,
WB-A80K,
WB-A80L,
WB-A80M,
WB-A80N,
WB-A80O,
WB-A80P,
WB-A80Q,
WB-A80V,
WB-A80Y,
WB-A814,
WB-A815,
WB-A816,
WB-A817,
WB-A818,
WB-A819,
WB-A81A,
WB-A81D,
WB-A81E,
WB-A81F,
WB-A81G,
WB-A81H,
WB-A81I,
WB-A81J,
WB-A81K,
WB-A81M,
WB-A81N,
WB-A81P,
WB-A81Q,
WB-A81R,
WB-A81S,
WB-A81T,
WB-A81V,
WB-A81W,
WB-A820,
WB-A821,
WB-A822,
XG-A823.\\
$\blacklozenge$ {\bf Prostate Cancer} (480 samples):\\
$\bullet$ Source B2 = \cite{Barbieri}. Sample IDs are of the form P0*-Tumor, where * is:\\
0-000450,
1-28,
2-1562,
2-2035,
3-1334,
3-1426,
3-1906,
3-2345,
3-2620,
3-3391,
3-595,
3-871,
4-1084,
4-1243,
4-1421,
4-1790,
4-2599,
4-2641,
4-2666,
4-2740,
4-47,
4-594,
5-2212,
5-2594,
5-3436,
5-3829,
5-3852,
5-3859,
5-620,
6-1125,
6-1696,
6-2325,
6-3676,
6-3939,
6-4428,
7-144,
7-360,
7-5036,
7-684,
7-718,
7-837,
8-2516,
8-590,
9-120,
9-1372,
9-1580,
9-2497,
9-649.\\
$\bullet$ Source B3 = \cite{Berger1}. Sample IDs are of the form PR-*, where * is:\\
0508,
0581,
1701,
1783,
2832,
3027,
3043.\\
Remaining sample IDs are of the form PR-*-Tumor, where * is:\\
00-1165,
00-160,
00-1823,
0099,
01-1934,
01-2382,
01-2492,
01-2554,
02-1082,
02-169,
02-1736,
02-1899,
02-2072,
02-2480,
02-254,
03-022,
03-1026,
03-870,
04-1367,
04-194,
04-3113,
04-3222,
04-3347,
04-639,
04-903,
0415,
0427,
05-3440,
05-3595,
05-839,
06-1651,
06-1749,
06-1999,
09-2517,
09-2744,
09-2767,
09-3421,
09-3566,
09-3687,
09-5094,
09-5245,
09-5446,
09-5630,
09-5700,
09-5702,
1024,
1043,
2661,
2682,
2740,
2761,
2762,
2858,
2872,
2915,
2916,
3023,
3026,
3034,
3035,
3036,
3048,
3051,
3127.\\
$\bullet$ Source G2 = \cite{Grasso1} (below * stands for WA, e.g., *10 = WA10):\\
T12,
T32,
T8,
T90,
T91,
T92,
T93,
T94,
T95,
T96,
T97,
*10,
*11,
*12,
*13,
*14,
*15,
*16,
*17,
*18,
*19,
*20,
*22,
*23,
*24,
*25,
*26,
*27,
*28,
*29,
*3,
*30,
*31,
*32,
*33,
*35,
*37,
*38,
*39,
*40,
*41,
*42,
*43-27,
*43-44,
*43-71,
*46,
*47,
*48,
*49,
*50,
*51,
*52,
*53,
*54,
*55,
*56,
*57,
*58,
*59,
*60,
*7.\\
$\bullet$ Source T19 = TCGA (see Acknowledgments). Sample IDs are of the form TCGA-*, where * is:\\
CH-5737,
CH-5738,
CH-5739,
CH-5740,
CH-5741,
CH-5743,
CH-5744,
CH-5745,
CH-5746,
CH-5748,
CH-5750,
CH-5751,
CH-5752,
CH-5753,
CH-5754,
CH-5761,
CH-5762,
CH-5763,
CH-5764,
CH-5765,
CH-5766,
CH-5767,
CH-5768,
CH-5769,
CH-5771,
CH-5772,
CH-5788,
CH-5789,
CH-5790,
CH-5791,
CH-5792,
CH-5794,
EJ-5494,
EJ-5495,
EJ-5496,
EJ-5497,
EJ-5498,
EJ-5499,
EJ-5501,
EJ-5502,
EJ-5503,
EJ-5504,
EJ-5505,
EJ-5506,
EJ-5507,
EJ-5508,
EJ-5509,
EJ-5510,
EJ-5511,
EJ-5512,
EJ-5514,
EJ-5515,
EJ-5516,
EJ-5517,
EJ-5518,
EJ-5519,
EJ-5521,
EJ-5522,
EJ-5524,
EJ-5525,
EJ-5526,
EJ-5527,
EJ-5530,
EJ-5531,
EJ-5532,
EJ-5542,
EJ-7115,
EJ-7123,
EJ-7125,
EJ-7218,
EJ-7312,
EJ-7314,
EJ-7315,
EJ-7317,
EJ-7318,
EJ-7321,
EJ-7325,
EJ-7327,
EJ-7328,
EJ-7330,
EJ-7331,
EJ-7781,
EJ-7782,
EJ-7783,
EJ-7784,
EJ-7785,
EJ-7786,
EJ-7788,
EJ-7789,
EJ-7791,
EJ-7792,
EJ-7793,
EJ-7794,
EJ-7797,
EJ-8468,
EJ-8469,
EJ-8470,
EJ-8472,
EJ-8474,
EJ-A46B,
EJ-A46D,
EJ-A46E,
EJ-A46F,
EJ-A46G,
EJ-A46H,
EJ-A46I,
EJ-A65B,
EJ-A65D,
EJ-A65E,
EJ-A65F,
EJ-A65G,
EJ-A65J,
EJ-A65M,
EJ-A6RA,
EJ-A6RC,
EJ-A7NF,
EJ-A7NG,
EJ-A7NH,
EJ-A7NM,
EJ-A7NN,
FC-7708,
FC-7961,
FC-A4JI,
FC-A5OB,
FC-A66V,
FC-A6HD,
G9-6329,
G9-6332,
G9-6333,
G9-6336,
G9-6338,
G9-6339,
G9-6342,
G9-6343,
G9-6347,
G9-6348,
G9-6351,
G9-6353,
G9-6354,
G9-6356,
G9-6361,
G9-6362,
G9-6363,
G9-6364,
G9-6365,
G9-6366,
G9-6367,
G9-6369,
G9-6370,
G9-6371,
G9-6373,
G9-6377,
G9-6378,
G9-6379,
G9-6384,
G9-6385,
G9-6494,
G9-6496,
G9-6498,
G9-6499,
G9-7510,
G9-7519,
G9-7521,
G9-7522,
G9-7523,
G9-7525,
H9-7775,
H9-A6BX,
H9-A6BY,
HC-7075,
HC-7077,
HC-7078,
HC-7079,
HC-7080,
HC-7081,
HC-7209,
HC-7210,
HC-7211,
HC-7212,
HC-7213,
HC-7230,
HC-7231,
HC-7232,
HC-7233,
HC-7736,
HC-7737,
HC-7738,
HC-7740,
HC-7742,
HC-7744,
HC-7745,
HC-7747,
HC-7748,
HC-7749,
HC-7750,
HC-7752,
HC-7817,
HC-7818,
HC-7819,
HC-7820,
HC-7821,
HC-8213,
HC-8216,
HC-8256,
HC-8257,
HC-8258,
HC-8259,
HC-8260,
HC-8261,
HC-8262,
HC-8264,
HC-8265,
HC-8266,
HC-A48F,
HC-A4ZV,
HC-A631,
HC-A632,
HC-A6AL,
HC-A6AN,
HC-A6AO,
HC-A6AP,
HC-A6AQ,
HC-A6AS,
HC-A6HX,
HC-A6HY,
HC-A76W,
HC-A76X,
HI-7168,
HI-7169,
HI-7170,
HI-7171,
J4-8198,
J4-8200,
J4-A67K,
J4-A67L,
J4-A67M,
J4-A67N,
J4-A67O,
J4-A67Q,
J4-A67R,
J4-A67S,
J4-A67T,
J4-A6G1,
J4-A6G3,
J4-A6M7,
J9-A52B,
J9-A52C,
J9-A52D,
J9-A52E,
KC-A4BL,
KC-A4BN,
KC-A4BO,
KC-A4BR,
KC-A4BV,
KC-A7F3,
KC-A7F5,
KC-A7F6,
KC-A7FA,
KC-A7FD,
KC-A7FE,
KK-A59V,
KK-A59X,
KK-A59Y,
KK-A59Z,
KK-A5A1,
KK-A6DY,
KK-A6E0,
KK-A6E1,
KK-A6E2,
KK-A6E3,
KK-A6E4,
KK-A6E5,
KK-A6E6,
KK-A6E7,
KK-A6E8,
KK-A7AP,
KK-A7AQ,
KK-A7AU,
KK-A7AV,
KK-A7AW,
KK-A7AY,
KK-A7AZ,
KK-A7B0,
KK-A7B1,
KK-A7B2,
KK-A7B3,
KK-A7B4,
M7-A71Y,
M7-A71Z,
M7-A720,
M7-A721,
M7-A723,
M7-A724,
M7-A725,
QU-A6IL,
QU-A6IM,
QU-A6IN,
QU-A6IO,
QU-A6IP,
SU-A7E7.\\
$\blacklozenge$ {\bf Rectum Adenocarcinoma} (115 samples):\\
$\bullet$ Source T20 = TCGA (see Acknowledgments). Sample IDs are of the form TCGA-*, where * is:\\
AF-2687,
AF-2689,
AF-2690,
AF-2691,
AF-2692,
AF-2693,
AF-3400,
AF-3911,
AF-4110,
AF-5654,
AF-6136,
AF-6655,
AF-6672,
AG-3574,
AG-3575,
AG-3578,
AG-3580,
AG-3581,
AG-3582,
AG-3583,
AG-3584,
AG-3586,
AG-3587,
AG-3591,
AG-3592,
AG-3593,
AG-3594,
AG-3598,
AG-3599,
AG-3600,
AG-3601,
AG-3602,
AG-3605,
AG-3608,
AG-3609,
AG-3611,
AG-3612,
AG-3725,
AG-3731,
AG-3732,
AG-3742,
AG-4021,
AG-4022,
AG-A002,
AG-A008,
AG-A00C,
AG-A00Y,
AG-A011,
AG-A014,
AG-A015,
AG-A016,
AG-A01L,
AH-6544,
AH-6643,
AH-6644,
AH-6897,
AH-6903,
BM-6198,
CI-6619,
CI-6620,
CI-6621,
CI-6622,
CI-6624,
CL-4957,
CL-5917,
CL-5918,
DC-4749,
DC-5337,
DC-5869,
DC-6154,
DC-6155,
DC-6157,
DC-6158,
DC-6681,
DC-6682,
DC-6683,
DT-5265,
DY-A0XA,
DY-A1DC,
DY-A1DD,
DY-A1DF,
DY-A1DG,
DY-A1H8,
EF-5830,
EI-6506,
EI-6507,
EI-6508,
EI-6509,
EI-6510,
EI-6511,
EI-6512,
EI-6513,
EI-6514,
EI-6881,
EI-6882,
EI-6883,
EI-6884,
EI-6885,
EI-6917,
EI-7002,
EI-7004,
F5-6464,
F5-6465,
F5-6571,
F5-6702,
F5-6812,
F5-6813,
F5-6814,
F5-6861,
F5-6863,
F5-6864,
G5-6233,
G5-6235,
G5-6572,
G5-6641.\\
$\blacklozenge$ {\bf Renal Cell Carcinoma} (709 samples):\\
$\bullet$ Source G3 = \cite{Guo1}:\\
K1,
K20,
K27,
K29,
K3,
K31,
K32,
K38,
K44,
K48,
T127,
T142,
T144,
T163,
T164,
T166,
T183M.\\
$\bullet$ Source T21 = TCGA (see Acknowledgments). Sample IDs are of the form TCGA-*, where * is:\\
A3-3308,
A3-3311,
A3-3313,
A3-3316,
A3-3317,
A3-3319,
A3-3320,
A3-3322,
A3-3323,
A3-3324,
A3-3326,
A3-3331,
A3-3346,
A3-3347,
A3-3349,
A3-3351,
A3-3357,
A3-3358,
A3-3362,
A3-3363,
A3-3365,
A3-3367,
A3-3370,
A3-3372,
A3-3373,
A3-3374,
A3-3376,
A3-3378,
A3-3380,
A3-3382,
A3-3383,
A3-3385,
A3-3387,
A4-7286,
A4-7287,
A4-7288,
A4-7583,
A4-7584,
A4-7585,
A4-7732,
A4-7734,
A4-7828,
A4-7915,
A4-7996,
A4-7997,
A4-8098,
A4-8310,
A4-8311,
A4-8312,
A4-8515,
A4-8516,
A4-8517,
A4-8518,
A4-8630,
A4-A48D,
A4-A4ZT,
A4-A57E,
A4-A5DU,
A4-A5XZ,
A4-A5Y0,
A4-A5Y1,
A4-A6HP,
AK-3425,
AK-3427,
AK-3428,
AK-3429,
AK-3430,
AK-3431,
AK-3434,
AK-3436,
AK-3440,
AK-3443,
AK-3444,
AK-3445,
AK-3447,
AK-3450,
AK-3451,
AK-3453,
AK-3454,
AK-3455,
AK-3456,
AK-3458,
AK-3460,
AK-3461,
AK-3465,
AL-3466,
AL-3467,
AL-3468,
AL-3472,
AL-3473,
AL-7173,
AL-A5DJ,
AS-3777,
AS-3778,
AT-A5NU,
B0-4690,
B0-4691,
B0-4693,
B0-4694,
B0-4697,
B0-4700,
B0-4703,
B0-4706,
B0-4707,
B0-4710,
B0-4712,
B0-4713,
B0-4714,
B0-4718,
B0-4810,
B0-4811,
B0-4813,
B0-4814,
B0-4815,
B0-4816,
B0-4817,
B0-4818,
B0-4819,
B0-4822,
B0-4823,
B0-4824,
B0-4827,
B0-4828,
B0-4833,
B0-4836,
B0-4837,
B0-4838,
B0-4839,
B0-4841,
B0-4842,
B0-4843,
B0-4844,
B0-4845,
B0-4846,
B0-4847,
B0-4848,
B0-4849,
B0-4852,
B0-4945,
B0-5075,
B0-5077,
B0-5080,
B0-5081,
B0-5083,
B0-5084,
B0-5085,
B0-5088,
B0-5092,
B0-5094,
B0-5095,
B0-5096,
B0-5097,
B0-5098,
B0-5099,
B0-5100,
B0-5102,
B0-5104,
B0-5106,
B0-5107,
B0-5108,
B0-5109,
B0-5110,
B0-5113,
B0-5115,
B0-5116,
B0-5117,
B0-5119,
B0-5120,
B0-5121,
B0-5399,
B0-5400,
B0-5402,
B0-5691,
B0-5692,
B0-5693,
B0-5694,
B0-5695,
B0-5696,
B0-5697,
B0-5698,
B0-5699,
B0-5701,
B0-5702,
B0-5703,
B0-5705,
B0-5706,
B0-5707,
B0-5709,
B0-5710,
B0-5711,
B0-5712,
B0-5713,
B0-5812,
B1-5398,
B1-A47M,
B1-A47N,
B1-A47O,
B1-A654,
B1-A655,
B1-A656,
B1-A657,
B2-3923,
B2-3924,
B2-4098,
B2-4099,
B2-4101,
B2-4102,
B2-5633,
B2-5635,
B2-5641,
B3-3925,
B3-3926,
B3-4103,
B3-4104,
B3-8121,
B4-5377,
B4-5832,
B4-5834,
B4-5835,
B4-5836,
B4-5838,
B4-5843,
B4-5844,
B8-4143,
B8-4146,
B8-4148,
B8-4151,
B8-4153,
B8-4154,
B8-4619,
B8-4620,
B8-4621,
B8-4622,
B8-5158,
B8-5159,
B8-5162,
B8-5163,
B8-5164,
B8-5165,
B8-5545,
B8-5546,
B8-5549,
B8-5550,
B8-5551,
B8-5552,
B8-5553,
B9-4113,
B9-4114,
B9-4115,
B9-4116,
B9-4117,
B9-4617,
B9-5155,
B9-5156,
B9-7268,
B9-A44B,
B9-A5W7,
B9-A5W8,
B9-A5W9,
B9-A69E,
BP-4158,
BP-4159,
BP-4160,
BP-4161,
BP-4162,
BP-4163,
BP-4164,
BP-4165,
BP-4166,
BP-4167,
BP-4169,
BP-4170,
BP-4173,
BP-4174,
BP-4176,
BP-4177,
BP-4326,
BP-4329,
BP-4330,
BP-4331,
BP-4337,
BP-4338,
BP-4340,
BP-4341,
BP-4342,
BP-4343,
BP-4345,
BP-4346,
BP-4347,
BP-4349,
BP-4351,
BP-4352,
BP-4354,
BP-4355,
BP-4756,
BP-4758,
BP-4759,
BP-4760,
BP-4761,
BP-4762,
BP-4763,
BP-4765,
BP-4766,
BP-4768,
BP-4770,
BP-4771,
BP-4774,
BP-4775,
BP-4777,
BP-4781,
BP-4782,
BP-4787,
BP-4789,
BP-4790,
BP-4795,
BP-4797,
BP-4798,
BP-4799,
BP-4801,
BP-4803,
BP-4804,
BP-4807,
BP-4960,
BP-4961,
BP-4962,
BP-4963,
BP-4964,
BP-4965,
BP-4967,
BP-4968,
BP-4969,
BP-4970,
BP-4971,
BP-4972,
BP-4973,
BP-4974,
BP-4975,
BP-4976,
BP-4977,
BP-4981,
BP-4982,
BP-4983,
BP-4985,
BP-4986,
BP-4987,
BP-4988,
BP-4989,
BP-4991,
BP-4992,
BP-4993,
BP-4994,
BP-4995,
BP-4998,
BP-4999,
BP-5000,
BP-5001,
BP-5004,
BP-5006,
BP-5007,
BP-5008,
BP-5009,
BP-5010,
BP-5168,
BP-5169,
BP-5170,
BP-5173,
BP-5174,
BP-5175,
BP-5176,
BP-5177,
BP-5178,
BP-5180,
BP-5181,
BP-5182,
BP-5183,
BP-5184,
BP-5185,
BP-5186,
BP-5187,
BP-5189,
BP-5190,
BP-5191,
BP-5192,
BP-5194,
BP-5195,
BP-5196,
BP-5198,
BP-5199,
BP-5200,
BP-5201,
BP-5202,
BQ-5875,
BQ-5876,
BQ-5877,
BQ-5878,
BQ-5879,
BQ-5880,
BQ-5881,
BQ-5882,
BQ-5883,
BQ-5884,
BQ-5885,
BQ-5886,
BQ-5887,
BQ-5888,
BQ-5889,
BQ-5890,
BQ-5891,
BQ-5892,
BQ-5893,
BQ-5894,
BQ-7044,
BQ-7045,
BQ-7046,
BQ-7048,
BQ-7049,
BQ-7050,
BQ-7051,
BQ-7053,
BQ-7055,
BQ-7056,
BQ-7058,
BQ-7059,
BQ-7060,
BQ-7061,
BQ-7062,
CJ-4634,
CJ-4635,
CJ-4636,
CJ-4637,
CJ-4638,
CJ-4639,
CJ-4640,
CJ-4641,
CJ-4643,
CJ-4644,
CJ-4868,
CJ-4869,
CJ-4870,
CJ-4871,
CJ-4872,
CJ-4873,
CJ-4874,
CJ-4875,
CJ-4876,
CJ-4878,
CJ-4881,
CJ-4882,
CJ-4884,
CJ-4885,
CJ-4886,
CJ-4887,
CJ-4888,
CJ-4889,
CJ-4890,
CJ-4891,
CJ-4892,
CJ-4893,
CJ-4894,
CJ-4895,
CJ-4897,
CJ-4899,
CJ-4900,
CJ-4901,
CJ-4902,
CJ-4903,
CJ-4904,
CJ-4905,
CJ-4907,
CJ-4908,
CJ-4912,
CJ-4913,
CJ-4916,
CJ-4918,
CJ-4920,
CJ-4923,
CJ-5671,
CJ-5672,
CJ-5675,
CJ-5676,
CJ-5677,
CJ-5678,
CJ-5679,
CJ-5680,
CJ-5681,
CJ-5682,
CJ-5683,
CJ-5684,
CJ-5686,
CJ-6027,
CJ-6028,
CJ-6030,
CJ-6031,
CJ-6032,
CJ-6033,
CW-5580,
CW-5581,
CW-5583,
CW-5584,
CW-5585,
CW-5588,
CW-5589,
CW-5591,
CW-6087,
CW-6090,
CW-6093,
CW-6097,
CZ-4853,
CZ-4854,
CZ-4856,
CZ-4857,
CZ-4858,
CZ-4859,
CZ-4861,
CZ-4862,
CZ-4863,
CZ-4865,
CZ-4866,
CZ-5451,
CZ-5452,
CZ-5453,
CZ-5454,
CZ-5455,
CZ-5456,
CZ-5457,
CZ-5458,
CZ-5459,
CZ-5460,
CZ-5461,
CZ-5462,
CZ-5463,
CZ-5464,
CZ-5465,
CZ-5466,
CZ-5467,
CZ-5468,
CZ-5469,
CZ-5470,
CZ-5982,
CZ-5984,
CZ-5985,
CZ-5986,
CZ-5987,
CZ-5988,
CZ-5989,
DV-5565,
DV-5566,
DV-5568,
DV-5569,
DV-5574,
DV-5575,
DV-5576,
DW-5560,
DW-5561,
DW-7834,
DW-7837,
DW-7838,
DW-7839,
DW-7840,
DW-7841,
DW-7842,
DW-7963,
DZ-6131,
DZ-6132,
DZ-6133,
DZ-6134,
DZ-6135,
EU-5904,
EU-5905,
EU-5906,
EU-5907,
EV-5901,
EV-5902,
EV-5903,
F9-A4JJ,
G7-6789,
G7-6790,
G7-6792,
G7-6793,
G7-6795,
G7-6796,
G7-6797,
G7-7501,
G7-7502,
G7-A4TM,
GL-6846,
GL-7773,
GL-7966,
GL-8500,
GL-A4EM,
GL-A59R,
GL-A59T,
HE-7128,
HE-7129,
HE-7130,
HE-A5NF,
HE-A5NH,
HE-A5NI,
HE-A5NJ,
HE-A5NK,
HE-A5NL,
IA-A40U,
IA-A40X,
IA-A40Y,
IZ-8195,
IZ-8196,
IZ-A6M8,
IZ-A6M9,
J7-6720,
J7-8537,
KL-8323,
KL-8324,
KL-8325,
KL-8326,
KL-8327,
KL-8328,
KL-8329,
KL-8330,
KL-8331,
KL-8332,
KL-8333,
KL-8334,
KL-8335,
KL-8336,
KL-8337,
KL-8338,
KL-8339,
KL-8340,
KL-8341,
KL-8342,
KL-8343,
KL-8344,
KL-8345,
KL-8346,
KM-8438,
KM-8439,
KM-8440,
KM-8441,
KM-8442,
KM-8443,
KM-8476,
KM-8477,
KM-8639,
KN-8418,
KN-8419,
KN-8421,
KN-8422,
KN-8423,
KN-8424,
KN-8425,
KN-8426,
KN-8427,
KN-8428,
KN-8429,
KN-8430,
KN-8431,
KN-8432,
KN-8433,
KN-8434,
KN-8435,
KN-8436,
KN-8437,
KO-8403,
KO-8404,
KO-8405,
KO-8406,
KO-8407,
KO-8408,
KO-8409,
KO-8410,
KO-8411,
KO-8413,
KO-8414,
KO-8415,
KO-8416,
KO-8417,
KV-A6GD,
KV-A6GE,
MH-A55W,
MH-A55Z,
MH-A560,
MH-A561,
MH-A562,
P4-A5E6,
P4-A5E7,
P4-A5E8,
P4-A5EA,
P4-A5EB,
P4-A5ED,
PJ-A5Z8,
PJ-A5Z9,
Q2-A5QZ.\\
$\blacklozenge$ {\bf Sarcoma} (255 samples):\\
$\bullet$ Source T22 = TCGA (see Acknowledgments). Sample IDs are of the form TCGA-*, where * is:\\
3B-A9HI,
3B-A9HJ,
3B-A9HL,
3B-A9HO,
3B-A9HP,
3B-A9HQ,
3B-A9HR,
3B-A9HS,
3B-A9HT,
3B-A9HU,
3B-A9HV,
3B-A9HX,
3B-A9HY,
3B-A9HZ,
3B-A9I0,
3B-A9I1,
3B-A9I3,
3R-A8YX,
DX-A1KU,
DX-A1KW,
DX-A1KX,
DX-A1KY,
DX-A1KZ,
DX-A1L0,
DX-A1L1,
DX-A1L2,
DX-A1L3,
DX-A1L4,
DX-A23R,
DX-A23T,
DX-A23U,
DX-A23V,
DX-A23Y,
DX-A240,
DX-A2IZ,
DX-A2J0,
DX-A2J1,
DX-A2J4,
DX-A3LS,
DX-A3LT,
DX-A3LU,
DX-A3LW,
DX-A3LY,
DX-A3M1,
DX-A3M2,
DX-A3U5,
DX-A3U6,
DX-A3U7,
DX-A3U8,
DX-A3U9,
DX-A3UA,
DX-A3UB,
DX-A3UC,
DX-A3UD,
DX-A3UE,
DX-A3UF,
DX-A48J,
DX-A48K,
DX-A48L,
DX-A48N,
DX-A48O,
DX-A48P,
DX-A48R,
DX-A48U,
DX-A48V,
DX-A6B7,
DX-A6B8,
DX-A6B9,
DX-A6BA,
DX-A6BB,
DX-A6BE,
DX-A6BF,
DX-A6BG,
DX-A6BH,
DX-A6BK,
DX-A6YQ,
DX-A6YR,
DX-A6YS,
DX-A6YT,
DX-A6YU,
DX-A6YV,
DX-A6YX,
DX-A6YZ,
DX-A6Z0,
DX-A6Z2,
DX-A7EF,
DX-A7EI,
DX-A7EL,
DX-A7EM,
DX-A7EN,
DX-A7EO,
DX-A7EQ,
DX-A7ER,
DX-A7ES,
DX-A7ET,
DX-A7EU,
DX-A8BG,
DX-A8BH,
DX-A8BJ,
DX-A8BK,
DX-A8BL,
DX-A8BM,
DX-A8BN,
DX-A8BO,
DX-A8BP,
DX-A8BR,
DX-A8BT,
DX-A8BU,
DX-A8BV,
DX-A8BX,
DX-A8BZ,
DX-AATS,
DX-AB2E,
DX-AB2F,
DX-AB2G,
DX-AB2H,
DX-AB2J,
DX-AB2L,
DX-AB2O,
DX-AB2P,
DX-AB2Q,
DX-AB2S,
DX-AB2T,
DX-AB2V,
DX-AB2W,
DX-AB2X,
DX-AB2Z,
DX-AB30,
DX-AB32,
DX-AB35,
DX-AB36,
DX-AB37,
DX-AB3A,
DX-AB3B,
DX-AB3C,
FX-A2QS,
FX-A3NJ,
FX-A3NK,
FX-A3RE,
FX-A3TO,
FX-A48G,
FX-A76Y,
FX-A8OO,
HB-A2OT,
HB-A3L4,
HB-A3YV,
HB-A43Z,
HB-A5W3,
HS-A5N7,
HS-A5N8,
HS-A5N9,
IE-A3OV,
IE-A4EH,
IE-A4EI,
IE-A4EJ,
IE-A4EK,
IE-A6BZ,
IF-A4AJ,
IF-A4AK,
IS-A3K6,
IS-A3K7,
IS-A3K8,
IS-A3KA,
IW-A3M4,
IW-A3M5,
IW-A3M6,
JV-A5VE,
JV-A5VF,
JV-A75J,
K1-A3PN,
K1-A3PO,
K1-A42W,
K1-A42X,
K1-A6RT,
K1-A6RU,
K1-A6RV,
KD-A5QS,
KD-A5QT,
KD-A5QU,
KF-A41W,
LI-A67I,
LI-A9QH,
MB-A5Y8,
MB-A5Y9,
MB-A5YA,
MB-A8JK,
MB-A8JL,
MJ-A68H,
MJ-A68J,
MJ-A850,
MO-A47P,
MO-A47R,
N1-A6IA,
PC-A5DK,
PC-A5DL,
PC-A5DM,
PC-A5DN,
PC-A5DO,
PC-A5DP,
QC-A6FX,
QC-A7B5,
QC-AA9N,
QQ-A5V2,
QQ-A5V9,
QQ-A5VA,
QQ-A5VB,
QQ-A5VC,
QQ-A5VD,
QQ-A8VB,
QQ-A8VD,
QQ-A8VF,
QQ-A8VG,
QQ-A8VH,
RN-A68Q,
RN-AAAQ,
SG-A6Z4,
SG-A6Z7,
SG-A849,
SI-A71O,
SI-A71P,
SI-A71Q,
SI-AA8B,
SI-AA8C,
UE-A6QT,
UE-A6QU,
VT-A80G,
VT-A80J,
VT-AB3D,
WK-A8XO,
WK-A8XQ,
WK-A8XS,
WK-A8XT,
WK-A8XX,
WK-A8XY,
WK-A8XZ,
WK-A8Y0,
WP-A9GB,
X2-A95T,
X6-A7W8,
X6-A7WA,
X6-A7WB,
X6-A7WC,
X6-A7WD,
X6-A8C2,
X6-A8C3,
X6-A8C4,
X6-A8C5,
X6-A8C6,
X6-A8C7,
X9-A971,
X9-A973,
Z4-A8JB,
Z4-A9VC,
Z4-AAPF,
Z4-AAPG.\\
$\blacklozenge$ {\bf Testicular Germ Cell Tumors} (150 samples):\\
$\bullet$ Source T23 = TCGA (see Acknowledgments). Sample IDs are of the form TCGA-*, where * is:\\
2G-AAEW,
2G-AAEX,
2G-AAF1,
2G-AAF4,
2G-AAF6,
2G-AAF8,
2G-AAFE,
2G-AAFG,
2G-AAFH,
2G-AAFI,
2G-AAFJ,
2G-AAFL,
2G-AAFM,
2G-AAFN,
2G-AAFO,
2G-AAFV,
2G-AAFY,
2G-AAFZ,
2G-AAG0,
2G-AAG3,
2G-AAG5,
2G-AAG6,
2G-AAG7,
2G-AAG8,
2G-AAG9,
2G-AAGA,
2G-AAGC,
2G-AAGE,
2G-AAGF,
2G-AAGG,
2G-AAGI,
2G-AAGJ,
2G-AAGK,
2G-AAGM,
2G-AAGN,
2G-AAGO,
2G-AAGP,
2G-AAGS,
2G-AAGT,
2G-AAGV,
2G-AAGW,
2G-AAGX,
2G-AAGY,
2G-AAGZ,
2G-AAH0,
2G-AAH2,
2G-AAH3,
2G-AAH4,
2G-AAH8,
2G-AAHA,
2G-AAHC,
2G-AAHG,
2G-AAHL,
2G-AAHN,
2G-AAHP,
2G-AAHT,
2G-AAKD,
2G-AAKG,
2G-AAKH,
2G-AAKL,
2G-AAKM,
2G-AAKO,
2G-AAL5,
2G-AAL7,
2G-AALF,
2G-AALG,
2G-AALN,
2G-AALO,
2G-AALP,
2G-AALQ,
2G-AALR,
2G-AALS,
2G-AALT,
2G-AALW,
2G-AALX,
2G-AALY,
2G-AALZ,
2G-AAM2,
2G-AAM3,
2G-AAM4,
2X-A9D5,
2X-A9D6,
4K-AA1G,
4K-AA1H,
4K-AA1I,
4K-AAAL,
S6-A8JW,
S6-A8JX,
S6-A8JY,
SB-A6J6,
SB-A76C,
SN-A6IS,
SN-A84W,
SN-A84X,
SN-A84Y,
SO-A8JP,
VF-A8A8,
VF-A8A9,
VF-A8AA,
VF-A8AB,
VF-A8AC,
VF-A8AD,
VF-A8AE,
W4-A7U2,
W4-A7U3,
W4-A7U4,
WZ-A7V3,
WZ-A7V4,
WZ-A7V5,
WZ-A8D5,
X3-A8G4,
XE-A8H1,
XE-A8H4,
XE-A8H5,
XE-A9SE,
XE-AANI,
XE-AANJ,
XE-AANR,
XE-AANV,
XE-AAO3,
XE-AAO4,
XE-AAO6,
XE-AAOB,
XE-AAOC,
XE-AAOD,
XE-AAOF,
XE-AAOJ,
XE-AAOL,
XY-A89B,
XY-A8S2,
XY-A8S3,
XY-A9T9,
YU-A90P,
YU-A90Q,
YU-A90S,
YU-A90W,
YU-A90Y,
YU-A912,
YU-A94D,
YU-A94I,
YU-AA4L,
YU-AA61,
ZM-AA05,
ZM-AA06,
ZM-AA0B,
ZM-AA0D,
ZM-AA0E,
ZM-AA0F,
ZM-AA0H,
ZM-AA0N.\\
$\blacklozenge$ {\bf Thymoma} (123 samples):\\
$\bullet$ Source T24 = TCGA (see Acknowledgments). Sample IDs are of the form TCGA-*, where * is:\\
3G-AB0O,
3G-AB0Q,
3G-AB0T,
3G-AB14,
3G-AB19,
3Q-A9WF,
3S-A8YW,
3S-AAYX,
3T-AA9L,
4V-A9QI,
4V-A9QJ,
4V-A9QL,
4V-A9QM,
4V-A9QN,
4V-A9QQ,
4V-A9QR,
4V-A9QS,
4V-A9QT,
4V-A9QU,
4V-A9QW,
4V-A9QX,
4X-A9F9,
4X-A9FA,
4X-A9FB,
4X-A9FC,
4X-A9FD,
5G-A9ZZ,
5K-AAAP,
5U-AB0D,
5U-AB0E,
5U-AB0F,
5V-A9RR,
X7-A8D6,
X7-A8D7,
X7-A8D8,
X7-A8D9,
X7-A8DB,
X7-A8DC,
X7-A8DD,
X7-A8DE,
X7-A8DF,
X7-A8DG,
X7-A8DI,
X7-A8DJ,
X7-A8M0,
X7-A8M1,
X7-A8M3,
X7-A8M4,
X7-A8M5,
X7-A8M6,
X7-A8M7,
X7-A8M8,
XH-A853,
XM-A8R8,
XM-A8R9,
XM-A8RB,
XM-A8RC,
XM-A8RD,
XM-A8RE,
XM-A8RF,
XM-A8RG,
XM-A8RH,
XM-A8RI,
XM-A8RL,
XM-AAZ1,
XM-AAZ2,
XM-AAZ3,
XU-A92O,
XU-A92Q,
XU-A92R,
XU-A92T,
XU-A92U,
XU-A92V,
XU-A92W,
XU-A92X,
XU-A92Y,
XU-A92Z,
XU-A930,
XU-A931,
XU-A932,
XU-A933,
XU-A936,
XU-AAXW,
XU-AAXX,
XU-AAXY,
XU-AAXZ,
XU-AAY0,
XU-AAY1,
YT-A95D,
YT-A95E,
YT-A95F,
YT-A95G,
YT-A95H,
ZB-A961,
ZB-A962,
ZB-A963,
ZB-A964,
ZB-A965,
ZB-A966,
ZB-A969,
ZB-A96A,
ZB-A96B,
ZB-A96C,
ZB-A96D,
ZB-A96E,
ZB-A96F,
ZB-A96G,
ZB-A96H,
ZB-A96I,
ZB-A96K,
ZB-A96L,
ZB-A96M,
ZB-A96O,
ZB-A96P,
ZB-A96Q,
ZB-A96R,
ZB-A96V,
ZC-AAA7,
ZC-AAAA,
ZC-AAAF,
ZC-AAAH,
ZL-A9V6,
ZT-A8OM.\\
$\blacklozenge$ {\bf Thyroid Cancer} (409 samples):\\
$\bullet$ Source T25 = TCGA (see Acknowledgments). Sample IDs are of the form TCGA-*, where * is:\\
BJ-A0YZ,
BJ-A0Z0,
BJ-A0Z2,
BJ-A0Z3,
BJ-A0Z5,
BJ-A0Z9,
BJ-A0ZA,
BJ-A0ZB,
BJ-A0ZC,
BJ-A0ZE,
BJ-A0ZF,
BJ-A0ZG,
BJ-A0ZH,
BJ-A0ZJ,
BJ-A18Y,
BJ-A18Z,
BJ-A190,
BJ-A191,
BJ-A192,
BJ-A28R,
BJ-A28S,
BJ-A28T,
BJ-A28V,
BJ-A28X,
BJ-A28Z,
BJ-A290,
BJ-A2N7,
BJ-A2N8,
BJ-A2N9,
BJ-A2NA,
BJ-A2P4,
BJ-A3EZ,
BJ-A3F0,
BJ-A3PR,
BJ-A3PT,
BJ-A3PU,
BJ-A45D,
BJ-A45E,
BJ-A45F,
BJ-A45G,
BJ-A45I,
BJ-A45J,
BJ-A45K,
BJ-A4O8,
BJ-A4O9,
CE-A13K,
CE-A27D,
CE-A3MD,
CE-A3ME,
CE-A482,
CE-A484,
CE-A485,
DE-A0XZ,
DE-A0Y2,
DE-A0Y3,
DE-A2OL,
DE-A3KN,
DE-A4M8,
DE-A4M9,
DJ-A13L,
DJ-A13M,
DJ-A13O,
DJ-A13P,
DJ-A13R,
DJ-A13S,
DJ-A13T,
DJ-A13U,
DJ-A13V,
DJ-A13W,
DJ-A13X,
DJ-A1QD,
DJ-A1QE,
DJ-A1QF,
DJ-A1QG,
DJ-A1QH,
DJ-A1QI,
DJ-A1QL,
DJ-A1QM,
DJ-A1QN,
DJ-A1QO,
DJ-A1QQ,
DJ-A2PN,
DJ-A2PO,
DJ-A2PP,
DJ-A2PQ,
DJ-A2PR,
DJ-A2PS,
DJ-A2PT,
DJ-A2PU,
DJ-A2PV,
DJ-A2PW,
DJ-A2PX,
DJ-A2PY,
DJ-A2PZ,
DJ-A2Q0,
DJ-A2Q1,
DJ-A2Q2,
DJ-A2Q3,
DJ-A2Q4,
DJ-A2Q5,
DJ-A2Q6,
DJ-A2Q7,
DJ-A2Q8,
DJ-A2Q9,
DJ-A2QA,
DJ-A2QB,
DJ-A2QC,
DJ-A3UK,
DJ-A3UM,
DJ-A3UN,
DJ-A3UO,
DJ-A3UP,
DJ-A3UQ,
DJ-A3UR,
DJ-A3US,
DJ-A3UT,
DJ-A3UU,
DJ-A3UV,
DJ-A3UW,
DJ-A3UX,
DJ-A3UY,
DJ-A3V7,
DJ-A3VA,
DJ-A3VB,
DJ-A3VE,
DJ-A3VF,
DJ-A3VJ,
DJ-A3VK,
DJ-A3VL,
DJ-A3VM,
DJ-A4UL,
DJ-A4UP,
DJ-A4UT,
DJ-A4UW,
DJ-A4V0,
DJ-A4V2,
DJ-A4V4,
DJ-A4V5,
DO-A1JZ,
DO-A1K0,
DO-A2HM,
E3-A3DY,
E3-A3DZ,
E3-A3E0,
E3-A3E1,
E3-A3E2,
E3-A3E3,
E3-A3E5,
E8-A242,
E8-A2EA,
E8-A413,
E8-A415,
E8-A418,
E8-A419,
E8-A433,
E8-A436,
E8-A437,
E8-A44K,
E8-A44M,
EL-A3CL,
EL-A3CM,
EL-A3CN,
EL-A3CO,
EL-A3CP,
EL-A3CR,
EL-A3CS,
EL-A3CT,
EL-A3CU,
EL-A3CV,
EL-A3CW,
EL-A3CX,
EL-A3CY,
EL-A3CZ,
EL-A3D0,
EL-A3D1,
EL-A3D4,
EL-A3D5,
EL-A3D6,
EL-A3GO,
EL-A3GP,
EL-A3GQ,
EL-A3GR,
EL-A3GS,
EL-A3GU,
EL-A3GV,
EL-A3GW,
EL-A3GX,
EL-A3GY,
EL-A3GZ,
EL-A3H1,
EL-A3H2,
EL-A3H3,
EL-A3H4,
EL-A3H5,
EL-A3H7,
EL-A3H8,
EL-A3MW,
EL-A3MX,
EL-A3MY,
EL-A3MZ,
EL-A3N2,
EL-A3N3,
EL-A3T0,
EL-A3T1,
EL-A3T2,
EL-A3T3,
EL-A3T6,
EL-A3T7,
EL-A3T8,
EL-A3T9,
EL-A3TA,
EL-A3TB,
EL-A3ZH,
EL-A3ZK,
EL-A3ZN,
EL-A3ZQ,
EL-A3ZR,
EL-A3ZT,
EL-A4JV,
EL-A4JW,
EL-A4JX,
EL-A4JZ,
EL-A4K0,
EL-A4K2,
EL-A4K4,
EL-A4K6,
EL-A4KD,
EL-A4KG,
EL-A4KH,
EL-A4KI,
EM-A1CS,
EM-A1CT,
EM-A1CU,
EM-A1CV,
EM-A1CW,
EM-A1YA,
EM-A1YB,
EM-A1YC,
EM-A1YD,
EM-A1YE,
EM-A22I,
EM-A22J,
EM-A22K,
EM-A22L,
EM-A22M,
EM-A22N,
EM-A22O,
EM-A22P,
EM-A22Q,
EM-A2CJ,
EM-A2CK,
EM-A2CL,
EM-A2CM,
EM-A2CN,
EM-A2CO,
EM-A2CP,
EM-A2CQ,
EM-A2CR,
EM-A2CT,
EM-A2CU,
EM-A2OV,
EM-A2OW,
EM-A2OX,
EM-A2OY,
EM-A2OZ,
EM-A2P0,
EM-A2P1,
EM-A2P2,
EM-A2P3,
EM-A3AI,
EM-A3AJ,
EM-A3AK,
EM-A3AL,
EM-A3AN,
EM-A3AO,
EM-A3AP,
EM-A3AQ,
EM-A3AR,
EM-A3FJ,
EM-A3FK,
EM-A3FL,
EM-A3FM,
EM-A3FN,
EM-A3FO,
EM-A3FP,
EM-A3FQ,
EM-A3FR,
EM-A3O3,
EM-A3O6,
EM-A3O7,
EM-A3O8,
EM-A3O9,
EM-A3OA,
EM-A3OB,
EM-A4FK,
EM-A4FM,
EM-A4FO,
EM-A4FQ,
EM-A4FR,
EM-A4FV,
EM-A4G1,
ET-A25G,
ET-A25I,
ET-A25J,
ET-A25K,
ET-A25O,
ET-A25R,
ET-A2MY,
ET-A2MZ,
ET-A2N0,
ET-A2N1,
ET-A2N4,
ET-A2N5,
ET-A39I,
ET-A39J,
ET-A39K,
ET-A39L,
ET-A39M,
ET-A39N,
ET-A39O,
ET-A39P,
ET-A39R,
ET-A39S,
ET-A39T,
ET-A3BN,
ET-A3BO,
ET-A3BP,
ET-A3BQ,
ET-A3BS,
ET-A3BT,
ET-A3BU,
ET-A3BV,
ET-A3BW,
ET-A3BX,
ET-A3DO,
ET-A3DP,
ET-A3DQ,
ET-A3DR,
ET-A3DS,
ET-A3DT,
ET-A3DU,
ET-A3DV,
ET-A3DW,
ET-A40S,
ET-A4KN,
FE-A22Z,
FE-A230,
FE-A231,
FE-A232,
FE-A233,
FE-A234,
FE-A235,
FE-A236,
FE-A237,
FE-A238,
FE-A23A,
FE-A3PA,
FE-A3PB,
FE-A3PC,
FE-A3PD,
FK-A3S3,
FK-A3SB,
FK-A3SD,
FK-A3SE,
FK-A3SG,
FK-A3SH,
FY-A2QD,
FY-A3BL,
FY-A3I4,
FY-A3I5,
FY-A3NM,
FY-A3NN,
FY-A3NP,
FY-A3ON,
FY-A3R6,
FY-A3R7,
FY-A3R8,
FY-A3R9,
FY-A3RA,
FY-A3W9,
FY-A3WA,
FY-A40K,
FY-A4B3,
GE-A2C6,
H2-A26U,
H2-A2K9,
H2-A3RH,
H2-A3RI,
H2-A421,
IM-A3EB,
IM-A3ED,
IM-A3U2,
IM-A3U3,
J8-A3NZ,
J8-A3O0,
J8-A3O1,
J8-A3YE,
J8-A3YH,
J8-A4HW,
KS-A41J,
KS-A4I5,
KS-A4I9,
KS-A4IB,
L6-A4EP,
L6-A4ET,
L6-A4EU,
MK-A4N6,
MK-A4N7,
MK-A4N9.\\
$\blacklozenge$ {\bf Uterine Cancer} (305 samples):\\
$\bullet$ Source T26 = TCGA (see Acknowledgments). Sample IDs are of the form TCGA-*, where * is:\\
A5-A0G3,
A5-A0G5,
A5-A0G9,
A5-A0GA,
A5-A0GB,
A5-A0GD,
A5-A0GE,
A5-A0GH,
A5-A0GI,
A5-A0GJ,
A5-A0GM,
A5-A0GN,
A5-A0GP,
A5-A0GQ,
A5-A0GU,
A5-A0GV,
A5-A0GW,
A5-A0GX,
A5-A0R6,
A5-A0R7,
A5-A0R8,
A5-A0R9,
A5-A0RA,
A5-A0VO,
A5-A0VP,
A5-A0VQ,
AJ-A23M,
AP-A051,
AP-A052,
AP-A053,
AP-A054,
AP-A056,
AP-A059,
AP-A05A,
AP-A05D,
AP-A05H,
AP-A05J,
AP-A05N,
AP-A05P,
AP-A0L8,
AP-A0L9,
AP-A0LD,
AP-A0LE,
AP-A0LF,
AP-A0LG,
AP-A0LH,
AP-A0LI,
AP-A0LJ,
AP-A0LL,
AP-A0LM,
AP-A0LN,
AP-A0LO,
AP-A0LP,
AP-A0LQ,
AP-A0LT,
AP-A0LV,
AP-A1DQ,
AX-A05S,
AX-A05T,
AX-A05U,
AX-A05W,
AX-A05Y,
AX-A05Z,
AX-A060,
AX-A062,
AX-A063,
AX-A064,
AX-A06B,
AX-A06H,
AX-A06L,
AX-A0IS,
AX-A0IU,
AX-A0IW,
AX-A0J0,
AX-A0J1,
AX-A1C7,
AX-A1C8,
AX-A1CP,
AX-A2H5,
AX-A2HF,
B5-A0JN,
B5-A0JR,
B5-A0JS,
B5-A0JT,
B5-A0JV,
B5-A0JY,
B5-A0JZ,
B5-A0K0,
B5-A0K1,
B5-A0K2,
B5-A0K3,
B5-A0K4,
B5-A0K6,
B5-A0K7,
B5-A0K8,
B5-A0K9,
B5-A11E,
B5-A11F,
B5-A11G,
B5-A11H,
B5-A11I,
B5-A11J,
B5-A11M,
B5-A11N,
B5-A11O,
B5-A11Q,
B5-A11R,
B5-A11S,
B5-A11U,
B5-A11V,
B5-A11W,
B5-A11X,
B5-A11Y,
B5-A11Z,
B5-A121,
B5-A1MU,
B5-A1MY,
BG-A0LW,
BG-A0LX,
BG-A0M0,
BG-A0M2,
BG-A0M3,
BG-A0M4,
BG-A0M6,
BG-A0M7,
BG-A0M8,
BG-A0M9,
BG-A0MC,
BG-A0MG,
BG-A0MI,
BG-A0MO,
BG-A0MQ,
BG-A0MS,
BG-A0MT,
BG-A0MU,
BG-A0RY,
BG-A0VT,
BG-A0VV,
BG-A0VW,
BG-A0VX,
BG-A0VZ,
BG-A0W1,
BG-A0W2,
BG-A0YU,
BG-A0YV,
BG-A186,
BG-A187,
BG-A18A,
BG-A18B,
BG-A18C,
BG-A2AE,
BK-A0C9,
BK-A0CA,
BK-A0CB,
BK-A0CC,
BK-A139,
BK-A13C,
BS-A0T9,
BS-A0TA,
BS-A0TC,
BS-A0TD,
BS-A0TE,
BS-A0TG,
BS-A0TI,
BS-A0TJ,
BS-A0U5,
BS-A0U7,
BS-A0U8,
BS-A0U9,
BS-A0UA,
BS-A0UF,
BS-A0UJ,
BS-A0UL,
BS-A0UM,
BS-A0UT,
BS-A0UV,
BS-A0V6,
BS-A0V7,
BS-A0V8,
BS-A0WQ,
D1-A0ZN,
D1-A0ZO,
D1-A0ZP,
D1-A0ZQ,
D1-A0ZR,
D1-A0ZS,
D1-A0ZU,
D1-A0ZV,
D1-A0ZZ,
D1-A101,
D1-A102,
D1-A103,
D1-A15V,
D1-A15W,
D1-A15X,
D1-A15Z,
D1-A160,
D1-A161,
D1-A163,
D1-A165,
D1-A167,
D1-A168,
D1-A169,
D1-A16B,
D1-A16D,
D1-A16E,
D1-A16F,
D1-A16G,
D1-A16I,
D1-A16J,
D1-A16N,
D1-A16O,
D1-A16Q,
D1-A16R,
D1-A16S,
D1-A16X,
D1-A16Y,
D1-A174,
D1-A176,
D1-A177,
D1-A17A,
D1-A17B,
D1-A17C,
D1-A17D,
D1-A17F,
D1-A17H,
D1-A17K,
D1-A17L,
D1-A17M,
D1-A17N,
D1-A17Q,
D1-A17R,
D1-A17S,
D1-A17T,
D1-A17U,
D1-A1NU,
D1-A1NX,
DI-A0WH,
DI-A1NN,
E6-A1LZ,
EO-A1Y5,
EO-A1Y8,
EY-A1GS,
EY-A212,
FI-A2D2,
FI-A2EW,
FI-A2EX,
FI-A2F8,
N5-A4R8,
N5-A4RA,
N5-A4RD,
N5-A4RF,
N5-A4RJ,
N5-A4RM,
N5-A4RN,
N5-A4RO,
N5-A4RS,
N5-A4RT,
N5-A4RU,
N5-A4RV,
N5-A59E,
N5-A59F,
N6-A4V9,
N6-A4VC,
N6-A4VD,
N6-A4VE,
N6-A4VF,
N6-A4VG,
N7-A4Y0,
N7-A4Y5,
N7-A4Y8,
N7-A59B,
N8-A4PI,
N8-A4PL,
N8-A4PM,
N8-A4PN,
N8-A4PO,
N8-A4PP,
N8-A4PQ,
N8-A56S,
N9-A4PZ,
N9-A4Q1,
N9-A4Q3,
N9-A4Q4,
N9-A4Q7,
N9-A4Q8,
NA-A4QV,
NA-A4QW,
NA-A4QX,
NA-A4QY,
NA-A4R0,
NA-A4R1,
NA-A5I1,
ND-A4W6,
ND-A4WA,
ND-A4WC,
ND-A4WF,
NF-A4WU,
NF-A4WX,
NF-A4X2,
NF-A5CP,
NG-A4VU,
NG-A4VW,
QM-A5NM,
QN-A5NN.
}

\newpage\clearpage
\begin{landscape}
\begin{table}[ht]
\noindent
\caption{Exome data summary. See Appendix \ref{app.IDs} for the data source definitions. Here we label cancer types via X1 through X32 for use in Tables below.}
{\scriptsize
\begin{tabular}{l l l l l} 
\\
\hline\hline 
Label & Cancer & Total  & \# of   & Source\\
      & Type   & Counts & Samples & \\[0.5ex] 
\hline 
X1 & Acute Lymphoblastic Leukemia & 938 & 86 & H1, Z1, D1\\
X2 & Acute Myeloid Leukemia & 1414 & 190 & T1\\
X3 & Adrenocortical Carcinoma & 11530 & 91 & T2\\
X4 & B-Cell Lymphoma & 706 & 24 & M1, L1\\
X5 & Benign Liver Tumor & 884 & 40 & P1\\
X6 & Bladder Cancer & 90121 & 341 & G1, T3\\
X7 & Brain Lower Grade Glioma & 38041 & 465 & T4\\
X8 & Breast Cancer & 201555 & 1182 & N1, S1, S2, T5\\
X9 & Cervical Cancer & 47715 & 197 & T6\\
X10 & Cholangiocarcinoma & 12156 & 139 & Z2, T7\\
X11 & Chronic Lymphocytic Leukemia & 975 & 80 & Q1\\
X12 & Colorectal Cancer & 214814 & 581 & S3, T8\\
X13 & Esophageal Cancer & 59088 & 329 & D2, T9\\
X14 & Gastric Cancer & 161078 & 401 & Z3, W1, T10\\
X15 & Glioblastoma Multiforme & 23230 & 359 & P2, T11\\
X16 & Head and Neck Cancer & 96816 & 591 & A1, S4, T12\\
X17 & Liver Cancer & 252755 & 452 & S5, H2, T13\\
X18 & Lung Cancer & 306071 & 1018 & D3, R1, P3, S6, I1, T14\\
X19 & Melanoma & 357060 & 594 & S7, D4, B1, A2, H3, T15\\
X20 & Nasopharyngeal Cancer & 2241 & 11 & L2\\
X21 & Oral Cancer & 13462 & 106 & I2\\
X22 & Ovarian Cancer & 20610 & 471 & J1, T16\\
X23 & Pancreatic Cancer & 39788 & 184 & W2, J2, T17\\
X24 & Pheochromocytoma and Paraganglioma & 3709 & 178 & T18\\
X25 & Prostate Cancer & 22808 & 480 & B2, B3, G2, T19\\
X26 & Rectum Adenocarcinoma & 32797 & 115 & T20\\
X27 & Renal Cell Carcinoma & 47635 & 709 & G3, T21\\
X28 & Sarcoma & 28256 & 255 & T22\\
X29 & Testicular Germ Cell Tumor & 6064 & 150 & T23\\
X30 & Thymoma & 4444 & 123 & T24\\
X31 & Thyroid Carcinoma & 6833 & 409 & T25\\
X32 & Uterine Cancer & 164211 & 305 & T26\\
--- & All Cancer Types & 2269805 & 10656 & Above\\ [1ex] 
\hline 
\end{tabular}
\label{table.exome.summary} 
}
\end{table}
\end{landscape}

\newpage\clearpage
\begin{landscape}
\begin{table}[ht]
\noindent
\caption{Occurrence counts for cancer types X1 through X16 for the first 48 mutation categories for the exome data summarized in Table \ref{table.exome.summary} aggregated by cancer types. Here and in Tables below the mutations are encoded as follows: XYZW = Y $>$ W: XYZ.}
{\tiny
\begin{tabular}{l l l l l l l l l l l l l l l l l} 
\\
\hline\hline 
Mutation & X1 & X2 & X3 & X4 & X5 & X6 & X7 & X8 & X9 & X10 & X11 & X12 & X13 & X14 & X15 & X16 \\[0.5ex] 
\hline 
\\
ACAA &  9 &  11 & 156 &  8 &  8 &   345 &  195 &  1823 &  190 & 136 &  9 &   681 &  513 &   493 &  178 &  892 \\
ACCA &  7 &  17 & 148 &  6 & 13 &   371 &  248 &  1775 &  177 & 112 & 11 &   726 &  465 &   561 &  214 &  712 \\
ACGA &  0 &   5 & 100 &  1 &  4 &   252 &   58 &   431 &   88 &  55 &  2 &   321 &  203 &   147 &   54 &  446 \\
ACTA &  6 &   6 & 105 &  6 &  6 &   259 &  436 &  1411 &  164 &  51 & 10 &   981 &  277 &   438 &  132 &  514 \\
CCAA & 13 &  13 & 333 &  5 & 18 &   520 &  258 &  1750 &  322 & 132 &  9 &  2019 &  823 &  1145 &  171 & 1426 \\
CCCA & 12 &   9 & 279 &  2 & 10 &   388 &  273 &  1506 &  222 & 134 &  6 &  2246 &  570 &  1681 &  172 & 1099 \\
CCGA &  3 &  10 & 180 &  3 &  9 &   373 &  127 &   541 &  175 & 102 &  7 &  1427 &  424 &   882 &   87 &  836 \\
CCTA & 10 &  14 & 292 &  2 &  9 &   403 & 1771 &  1804 &  298 & 121 & 11 &  5562 &  844 &  4206 &  191 & 1153 \\
GCAA & 12 &  12 & 278 & 11 & 15 &   356 &  127 &  1437 &  245 & 134 & 10 &  1258 &  824 &   816 &  118 & 1146 \\
GCCA &  7 &   8 & 270 &  5 & 13 &   382 &  218 &  1281 &  245 & 190 &  9 &  1265 &  486 &   772 &  138 &  842 \\
GCGA &  6 &   7 & 163 &  4 & 17 &   262 &   99 &   372 &   95 & 142 &  1 &   698 &  331 &   357 &   60 &  656 \\
GCTA & 18 &   1 & 224 &  7 & 20 &   332 &  571 &  1216 &  233 & 111 &  6 &  2222 &  611 &  1447 &  112 &  700 \\
TCAA & 16 &   5 & 273 &  7 & 12 &  1800 &  147 &  4102 &  867 & 143 &  7 &  3007 &  718 &   688 &  160 & 1450 \\
TCCA & 18 &  17 & 282 &  9 & 20 &  1247 &  305 &  2813 &  601 & 159 & 10 &  1634 &  709 &   796 &  180 & 1490 \\
TCGA &  1 &   5 &  93 &  2 &  3 &   549 &   84 &   640 &  273 &  73 &  3 &   821 &  397 &   268 &   69 &  632 \\
TCTA & 25 &  13 & 281 & 13 &  8 &  2214 & 1292 &  4889 & 1150 & 126 & 13 & 15896 &  830 &  2202 &  157 & 1359 \\
ACAG &  5 &  13 &  50 &  5 &  7 &   272 &  147 &  1287 &  138 &  40 &  9 &   374 &  239 &   250 &  121 &  423 \\
ACCG &  3 &   3 &  50 &  3 &  4 &   190 &  115 &   968 &   84 &  51 &  3 &   289 &  221 &   350 &   95 &  381 \\
ACGG &  3 &   8 &  33 &  2 &  4 &   169 &   43 &   516 &   71 &  37 &  2 &   139 &  110 &   143 &   66 &  229 \\
ACTG &  3 &   6 &  55 &  3 &  3 &   275 &  169 &  1363 &   88 &  52 &  6 &   446 &  268 &   374 &  129 &  440 \\
CCAG &  4 &  12 &  71 &  1 &  3 &   406 &   89 &  1186 &  162 &  63 &  6 &   272 &  216 &   270 &  109 &  517 \\
CCCG &  1 &   2 &  66 &  5 &  8 &   277 &   89 &   876 &  102 &  63 &  4 &   249 &  212 &   202 &   86 &  426 \\
CCGG &  0 &  11 &  48 &  1 &  8 &   290 &   40 &   569 &   93 &  52 &  2 &   230 &  140 &   148 &   55 &  397 \\
CCTG &  2 &   3 &  97 &  2 &  2 &   519 &  110 &  1594 &  190 &  73 & 11 &   335 &  329 &   341 &  129 &  636 \\
GCAG &  3 &   7 &  52 &  8 &  3 &   256 &   94 &   720 &   85 &  32 &  3 &   251 &  170 &   227 &   87 &  310 \\
GCCG &  3 &   5 &  83 &  4 &  4 &   219 &  136 &   791 &   97 &  46 &  3 &   493 &  220 &   470 &   92 &  387 \\
GCGG &  7 &   7 &  42 &  2 &  0 &   183 &   39 &   335 &   34 &  40 &  3 &   144 &  109 &   100 &   29 &  260 \\
GCTG &  3 &   1 &  64 &  9 &  8 &   273 &  133 &  1106 &  118 &  52 &  4 &   383 &  248 &   314 &   93 &  371 \\
TCAG &  5 &   5 & 151 &  8 &  6 &  6944 &  164 & 13309 & 4554 &  96 &  7 &   478 & 1155 &   539 &  179 & 4640 \\
TCCG &  4 &   5 & 135 &  5 &  8 &  2539 &  134 &  4348 & 1400 &  74 &  4 &   415 &  546 &   453 &  159 & 1864 \\
TCGG &  4 &   8 &  43 &  0 &  1 &   995 &   71 &   997 &  649 &  46 &  2 &   159 &  225 &   138 &   46 &  706 \\
TCTG &  8 &   6 & 216 &  6 &  6 &  8468 &  332 & 17536 & 4883 & 126 & 14 &   766 & 1446 &   900 &  250 & 5340 \\
ACAT & 13 &  43 & 114 & 10 & 10 &   663 &  739 &  2293 &  292 & 132 & 21 &  2578 &  665 &  1964 &  339 & 1017 \\
ACCT & 20 &  28 &  99 & 10 & 15 &   626 &  582 &  1511 &  235 & 154 & 15 &  2141 &  589 &  1775 &  438 & 1081 \\
ACGT & 77 & 155 & 650 & 27 & 26 &  2163 & 3790 &  4880 & 1264 & 442 & 61 & 12877 & 3611 & 15083 & 3208 & 3132 \\
ACTT &  4 &  20 &  69 &  4 &  9 &   519 &  594 &  1625 &  190 &  97 & 15 &  1672 &  505 &  1204 &  294 &  765 \\
CCAT & 28 &  28 & 158 & 15 & 19 &  1363 &  451 &  2613 &  597 & 198 & 12 &  2612 &  910 &  1442 &  428 & 2033 \\
CCCT & 24 &  21 & 154 &  6 & 18 &   930 &  651 &  1735 &  340 & 188 & 17 &  2422 &  941 &  1508 &  729 & 2167 \\
CCGT & 71 & 158 & 586 & 32 & 36 &  2541 & 2669 &  4215 & 1538 & 502 & 45 & 13132 & 3976 & 14824 & 1933 & 4047 \\
CCTT & 26 &  54 & 174 &  9 & 28 &  1196 &  582 &  2496 &  444 & 220 & 19 &  2210 &  956 &  1604 &  497 & 2161 \\
GCAT & 10 &  19 & 213 & 22 & 23 &   902 & 1019 &  2266 &  445 & 173 & 20 &  4317 &  784 &  3857 &  316 & 1392 \\
GCCT & 19 &  34 & 231 & 16 & 23 &  1197 & 1447 &  2181 &  470 & 232 & 15 &  7695 & 1094 &  6834 &  985 & 1864 \\
GCGT & 80 & 131 & 662 & 29 & 40 &  2440 & 4477 &  4580 & 1665 & 452 & 44 & 19823 & 4183 & 21712 & 2596 & 3885 \\
GCTT &  8 &  37 & 234 & 16 & 12 &   907 & 1128 &  2095 &  432 & 194 & 18 &  5329 &  888 &  4192 &  406 & 1269 \\
TCAT & 29 &  30 & 252 & 13 &  7 & 12157 &  353 & 21889 & 7214 & 226 & 19 &  2277 & 1942 &  1756 &  340 & 7315 \\
TCCT & 43 &  33 & 227 & 16 & 14 &  4126 &  741 &  6715 & 2397 & 204 & 25 &  3327 & 1302 &  1988 &  594 & 4506 \\
TCGT & 58 &  73 & 356 & 19 & 21 &  6568 & 1826 &  7512 & 3602 & 293 & 27 & 17135 & 2630 &  9312 & 1178 & 4328 \\
TCTT & 25 &  19 & 193 &  8 & 17 &  6716 &  611 & 12470 & 3850 & 166 & 18 &  3770 & 1474 &  1666 &  413 & 4591 \\
[1ex] 
\hline 
\end{tabular}
}
\label{table.aggr.data.11} 
\end{table}
\end{landscape}

\newpage\clearpage
\begin{landscape}
\begin{table}[ht]
\noindent
\caption{Table \ref{table.aggr.data.11} continued: occurrence counts (aggregated by cancer types) for the
next 48 mutation categories for cancer types X1 through X16.}
{\tiny
\begin{tabular}{l l l l l l l l l l l l l l l l l} 
\\
\hline\hline 
Mutation & X1 & X2 & X3 & X4 & X5 & X6 & X7 & X8 & X9 & X10 & X11 & X12 & X13 & X14 & X15 & X16 \\[0.5ex] 
\hline 
\\
ATAA &  1 &   0 &  44 &  9 &  4 &   125 &   48 &   729 &   34 &  85 &  4 &   279 &  123 &   261 &   50 &  293 \\
ATCA &  3 &   8 &  56 & 10 &  5 &   162 &   90 &   861 &   64 &  70 &  7 &   564 &  224 &   791 &   74 &  407 \\
ATGA &  0 &   3 &  48 &  8 &  6 &   234 &   78 &   893 &   71 & 202 & 14 &   280 &  206 &   296 &   83 &  424 \\
ATTA &  8 &   2 &  21 &  4 &  3 &   151 &   55 &  1087 &   49 &  44 &  8 &   603 &  171 &   529 &   58 &  295 \\
CTAA &  2 &   2 &  29 &  5 &  0 &   132 &   34 &   525 &   29 & 192 &  2 &   242 &   82 &   131 &   41 &  209 \\
CTCA &  2 &   7 &  67 &  6 &  6 &   221 &  139 &   866 &   70 & 200 &  5 &   530 &  328 &   529 &   87 &  481 \\
CTGA &  8 &  10 & 128 &  5 & 11 &   480 &  132 &   888 &   93 & 691 & 14 &   606 &  310 &   430 &  154 &  845 \\
CTTA &  3 &   2 &  50 &  5 &  6 &   221 &  110 &   981 &   80 & 179 &  6 &   560 &  535 &   714 &  103 &  507 \\
GTAA &  0 &   2 &  41 &  5 &  0 &   129 &   45 &   412 &   34 & 119 &  6 &   315 &   79 &   173 &   40 &  232 \\
GTCA &  3 &   3 &  43 &  7 &  2 &   121 &   75 &   544 &   40 &  62 &  4 &   443 &  134 &   575 &   45 &  263 \\
GTGA &  2 &   3 &  42 &  4 &  4 &   166 &   44 &   537 &   44 & 194 & 10 &   302 &  117 &   239 &   60 &  379 \\
GTTA &  1 &   5 &  20 &  0 &  3 &   117 &   65 &   666 &   37 &  44 &  6 &   326 &  184 &   337 &   55 &  219 \\
TTAA &  0 &   1 &  20 &  3 &  2 &   105 &   36 &   905 &   35 &  75 &  7 &   281 &   83 &   187 &   32 &  199 \\
TTCA &  5 &   4 &  33 &  3 &  4 &   161 &   65 &   660 &   53 & 107 &  8 &   363 &  155 &   281 &   65 &  278 \\
TTGA &  4 &   3 &  30 &  2 &  5 &   163 &   48 &   502 &   31 & 188 &  5 &   231 &   99 &   140 &   53 &  218 \\
TTTA &  2 &   3 &  26 &  3 &  3 &   141 &   93 &  1054 &   45 &  67 &  9 &   459 &  194 &   261 &   58 &  258 \\
ATAC &  5 &  17 & 104 &  4 & 14 &   632 &  350 &  1677 &  138 & 174 & 17 &  1585 &  541 &  1300 &  160 & 1137 \\
ATCC & 11 &  12 &  63 &  6 &  4 &   319 &  349 &   964 &  110 &  93 & 15 &  2728 &  330 &  1306 &  143 &  450 \\
ATGC & 10 &  21 & 104 &  7 & 21 &   731 &  493 &  1781 &  188 & 234 & 14 &  3410 &  521 &  2601 &  208 &  876 \\
ATTC &  3 &  19 &  90 &  7 & 11 &   535 &  459 &  1762 &  153 & 118 & 18 &  1670 &  484 &   977 &  236 &  834 \\
CTAC &  5 &   6 &  48 &  4 &  7 &   267 &  175 &   808 &   88 & 100 &  5 &  1110 &  273 &  1092 &  117 &  403 \\
CTCC &  3 &  14 &  68 &  9 & 13 &   459 &  406 &  1316 &  154 & 125 & 11 &  4194 &  724 &  1625 &  252 &  526 \\
CTGC & 15 &  15 & 130 &  9 & 19 &   692 &  450 &  1464 &  249 & 245 & 14 &  4936 &  842 &  3653 &  324 &  863 \\
CTTC &  3 &  10 &  75 &  8 &  9 &   427 &  337 &  1148 &  162 & 142 & 12 &  3577 & 1673 &  2424 &  222 &  780 \\
GTAC & 10 &   4 & 112 &  8 & 10 &   426 &  354 &  1164 &  163 & 113 & 10 &  2598 &  434 &  3655 &  126 &  808 \\
GTCC &  8 &   8 &  73 & 14 &  9 &   373 &  443 &   780 &  227 &  90 & 15 &  4544 &  483 &  2647 &  169 &  575 \\
GTGC &  8 &   7 &  83 &  7 & 13 &   413 &  331 &  1004 &  204 & 143 &  9 &  3583 &  456 &  3093 &  140 &  776 \\
GTTC &  4 &   8 &  66 &  3 & 11 &   391 &  379 &  1134 &  194 & 124 & 16 &  3277 &  459 &  2338 &  179 &  773 \\
TTAC &  3 &   0 &  54 &  5 &  6 &   288 &  150 &   846 &   77 &  82 &  9 &   941 &  196 &  1012 &   78 &  353 \\
TTCC & 11 &   9 &  63 & 11 & 11 &   290 &  339 &  1048 &  165 &  93 &  6 &  2964 &  393 &  1362 &  155 &  428 \\
TTGC &  3 &   4 &  77 &  4 &  8 &   352 &  342 &   721 &  141 & 113 &  7 &  2088 &  283 &  1444 &   99 &  456 \\
TTTC &  5 &  15 &  52 &  7 & 11 &   267 &  318 &  1037 &  152 &  82 & 10 &  1910 &  282 &   899 &  142 &  423 \\
ATAG &  2 &   1 &  12 &  6 &  4 &    63 &   43 &   423 &   33 &  25 &  4 &   208 &   62 &   104 &   24 &  110 \\
ATCG &  2 &   2 &  14 &  6 &  0 &    93 &   85 &   320 &   43 &  26 &  4 &   446 &   86 &   258 &   41 &  115 \\
ATGG &  0 &   2 &  26 &  6 &  4 &   128 &   77 &   588 &   50 &  44 &  2 &   276 &  130 &   205 &   55 &  156 \\
ATTG &  1 &   1 &  19 &  5 &  2 &   132 &  136 &   564 &  112 &  43 &  2 &  1193 &  322 &   588 &   38 &  123 \\
CTAG &  1 &   1 &  14 &  2 &  1 &    74 &   29 &   318 &   19 &  36 &  6 &   238 &   70 &   162 &   21 &   77 \\
CTCG &  2 &   5 &  28 &  5 &  4 &   129 &   99 &   519 &   49 &  74 &  6 &   624 &  271 &   737 &   50 &  149 \\
CTGG &  3 &  10 &  60 &  6 &  9 &   271 &   93 &   826 &   87 &  86 &  8 &   863 &  339 &   934 &   95 &  317 \\
CTTG &  2 &   4 &  19 & 32 &  6 &   266 &  176 &   994 &  123 & 129 & 14 &  2194 & 3027 &  3903 &   59 &  313 \\
GTAG &  0 &   3 &  25 &  4 &  1 &   138 &   31 &   544 &   31 &  24 &  3 &   161 &   75 &   125 &   23 &  102 \\
GTCG &  2 &   2 &  28 &  2 &  4 &   159 &   67 &   464 &   50 &  19 &  4 &   362 &  138 &   334 &   49 &   93 \\
GTGG &  5 &   4 & 133 &  2 & 10 &   656 &  116 &  5379 &   58 &  54 &  2 &   368 &  155 &   314 &  118 &  243 \\
GTTG &  2 &   1 &  27 & 13 &  7 &   152 &   84 &   683 &   85 &  31 &  4 &   714 &  727 &  1161 &   48 &  109 \\
TTAG &  1 &   3 &   9 &  3 &  1 &   102 &   35 &   503 &   59 &  21 &  6 &   466 &   84 &   137 &   35 &   96 \\
TTCG &  2 &   4 &  29 &  4 &  2 &   159 &   86 &   474 &   98 &  65 &  1 &   778 &  179 &   367 &   30 &  172 \\
TTGG &  3 &   1 &  20 &  3 &  6 &   151 &   82 &   635 &   59 &  63 &  3 &   351 &  141 &   247 &   61 &  226 \\
TTTG &  3 &   5 &  22 &  8 &  4 &   415 &  196 &  1189 &  317 &  89 &  5 &  4569 &  555 &  1361 &   63 &  214 \\
[1ex] 
\hline 
\end{tabular}
}
\label{table.aggr.data.21} 
\end{table}
\end{landscape}

\newpage\clearpage
\begin{landscape}
\begin{table}[ht]
\noindent
\caption{Occurrence counts for cancer types X17 through X32 for the first 48 mutation categories for the exome data summarized in Table \ref{table.exome.summary} aggregated by cancer types.}
{\tiny
\begin{tabular}{l l l l l l l l l l l l l l l l l} 
\\
\hline\hline 
Mutation & X17 & X18 & X19 & X20 & X21 & X22 & X23 & X24 & X25 & X26 & X27 & X28 & X29 & X30 & X31 & X32 \\[0.5ex] 
\hline 
\\
ACAA & 2813 &  6149 &   562 &  63 & 323 & 308 &  147 &  36 &  264 &  104 &  555 &  273 &  58 &  64 &  56 &   457 \\
ACCA & 2218 &  6108 &   394 &   8 & 340 & 352 &  214 &  28 &  216 &  131 &  511 &  256 & 111 &  55 &  51 &   554 \\
ACGA & 1014 &  3979 &   157 &   2 & 193 &  67 &   68 &  21 &   64 &   41 &  202 &  127 &  67 &  36 &  34 &   185 \\
ACTA & 1623 &  4089 &   350 &  11 & 187 & 217 &  351 &  10 &  129 &  212 &  360 &  189 &  52 &  36 &  43 &  1594 \\
CCAA & 2593 & 13874 &  2172 &  40 & 379 & 326 &  291 &  38 &  259 &  165 &  673 &  304 & 109 &  75 &  68 &  1094 \\
CCCA & 2454 & 12582 &   851 &  18 & 312 & 317 &  257 &  33 &  189 &  125 &  645 &  261 &  95 &  62 &  42 &  1238 \\
CCGA & 1644 &  8004 &  1897 &  16 & 225 &  82 &  296 &  22 &  145 &  132 &  338 &  198 & 128 &  66 &  53 &   548 \\
CCTA & 2054 & 10717 &  1243 &  14 & 350 & 287 & 1368 &  17 &  233 &  469 &  704 &  323 & 114 &  83 &  52 &  6273 \\
GCAA & 2244 &  6380 &   358 &  30 & 662 & 260 &  147 &  28 &  273 &  162 &  517 &  959 & 104 &  44 &  55 &   586 \\
GCCA & 2318 &  7537 &   333 &  22 & 756 & 292 &  322 &  22 &  234 &  149 &  531 &  943 & 195 &  37 &  32 &   899 \\
GCGA & 1534 &  5020 &   153 &  14 & 375 &  80 &  124 &  15 &  135 &   85 &  261 &  571 & 146 &  49 &  30 &   239 \\
GCTA & 1597 &  4815 &   286 &  23 & 610 & 245 & 1352 &  18 &  159 &  297 &  482 & 2182 & 136 &  44 &  36 &  3015 \\
TCAA & 2141 &  6882 &   941 &  26 & 309 & 259 &  399 &  21 &  214 &  782 &  696 &  242 &  67 &  48 &  57 &  2984 \\
TCCA & 2380 & 10492 &  1213 &  16 & 319 & 373 &  404 &  30 &  297 &  262 &  727 &  248 &  83 &  45 &  52 &  1575 \\
TCGA &  926 &  3791 &   429 &   4 & 204 &  88 &  240 &  16 &  120 &  141 &  255 &  137 &  69 &  36 &  40 &   409 \\
TCTA & 2464 &  7464 &   864 &  14 & 256 & 306 & 1254 &  32 &  261 & 5224 &  703 &  278 &  82 &  40 &  41 & 22736 \\
ACAG & 1852 &  1947 &   301 &   8 &  44 & 276 &   83 &  23 &  130 &   55 &  284 &  111 &  39 &  15 &  20 &   168 \\
ACCG & 1075 &  1377 &   284 &   3 &  33 & 166 &   39 &  30 &   95 &   42 &  309 &   98 &  30 &  21 &  28 &   181 \\
ACGG &  824 &   901 &   214 &  16 &  22 & 110 &   51 &  10 &   48 &   21 &  148 &   70 &  51 &  16 &  21 &   140 \\
ACTG & 1621 &  1467 &   319 &   5 &  31 & 247 &   65 &  14 &  118 &   50 &  324 &   92 &  39 &  13 &  33 &   213 \\
CCAG & 1139 &  2373 &   735 &  12 &  51 & 257 &   83 &  26 &  109 &   49 &  279 &  104 &  24 &  11 &  38 &    93 \\
CCCG & 1417 &  2025 &   684 &  11 &  69 & 238 &   54 &  22 &   91 &   35 &  355 &   92 &  19 &  13 &  43 &    97 \\
CCGG &  857 &  1762 &   823 &  26 &  41 & 120 &   44 &   8 &   83 &   53 &  223 &   73 &  28 &  18 &  29 &   128 \\
CCTG & 1715 &  2383 &   496 &   6 &  88 & 315 &   82 &  28 &  158 &   38 &  506 &  154 &  51 &  23 &  45 &   154 \\
GCAG &  958 &  1595 &   172 &  31 &  28 & 176 &   61 &  18 &   76 &   50 &  222 &   84 &  36 &  13 &  21 &   122 \\
GCCG & 1091 &  1993 &   337 &  88 &  29 & 165 &   66 &  27 &  131 &   62 &  241 &  104 &  32 &  20 &  43 &   215 \\
GCGG &  873 &  1350 &   181 &  90 &  12 &  49 &   30 &  12 &   37 &   35 &  135 &   51 &  25 &  12 &  15 &    62 \\
GCTG & 1143 &  1498 &   226 &  31 &  41 & 293 &  101 &  19 &  136 &   55 &  279 &  124 &  66 &  23 &  43 &   201 \\
TCAG & 1485 &  7479 &   480 &  18 & 396 & 396 &  175 &  37 &  189 &   87 &  429 &  191 &  43 &  37 & 129 &   548 \\
TCCG & 1511 &  4330 &   676 &  12 & 217 & 371 &  108 &  36 &  173 &   72 &  496 &  197 &  37 &  23 &  82 &   253 \\
TCGG &  745 &  1559 &   264 &  13 &  64 &  82 &   62 &  13 &   45 &   33 &  173 &   52 &  30 &  12 &  23 &   118 \\
TCTG & 2700 &  8876 &   790 &  17 & 437 & 598 &  196 &  39 &  304 &  141 &  714 &  341 &  58 &  24 & 173 &   653 \\
ACAT & 6483 &  3019 &   839 &  12 &  91 & 333 &  446 &  52 &  321 &  173 &  667 &  294 &  85 &  66 & 109 &  2148 \\
ACCT & 4063 &  2437 &  9149 &  14 & 108 & 247 &  704 &  61 &  243 &  222 &  616 &  398 &  65 &  42 & 139 &  1921 \\
ACGT & 8931 &  4557 &  1387 &  58 & 454 & 659 & 2672 & 222 & 1955 & 2009 & 1758 & 1318 & 199 & 310 & 360 &  7951 \\
ACTT & 4600 &  1931 &  3358 &  14 &  54 & 253 &  609 &  55 &  240 &  207 &  545 &  256 &  76 &  44 &  93 &  1766 \\
CCAT & 4593 &  6129 & 23356 &  16 & 236 & 400 &  367 &  70 &  341 &  201 &  910 &  734 &  98 &  49 & 136 &   937 \\
CCCT & 5514 &  4925 & 35156 &  31 & 245 & 346 &  575 &  62 &  341 &  217 &  860 & 1061 &  94 &  52 & 132 &  1405 \\
CCGT & 9037 &  6230 &  9358 &  61 & 616 & 690 & 2479 & 167 & 1900 & 1747 & 1671 & 1382 & 194 & 275 & 368 &  7385 \\
CCTT & 6055 &  5475 & 26855 &  22 & 178 & 366 &  752 &  82 &  390 &  259 &  913 &  960 & 138 &  83 & 138 &  1599 \\
GCAT & 3913 &  3263 &   895 &  18 & 126 & 454 &  962 &  79 &  430 &  244 &  917 &  304 & 147 &  89 & 125 &  4051 \\
GCCT & 5014 &  3961 &  8614 &  38 & 183 & 380 & 2130 & 173 &  742 &  693 & 1161 &  553 & 197 & 125 & 242 &  6013 \\
GCGT & 7796 &  5744 &  1491 &  64 & 558 & 610 & 5591 & 312 & 2235 & 3018 & 1719 & 1415 & 329 & 410 & 365 & 12763 \\
GCTT & 4278 &  3066 &  6132 &  16 & 116 & 350 & 1770 & 112 &  453 &  652 &  890 &  388 & 151 & 100 & 175 &  5056 \\
TCAT & 3855 & 11556 & 34755 &  48 & 595 & 534 &  499 &  53 &  405 &  268 & 1142 &  800 &  68 &  56 & 299 &  1758 \\
TCCT & 4397 &  7323 & 84453 &  28 & 466 & 449 &  866 &  95 &  447 &  571 & 1017 & 1728 &  67 &  88 & 212 &  3178 \\
TCGT & 5045 &  5329 & 33171 &  40 & 595 & 456 & 2073 & 111 & 1022 & 6497 & 1025 & 1086 &  96 & 138 & 235 & 18507 \\
TCTT & 4336 &  7487 & 27330 &  69 & 326 & 376 &  932 &  64 &  355 &  789 &  752 &  842 &  58 &  62 & 155 &  4029 \\
[1ex] 
\hline 
\end{tabular}
}
\label{table.aggr.data.12} 
\end{table}
\end{landscape}

\newpage\clearpage
\begin{landscape}
\begin{table}[ht]
\noindent
\caption{Table \ref{table.aggr.data.12} continued: occurrence counts (aggregated by cancer types) for the
next 48 mutation categories for cancer types X17 through X32.}
{\tiny
\begin{tabular}{l l l l l l l l l l l l l l l l l l} 
\\
\hline\hline 
Mutation & X17 & X18 & X19 & X20 & X21 & X22 & X23 & X24 & X25 & X26 & X27 & X28 & X29 & X30 & X31 & X32 \\[0.5ex] 
\hline 
\\
ATAA & 2099 &  1060 &   431 &  31 &   9 &  73 &   52 &   9 &   76 &   34 &  390 &   54 &   9 &   4 &  17 &   119 \\
ATCA & 1201 &  1112 &   418 &   4 &  22 & 151 &   75 &  29 &   97 &   54 &  238 &   93 &  24 &  16 &  41 &   256 \\
ATGA & 1558 &  2176 &   451 &   6 &  26 & 148 &   55 &  19 &   77 &   40 &  300 &  108 &  23 &  14 &  40 &   119 \\
ATTA & 1501 &   873 &  1059 &   2 &  17 & 116 &   98 &  12 &   91 &   62 &  265 &   99 &  12 &  18 &  21 &   424 \\
CTAA & 1106 &  1366 &   219 &   9 &  16 &  60 &   33 &  11 &   43 &   22 &  914 &   49 &   5 &   8 &   9 &    62 \\
CTCA & 1580 &  2157 &   567 &   3 &  26 & 188 &   99 &  25 &   90 &   56 &  443 &  126 &  16 &  21 &  25 &   252 \\
CTGA & 3163 &  6266 &   518 &   2 &  30 & 191 &  122 &  29 &  132 &  100 &  540 &  140 &  44 &  31 &  42 &   212 \\
CTTA & 1253 &  2176 &   762 &   2 &  17 & 194 &   80 &  10 &  104 &   80 &  377 &  110 &  23 &  70 &  19 &   271 \\
GTAA &  737 &  1419 &   211 &  24 &  22 &  73 &   31 &  11 &   65 &   20 &  377 &   60 &   8 &   5 &  10 &   103 \\
GTCA &  808 &  1125 &   248 &   5 &   9 & 111 &   72 &  12 &   65 &   36 &  204 &   91 &  22 &  16 &  12 &   218 \\
GTGA & 1232 &  2391 &   395 &   6 &  31 & 126 &   38 &  12 &   72 &   35 &  243 &   82 &  15 &  17 & 255 &   131 \\
GTTA &  888 &  1029 &   356 &   3 &  11 & 102 &   64 &  12 &   52 &   38 &  143 &   67 &   9 &   3 &  17 &   185 \\
TTAA & 1735 &   827 &   424 &   6 &  17 &  49 &   23 &   6 &   71 &   39 &  264 &   57 &  18 &   2 &   8 &    90 \\
TTCA &  897 &  1062 &   577 &   1 &  26 & 145 &   50 &   7 &   66 &   38 &  327 &   96 &  22 &   8 &  16 &   217 \\
TTGA & 1148 &  1623 &   379 &   1 &  16 &  82 &   21 &   4 &   49 &   36 &  251 &   61 &  12 &   6 &  17 &   113 \\
TTTA & 1985 &   940 &  1538 &   4 &  12 & 131 &   47 &   6 &   67 &   80 &  329 &   94 &  11 &   7 &  20 &   299 \\
ATAC & 9096 &  2438 &   512 &   9 &  39 & 189 &  243 &  66 &  251 &  135 &  530 &  168 &  43 &  54 & 101 &   922 \\
ATCC & 2835 &  1074 &   467 &  24 &  40 & 183 &  266 &  43 &  171 &  125 &  498 &  137 &  62 &  40 &  68 &  1163 \\
ATGC & 8618 &  2982 &   746 &  10 &  45 & 248 &  363 &  93 &  252 &  122 &  761 &  213 &  49 &  56 &  92 &  1716 \\
ATTC & 5928 &  1716 &   776 &  10 &  39 & 261 &  298 &  84 &  312 &  165 &  566 &  213 &  35 &  57 & 133 &   788 \\
CTAC & 2951 &  1387 &   613 &   4 &  16 &  87 &   94 &  26 &  108 &   85 &  432 &   93 &  34 &  23 &  54 &   557 \\
CTCC & 3360 &  1350 &   880 &  92 &  32 & 220 &  290 &  30 &  202 &  127 &  727 &  280 & 180 &  66 &  96 &   957 \\
CTGC & 6754 &  3236 &  1008 &  28 &  64 & 209 &  414 &  90 &  284 &  117 &  910 &  257 &  89 &  78 & 126 &  2141 \\
CTTC & 4047 &  1913 &  1913 &   9 &  33 & 224 &  391 &  50 &  231 &  127 &  730 &  347 & 255 & 117 & 159 &   812 \\
GTAC & 3048 &  1831 &   513 &   2 &  44 & 123 &  390 &  38 &  229 &  204 &  422 &  157 &  38 &  47 &  56 &  1717 \\
GTCC & 2708 &  1127 &   591 &  28 &  39 & 131 &  445 &  29 &  237 &  311 &  404 &  151 &  54 &  52 &  35 &  1844 \\
GTGC & 4987 &  1824 &   626 &  13 &  48 & 140 &  451 &  39 &  204 &  166 &  515 &  198 &  45 &  49 &  53 &  1871 \\
GTTC & 3124 &  1410 &  3319 &   6 &  40 & 163 &  400 &  50 &  255 &  267 &  494 &  236 &  51 &  77 &  76 &  1833 \\
TTAC & 3220 &   921 &   934 &   6 &  19 &  91 &  106 &  22 &  138 &  108 &  272 &  101 &  25 &  21 &  24 &   782 \\
TTCC & 2872 &   962 &   930 &  22 &  36 & 176 &  207 &  32 &  166 &  163 &  396 &  180 &  46 &  43 &  40 &  1411 \\
TTGC & 3895 &  1271 &   903 &   7 &  18 & 114 &  254 &  54 &  159 &  110 &  426 &  143 &  51 &  36 &  59 &  1429 \\
TTTC & 3660 &   994 &   986 &   3 &  22 & 148 &  245 &  58 &  168 &  175 &  417 &  140 &  37 &  37 &  45 &  1159 \\
ATAG & 1333 &   325 &   200 &   1 &   6 &  42 &   39 &   7 &   52 &   45 &  122 &   21 &  13 &  10 &   4 &   138 \\
ATCG &  705 &   295 &   272 &   1 &  14 &  58 &   97 &   6 &   55 &   62 &  230 &   35 &  13 &  18 &   5 &   422 \\
ATGG & 1122 &   522 &   549 &   4 &  13 &  87 &   76 &  12 &   49 &   40 &  229 &   78 &  21 &  14 &   8 &   215 \\
ATTG & 1101 &   348 &   334 &   0 &   7 &  55 &  182 &   7 &   95 &  239 &  272 &   30 &  18 &   8 &  19 &  1241 \\
CTAG &  503 &   238 &   119 &   0 &  11 &  43 &   13 &   3 &   36 &   37 &   79 &   22 &   3 &   4 &   8 &   184 \\
CTCG & 1030 &   504 &   443 &   6 &   9 &  84 &  100 &  11 &   92 &   42 &  210 &   70 &  11 &  16 &  24 &   372 \\
CTGG & 1519 &  1384 &   606 &  30 &  22 & 164 &  128 &  21 &  112 &   64 &  291 &  101 &  15 &  14 &  33 &   577 \\
CTTG & 1240 &  1100 &  1090 &   4 &  29 & 141 &  280 &  19 &  154 &  271 &  336 &   88 &  15 &  17 &  31 &  1741 \\
GTAG &  643 &   319 &   128 &  61 &   9 &  44 &   39 &   6 &   48 &   24 &  102 &   24 &  21 &   9 &   6 &    78 \\
GTCG &  658 &   311 &   202 &  69 &   5 &  52 &  100 &  10 &   65 &   55 &  127 &   34 &  58 &  19 &   8 &   259 \\
GTGG & 1314 &   835 &   904 & 384 &  33 & 121 &  190 &  15 &  128 &   33 &  350 &  103 & 124 &  24 &  15 &   181 \\
GTTG & 1097 &   478 &   467 &  35 &  12 &  67 &  178 &   9 &   92 &   67 &  171 &   62 &  36 &  19 &   7 &   408 \\
TTAG &  960 &   272 &   237 &   1 &  10 &  54 &   31 &   7 &   48 &  120 &  117 &   26 &   9 &   2 &   5 &   390 \\
TTCG & 1040 &   462 &   454 &   0 &  12 &  93 &  116 &   9 &   86 &  148 &  315 &   82 &  11 &  52 &  19 &   754 \\
TTGG & 1331 &   733 &   419 &   4 &  14 & 119 &   62 &  18 &   82 &   64 &  267 &   65 &  29 &   6 &  18 &   264 \\
TTTG & 2232 &   677 &   882 &   2 &  25 & 118 &  379 &  12 &   95 & 1083 &  472 &   66 &  14 &  10 &  26 &  4615 \\
[1ex] 
\hline 
\end{tabular}
}
\label{table.aggr.data.22} 
\end{table}
\end{landscape}

\newpage\clearpage
\begin{landscape}
\begin{table}[ht]
\noindent
\caption{Top-10 clusterings (Clustering-E1 through Clustering-E10) by occurrence counts (second column) in 30,000 runs (performed as 3 consecutive batches of 10,000 runs in each batch). Each run is based on 1,000 samplings (i.e., {\tt\small num.try = 1000} in the R function {\tt\small qrm.stat.ind.class()} in Appendix A of \cite{*K-means}); also, the {\em target} number of clusters is {\tt\small k = 13}, which is based on the effective rank a.k.a. eRank (and is computed using the R function {\tt{\small bio.erank.pc()}} in Appendix B of \cite{BioFM}). The columns ``Cl-1" through ``Cl-13" give the numbers of mutations in each cluster (the total number of mutations in each clustering is 96). The entries ``--" correspond to clusterings with fewer than 13 (the target number) of clusters. (Note that top-10 clusterings have either 11 or 12 clusters; however, there are other, less frequently occurring clusterings with 13 clusters.) While there was variability in the placement (by occurrence counts) of the top-10 clusterings within the aforesaid 3 batches of 10,000 runs, in each batch Clustering-E1 invariably had the highest count by a large margin: Batch1,  Clustering-E1 count = 95, second place count = 47; Batch2, Clustering-E1 count = 124, second place count = 46; Batch3, Clustering-E1 count = 115, second place count = 49.}
\begin{tabular}{l l l l l l l l l l l l l l l} 
\\
\hline\hline 
Name & Count & Cl-1 & Cl-2 & Cl-3 & Cl-4 & Cl-5 & Cl-6 & Cl-7 & Cl-8 & Cl-9 & Cl-10 & Cl-11 & Cl-12 & Cl-13\\[0.5ex] 
\hline 
Clustering-E1 & 334 & 3 & 4 & 6 & 6 & 6 & 7 & 7 & 9 & 16 & 16 & 16 & -- & -- \\
Clustering-E2 & 134 & 3 & 4 & 6 & 6 & 6 & 7 & 7 & 7 & 8 & 9 & 16 & 17 & -- \\
Clustering-E3 & 126 & 2 & 4 & 6 & 6 & 6 & 7 & 7 & 9 & 16 & 16 & 17 & -- & -- \\
Clustering-E4 & 120 & 2 & 4 & 6 & 6 & 6 & 7 & 7 & 8 & 9 & 9 & 16 & 16 & -- \\
Clustering-E5 & 112 & 3 & 4 & 6 & 6 & 6 & 7 & 7 & 9 & 15 & 16 & 17 & -- & -- \\
Clustering-E6 & 109 & 1 & 3 & 3 & 6 & 6 & 6 & 7 & 7 & 9 & 16 & 16 & 16 & -- \\
Clustering-E7 & 109 & 3 & 3 & 6 & 6 & 6 & 7 & 7 & 8 & 9 & 10 & 15 & 16 & -- \\
Clustering-E8 & 105 & 3 & 4 & 6 & 6 & 6 & 7 & 7 & 7 & 9 & 10 & 15 & 16 & -- \\
Clustering-E9 & 78 & 2 & 4 & 6 & 6 & 6 & 6 & 7 & 7 & 9 & 11 & 16 & 16 & -- \\
Clustering-E10 & 76 & 3 & 3 & 6 & 6 & 6 & 7 & 7 & 8 & 8 & 9 & 16 & 17 & -- \\
 [1ex] 
\hline 
\end{tabular}
\label{table.occurrence.cts} 
\end{table}
\end{landscape}

\newpage\clearpage
\begin{table}[ht]
\noindent
\caption{Weights (in the units of 1\%, rounded to 2 digits) for the first 48 mutation categories for the 11 clusters in Clustering-E1 (see Table \ref{table.occurrence.cts}) based on unnormalized regressions (see Subsection \ref{sub.res} for details). Each cluster is defined as containing the mutations with nonzero weights. For instance, cluster Cl-1 contains 3 mutations GCGA, TCGA, CTGA (also see Table \ref{table.weights.E.2}). In each cluster the weights are normalized to add up to 100\% (up to 2 digits due to the aforesaid rounding).}
{\scriptsize
\begin{tabular}{l l l l l l l l l l l l} 
\\
\hline\hline 
Mutation & Cl-1 & Cl-2 & Cl-3 & Cl-4 & Cl-5 & Cl-6 & Cl-7 & Cl-8 & Cl-9 & Cl-10 & Cl-11\\[0.5ex] 
\hline 
\\
ACAA &  0.00 &  0.00 &  0.00 &  0.00 &  0.00 &  0.00 &  0.00 &  0.00 &  3.51 &  0.00 & 0.00 \\
ACCA &  0.00 &  0.00 &  0.00 &  0.00 &  0.00 &  0.00 &  0.00 &  0.00 &  3.45 &  0.00 & 0.00 \\
ACGA &  0.00 &  0.00 &  0.00 &  0.00 &  0.00 &  0.00 & 14.21 &  0.00 &  0.00 &  0.00 & 0.00 \\
ACTA &  0.00 &  0.00 &  0.00 &  0.00 &  0.00 &  0.00 &  0.00 &  0.00 &  2.78 &  0.00 & 0.00 \\
CCAA &  0.00 &  0.00 &  0.00 &  0.00 &  0.00 &  0.00 &  0.00 &  0.00 &  5.01 &  0.00 & 0.00 \\
CCCA &  0.00 &  0.00 &  0.00 &  0.00 &  0.00 &  0.00 &  0.00 &  0.00 &  4.19 &  0.00 & 0.00 \\
CCGA &  0.00 &  0.00 &  0.00 &  0.00 &  0.00 &  0.00 &  0.00 &  0.00 &  3.05 &  0.00 & 0.00 \\
CCTA &  0.00 &  0.00 &  0.00 &  0.00 &  0.00 &  0.00 &  0.00 &  0.00 &  6.91 &  0.00 & 0.00 \\
GCAA &  0.00 &  0.00 &  0.00 &  0.00 &  0.00 &  0.00 &  0.00 &  0.00 &  5.05 &  0.00 & 0.00 \\
GCCA &  0.00 &  0.00 &  0.00 &  0.00 &  0.00 &  0.00 &  0.00 &  0.00 &  5.39 &  0.00 & 0.00 \\
GCGA & 39.73 &  0.00 &  0.00 &  0.00 &  0.00 &  0.00 &  0.00 &  0.00 &  0.00 &  0.00 & 0.00 \\
GCTA &  0.00 &  0.00 &  0.00 &  0.00 &  0.00 &  0.00 &  0.00 &  0.00 &  6.39 &  0.00 & 0.00 \\
TCAA &  0.00 &  0.00 &  0.00 &  0.00 &  0.00 &  0.00 &  0.00 &  0.00 &  6.05 &  0.00 & 0.00 \\
TCCA &  0.00 &  0.00 &  0.00 &  0.00 &  0.00 &  0.00 &  0.00 &  0.00 &  5.20 &  0.00 & 0.00 \\
TCGA & 25.16 &  0.00 &  0.00 &  0.00 &  0.00 &  0.00 &  0.00 &  0.00 &  0.00 &  0.00 & 0.00 \\
TCTA &  0.00 &  0.00 &  0.00 &  0.00 &  0.00 &  0.00 &  0.00 &  0.00 & 15.88 &  0.00 & 0.00 \\
ACAG &  0.00 &  0.00 &  0.00 &  0.00 &  0.00 &  0.00 & 14.55 &  0.00 &  0.00 &  0.00 & 0.00 \\
ACCG &  0.00 &  0.00 & 13.07 &  0.00 &  0.00 &  0.00 &  0.00 &  0.00 &  0.00 &  0.00 & 0.00 \\
ACGG &  0.00 &  0.00 &  0.00 &  0.00 &  0.00 &  9.41 &  0.00 &  0.00 &  0.00 &  0.00 & 0.00 \\
ACTG &  0.00 &  0.00 & 15.05 &  0.00 &  0.00 &  0.00 &  0.00 &  0.00 &  0.00 &  0.00 & 0.00 \\
CCAG &  0.00 &  0.00 &  0.00 &  0.00 &  0.00 &  0.00 & 14.65 &  0.00 &  0.00 &  0.00 & 0.00 \\
CCCG &  0.00 &  0.00 &  0.00 &  0.00 &  0.00 &  0.00 & 12.89 &  0.00 &  0.00 &  0.00 & 0.00 \\
CCGG &  0.00 &  0.00 &  0.00 &  0.00 &  0.00 &  0.00 & 11.72 &  0.00 &  0.00 &  0.00 & 0.00 \\
CCTG &  0.00 &  0.00 &  0.00 &  0.00 &  0.00 &  0.00 & 17.13 &  0.00 &  0.00 &  0.00 & 0.00 \\
GCAG &  0.00 &  0.00 &  0.00 &  0.00 &  0.00 & 14.48 &  0.00 &  0.00 &  0.00 &  0.00 & 0.00 \\
GCCG &  0.00 &  0.00 &  0.00 & 14.07 &  0.00 &  0.00 &  0.00 &  0.00 &  0.00 &  0.00 & 0.00 \\
GCGG &  0.00 &  0.00 &  0.00 &  0.00 &  0.00 & 25.13 &  0.00 &  0.00 &  0.00 &  0.00 & 0.00 \\
GCTG &  0.00 &  0.00 &  0.00 &  7.35 &  0.00 &  0.00 &  0.00 &  0.00 &  0.00 &  0.00 & 0.00 \\
TCAG &  0.00 &  0.00 &  0.00 &  0.00 &  0.00 &  0.00 &  0.00 &  0.00 & 10.41 &  0.00 & 0.00 \\
TCCG &  0.00 &  0.00 &  0.00 &  0.00 &  0.00 &  0.00 &  0.00 &  0.00 &  4.51 &  0.00 & 0.00 \\
TCGG &  0.00 &  0.00 &  0.00 &  0.00 &  0.00 &  0.00 & 14.86 &  0.00 &  0.00 &  0.00 & 0.00 \\
TCTG &  0.00 &  0.00 &  0.00 &  0.00 &  0.00 &  0.00 &  0.00 &  0.00 & 12.23 &  0.00 & 0.00 \\
ACAT &  0.00 &  0.00 &  0.00 &  0.00 &  0.00 &  0.00 &  0.00 &  0.00 &  0.00 &  1.63 & 0.00 \\
ACCT &  0.00 &  0.00 &  0.00 &  0.00 &  0.00 &  0.00 &  0.00 &  0.00 &  0.00 &  2.62 & 0.00 \\
ACGT &  0.00 &  0.00 &  0.00 &  0.00 &  0.00 &  0.00 &  0.00 &  0.00 &  0.00 &  7.61 & 0.00 \\
ACTT &  0.00 &  0.00 &  0.00 &  0.00 &  0.00 &  0.00 &  0.00 &  0.00 &  0.00 &  0.00 & 6.83 \\
CCAT &  0.00 &  0.00 &  0.00 &  0.00 &  0.00 &  0.00 &  0.00 &  0.00 &  0.00 &  4.65 & 0.00 \\
CCCT &  0.00 &  0.00 &  0.00 &  0.00 &  0.00 &  0.00 &  0.00 &  0.00 &  0.00 &  6.18 & 0.00 \\
CCGT &  0.00 &  0.00 &  0.00 &  0.00 &  0.00 &  0.00 &  0.00 &  0.00 &  0.00 &  8.07 & 0.00 \\
CCTT &  0.00 &  0.00 &  0.00 &  0.00 &  0.00 &  0.00 &  0.00 &  0.00 &  0.00 &  5.36 & 0.00 \\
GCAT &  0.00 &  0.00 &  0.00 &  0.00 &  0.00 &  0.00 &  0.00 &  0.00 &  0.00 &  2.17 & 0.00 \\
GCCT &  0.00 &  0.00 &  0.00 &  0.00 &  0.00 &  0.00 &  0.00 &  0.00 &  0.00 &  4.48 & 0.00 \\
GCGT &  0.00 &  0.00 &  0.00 &  0.00 &  0.00 &  0.00 &  0.00 &  0.00 &  0.00 &  9.57 & 0.00 \\
GCTT &  0.00 &  0.00 &  0.00 &  0.00 &  0.00 &  0.00 &  0.00 &  0.00 &  0.00 &  3.32 & 0.00 \\
TCAT &  0.00 &  0.00 &  0.00 &  0.00 &  0.00 &  0.00 &  0.00 &  0.00 &  0.00 &  9.26 & 0.00 \\
TCCT &  0.00 &  0.00 &  0.00 &  0.00 &  0.00 &  0.00 &  0.00 &  0.00 &  0.00 & 13.86 & 0.00 \\
TCGT &  0.00 &  0.00 &  0.00 &  0.00 &  0.00 &  0.00 &  0.00 &  0.00 &  0.00 & 12.42 & 0.00 \\
TCTT &  0.00 &  0.00 &  0.00 &  0.00 &  0.00 &  0.00 &  0.00 &  0.00 &  0.00 &  7.13 & 0.00 \\
[1ex] 
\hline 
\end{tabular}
}
\label{table.weights.E.1} 
\end{table}

\newpage\clearpage
\begin{table}[ht]
\noindent
\caption{Table \ref{table.weights.E.1} continued: weights for the next 48 mutation categories.}
{\scriptsize
\begin{tabular}{l l l l l l l l l l l l} 
\\
\hline\hline 
Mutation & Cl-1 & Cl-2 & Cl-3 & Cl-4 & Cl-5 & Cl-6 & Cl-7 & Cl-8 & Cl-9 & Cl-10 & Cl-11\\[0.5ex] 
\hline 
\\
ATAA &  0.00 &  0.00 &  0.00 &  0.00 &  0.00 & 13.50 &  0.00 &  0.00 &  0.00 &  0.00 & 0.00 \\
ATCA &  0.00 &  0.00 & 15.56 &  0.00 &  0.00 &  0.00 &  0.00 &  0.00 &  0.00 &  0.00 & 0.00 \\
ATGA &  0.00 &  0.00 & 16.90 &  0.00 &  0.00 &  0.00 &  0.00 &  0.00 &  0.00 &  0.00 & 0.00 \\
ATTA &  0.00 &  0.00 &  0.00 &  0.00 & 14.10 &  0.00 &  0.00 &  0.00 &  0.00 &  0.00 & 0.00 \\
CTAA &  0.00 &  0.00 &  0.00 &  0.00 &  0.00 &  9.12 &  0.00 &  0.00 &  0.00 &  0.00 & 0.00 \\
CTCA &  0.00 &  0.00 & 17.42 &  0.00 &  0.00 &  0.00 &  0.00 &  0.00 &  0.00 &  0.00 & 0.00 \\
CTGA & 35.11 &  0.00 &  0.00 &  0.00 &  0.00 &  0.00 &  0.00 &  0.00 &  0.00 &  0.00 & 0.00 \\
CTTA &  0.00 &  0.00 &  0.00 &  0.00 & 19.22 &  0.00 &  0.00 &  0.00 &  0.00 &  0.00 & 0.00 \\
GTAA &  0.00 &  0.00 &  0.00 &  0.00 &  0.00 & 11.14 &  0.00 &  0.00 &  0.00 &  0.00 & 0.00 \\
GTCA &  0.00 &  0.00 &  0.00 &  0.00 &  0.00 &  0.00 &  0.00 & 13.54 &  0.00 &  0.00 & 0.00 \\
GTGA &  0.00 &  0.00 & 22.00 &  0.00 &  0.00 &  0.00 &  0.00 &  0.00 &  0.00 &  0.00 & 0.00 \\
GTTA &  0.00 &  0.00 &  0.00 &  0.00 &  0.00 &  0.00 &  0.00 & 10.69 &  0.00 &  0.00 & 0.00 \\
TTAA &  0.00 & 19.89 &  0.00 &  0.00 &  0.00 &  0.00 &  0.00 &  0.00 &  0.00 &  0.00 & 0.00 \\
TTCA &  0.00 & 26.53 &  0.00 &  0.00 &  0.00 &  0.00 &  0.00 &  0.00 &  0.00 &  0.00 & 0.00 \\
TTGA &  0.00 & 24.44 &  0.00 &  0.00 &  0.00 &  0.00 &  0.00 &  0.00 &  0.00 &  0.00 & 0.00 \\
TTTA &  0.00 & 29.14 &  0.00 &  0.00 &  0.00 &  0.00 &  0.00 &  0.00 &  0.00 &  0.00 & 0.00 \\
ATAC &  0.00 &  0.00 &  0.00 &  0.00 &  0.00 &  0.00 &  0.00 &  0.00 &  0.00 &  0.00 & 6.02 \\
ATCC &  0.00 &  0.00 &  0.00 &  0.00 &  0.00 &  0.00 &  0.00 &  0.00 &  0.00 &  0.00 & 5.08 \\
ATGC &  0.00 &  0.00 &  0.00 &  0.00 &  0.00 &  0.00 &  0.00 &  0.00 &  0.00 &  0.00 & 7.83 \\
ATTC &  0.00 &  0.00 &  0.00 &  0.00 &  0.00 &  0.00 &  0.00 &  0.00 &  0.00 &  0.00 & 6.08 \\
CTAC &  0.00 &  0.00 &  0.00 &  0.00 & 19.38 &  0.00 &  0.00 &  0.00 &  0.00 &  0.00 & 0.00 \\
CTCC &  0.00 &  0.00 &  0.00 &  0.00 &  0.00 &  0.00 &  0.00 &  0.00 &  0.00 &  0.00 & 7.15 \\
CTGC &  0.00 &  0.00 &  0.00 &  0.00 &  0.00 &  0.00 &  0.00 &  0.00 &  0.00 &  1.67 & 0.00 \\
CTTC &  0.00 &  0.00 &  0.00 &  0.00 &  0.00 &  0.00 &  0.00 &  0.00 &  0.00 &  0.00 & 7.81 \\
GTAC &  0.00 &  0.00 &  0.00 &  0.00 &  0.00 &  0.00 &  0.00 &  0.00 &  0.00 &  0.00 & 6.00 \\
GTCC &  0.00 &  0.00 &  0.00 &  0.00 &  0.00 &  0.00 &  0.00 &  0.00 &  0.00 &  0.00 & 6.68 \\
GTGC &  0.00 &  0.00 &  0.00 &  0.00 &  0.00 &  0.00 &  0.00 &  0.00 &  0.00 &  0.00 & 6.45 \\
GTTC &  0.00 &  0.00 &  0.00 &  0.00 &  0.00 &  0.00 &  0.00 &  0.00 &  0.00 &  0.00 & 6.71 \\
TTAC &  0.00 &  0.00 &  0.00 &  0.00 & 18.54 &  0.00 &  0.00 &  0.00 &  0.00 &  0.00 & 0.00 \\
TTCC &  0.00 &  0.00 &  0.00 &  0.00 &  0.00 &  0.00 &  0.00 &  0.00 &  0.00 &  0.00 & 5.16 \\
TTGC &  0.00 &  0.00 &  0.00 &  0.00 &  0.00 &  0.00 &  0.00 &  0.00 &  0.00 &  0.00 & 4.64 \\
TTTC &  0.00 &  0.00 &  0.00 &  0.00 &  0.00 &  0.00 &  0.00 &  0.00 &  0.00 &  0.00 & 4.63 \\
ATAG &  0.00 &  0.00 &  0.00 &  0.00 &  0.00 &  0.00 &  0.00 &  9.17 &  0.00 &  0.00 & 0.00 \\
ATCG &  0.00 &  0.00 &  0.00 &  0.00 &  0.00 &  0.00 &  0.00 & 10.65 &  0.00 &  0.00 & 0.00 \\
ATGG &  0.00 &  0.00 &  0.00 &  0.00 &  0.00 &  0.00 &  0.00 & 11.82 &  0.00 &  0.00 & 0.00 \\
ATTG &  0.00 &  0.00 &  0.00 &  0.00 & 14.65 &  0.00 &  0.00 &  0.00 &  0.00 &  0.00 & 0.00 \\
CTAG &  0.00 &  0.00 &  0.00 &  0.00 &  0.00 &  0.00 &  0.00 &  6.72 &  0.00 &  0.00 & 0.00 \\
CTCG &  0.00 &  0.00 &  0.00 &  0.00 &  0.00 &  0.00 &  0.00 & 14.76 &  0.00 &  0.00 & 0.00 \\
CTGG &  0.00 &  0.00 &  0.00 &  7.09 &  0.00 &  0.00 &  0.00 &  0.00 &  0.00 &  0.00 & 0.00 \\
CTTG &  0.00 &  0.00 &  0.00 &  0.00 &  0.00 &  0.00 &  0.00 &  0.00 &  0.00 &  0.00 & 6.31 \\
GTAG &  0.00 &  0.00 &  0.00 &  0.00 &  0.00 & 17.22 &  0.00 &  0.00 &  0.00 &  0.00 & 0.00 \\
GTCG &  0.00 &  0.00 &  0.00 & 10.70 &  0.00 &  0.00 &  0.00 &  0.00 &  0.00 &  0.00 & 0.00 \\
GTGG &  0.00 &  0.00 &  0.00 & 53.39 &  0.00 &  0.00 &  0.00 &  0.00 &  0.00 &  0.00 & 0.00 \\
GTTG &  0.00 &  0.00 &  0.00 &  7.38 &  0.00 &  0.00 &  0.00 &  0.00 &  0.00 &  0.00 & 0.00 \\
TTAG &  0.00 &  0.00 &  0.00 &  0.00 &  0.00 &  0.00 &  0.00 &  9.41 &  0.00 &  0.00 & 0.00 \\
TTCG &  0.00 &  0.00 &  0.00 &  0.00 & 14.11 &  0.00 &  0.00 &  0.00 &  0.00 &  0.00 & 0.00 \\
TTGG &  0.00 &  0.00 &  0.00 &  0.00 &  0.00 &  0.00 &  0.00 & 13.24 &  0.00 &  0.00 & 0.00 \\
TTTG &  0.00 &  0.00 &  0.00 &  0.00 &  0.00 &  0.00 &  0.00 &  0.00 &  0.00 &  0.00 & 6.63 \\
[1ex] 
\hline 
\end{tabular}
}
\label{table.weights.E.2} 
\end{table}

\newpage\clearpage
\begin{table}[ht]
\noindent
\caption{Weights for the first 48 mutation categories for the 11 clusters in Clustering-E1 (see Table \ref{table.occurrence.cts}) based on normalized regressions (see Subsection \ref{sub.res} for details). The conventions are the same as in Table \ref{table.weights.E.1}.}
{\scriptsize
\begin{tabular}{l l l l l l l l l l l l} 
\\
\hline\hline 
Mutation & Cl-1 & Cl-2 & Cl-3 & Cl-4 & Cl-5 & Cl-6 & Cl-7 & Cl-8 & Cl-9 & Cl-10 & Cl-11\\[0.5ex] 
\hline 
\\
ACAA &  0.00 &  0.00 &  0.00 &  0.00 &  0.00 &  0.00 &  0.00 &  0.00 &  4.05 &  0.00 & 0.00 \\
ACCA &  0.00 &  0.00 &  0.00 &  0.00 &  0.00 &  0.00 &  0.00 &  0.00 &  3.94 &  0.00 & 0.00 \\
ACGA &  0.00 &  0.00 &  0.00 &  0.00 &  0.00 &  0.00 & 13.92 &  0.00 &  0.00 &  0.00 & 0.00 \\
ACTA &  0.00 &  0.00 &  0.00 &  0.00 &  0.00 &  0.00 &  0.00 &  0.00 &  2.98 &  0.00 & 0.00 \\
CCAA &  0.00 &  0.00 &  0.00 &  0.00 &  0.00 &  0.00 &  0.00 &  0.00 &  5.71 &  0.00 & 0.00 \\
CCCA &  0.00 &  0.00 &  0.00 &  0.00 &  0.00 &  0.00 &  0.00 &  0.00 &  4.76 &  0.00 & 0.00 \\
CCGA &  0.00 &  0.00 &  0.00 &  0.00 &  0.00 &  0.00 &  0.00 &  0.00 &  3.49 &  0.00 & 0.00 \\
CCTA &  0.00 &  0.00 &  0.00 &  0.00 &  0.00 &  0.00 &  0.00 &  0.00 &  7.19 &  0.00 & 0.00 \\
GCAA &  0.00 &  0.00 &  0.00 &  0.00 &  0.00 &  0.00 &  0.00 &  0.00 &  5.78 &  0.00 & 0.00 \\
GCCA &  0.00 &  0.00 &  0.00 &  0.00 &  0.00 &  0.00 &  0.00 &  0.00 &  6.17 &  0.00 & 0.00 \\
GCGA & 39.97 &  0.00 &  0.00 &  0.00 &  0.00 &  0.00 &  0.00 &  0.00 &  0.00 &  0.00 & 0.00 \\
GCTA &  0.00 &  0.00 &  0.00 &  0.00 &  0.00 &  0.00 &  0.00 &  0.00 &  6.96 &  0.00 & 0.00 \\
TCAA &  0.00 &  0.00 &  0.00 &  0.00 &  0.00 &  0.00 &  0.00 &  0.00 &  5.91 &  0.00 & 0.00 \\
TCCA &  0.00 &  0.00 &  0.00 &  0.00 &  0.00 &  0.00 &  0.00 &  0.00 &  5.56 &  0.00 & 0.00 \\
TCGA & 26.06 &  0.00 &  0.00 &  0.00 &  0.00 &  0.00 &  0.00 &  0.00 &  0.00 &  0.00 & 0.00 \\
TCTA &  0.00 &  0.00 &  0.00 &  0.00 &  0.00 &  0.00 &  0.00 &  0.00 & 13.30 &  0.00 & 0.00 \\
ACAG &  0.00 &  0.00 &  0.00 &  0.00 &  0.00 &  0.00 & 14.83 &  0.00 &  0.00 &  0.00 & 0.00 \\
ACCG &  0.00 &  0.00 & 13.73 &  0.00 &  0.00 &  0.00 &  0.00 &  0.00 &  0.00 &  0.00 & 0.00 \\
ACGG &  0.00 &  0.00 &  0.00 &  0.00 &  0.00 & 10.02 &  0.00 &  0.00 &  0.00 &  0.00 & 0.00 \\
ACTG &  0.00 &  0.00 & 15.79 &  0.00 &  0.00 &  0.00 &  0.00 &  0.00 &  0.00 &  0.00 & 0.00 \\
CCAG &  0.00 &  0.00 &  0.00 &  0.00 &  0.00 &  0.00 & 14.81 &  0.00 &  0.00 &  0.00 & 0.00 \\
CCCG &  0.00 &  0.00 &  0.00 &  0.00 &  0.00 &  0.00 & 13.10 &  0.00 &  0.00 &  0.00 & 0.00 \\
CCGG &  0.00 &  0.00 &  0.00 &  0.00 &  0.00 &  0.00 & 11.85 &  0.00 &  0.00 &  0.00 & 0.00 \\
CCTG &  0.00 &  0.00 &  0.00 &  0.00 &  0.00 &  0.00 & 17.23 &  0.00 &  0.00 &  0.00 & 0.00 \\
GCAG &  0.00 &  0.00 &  0.00 &  0.00 &  0.00 & 14.97 &  0.00 &  0.00 &  0.00 &  0.00 & 0.00 \\
GCCG &  0.00 &  0.00 &  0.00 & 14.36 &  0.00 &  0.00 &  0.00 &  0.00 &  0.00 &  0.00 & 0.00 \\
GCGG &  0.00 &  0.00 &  0.00 &  0.00 &  0.00 & 23.52 &  0.00 &  0.00 &  0.00 &  0.00 & 0.00 \\
GCTG &  0.00 &  0.00 &  0.00 &  9.16 &  0.00 &  0.00 &  0.00 &  0.00 &  0.00 &  0.00 & 0.00 \\
TCAG &  0.00 &  0.00 &  0.00 &  0.00 &  0.00 &  0.00 &  0.00 &  0.00 &  9.11 &  0.00 & 0.00 \\
TCCG &  0.00 &  0.00 &  0.00 &  0.00 &  0.00 &  0.00 &  0.00 &  0.00 &  4.30 &  0.00 & 0.00 \\
TCGG &  0.00 &  0.00 &  0.00 &  0.00 &  0.00 &  0.00 & 14.26 &  0.00 &  0.00 &  0.00 & 0.00 \\
TCTG &  0.00 &  0.00 &  0.00 &  0.00 &  0.00 &  0.00 &  0.00 &  0.00 & 10.79 &  0.00 & 0.00 \\
ACAT &  0.00 &  0.00 &  0.00 &  0.00 &  0.00 &  0.00 &  0.00 &  0.00 &  0.00 &  1.97 & 0.00 \\
ACCT &  0.00 &  0.00 &  0.00 &  0.00 &  0.00 &  0.00 &  0.00 &  0.00 &  0.00 &  2.65 & 0.00 \\
ACGT &  0.00 &  0.00 &  0.00 &  0.00 &  0.00 &  0.00 &  0.00 &  0.00 &  0.00 &  9.10 & 0.00 \\
ACTT &  0.00 &  0.00 &  0.00 &  0.00 &  0.00 &  0.00 &  0.00 &  0.00 &  0.00 &  0.00 & 6.90 \\
CCAT &  0.00 &  0.00 &  0.00 &  0.00 &  0.00 &  0.00 &  0.00 &  0.00 &  0.00 &  4.19 & 0.00 \\
CCCT &  0.00 &  0.00 &  0.00 &  0.00 &  0.00 &  0.00 &  0.00 &  0.00 &  0.00 &  5.37 & 0.00 \\
CCGT &  0.00 &  0.00 &  0.00 &  0.00 &  0.00 &  0.00 &  0.00 &  0.00 &  0.00 &  9.13 & 0.00 \\
CCTT &  0.00 &  0.00 &  0.00 &  0.00 &  0.00 &  0.00 &  0.00 &  0.00 &  0.00 &  4.90 & 0.00 \\
GCAT &  0.00 &  0.00 &  0.00 &  0.00 &  0.00 &  0.00 &  0.00 &  0.00 &  0.00 &  2.64 & 0.00 \\
GCCT &  0.00 &  0.00 &  0.00 &  0.00 &  0.00 &  0.00 &  0.00 &  0.00 &  0.00 &  4.98 & 0.00 \\
GCGT &  0.00 &  0.00 &  0.00 &  0.00 &  0.00 &  0.00 &  0.00 &  0.00 &  0.00 & 11.43 & 0.00 \\
GCTT &  0.00 &  0.00 &  0.00 &  0.00 &  0.00 &  0.00 &  0.00 &  0.00 &  0.00 &  3.69 & 0.00 \\
TCAT &  0.00 &  0.00 &  0.00 &  0.00 &  0.00 &  0.00 &  0.00 &  0.00 &  0.00 &  8.08 & 0.00 \\
TCCT &  0.00 &  0.00 &  0.00 &  0.00 &  0.00 &  0.00 &  0.00 &  0.00 &  0.00 & 11.38 & 0.00 \\
TCGT &  0.00 &  0.00 &  0.00 &  0.00 &  0.00 &  0.00 &  0.00 &  0.00 &  0.00 & 12.03 & 0.00 \\
TCTT &  0.00 &  0.00 &  0.00 &  0.00 &  0.00 &  0.00 &  0.00 &  0.00 &  0.00 &  6.42 & 0.00 \\
[1ex] 
\hline 
\end{tabular}
}
\label{table.weights.E.N1} 
\end{table}

\newpage\clearpage
\begin{table}[ht]
\noindent
\caption{Table \ref{table.weights.E.N1} continued: weights for the next 48 mutation categories.}
{\scriptsize
\begin{tabular}{l l l l l l l l l l l l} 
\\
\hline\hline 
Mutation & Cl-1 & Cl-2 & Cl-3 & Cl-4 & Cl-5 & Cl-6 & Cl-7 & Cl-8 & Cl-9 & Cl-10 & Cl-11 \\[0.5ex] 
\hline 
\\
ATAA &  0.00 &  0.00 &  0.00 &  0.00 &  0.00 & 13.78 &  0.00 &  0.00 &  0.00 &  0.00 & 0.00 \\
ATCA &  0.00 &  0.00 & 16.08 &  0.00 &  0.00 &  0.00 &  0.00 &  0.00 &  0.00 &  0.00 & 0.00 \\
ATGA &  0.00 &  0.00 & 16.98 &  0.00 &  0.00 &  0.00 &  0.00 &  0.00 &  0.00 &  0.00 & 0.00 \\
ATTA &  0.00 &  0.00 &  0.00 &  0.00 & 14.23 &  0.00 &  0.00 &  0.00 &  0.00 &  0.00 & 0.00 \\
CTAA &  0.00 &  0.00 &  0.00 &  0.00 &  0.00 & 10.07 &  0.00 &  0.00 &  0.00 &  0.00 & 0.00 \\
CTCA &  0.00 &  0.00 & 18.00 &  0.00 &  0.00 &  0.00 &  0.00 &  0.00 &  0.00 &  0.00 & 0.00 \\
CTGA & 33.97 &  0.00 &  0.00 &  0.00 &  0.00 &  0.00 &  0.00 &  0.00 &  0.00 &  0.00 & 0.00 \\
CTTA &  0.00 &  0.00 &  0.00 &  0.00 & 19.11 &  0.00 &  0.00 &  0.00 &  0.00 &  0.00 & 0.00 \\
GTAA &  0.00 &  0.00 &  0.00 &  0.00 &  0.00 & 11.46 &  0.00 &  0.00 &  0.00 &  0.00 & 0.00 \\
GTCA &  0.00 &  0.00 &  0.00 &  0.00 &  0.00 &  0.00 &  0.00 & 13.53 &  0.00 &  0.00 & 0.00 \\
GTGA &  0.00 &  0.00 & 19.41 &  0.00 &  0.00 &  0.00 &  0.00 &  0.00 &  0.00 &  0.00 & 0.00 \\
GTTA &  0.00 &  0.00 &  0.00 &  0.00 &  0.00 &  0.00 &  0.00 & 10.75 &  0.00 &  0.00 & 0.00 \\
TTAA &  0.00 & 20.00 &  0.00 &  0.00 &  0.00 &  0.00 &  0.00 &  0.00 &  0.00 &  0.00 & 0.00 \\
TTCA &  0.00 & 26.57 &  0.00 &  0.00 &  0.00 &  0.00 &  0.00 &  0.00 &  0.00 &  0.00 & 0.00 \\
TTGA &  0.00 & 24.38 &  0.00 &  0.00 &  0.00 &  0.00 &  0.00 &  0.00 &  0.00 &  0.00 & 0.00 \\
TTTA &  0.00 & 29.05 &  0.00 &  0.00 &  0.00 &  0.00 &  0.00 &  0.00 &  0.00 &  0.00 & 0.00 \\
ATAC &  0.00 &  0.00 &  0.00 &  0.00 &  0.00 &  0.00 &  0.00 &  0.00 &  0.00 &  0.00 & 6.14 \\
ATCC &  0.00 &  0.00 &  0.00 &  0.00 &  0.00 &  0.00 &  0.00 &  0.00 &  0.00 &  0.00 & 5.13 \\
ATGC &  0.00 &  0.00 &  0.00 &  0.00 &  0.00 &  0.00 &  0.00 &  0.00 &  0.00 &  0.00 & 7.96 \\
ATTC &  0.00 &  0.00 &  0.00 &  0.00 &  0.00 &  0.00 &  0.00 &  0.00 &  0.00 &  0.00 & 6.21 \\
CTAC &  0.00 &  0.00 &  0.00 &  0.00 & 19.54 &  0.00 &  0.00 &  0.00 &  0.00 &  0.00 & 0.00 \\
CTCC &  0.00 &  0.00 &  0.00 &  0.00 &  0.00 &  0.00 &  0.00 &  0.00 &  0.00 &  0.00 & 7.14 \\
CTGC &  0.00 &  0.00 &  0.00 &  0.00 &  0.00 &  0.00 &  0.00 &  0.00 &  0.00 &  2.04 & 0.00 \\
CTTC &  0.00 &  0.00 &  0.00 &  0.00 &  0.00 &  0.00 &  0.00 &  0.00 &  0.00 &  0.00 & 7.76 \\
GTAC &  0.00 &  0.00 &  0.00 &  0.00 &  0.00 &  0.00 &  0.00 &  0.00 &  0.00 &  0.00 & 5.99 \\
GTCC &  0.00 &  0.00 &  0.00 &  0.00 &  0.00 &  0.00 &  0.00 &  0.00 &  0.00 &  0.00 & 6.64 \\
GTGC &  0.00 &  0.00 &  0.00 &  0.00 &  0.00 &  0.00 &  0.00 &  0.00 &  0.00 &  0.00 & 6.46 \\
GTTC &  0.00 &  0.00 &  0.00 &  0.00 &  0.00 &  0.00 &  0.00 &  0.00 &  0.00 &  0.00 & 6.72 \\
TTAC &  0.00 &  0.00 &  0.00 &  0.00 & 18.66 &  0.00 &  0.00 &  0.00 &  0.00 &  0.00 & 0.00 \\
TTCC &  0.00 &  0.00 &  0.00 &  0.00 &  0.00 &  0.00 &  0.00 &  0.00 &  0.00 &  0.00 & 5.18 \\
TTGC &  0.00 &  0.00 &  0.00 &  0.00 &  0.00 &  0.00 &  0.00 &  0.00 &  0.00 &  0.00 & 4.68 \\
TTTC &  0.00 &  0.00 &  0.00 &  0.00 &  0.00 &  0.00 &  0.00 &  0.00 &  0.00 &  0.00 & 4.69 \\
ATAG &  0.00 &  0.00 &  0.00 &  0.00 &  0.00 &  0.00 &  0.00 &  9.14 &  0.00 &  0.00 & 0.00 \\
ATCG &  0.00 &  0.00 &  0.00 &  0.00 &  0.00 &  0.00 &  0.00 & 10.60 &  0.00 &  0.00 & 0.00 \\
ATGG &  0.00 &  0.00 &  0.00 &  0.00 &  0.00 &  0.00 &  0.00 & 11.81 &  0.00 &  0.00 & 0.00 \\
ATTG &  0.00 &  0.00 &  0.00 &  0.00 & 14.48 &  0.00 &  0.00 &  0.00 &  0.00 &  0.00 & 0.00 \\
CTAG &  0.00 &  0.00 &  0.00 &  0.00 &  0.00 &  0.00 &  0.00 &  6.74 &  0.00 &  0.00 & 0.00 \\
CTCG &  0.00 &  0.00 &  0.00 &  0.00 &  0.00 &  0.00 &  0.00 & 14.76 &  0.00 &  0.00 & 0.00 \\
CTGG &  0.00 &  0.00 &  0.00 &  9.04 &  0.00 &  0.00 &  0.00 &  0.00 &  0.00 &  0.00 & 0.00 \\
CTTG &  0.00 &  0.00 &  0.00 &  0.00 &  0.00 &  0.00 &  0.00 &  0.00 &  0.00 &  0.00 & 6.03 \\
GTAG &  0.00 &  0.00 &  0.00 &  0.00 &  0.00 & 16.18 &  0.00 &  0.00 &  0.00 &  0.00 & 0.00 \\
GTCG &  0.00 &  0.00 &  0.00 & 10.55 &  0.00 &  0.00 &  0.00 &  0.00 &  0.00 &  0.00 & 0.00 \\
GTGG &  0.00 &  0.00 &  0.00 & 47.90 &  0.00 &  0.00 &  0.00 &  0.00 &  0.00 &  0.00 & 0.00 \\
GTTG &  0.00 &  0.00 &  0.00 &  8.98 &  0.00 &  0.00 &  0.00 &  0.00 &  0.00 &  0.00 & 0.00 \\
TTAG &  0.00 &  0.00 &  0.00 &  0.00 &  0.00 &  0.00 &  0.00 &  9.40 &  0.00 &  0.00 & 0.00 \\
TTCG &  0.00 &  0.00 &  0.00 &  0.00 & 13.98 &  0.00 &  0.00 &  0.00 &  0.00 &  0.00 & 0.00 \\
TTGG &  0.00 &  0.00 &  0.00 &  0.00 &  0.00 &  0.00 &  0.00 & 13.26 &  0.00 &  0.00 & 0.00 \\
TTTG &  0.00 &  0.00 &  0.00 &  0.00 &  0.00 &  0.00 &  0.00 &  0.00 &  0.00 &  0.00 & 6.37 \\
[1ex] 
\hline 
\end{tabular}
}
\label{table.weights.E.N2} 
\end{table}

\newpage\clearpage
\begin{landscape}
\begin{table}[ht]
\noindent
\caption{The within-cluster cross-sectional correlations $\Theta_{sA}$ (columns 2-12), the overall correlations $\Xi_s$ (column 15) based on the overall cross-sectional regressions, and multiple $R^2$ and adjusted $R^2$ of these regressions (columns 13 and 14). The cluster weights are based on unnormalized regressions (see Subsections \ref{sub.res} and \ref{sub.within} for details). All quantities are in the units of 1\% rounded to 2 digits. The values above 80\% are given in bold font. The values above 70\% are underlined.
}
{\scriptsize
\begin{tabular}{l l l l l l l l l l l l l l l} 
\\
\hline\hline 
Type & Cl-1 & Cl-2 & Cl-3 & Cl-4 & Cl-5 & Cl-6 & Cl-7 & Cl-8 & Cl-9 & Cl-10 & Cl-11 & $R^2$ & adj-$R^2$ & Cor\\[0.5ex] 
\hline 
X1 & {\bf 82.73} & 52.61 & -38.35 & {\bf 88.99} & -2.48 & 62.31 & 44.74 & 47.99 & 46.96 & 68.29 & -17.3 & {\bf 81.53} & {\underline{79.14}} & {\bf 86.03} \\
X2 & 57.84 & {\underline{79.57}} & -21.35 & 2.46 & 14.07 & 29 & -23.27 & 45.7 & -9.06 & 37.86 & 23.69 & 61.55 & 56.57 & {\underline{70.97}} \\
X3 & {\bf 97.84} & 59.33 & -34.88 & {\bf 85.84} & {\bf 93.71} & 20.49 & 49.36 & {\underline{72.28}} & 24.43 & 48.92 & 13.55 & {\underline{75.84}} & {\underline{72.71}} & {\underline{75.03}} \\
X4 & {\underline{79.67}} & 9.54 & 2.33 & -53.43 & 33.46 & -25.78 & -10.98 & 37.47 & 49.42 & 35.11 & 6.69 & {\underline{70.35}} & 66.51 & 60.11 \\
X5 & {\bf 99.21} & 36.43 & 13.54 & 46.65 & {\bf 96.99} & -30.2 & -76.87 & 51.58 & -27.23 & 18.49 & 37.03 & {\underline{70.96}} & 67.21 & 61.21 \\
X6 & -87.79 & 64.06 & -30.37 & {\bf 93.94} & {\bf 89.43} & 27.25 & 41.11 & {\bf 81.67} & 66.06 & 61.77 & 57.68 & 64.09 & 59.44 & {\underline{74.35}} \\
X7 & 49.56 & {\bf 94.33} & -63.27 & 23.6 & 69.48 & -4.55 & 53.97 & {\bf 88.73} & 45.59 & 34.97 & 28.95 & 59.28 & 54.01 & 68 \\
X8 & -33.14 & 16.16 & -72.5 & {\bf 97.72} & 36.79 & -35.38 & {\underline{76.35}} & 58.44 & 67.06 & 51.8 & 44.57 & 65.49 & 61.02 & {\underline{72.05}} \\
X9 & -94.76 & 61.06 & -88.86 & -49.91 & -5.3 & -20.18 & 30.47 & 59.55 & 64.49 & 62.66 & 37.67 & 61.25 & 56.24 & {\underline{73.38}} \\
X10 & 30.52 & -7.31 & {\underline{75.57}} & 7.44 & {\underline{77.24}} & -53.34 & 36.34 & {\bf 82.27} & 10.41 & 42.52 & 54.56 & {\underline{74.72}} & {\underline{71.45}} & 65.04 \\
X11 & 6.48 & 54.54 & 53.91 & -58.77 & 44.42 & -0.39 & {\underline{70.02}} & -30.55 & 49.29 & 42.58 & 28.14 & {\underline{77.46}} & {\underline{74.55}} & {\underline{72.98}} \\
X12 & -72.76 & {\underline{76.02}} & -15.94 & -43.69 & 7.61 & -44.71 & 37.64 & 64.45 & 67.75 & 47.08 & 50.24 & 67.32 & 63.09 & {\underline{73.76}} \\
X13 & -85.31 & {\bf 93.35} & -52.58 & -40.15 & 50.94 & 11.64 & {\bf 93.66} & {\underline{76.36}} & {\underline{73.57}} & 58.14 & 31.23 & {\underline{73.99}} & {\underline{70.63}} & {\underline{76.38}} \\
X14 & {\underline{70.94}} & 62.01 & -32.58 & -42.79 & {\bf 85.81} & -31.98 & 69.19 & {\underline{77.94}} & 31.37 & 35.25 & 38.43 & 55.44 & 49.67 & 65.84 \\
X15 & 12.1 & {\bf 87.76} & -64.16 & 62.01 & {\bf 92.19} & -40.36 & 56.64 & {\underline{77.94}} & 34.37 & 39.43 & 46.34 & 60.01 & 54.84 & {\underline{70.44}} \\
X16 & 30.62 & {\bf 83.56} & -8.44 & -1.16 & {\bf 84.79} & 1 & {\underline{71.1}} & 69.7 & 60.24 & {\bf 80.7} & 37.18 & {\bf 85.79} & {\bf 83.95} & {\bf 87.99} \\
X17 & 45.65 & 1.38 & 8.66 & 23.53 & 66.13 & -13.07 & 45.08 & 40 & 17.91 & 8.63 & 23.89 & {\underline{75.75}} & {\underline{72.62}} & 65.94 \\
X18 & 66.52 & 9.2 & {\underline{79.62}} & -13.51 & {\underline{78.2}} & -5.95 & 16.95 & 56.68 & 8.34 & 65.83 & 51.17 & {\underline{76.87}} & {\underline{73.88}} & {\underline{74.51}} \\
X19 & -56.72 & {\underline{76.08}} & 41.95 & {\underline{77.89}} & 21.98 & -24.67 & -44.45 & 69.91 & -6.06 & 69.84 & 31.03 & {\underline{70.19}} & 66.33 & {\bf 81.77} \\
X20 & 63.1 & -45.68 & 59.23 & {\bf 99.95} & {\underline{71.77}} & {\bf 98.3} & -66.7 & {\bf 94.37} & -19.75 & 54.91 & 20.7 & {\bf 91.03} & {\bf 89.87} & {\bf 94.01} \\
X21 & 30.55 & -9.19 & -3.43 & 32.96 & 58.57 & -42.26 & 18.66 & 10.7 & 5.66 & {\bf 87.75} & 43.01 & {\underline{78.2}} & {\underline{75.37}} & {\bf 81.7} \\
X22 & 14.3 & {\bf 89.91} & -48.97 & -15.32 & 41.05 & -28.35 & 45.06 & {\underline{77.4}} & 49.01 & 57.61 & 46.71 & {\bf 82.71} & {\bf 80.48} & {\underline{73.91}} \\
X23 & -94.6 & {\underline{78.61}} & -10.88 & 54.36 & -54.29 & -25.86 & {\bf 80.35} & {\underline{79.53}} & 41.25 & 36.57 & 56.69 & 59.87 & 54.68 & 69.15 \\
X24 & 14.36 & 17.95 & -64.97 & -6.44 & 67.95 & 3.25 & 68.4 & {\underline{77.5}} & 33.97 & 30.86 & 8.48 & 69.67 & 65.74 & {\underline{71.2}} \\
X25 & {\bf 99.22} & -10.4 & -68.67 & 31.16 & {\underline{70.04}} & -30.46 & 51.85 & {\bf 88.25} & 39.8 & 42.09 & 38.82 & 65.17 & 60.66 & {\underline{70.86}} \\
X26 & -99.86 & 68.28 & -42.04 & -91.74 & -44.57 & 41.96 & -32.18 & -17.2 & {\underline{71.07}} & 59.33 & 12.88 & 51.13 & 44.8 & 67.02 \\
X27 & 22.46 & {\underline{77.37}} & -17.66 & 60.25 & 67.81 & -54.02 & 54.78 & {\bf 81} & 43.57 & 45.36 & 69.68 & {\bf 81.75} & {\underline{79.39}} & {\underline{71.86}} \\
X28 & {\underline{74.8}} & {\bf 86.01} & -20.06 & 20.25 & 52.18 & -29.64 & 60.87 & {\bf 82.46} & -0.92 & {\bf 87.42} & 56.07 & {\underline{74.64}} & {\underline{71.36}} & {\underline{78.39}} \\
X29 & 56.6 & -32.03 & -73.41 & {\bf 86.76} & {\bf 89.79} & 5.85 & 45.1 & 65.9 & -19.74 & 8.92 & 48.8 & 63.01 & 58.22 & 52.27 \\
X30 & 53.68 & {\bf 89.56} & -8.73 & 59.42 & 27.29 & 14.29 & 14.57 & 55.29 & -34.73 & 35.97 & 50.84 & 63.94 & 59.27 & 66.09 \\
X31 & -63.1 & {\bf 94.6} & {\bf 86.58} & -25.37 & 54.44 & -4.68 & 23.57 & {\underline{70.77}} & 45.75 & 58.82 & 45.69 & {\bf 80.63} & {\underline{78.12}} & {\bf 81.63} \\
X32 & -90.38 & {\bf 92.13} & -40.68 & -46.9 & -41.39 & -47.56 & 29.25 & 28.35 & {\underline{70.58}} & 53.25 & 14.12 & 60.58 & 55.47 & {\underline{71.25}} \\
 [1ex] 
\hline 
\end{tabular}
}
\label{table.fit.theta} 
\end{table}
\end{landscape}

\newpage\clearpage
\begin{landscape}
\begin{table}[ht]
\noindent
\caption{The within-cluster cross-sectional correlations $\Theta_{sA}$ (columns 2-12), the overall correlations $\Xi_s$ (column 15) based on the overall cross-sectional regressions, and multiple $R^2$ and adjusted $R^2$ of these regressions (columns 13 and 14). The cluster weights are based on normalized regressions (see Subsections \ref{sub.res} and \ref{sub.within} for details). All quantities are in the units of 1\% rounded to 2 digits. The values above 80\% are given in bold font. The values above 70\% are underlined.
}
{\scriptsize
\begin{tabular}{l l l l l l l l l l l l l l l} 
\\
\hline\hline 
Type & Cl-1 & Cl-2 & Cl-3 & Cl-4 & Cl-5 & Cl-6 & Cl-7 & Cl-8 & Cl-9 & Cl-10 & Cl-11 & $R^2$ & adj-$R^2$ & Cor\\[0.5ex] 
\hline 
X1 & {\underline{74.82}} & 52.72 & -43.94 & {\bf 90.19} & 1.85 & 65.36 & 45.69 & 47.77 & 51.21 & {\bf 84.05} & -13.31 & {\bf 89.21} & {\bf 87.81} & {\bf 92.03} \\
X2 & 46.86 & {\underline{79.8}} & -0.94 & 3.68 & 14.82 & 30.35 & -20.14 & 46.18 & -6.97 & 59.59 & 32.36 & {\underline{72.73}} & 69.2 & {\bf 80.3} \\
X3 & {\bf 99.69} & 59.47 & -9.32 & {\bf 86.89} & {\bf 93.76} & 26.13 & 50.46 & {\underline{72.39}} & 34.47 & {\underline{70.55}} & 23.15 & {\bf 85.24} & {\bf 83.34} & {\bf 85.05} \\
X4 & {\underline{71.23}} & 10.76 & 15.04 & -51.44 & 30.56 & -21.88 & 1.98 & 36.41 & 47.97 & 55.7 & -1.58 & {\underline{75.85}} & {\underline{72.72}} & 67.89 \\
X5 & {\bf 100} & 35.83 & 30.3 & 49.26 & {\bf 98.08} & -28.3 & -65.45 & 51.98 & -17.43 & 38.98 & 44.82 & {\underline{77.99}} & {\underline{75.14}} & {\underline{70.79}} \\
X6 & -93.22 & 63.8 & -16.06 & {\bf 94.84} & {\bf 90.84} & 31.67 & 27.56 & {\bf 82.12} & 61.61 & 53.05 & 64.31 & 58.83 & 53.5 & 69.75 \\
X7 & 37.97 & {\bf 94.43} & -47.29 & 25.4 & 69.11 & 0.83 & 63.68 & {\bf 88.6} & 48.09 & 59.4 & 38.15 & {\underline{70.76}} & 66.97 & {\underline{78.2}} \\
X8 & -45.01 & 16.5 & -62.63 & {\bf 98.28} & 39.35 & -32.27 & {\bf 83.6} & 59.19 & 63.11 & 43.4 & 50.52 & 61.19 & 56.17 & 67.77 \\
X9 & -89.86 & 62.25 & -80.98 & -47.45 & -8.97 & -16.74 & 16.12 & 59.39 & 60.3 & 55 & 33.34 & 56.23 & 50.57 & 69.13 \\
X10 & 18.02 & -8.2 & {\bf 82.63} & 10.18 & {\underline{74.99}} & -52.09 & 41.77 & {\bf 82.73} & 20.02 & 63.78 & 58.84 & {\underline{79.37}} & {\underline{76.7}} & {\underline{71.53}} \\
X11 & -6.42 & 55.44 & 52.93 & -57.54 & 47.63 & -1.28 & {\bf 81.7} & -30.07 & 50.66 & 62.3 & 33.12 & {\bf 84.34} & {\bf 82.31} & {\bf 81.56} \\
X12 & -63.33 & {\underline{76.58}} & 4.49 & -41.77 & 7.54 & -42.03 & 48.89 & 64.25 & 66.89 & 68.73 & 45.58 & {\underline{74.94}} & {\underline{71.69}} & {\bf 80.69} \\
X13 & -77.89 & {\bf 93.71} & -27.8 & -38.68 & 47.29 & 17.44 & {\bf 95.87} & {\underline{76.87}} & {\underline{75.5}} & {\underline{77.51}} & 23.68 & {\bf 81.52} & {\underline{79.13}} & {\bf 83.69} \\
X14 & 61.29 & 63.15 & -18.3 & -41.27 & {\bf 87.58} & -28.2 & {\bf 81.25} & {\underline{78.09}} & 36.86 & 60.05 & 33.54 & 67.85 & 63.69 & {\underline{76.76}} \\
X15 & -0.78 & {\bf 87.99} & -55.94 & 63.9 & {\bf 93.15} & -35.35 & {\underline{70.96}} & {\underline{78.38}} & 33.18 & 61.53 & 54.21 & {\underline{71.73}} & 68.07 & {\bf 80.1} \\
X16 & 18.11 & {\bf 84.31} & 19.73 & 0.91 & {\bf 85.33} & 7.48 & 60.17 & {\underline{70.2}} & 57.35 & {\underline{76.08}} & 45.64 & {\bf 84.17} & {\bf 82.13} & {\bf 86.47} \\
X17 & 33.82 & 1.05 & 30.67 & 26.74 & 69.21 & -9.94 & 58.67 & 39.91 & 22.7 & 29.39 & 32.72 & {\bf 80.42} & {\underline{77.89}} & {\underline{72.46}} \\
X18 & 56.36 & 8.18 & {\bf 80.84} & -11.72 & {\underline{77.47}} & 0.12 & 14.08 & 57.32 & 14.3 & 55.78 & 58.66 & {\bf 81.32} & {\underline{78.91}} & {\underline{79.27}} \\
X19 & -66.84 & {\underline{76.03}} & 63.58 & {\underline{79.35}} & 25.45 & -22.45 & -33.4 & {\underline{70.1}} & -5.23 & 47.49 & 31.47 & 57.05 & 51.49 & {\underline{72.24}} \\
X20 & {\underline{72.56}} & -45.53 & 50.76 & {\bf 99.76} & {\underline{74.67}} & {\bf 96.7} & -64.71 & {\bf 94.63} & -15.57 & 67 & 21.85 & {\bf 90.27} & {\bf 89.01} & {\bf 93.47} \\
X21 & 42.55 & -8.08 & -21.72 & 35.24 & 61.4 & -38.89 & 10.53 & 10.83 & 17.15 & {\bf 92.87} & 48.29 & {\bf 84.49} & {\bf 82.49} & {\bf 87.07} \\
X22 & 1.45 & {\bf 90.5} & -35.62 & -13.13 & 39.71 & -23.66 & 60.34 & {\underline{77.98}} & 48.6 & {\underline{72.1}} & 53.62 & {\bf 87.25} & {\bf 85.6} & {\bf 80.31} \\
X23 & -89.64 & {\underline{79.52}} & 16.97 & 55.42 & -56.06 & -22.53 & {\bf 84.97} & {\underline{79.22}} & 45.96 & 59.23 & 56.69 & {\underline{70.93}} & 67.17 & {\underline{78.71}} \\
X24 & 1.5 & 19.38 & -60 & -5.03 & {\underline{71.17}} & 9.51 & {\underline{76.81}} & {\underline{77.96}} & 32.31 & 54.11 & 18.55 & {\underline{79.13}} & {\underline{76.43}} & {\bf 80.84} \\
X25 & {\bf 96.78} & -9.27 & -60.49 & 33.73 & {\underline{71.27}} & -27.25 & 66.72 & {\bf 88.29} & 45.16 & 65.44 & 47.19 & {\underline{76.9}} & {\underline{73.91}} & {\bf 81.59} \\
X26 & -98.33 & 68.09 & -18.47 & -91.28 & -47.48 & 47.37 & -21.32 & -17.49 & 68.11 & {\underline{70.34}} & 4.83 & 52.9 & 46.8 & 68.69 \\
X27 & 9.74 & {\underline{78.26}} & 3.2 & 62.7 & 67.15 & -51.82 & 65.98 & {\bf 80.74} & 48.21 & 65.36 & {\underline{74.89}} & {\bf 87.38} & {\bf 85.74} & {\bf 80.4} \\
X28 & {\bf 82.72} & {\bf 86.77} & 3.18 & 23.06 & 54.14 & -24.21 & 69.04 & {\bf 82.83} & 8.5 & {\bf 88.99} & 61.88 & {\underline{77.71}} & {\underline{74.83}} & {\bf 81.01} \\
X29 & 66.74 & -30.7 & -75.41 & {\bf 86.26} & {\bf 90.39} & 6.19 & 40.05 & 65.75 & -9.15 & 33.08 & 50 & {\underline{70.78}} & 67 & 63.15 \\
X30 & 64.09 & {\bf 89.73} & -9.78 & 59.19 & 24.42 & 15.12 & 10.09 & 54.47 & -29.33 & 60 & 55.91 & {\underline{74.91}} & {\underline{71.66}} & {\underline{77.47}} \\
X31 & -72.56 & {\bf 94.15} & 68.86 & -23.61 & 56.73 & -0.07 & 29.76 & {\underline{71.5}} & 42.07 & {\underline{75.55}} & 52.13 & {\bf 85.8} & {\bf 83.96} & {\bf 86.69} \\
X32 & -84.12 & {\bf 92.48} & -23.26 & -45.55 & -42.69 & -46.15 & 27.16 & 27.83 & 68.89 & 68.65 & 6.59 & 62.67 & 57.84 & {\underline{73.22}} \\
 [1ex] 
\hline 
\end{tabular}
}
\label{table.fit.theta.N} 
\end{table}
\end{landscape}

\newpage\clearpage
\begin{landscape}
\begin{table}[ht]
\noindent
\caption{Cross-sectional correlations between 30 COSMIC signatures and cancer types X1 through X16 for the exome data summarized in Table \ref{table.exome.summary} aggregated by cancer types. The weights for COSMIC signatures are available from http://cancer.sanger.ac.uk/cancergenome/assets/signatures\_probabilities.txt. The values above 80\% are given in bold font. The values above 70\% are underlined.}
{\tiny
\begin{tabular}{l l l l l l l l l l l l l l l l l} 
\\
\hline\hline 
Signature & X1 & X2 & X3 & X4 & X5 & X6 & X7 & X8 & X9 & X10 & X11 & X12 & X13 & X14 & X15 & X16 \\[0.5ex] 
\hline 
\\
COSMIC1 & {\bf 89.45} & {\bf 94.29} & {\bf 80.44} & {\underline{70.24}} & 67.36 & 27.61 & {\bf 90.83} & 23.73 & 27.58 & 63.3 & {\bf 90.12} & {\underline{78.51}} & {\bf 81.18} & {\bf 89.33} & {\bf 97.03} & 49.55 \\
COSMIC2 & 24.57 & 9.98 & 18.61 & 14.15 & 6.23 & {\bf 82.29} & 4.8 & {\bf 81.07} & {\bf 81.74} & 14.12 & 19.06 & 7.69 & 27.65 & 2.91 & 7.96 & {\underline{71.12}} \\
COSMIC3 & -12.79 & -15.52 & 3.13 & -8.38 & 2.4 & 17 & -12.56 & 28.33 & 15.5 & -6.34 & -2.18 & -15.44 & -6.58 & -22.57 & -15.97 & 15.47 \\
COSMIC4 & 8.78 & -2.48 & 40.12 & -7.33 & 26.25 & -5.15 & 2.42 & -1.54 & -5.65 & 22.44 & -1.49 & 4.36 & -0.21 & -7.05 & -5.54 & 6.11 \\
COSMIC5 & 31.29 & 30.66 & 27.6 & 36.36 & 52.7 & 7.3 & 33.06 & 9.53 & 3.63 & 38.61 & 51.32 & 34.3 & 28.43 & 27.87 & 27.64 & 23.26 \\
COSMIC6 & {\underline{75.67}} & {\bf 82.76} & {\underline{77.34}} & 69.83 & {\underline{72.74}} & 16.2 & {\bf 92.47} & 13.84 & 16.8 & 59.95 & {\underline{77.34}} & {\underline{78.66}} & {\underline{74.2}} & {\bf 92.46} & {\bf 84.35} & 40.69 \\
COSMIC7 & 44.18 & 22.95 & 22.21 & 25.98 & 26.29 & 51.13 & 15.5 & 43.93 & 49.06 & 22.89 & 31.67 & 18.79 & 28.88 & 10.83 & 21.58 & 59.96 \\
COSMIC8 & 12.82 & 8.08 & 29.72 & 1.54 & 16.56 & -4.54 & 14.91 & 2.71 & -4.65 & 18.14 & 18.4 & 15.45 & 7.78 & 3.28 & 6.33 & 2.89 \\
COSMIC9 & -3.53 & -3.98 & -10.19 & 22.79 & -2.13 & -10.04 & 1.52 & -7.89 & -9.83 & -3.79 & 8 & 15.48 & 14.23 & 4.86 & -3.85 & -12.77 \\
COSMIC10 & 34.34 & 17.64 & 28.23 & 26.03 & 14.07 & 27.2 & 30.53 & 20.53 & 25.27 & 13.36 & 19.77 & 64.86 & 25.88 & 23.06 & 16.43 & 22.68 \\
COSMIC11 & 35.2 & 23.69 & 19.25 & 23.9 & 36.72 & 27.19 & 21.53 & 24.41 & 24.5 & 22.07 & 32.83 & 15.31 & 21.76 & 13.07 & 26.44 & 42.54 \\
COSMIC12 & -1.61 & -2.34 & -5.85 & 4.67 & 23.61 & -5.74 & 0.37 & -5.46 & -7.78 & 9.77 & 14.57 & 9.17 & -0.39 & 6.51 & -4.08 & -1.68 \\
COSMIC13 & 2.87 & -3.44 & 13.74 & 1.98 & -4.91 & {\underline{71.24}} & -2.55 & {\underline{76.35}} & {\underline{72.85}} & 0.63 & 3.61 & -4.22 & 18.47 & -4.44 & 0.2 & 60.08 \\
COSMIC14 & 52.79 & 48.45 & 61.44 & 45.38 & 54.47 & 12.36 & {\underline{74.05}} & 13.53 & 13.19 & 35.41 & 51.75 & 61.67 & 47.65 & 60.64 & 51.76 & 31.07 \\
COSMIC15 & 51.28 & 50.19 & 53.9 & 54.12 & 57.75 & 10.52 & {\underline{70.48}} & 9.53 & 11.22 & 39.95 & 49.65 & 62.24 & 49.85 & {\underline{71.45}} & 58.85 & 28.77 \\
COSMIC16 & -2.28 & -2.35 & -0.5 & 7.28 & 22.17 & 8.9 & -1.97 & 11.41 & 5.71 & 16.17 & 18.76 & 4.3 & 6.41 & -3.4 & -6.17 & 11.89 \\
COSMIC17 & -1.47 & -0.75 & -9.35 & 44.18 & 0.18 & -0.04 & -0.7 & -0.65 & 0.14 & 3.85 & 8.05 & 7.81 & 39.42 & 11.2 & -0.52 & -2.05 \\
COSMIC18 & 19.89 & 3.83 & 43.34 & 10.48 & 24.38 & 6.55 & 12.58 & 9.73 & 6.77 & 8.44 & 5.53 & 25.92 & 12.85 & 4.01 & 1.93 & 12.79 \\
COSMIC19 & 42.96 & 45.14 & 38.47 & 37.84 & 56.04 & 19.22 & 41.38 & 18.9 & 16.2 & 36.72 & 48.65 & 31.6 & 34.49 & 33.11 & 42.37 & 39.03 \\
COSMIC20 & 31.7 & 35.8 & 41.02 & 30.54 & 48.6 & 5.94 & 47.07 & 5.42 & 4.3 & 34.21 & 40.46 & 42.75 & 30.67 & 41.04 & 32.11 & 20.18 \\
COSMIC21 & 9.79 & 9.96 & 4.77 & 16.56 & 20.54 & -3.68 & 14.23 & -5.29 & -4.34 & 11.95 & 22.03 & 19.9 & 8.42 & 23.88 & 11.67 & 2.69 \\
COSMIC22 & -10.52 & -9.55 & -11.9 & -10.11 & -11.87 & -11.33 & -12.09 & -12.28 & -12.25 & 52.49 & -2.78 & -14.12 & -14.12 & -11.19 & -8.46 & -11.76 \\
COSMIC23 & 27.16 & 25.81 & 19.44 & 25.44 & 42.9 & 9.57 & 24.3 & 7.91 & 6.39 & 23.54 & 27.9 & 17.59 & 17.94 & 17.64 & 27.65 & 26.29 \\
COSMIC24 & 10.03 & -0.84 & 43.35 & 2.42 & 27.86 & 7.29 & 3.83 & 9.45 & 8.17 & 10.73 & -0.15 & 5.55 & 6.34 & -1.74 & -1.97 & 14.26 \\
COSMIC25 & 12.26 & 11.02 & 11.7 & 10.08 & 13.83 & 6.66 & 11.16 & 8.12 & 4.93 & 57.53 & 24.31 & 13.54 & 8.95 & 9.18 & 9.94 & 10.21 \\
COSMIC26 & 13.37 & 13.38 & 8.26 & 17.37 & 29.37 & 0.77 & 20.94 & 0.15 & -0.69 & 18.19 & 28.82 & 25.28 & 14.23 & 28.18 & 14.85 & 7.4 \\
COSMIC27 & -4.19 & -4.22 & -5.91 & 1.02 & -11.83 & -2.66 & -4.31 & -2.91 & -2.63 & 8.24 & -7.21 & -4.96 & -6.82 & -4.21 & -3.09 & -4.06 \\
COSMIC28 & -10.61 & -8.93 & -19.23 & 17.31 & -11.5 & -8.21 & -7.57 & -8.29 & -6.85 & -5.6 & -7.6 & 4.11 & 9.38 & 0.1 & -8.41 & -14.27 \\
COSMIC29 & 26.94 & 17.44 & 53.8 & 14.87 & 37.87 & 3.04 & 23.71 & 6.25 & 4.06 & 19.4 & 19.72 & 24.28 & 21.17 & 16.51 & 16.68 & 14.59 \\
COSMIC30 & 44.21 & 32.58 & 25.34 & 39.34 & 37.09 & 44.06 & 26.27 & 41.01 & 41.95 & 30.01 & 41.72 & 19.63 & 31.65 & 19.97 & 31.06 & 55.25 \\
[1ex] 
\hline 
\end{tabular}
}
\label{table.cosmic.exome.cor.1} 
\end{table}
\end{landscape}

\newpage\clearpage
\begin{landscape}
\begin{table}[ht]
\noindent
\caption{Cross-sectional correlations between 30 COSMIC signatures and cancer types X17 through X32 for the exome data summarized in Table \ref{table.exome.summary} aggregated by cancer types. The weights for COSMIC signatures are available from http://cancer.sanger.ac.uk/cancergenome/assets/signatures\_probabilities.txt. The values above 80\% are given in bold font. The values above 70\% are underlined.}
{\tiny
\begin{tabular}{l l l l l l l l l l l l l l l l l l} 
\\
\hline\hline 
Signature & X17 & X18 & X19 & X20 & X21 & X22 & X23 & X24 & X25 & X26 & X27 & X28 & X29 & X30 & X31 & X32 \\[0.5ex] 
\hline 
\\
COSMIC1 & 66.66 & 22.83 & 19.86 & 14.47 & 48.08 & 66.58 & {\bf 80.24} & {\bf 83.31} & {\bf 94.16} & 58.88 & {\underline{78.37}} & 59.4 & 58.17 & {\bf 88.27} & {\underline{76.26}} & 59.17 \\
COSMIC2 & 13.37 & 40.04 & 50.81 & 10.46 & 36.67 & 35.99 & 8.13 & 10.26 & 10.38 & 9.99 & 28.05 & 27.39 & 2.04 & 5.91 & 38.8 & 10.5 \\
COSMIC3 & 5.83 & 31.25 & 0.86 & 4.42 & 9.65 & 41.57 & -16.05 & -8.32 & -13.9 & -12.48 & 9.73 & -2.39 & 1.36 & -14.16 & -0.95 & -7.39 \\
COSMIC4 & 1.81 & {\underline{75.05}} & -2.14 & -4.6 & 49.97 & 27.1 & 2.51 & -5.36 & -1.11 & 5.96 & 17.97 & 21.39 & 27.4 & 5.24 & -4.28 & 12.78 \\
COSMIC5 & {\underline{79.6}} & 15.5 & 26.99 & -7.31 & 10.73 & 42.45 & 29.35 & 49.26 & 30.41 & 15.31 & 53.95 & 31.44 & 31.66 & 34.09 & 42.47 & 27.19 \\
COSMIC6 & 63.89 & 25.94 & 8.82 & 12.01 & 41.32 & 62.45 & {\bf 91.06} & {\bf 87.74} & {\bf 88.24} & 46.22 & {\underline{77.96}} & 50.76 & {\underline{70.16}} & {\bf 88.86} & {\underline{72.53}} & 58.1 \\
COSMIC7 & 27.65 & 31.87 & {\bf 99.66} & 7 & 36.37 & 38.52 & 18.54 & 25.81 & 19.5 & 22.17 & 38.17 & 56.58 & 11.02 & 16.39 & 42.31 & 20.47 \\
COSMIC8 & 14.72 & 43.44 & -9.3 & -5.21 & 29.34 & 25.78 & 10.47 & 2.31 & 8.37 & 18 & 21.19 & 14.7 & 14.47 & 13.53 & 0.94 & 23.61 \\
COSMIC9 & 15.26 & -17.36 & -6.58 & -14.5 & -14.65 & -13.15 & 1.24 & -0.21 & -1.92 & 14.47 & 0.55 & -8.02 & -8.43 & -2.77 & -6.73 & 18.46 \\
COSMIC10 & 11.48 & 19.21 & 18.58 & 0.86 & 24.12 & 19.65 & 31.75 & 14.87 & 20.33 & {\bf 89.16} & 22.14 & 17.66 & 10.06 & 15.29 & 14.56 & {\bf 87.49} \\
COSMIC11 & 37.02 & 21.94 & {\underline{77.36}} & 5.72 & 21.56 & 36.81 & 26.72 & 35.61 & 20.25 & 6.45 & 40.26 & 48.93 & 23.07 & 19.11 & 41.85 & 14.48 \\
COSMIC12 & 50.27 & -7.26 & 1.58 & -9.28 & -16.64 & -2.1 & -2.52 & 16.72 & -0.01 & -9.43 & 16.39 & -2.97 & 8.68 & 8.56 & 9.4 & -2.31 \\
COSMIC13 & -2.04 & 35.65 & 6.66 & -0.38 & 31.29 & 38.21 & -2.83 & 1.34 & 2.46 & -0.11 & 11.23 & 4.34 & -2.61 & -3.21 & 22.91 & -0.89 \\
COSMIC14 & 38.55 & 36.5 & 14.94 & 5.86 & 43.08 & 49.41 & {\underline{76.65}} & 56.58 & 54.4 & 38.22 & 57.5 & 54.34 & 59.64 & 59.99 & 46.35 & 58.74 \\
COSMIC15 & 42.05 & 14.95 & 7.14 & 9.89 & 27.61 & 41.95 & {\bf 82.98} & {\underline{76}} & 63.73 & 32.73 & 56.75 & 40.14 & 62.71 & 68.42 & 53.87 & 46.38 \\
COSMIC16 & 56.26 & 10.38 & 9.23 & -13.64 & -4.46 & 13.69 & -4.71 & 15.75 & -1.59 & -3.7 & 20.56 & 2.25 & 4.06 & 3.41 & 15.09 & 1.88 \\
COSMIC17 & 1.26 & -11.27 & 4.05 & -2.77 & -8.56 & -3.3 & 0.91 & -0.68 & 0.94 & 3.56 & 3.22 & -0.83 & 4.77 & 3.02 & 3.12 & 3.93 \\
COSMIC18 & 1.08 & 55.29 & -1.04 & 6.43 & 53.97 & 23.12 & 15.21 & 0.28 & 8.82 & 32.61 & 18.54 & 33.36 & 24.02 & 10.97 & 1.16 & 38.89 \\
COSMIC19 & 55.06 & 26.57 & 48.76 & 5.01 & 23.54 & 50.79 & 44.61 & 54.38 & 39.57 & 14.42 & 57.89 & 47.64 & 45.4 & 40.3 & 52.16 & 25.02 \\
COSMIC20 & 55.9 & 31.1 & 9.54 & 2.85 & 19.92 & 37.82 & 40.79 & 43.95 & 34.15 & 17.31 & 52.43 & 24.8 & 42.43 & 40.32 & 39.66 & 35.22 \\
COSMIC21 & 33.38 & -12.86 & 0.05 & -6.89 & -9.66 & 0.23 & 12.35 & 21.42 & 15.34 & 2.9 & 15.33 & 2.96 & 6.55 & 18.55 & 14.91 & 8.78 \\
COSMIC22 & -7.62 & -1.77 & -7.78 & -12.32 & -18.87 & -15.29 & -12.86 & -12.47 & -11.18 & -8.83 & -2.13 & -13.97 & -20.14 & -11.76 & -10.2 & -12.77 \\
COSMIC23 & 37.74 & 14.67 & 47.26 & 2.88 & 12.21 & 33.83 & 30.51 & 39.51 & 22.27 & 4.53 & 40.81 & 36.52 & 33.83 & 21.75 & 38.16 & 14.34 \\
COSMIC24 & -3.48 & 64.69 & -1.2 & 1.23 & 67.23 & 26.96 & 9.63 & -0.1 & 5.17 & 6.47 & 15.79 & 36.14 & 40.4 & 8.82 & 0.64 & 11.09 \\
COSMIC25 & 26.48 & 14.87 & -0.68 & -6.78 & -0.38 & 12.12 & 6.88 & 11.3 & 9.67 & 13.23 & 23.94 & 4.69 & -3.35 & 9.51 & 18.18 & 14.51 \\
COSMIC26 & 54.08 & -8.65 & 0.08 & -6.4 & -9.68 & 6.33 & 17.87 & 33.03 & 18.87 & 3.97 & 25.39 & 4 & 12.13 & 24.86 & 22.75 & 11.68 \\
COSMIC27 & -5.21 & -5.53 & -0.83 & -1.13 & -5.61 & -10.71 & -3.71 & -4.84 & -4.08 & -3.2 & 15.48 & -4.27 & -10.11 & -6.96 & -6.57 & -4.13 \\
COSMIC28 & -5 & -21.81 & -5.79 & -8.13 & -18.66 & -17.36 & -5.6 & -10.39 & -8.43 & 3.09 & -5.96 & -13.41 & -15.94 & -10.54 & -11.47 & 3.99 \\
COSMIC29 & 14.75 & 60.17 & -4.79 & 5.76 & {\underline{71.38}} & 34.28 & 23.68 & 12.79 & 22.55 & 21.94 & 27.84 & 45.59 & 43.15 & 26.32 & 10.28 & 27.23 \\
COSMIC30 & 41.46 & 26.37 & {\underline{76.08}} & 5.97 & 30.05 & 45.82 & 27.99 & 38.08 & 27.75 & 8.55 & 48.65 & 51.88 & 23.83 & 23.84 & 51.58 & 15.89 \\
[1ex] 
\hline 
\end{tabular}
}
\label{table.cosmic.exome.cor.2} 
\end{table}
\end{landscape}

\newpage\clearpage
\begin{landscape}
\begin{table}[ht]
\noindent
\caption{Cross-sectional correlations between 30 COSMIC signatures and cancer types G.X1 through G.X14 for the genome data summarized in Table 1 of \cite{*K-means} aggregated by cancer types. G.X1 = B-Cell Lymphoma, G.X2 =
Bone Cancer, G.X3 = Brain Lower Grade Glioma, G.X4 = Breast Cancer, G.X5 = Chronic
Lymphocytic Leukemia, G.X6 = Esophageal Cancer, G.X7 = Gastric Cancer, G.X8 = Liver
Cancer, G.X9 = Lung Cancer, G.X10 = Medulloblastoma, G.X11 = Ovarian Cancer, G.X12
= Pancreatic Cancer, G.X13 = Prostate Cancer, G.X14 = Renal Cell Carcinoma. The weights for COSMIC signatures are available from http://cancer.sanger.ac.uk/cancergenome/assets/signatures\_probabilities.txt. The values above 80\% are given in bold font. The values above 70\% are underlined.}
{\tiny
\begin{tabular}{l l l l l l l l l l l l l l l} 
\\
\hline\hline 
Signature & G.X1 & G.X2 & G.X3 & G.X4 & G.X5 & G.X6 & G.X7 & G.X8 & G.X9 & G.X10 & G.X11 & G.X12 & G.X13 & G.X14 \\[0.5ex] 
\hline 
\\
COSMIC1 & 39.38 & {\bf 86.27} & {\bf 91.05} & 15.44 & 69.73 & {\underline{74.43}} & 48.71 & -2.93 & 6.59 & {\bf 94.86} & 46.94 & {\bf 95.31} & {\bf 83.27} & 19.48 \\
COSMIC2 & 47.64 & 22 & 18.39 & {\underline{79.91}} & 11.54 & 54.23 & 7.52 & -10.44 & 32.48 & 14.39 & 32.28 & 12.87 & 27.25 & 22.72 \\
COSMIC3 & 11.51 & 10.77 & 1.15 & 36 & 10.73 & 6.03 & 4.29 & 1.37 & 49.07 & 1.04 & 60.73 & -16.4 & 13.8 & 36.19 \\
COSMIC4 & -10.31 & 16.21 & 0.21 & 1.07 & 1.06 & 1.39 & -2.62 & 8.44 & {\bf 82.53} & 7.79 & 26.95 & -3.57 & 17.3 & 28.1 \\
COSMIC5 & 55.4 & 57.76 & 55.72 & 13.06 & {\underline{75.02}} & 22.07 & 46.35 & 1.32 & 29.05 & 48.62 & 63.51 & 24.5 & 51.15 & 54.23 \\
COSMIC6 & 33.24 & 66.83 & {\underline{71.52}} & 6.37 & 58.41 & 59.46 & 50.19 & -4.93 & 8.23 & {\underline{76.42}} & 37.65 & {\bf 86.72} & 66.87 & 11.58 \\
COSMIC7 & 42.62 & 26.5 & 30.74 & 39.25 & 22.51 & 36.55 & 7.21 & 0.22 & 18.84 & 23.95 & 22.82 & 19.38 & 25.4 & 21.84 \\
COSMIC8 & 15.64 & 46.16 & 22.68 & 7.68 & 37.63 & 9.85 & 20.6 & -4.54 & 68.63 & 31.67 & 57.31 & 7.37 & 44.86 & 54.24 \\
COSMIC9 & 62.32 & 14.33 & 12.05 & -3.79 & 51.08 & -6.83 & 60.78 & -13.48 & -0.57 & 9.82 & 16.12 & -1.57 & 18.48 & 31.51 \\
COSMIC10 & 27.68 & 28.48 & 22.43 & 11.9 & 23.97 & 30.68 & 25.05 & 1.74 & 20.12 & 30.22 & 18.54 & 21.41 & 35.38 & 13.93 \\
COSMIC11 & 36.78 & 28.03 & 31.95 & 22.49 & 29.37 & 22.68 & 8.2 & 6.52 & 13.79 & 25.95 & 22.77 & 17.48 & 21.58 & 20.6 \\
COSMIC12 & 23.88 & 14.26 & 16.17 & -1.85 & 34.02 & -6.63 & 23.78 & -3.79 & 3.57 & 7.7 & 23.78 & -4.48 & 9.78 & 18.13 \\
COSMIC13 & 13.66 & 4.28 & 2.33 & {\bf 80.9} & -4.65 & 47.36 & -1.17 & -14.68 & 28.78 & 0.59 & 43.56 & 3.99 & 16.71 & 10.38 \\
COSMIC14 & 30.05 & 52.91 & 44.83 & 9.22 & 44.04 & 39.25 & 39.49 & -2.03 & 30.51 & 53.35 & 36.4 & 53.72 & 52.42 & 16.22 \\
COSMIC15 & 25.67 & 42.18 & 41.37 & 4.53 & 39.78 & 35.89 & 34.97 & -0.46 & 2.31 & 48.67 & 24.49 & 58.99 & 46.22 & 5.51 \\
COSMIC16 & 45.02 & 27.55 & 25.39 & 18.48 & 50.43 & 4.96 & 32.15 & -1.05 & 27.99 & 14.62 & 49.19 & -6.52 & 24.2 & 44.42 \\
COSMIC17 & 54.33 & -2.69 & 2.27 & -0.59 & 19 & 0.43 & {\underline{73.87}} & -3.68 & -11.64 & -2.52 & -3 & 2.2 & -2.89 & -2.4 \\
COSMIC18 & 8.56 & 27.17 & 8.01 & 9.55 & 14.24 & 15.81 & 11.6 & 4.2 & 66.55 & 18.73 & 24.89 & 6.48 & 31.09 & 18.59 \\
COSMIC19 & 37.46 & 48.08 & 56.92 & 17.83 & 49.29 & 30.71 & 20.98 & 3.49 & 20.9 & 45.85 & 40.02 & 34.23 & 39.33 & 28.58 \\
COSMIC20 & 22.09 & 35.41 & 36.98 & 2.45 & 36.13 & 23.82 & 32.67 & -7.1 & 25.78 & 34.99 & 28.96 & 32.35 & 28.42 & 12.46 \\
COSMIC21 & 12.23 & 15.22 & 18.41 & -6.01 & 23.79 & 3.96 & 16.11 & 1.44 & -12.08 & 14.86 & 8.33 & 11.16 & 12.87 & -1.38 \\
COSMIC22 & -15.62 & -7.33 & -10.46 & -10.43 & -7 & -8.75 & -13.39 & -16.17 & 2.97 & -11.57 & -5.9 & -9.23 & -8.08 & 47.87 \\
COSMIC23 & 23.12 & 24.08 & 34.44 & 7.06 & 29.31 & 12.74 & 8.41 & 5.5 & 7.78 & 24.45 & 17.62 & 16.66 & 16.08 & 15.21 \\
COSMIC24 & -6.8 & 12.77 & -1.61 & 10.16 & -3.47 & 10.01 & -3.98 & 10.27 & 62.16 & 6.07 & 16.1 & 1.86 & 14.9 & 6.58 \\
COSMIC25 & 15.8 & 28.11 & 20.81 & 10.07 & 29.92 & 16.63 & 16.12 & -21.2 & 25.42 & 19.51 & 33.86 & 10.23 & 27.09 & 61.26 \\
COSMIC26 & 25 & 24.23 & 25.82 & 0.87 & 37.31 & 7.71 & 27.11 & 0.51 & -4.25 & 21.13 & 24.3 & 14.95 & 20.76 & 8.6 \\
COSMIC27 & -4.17 & -0.03 & -4.34 & -2.16 & -2.63 & -4.3 & -7.7 & -8.59 & 2.35 & -3.32 & -4.12 & -3.36 & 0.47 & 35.18 \\
COSMIC28 & 42.33 & -9.94 & -3.87 & -6.62 & 16.24 & -11.59 & 52.44 & 3.29 & -18.66 & -8.75 & -7.25 & -5.64 & -5.71 & 12.33 \\
COSMIC29 & 7.41 & 38.85 & 21.13 & 5.93 & 23.52 & 20.41 & 14.98 & 4.24 & 68.09 & 31.18 & 33.47 & 19.68 & 38.55 & 16.98 \\
COSMIC30 & 49.59 & 37.46 & 41.49 & 38.84 & 35.91 & 39.02 & 16.17 & 0.96 & 15.15 & 34.52 & 27.39 & 26.86 & 32.02 & 24.89 \\
[1ex] 
\hline 
\end{tabular}
}
\label{table.cosmic.fit.genome} 
\end{table}
\end{landscape}

\newpage\clearpage
\begin{landscape}
\begin{table}[ht]
\noindent
\caption{The within-cluster cross-sectional correlations $\Theta_{sA}$ (columns 2-12), the overall correlations $\Xi_s$ (column 15) based on the overall cross-sectional regressions, and multiple $R^2$ and adjusted $R^2$ of these regressions (columns 13 and 14). The cluster weights are based on normalized regressions (see Subsections \ref{sub.res} and \ref{sub.within} for details). The definitions of cancer types G.X1 through G.X14 for genome data are given in Table \ref{table.cosmic.fit.genome}. All quantities are in the units of 1\% rounded to 2 digits. The values above 80\% are given in bold font. The values above 70\% are underlined.}
{\scriptsize
\begin{tabular}{l l l l l l l l l l l l l l l} 
\\
\hline\hline 
Type & Cl-1 & Cl-2 & Cl-3 & Cl-4 & Cl-5 & Cl-6 & Cl-7 & Cl-8 & Cl-9 & Cl-10 & Cl-11 & $R^2$ & adj-$R^2$ & Cor\\[0.5ex] 
\hline 
G.X1 & 8.47 & -42.14 & -5.83 & -28.47 & 8.6 & -27.81 & 68.1 & -58.76 & {\underline{78.71}} & 43.09 & 1.98 & {\underline{76.74}} & {\underline{73.73}} & 64.82 \\
G.X2 & 12.58 & -8.39 & 5.78 & -17.48 & 36.44 & -39.46 & 65.49 & -12.25 & 32.07 & 52.76 & 18.76 & {\bf 80.04} & {\underline{77.46}} & {\underline{74.6}} \\
G.X3 & 7.9 & 17.51 & -12.85 & 37.46 & 63.86 & -48.79 & 43.86 & 40.63 & 20.77 & 53.57 & 10.21 & {\underline{78.96}} & {\underline{76.24}} & {\underline{79.3}} \\
G.X4 & -7.33 & -4.67 & -35.67 & {\bf 90.16} & 27.48 & -35.7 & {\underline{76.34}} & 21.87 & 59.39 & 29.95 & 18.74 & 57.38 & 51.87 & 60.47 \\
G.X5 & 8.64 & -2.63 & 4.76 & 13.57 & 18.72 & -19.86 & 48.29 & -54.11 & 38.75 & 38.46 & 5.22 & {\bf 80.96} & {\underline{78.49}} & 66.43 \\
G.X6 & 19.29 & {\bf 86.79} & 63.27 & -26.72 & -1.53 & -52.18 & {\bf 83.96} & 34.55 & 69.9 & {\underline{77.08}} & 56.94 & {\bf 83.66} & {\bf 81.54} & {\bf 86.8} \\
G.X7 & 0.1 & 15.21 & 40.26 & -28.56 & 3.03 & -38.4 & 60.42 & 63.09 & 56.62 & 42.03 & 9.24 & 68.45 & 64.37 & 62.44 \\
G.X8 & 58.39 & 25.87 & -1.42 & -17.3 & -83.31 & {\underline{75.44}} & -68.56 & 65.93 & -23.25 & -27.49 & 17.47 & 58.65 & 53.3 & 9.81 \\
G.X9 & 28.73 & -62.34 & {\underline{77.8}} & 64.57 & {\bf 87.7} & -47.09 & 49.71 & 17.1 & 15.13 & 3.02 & 29.05 & {\underline{76.2}} & {\underline{73.12}} & 69.99 \\
G.X10 & -20.84 & -15.96 & -61.68 & 24.6 & 17.53 & -33.44 & 39.85 & -5.4 & 34.93 & 58.49 & 4.99 & {\underline{78.48}} & {\underline{75.7}} & {\underline{78.18}} \\
G.X11 & 7.25 & 39.51 & -7.86 & 44.23 & 46 & -54.88 & 67.08 & 25.45 & 50.2 & 41.11 & 15.02 & {\bf 83.99} & {\bf 81.92} & 65.21 \\
G.X12 & 7.9 & -88.83 & -70.05 & 21.84 & {\bf 90.47} & -23.28 & 66.92 & 53.73 & 59.49 & 67.2 & {\underline{70.4}} & {\underline{73.55}} & {\underline{70.12}} & {\bf 81.42} \\
G.X13 & -5.33 & -30.41 & -61.53 & -56.72 & -11.91 & -37.94 & 64.53 & -20.61 & 60.22 & 66.73 & -4.61 & {\bf 84.31} & {\bf 82.28} & {\underline{79.24}} \\
G.X14 & 6.33 & -39.94 & 42.62 & -21 & 56.7 & -51.19 & 65.01 & 7 & 26.19 & 5.52 & -12.27 & {\underline{71.58}} & 67.9 & 39.79 \\
 [1ex] 
\hline 
\end{tabular}
}
\label{table.fit.theta.N.genome} 
\end{table}
\end{landscape}

\newpage\clearpage

\newpage\clearpage

\newpage\clearpage

\newpage\clearpage

\newpage\clearpage

\newpage\clearpage

\newpage\clearpage

\newpage\clearpage

\newpage\clearpage

\newpage\clearpage

\newpage\clearpage

\newpage\clearpage

\newpage\clearpage

\newpage\clearpage

\newpage\clearpage

\newpage\clearpage

\newpage\clearpage

\newpage\clearpage

\newpage\clearpage

\newpage\clearpage

\newpage\clearpage

\newpage\clearpage

\end{document}